\documentclass[11pt, a4paper]{etc/feereport}
\usepackage{subfiles}
\graphicspath{{img/}{../img/}{../../img/}}
\usepackage{hyperref}
\hypersetup{
	unicode=true, 
	pdfnewwindow=true, 
	colorlinks=true, 	% (false,true)
	pdfborder={0 0 0},	
	linkcolor=blue,
	linktoc=all, 		% (none,all) 
	citecolor=blue,
	urlcolor=blue,
	breaklinks=true
}
\newcounter{address}
\renewcommand{\pd}[2]{\frac{\partial #1}{\partial #2}}

\newcommand{\dd}[2]{\ensuremath{\frac{\mathrm{d} #1}{\mathrm{d} #2}}\xspace}
\newcommand{\ddn}[3]{\frac{\partial^{#3}#1}{\partial#2^{#3}}\xspace}

\newcommand{\avg}[1]{\left\langle #1 \right\rangle}
\newcommand{\pars}[1]{\left(#1 \right)}
\newcommand{\bracs}[1]{\left[#1 \right]}
\newcommand{\p}{%
    \mathchoice%
        {\turnbox{12}{$\displaystyle\,'$}\;}%
        {\turnbox{12}{$\textstyle\,'$}\;}%
        {\turnbox{12}{$\scriptstyle\,'$}\;}%
        {\turnbox{12}{$\scriptscriptstyle\,'$}\;}%
}%

\newcommand{\vect}[1]{\boldsymbol{#1}}
\newcommand{\matr}[1]{\underline{\bm{\mathit{#1}}}}

\newcommand{\smallgtrsim}{\smallsym{\mathrel}{\gtrsim}}
\makeatletter
\newcommand{\smallsym}[2]{#1{\mathpalette\make@small@sym{#2}}}
\newcommand{\make@small@sym}[2]{%
  \vcenter{\hbox{$\m@th\downgrade@style#1#2$}}%
}
\newcommand{\downgrade@style}[1]{%
  \ifx#1\displaystyle\scriptstyle\else
    \ifx#1\textstyle\scriptstyle\else
      \scriptscriptstyle
  \fi\fi
}
\makeatother

\newcolumntype{A}{ >{$} r <{$} @{} >{${}} l <{$} } % Align table cells for equations
\newcolumntype{L}[1]{>{\raggedright\let\newline\\\arraybackslash\hspace{0pt}}m{#1}}
\newcolumntype{C}[1]{>{\tiny\centering\let\newline\\\arraybackslash\hspace{0pt}}m{#1}}
\newcolumntype{R}[1]{>{\raggedleft\let\newline\\\arraybackslash\hspace{0pt}}m{#1}}
\newcolumntype{N}{@{}m{0pt}@{}}

\newcommand*{\ovB}[1]{%
  \m@th\overline{\mbox{#1\rule{0pt}{3mm}}}
}
\newcommand*{\ovA}[1]{%
  \m@th\overline{\mbox{#1}\raisebox{3mm}{}}
}
\newcommand{\specialcell}[2][c]{%
  \begin{tabular}[#1]{@{}c@{}}#2\end{tabular}}
  
\setlist[itemize]{itemsep=3pt, topsep=0pt}
\makeatletter
\def\ignorecitefornumbering#1{%
     \begingroup
         \@fileswfalse
         #1%                     % do \cite comand
    \endgroup
}
\makeatother

\newcommand{\BibTeX}{{\rm B\kern-.05em{\sc i\kern-.025em b}\kern-.08em T\kern-.1667em\lower.7ex\hbox{E}\kern-.125emX}}

\begin{document}
\frontmatter

\title{Modelling of Flow Past Long Cylindrical Structures}
\date{September 2020}
\author{Bernat Font Garcia}
\email{b.fontgarcia@soton.ac.uk}
\supervisor{Owen R. Tutty, Gabriel D. Weymouth, Vinh-Tan Nguyen}
\department{Aerodynamics and Flight Mechanics Group}
\group{Aerodynamics and Flight Mechanics Group}
\faculty{Faculty of Engineering and Physical Sciences}
\university{University of Southampton}

\maketitle

\begin{abstract}
Turbulent flows are fundamental in engineering and the environment, but their chaotic and three-dimensional (3-D) nature makes them computationally expensive to simulate.
In this work, a dimensionality reduction technique is investigated to exploit flows presenting an homogeneous direction, such as wake flows of extruded two-dimensional (2-D) geometries.
First, we examine the effect of the homogeneous direction span on the wake turbulence dynamics of incompressible flow past a circular cylinder at $Re=10^4$.
It is found that the presence of a solid wall induces 3-D structures even in highly constricted domains.
The 3-D structures are rapidly two-dimensionalised by the large-scale K\'{a}rm\'{a}n vortices if the cylinder span is 50\% of the diameter or less, as a result of the span being shorter than the natural wake Mode B instability wavelength.
It is also observed that 2-D and 3-D turbulence dynamics can coexist at certain points in the wake depending on the domain geometric anisotropy.
With this physical understanding, a 2-D data-driven model that incorporates 3-D effects, as found in the 3-D wake flow, is presented.
The 2-D model is derived from a novel flow decomposition based on a local spanwise average of the flow, yielding the spanwise-averaged Navier--Stokes (SANS) equations.
The 3-D effects included in the SANS equations are in the form of spanwise-stress residual (SSR) terms.
The inclusion of the SSR terms in 2-D systems modifies the flow dynamics from standard 2-D Navier--Stokes to spanwise-averaged dynamics.
A machine-learning (ML) model is employed to provide closure to the SANS equations.
In the a-priori framework, the ML model yields accurate predictions of the SSR terms, in contrast to a standard eddy-viscosity model which completely fails to capture the closure term structures.
The trained ML model is also assessed for different Reynolds regimes and body shapes to the training case where, despite some discrepancies in the shear-layer region, high correlation values are still observed.
In the a-posteriori analysis, while we find evidence of known stability issues with long-time ML predictions for dynamical systems, the closed SANS equations are still capable of predicting wake metrics and induced forces with errors from 1-10\%.
This results in approximately an order of magnitude improvement over standard 2-D simulations while reducing the computational cost of 3-D simulations by 99.5\%.
\end{abstract}

\begingroup 
\tableofcontents
\listoffigures
\listoftables
\endgroup

\declaration
\dedicatory{Dedicat a la meva família, la de naixement i l'escollida.}
\acknowledgements{
I am forever indebted with all my three supervisors: Gabriel D. Weymouth, Owen R. Tutty and Vinh-Tan Nguyen. 
I have learnt so much from all of you, and I could have never imagined working in a better team, both from a personal and a professional standpoint.
You have always been there for the good and (more importantly) for the bad. 
Thank you very much.

I also thank the University of Southampton and the Singapore Agency for Science, Technology and Research (A*STAR).
The collaboration between the A*STAR Institute of High Performance Computing and the Faculty of Engineering and Physical Sciences of the University of Southampton has made this work possible.

Last but not least, I want to thank all the good friends I have made in Southampton and Singapore.
Together we have supported each other and shared unforgettable moments.
Thank you for being there.
}
\nomenclature

%% ---------------------------------------------------------------
\mainmatter

\chapter{Introduction}

\begin{figure}[t]
\centering
\includegraphics[width=0.5\linewidth]{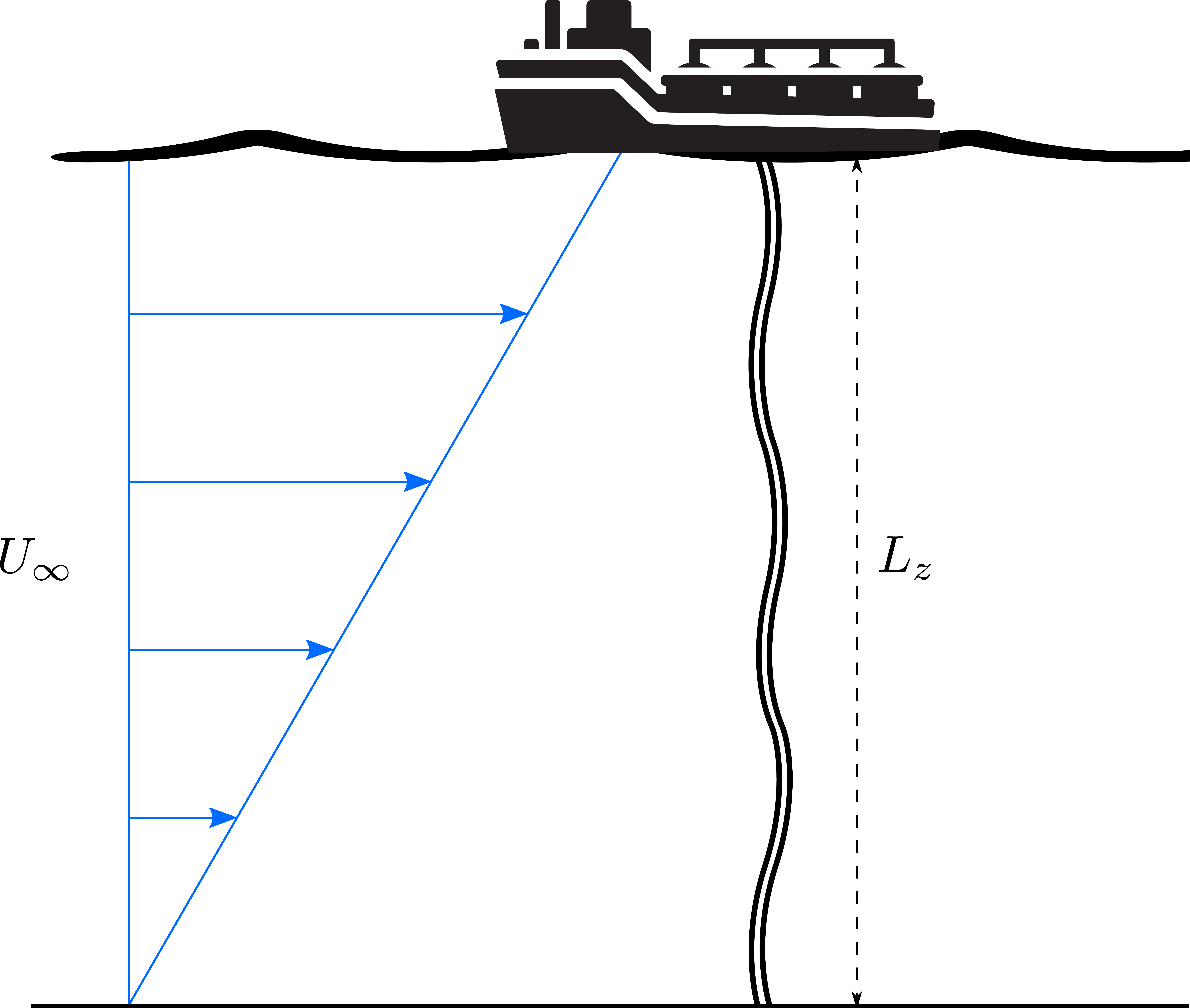}
\caption{Sketch of a marine riser exposed to a sheared onset current profile.}
\label{fig:riser}
\end{figure}

Marine risers are long flexible cylindrical structures that extract oil and gas from subsea oil fields to the surface (\fref{fig:riser}).
Under ocean currents, they are exposed to vortex-induced vibrations (VIV), a phenomenon arising in external flows past solid bodies which is particularly important for long, thin and flexible body shapes.
An alternate shedding of vortices from the upper and the lower sides of the riser circular section, a.k.a von-K\'{a}rm\'{a}n vortex street, creates an unsteady asymmetrical load pattern that stimulates a vibrational state of the riser.
This endangers the structure to severe fatigue reducing its lifespan significantly.
Besides the VIV phenomenon, wake-induced vibrations (WIV) also come into the picture for structures composed by multiple risers, adding notable complexity to the problem.
WIV, or wake galloping, takes place when one body is exposed to the non-uniform wake of another one.
This phenomenon can cause larger structural displacements compared to the VIV response of the isolated body \citep{Assi2014}.

In regard to the risers design, it is critical to correctly predict the \textit{lock-in} conditions of the structure.
A state of resonance can occur when the shedding frequency is close to the natural frequency of the riser yielding large vibration amplitudes.
From the offshore industry point of view, it is important to have reliable lock-in prediction models in order to prevent the failure of the structure.
To do so, experimental work is performed to provide representative databases of the forces acting on the cylinder at different Reynolds $(Re)$ regimes and onset current profiles.
From this data, semi-empirical structural and hydrodynamic models can be derived which provide an estimation of fatigue damage and predict the structural failure \citep{Mukundan2008}.

A disadvantage of semi-empirical models is that databases need to be repopulated every time new conditions are considered for a novel riser design.
Experimental work does not often provide sufficient data points on Reynolds regimes and single testing frequencies induce a reduced motion of the riser \citep{Mukundan2008}.
Additionally, vibrations might be restricted to the transverse (crossflow) direction alone since it is typically twice as large as the inline (streamwise) response.
However, omitting inline vibrations can yield significant differences to realistic flexible mounted cylinder systems \citep{Dahl2007}.
The reasons behind these limitations are the dimensions of the riser (a full scale experiment is very expensive and scaled experiments are hardly realistic), the number of available sensors, and the correct reconstruction of the riser motion once the data is processed.
It is because of this that numerical simulations can greatly improve the risers design process.

A complete approach to model flexible risers is to solve the incompressible Navier--Stokes momentum equations taking into account the fluid-structure interaction (FSI).
In short, the momentum equations provide the forces induced to the structure which are then used to calculate the structural displacements.
The new solid configuration is considered for the solution of the momentum equations at the next time step, and the process is repeated.
Multiple methods have been developed to couple the interaction between fluid and solid equations, typically solved by looping through both systems of equations alternately.
However, risers operate at high Reynolds regimes $(10^4<Re<10^6)$ and have a large extent ($L_z \sim10^3\,\mathrm{m}$) \citep{Willden2001,Willden2004}, making fully resolved three-dimensional (3-D) fluid dynamics simulations unaffordable nowadays.
Additionally, the FSI problem further increases the computational requirements.

\section{Background and motivation}

Turbulence plays a crucial role on long cylindrical structures such as risers.
As reported in \cite{Evangelinos1999}, the Reynolds number has a direct impact on the vibration amplitude of the cylinder.
Therefore, it is imperative to correctly capture the turbulence nature of the flow for a realistic prediction of VIV and the fatigue that the structure will suffer during its service time.
From a computational cost standpoint, it is impractical to fully resolve all fluid and structural scales inherent in the system because of the large extent of the structure and the high Reynolds regimes at which offshore structures operate.

\subsection{The computational cost of turbulence}\label{sec:turbulence_cost}

Consider homogeneous isotropic turbulence in a $L^3$ periodic box.
Following \cite[p. 345]{Pope2000}, the spectral (Fourier) representation of the flow requires $\kappa_{\max}=N \kappa_0/2=\pi N/L$ modes in each direction, where the lowest mode is $\kappa_0=2\pi/L$ and $N$ is the number of grid points per direction.
The resolved number of modes indicates the effective Reynolds number of the simulation.
In physical space, the grid spacing is $\delta x=L/N=\pi/\kappa_{\max}$.

To resolve all spatial scales, $\delta x$ must be in the order of the Kolmogorov lengthscale $\eta$, which is the smallest scale of turbulence where vortical structures are destroyed by viscous effects.
A typical choice is $\delta x/\eta=\pi/1.5$.
Also, $L$ must be at least 8 times the integral lengthscale $L_{11}$, $L=8L_{11}$.
From these constraints, it can be derived that the required number of grid points per direction in homogeneous isotropic turbulence is $N\sim 1.6Re_L^{3/4}$, totalling $N^3\sim 4.4Re_L^{9/4}$ in the cubic domain.
Similarly, time integration methods must sufficiently resolve the Kolmogorov timescale of motion.
This is ensured by the Courant--Friedrichs--Lewy (CFL) condition which imposes that fluid particles move less than one grid cell per time step.
Considering a simulation time of 4 turbulence timescales ($\tau$), the resulting total number of time steps is $M=4\tau/\delta t\sim38.2 Re_{L}^{3/4}$ \citep[p. 346]{Coleman2010,Pope2000}.

Combining space and time resolution requirements, we find that $N^3M\sim169.3 Re_{L}^{3}$ calculations are required to simulate 4 turbulence timescales of homogeneous isotropic turbulence in a $L^3$ box.
Depending on the numerical implementation, a different number of floating-point operations (FLOP) are required per pointwise calculation of the spatio-temporal domain.
Assuming 1000 FLOP per calculation \citep[p. 348]{Pope2000} and $Re_L=10^6$ (similar to offshore applications), this would require $1.69\times10^{23}$ FLOP.
Considering that the Iridis5 supercomputer at the University of Southampton offers 1.3 PFLOP/second peak, the simulation presented above would take more than 4 years of CPU time to complete employing all its 20,000 processor-cores simultaneously.

The analysis above highlights the impracticality of fully-resolved simulations of high-Reynolds flows, which are even less feasible for standard computers (GFLOP/second).
With the objective of reducing the computational cost, turbulence models are employed to allow a coarser spatio-temporal discretisation of the problem domain.
Briefly, the flow field is decomposed into resolved and unresolved scales.
The resolved scales are those captured by the arbitrary discretisation level and the unresolved scales are those smaller than the finest scale available in the discretised domain.
Turbulence models account for the physical effect of the unresolved scales by relating them to the resolved ones, thus lowering the computational requirements.
Popularly, the flow decomposition might be based in time averaging (a.k.a. Reynolds-averaged Navier--Stokes, RANS) or local spatial filtering (a.k.a. large-eddy simulation, LES).
The former models all turbulence scales while the latter only models the subgrid scales representing the viscous dissipative effects inherent in the high wavenumbers of the turbulence spectrum (for isotropic turbulence, dissipative scales represent 99.98\% of the spectrum as noted in \cite{Pope2000}, p. 350).

It is important to realise that turbulence models can decrease the accuracy of the solution when the physical hypothesis linking resolved and unresolved scales does not hold.
Additionally, it is also a non-trivial task to select the specific turbulence model which best fits the problem, or the semi-empirical constant values used to tune the model.

\subsection{The importance of three-dimensional effects}

As a result of the high computational cost of turbulence, fully resolved numerical investigations (a.k.a. direct numerical simulations, DNS) on VIV of long flexible cylinders have been limited to low Reynolds numbers and short spans.
\cite{Evangelinos1999} investigated rigid and flexible cylinders for $Re=10^3$ and aspect ratio $L_z/D=100$ (where $D$ is the cylinder diameter) finding travelling wave motions on the structural response when the cylinder was allowed to vibrate.
\cite{Lucor2001} studied the effect of the oncoming flow profile on a flexible cylinder in the transverse direction for an aspect ratio of $500<L_z/D<1000$ and $Re=10^3$.
\cite{Bourguet2011} studied a flexible cylinder for  $L_z/D=200$ allowed to vibrate in the inline and transverse directions within the $100<Re<1100$ range.

Attempting to study VIV for Reynolds regimes more relevant to offshore engineering applications, \cite{Huang2011} investigated risers at the order of $Re=10^4$ and $L_z/D=1000$.
However, a compromise in the spanwise resolution $(\delta z)$ was required to achieve feasible computational times, ranging from $\delta z/D\sim5$ for $L_z/D=482$, to $\delta z/D\sim34$ for $L_z/D=3350$.
Evidently, most of the 3-D turbulence scales are missed with this spanwise discretisation resulting in a two-dimensionalised wake.
In this fashion, \cite{Holmes2006} studied a similar case and concluded that a coarse spanwise resolution yields discrepancies in the prediction of the forces induced to the cylinder.
Authors often justify the use of a coarse spanwise resolution pointing at the increase of the spanwise correlation length $(\lambda_z)$ of the turbulent wake when the cylinder is allowed to vibrate in the transverse direction \citep{Toebes1969,Blevins1974,Wu1994}.
Although it has been demonstrated that the structural response can display lower spanwise modes under transverse vibrations, the turbulent wake still presents 3-D small-scale structures along the homogeneous direction.
As mentioned in \cite{Toebes1969}: ``At the same time it is clear that some distortion in the axial direction remains present.
It is plausible to regard this distortion as being ``fed back'' from the wake structure whose large features must sooner or later be deformed by instability and
diffusing turbulence''.
Hence, not accounting for the physical effect of 3-D small-scale structures (i.e. diffusing turbulence) means to disregard the natural dissipation mechanism in the turbulent wake.

\cite{Mittal1995} investigated the physical traits of 2-D and 3-D wakes at $Re=525$ to explain why different forces are captured on both systems.
Higher in-plane $(x$--$y)$ Reynolds stresses were found in the 2-D simulation (these being more pronounced for bluffer bodies) as well as a shorter distance from the vortices roll-up to the cylinder.
Although larger turbulent fluctuations have been found in 2-D simulations, these do not explain why the 3-D behaviour is lost when the spanwise direction is constrained.

An important note also provided by \cite{Mittal1995} states that:
\begin{quote}``An understanding of the basic mechanisms that result in the discrepancy between 2-D and 3-D results could eventually lead to the incorporation of additional physics, which would allow 2-D simulations to predict the aerodynamic forces more accurately.''
\end{quote}
With this purpose, the first part of the thesis investigates the different turbulence dynamics encountered in 2-D and 3-D wakes of flow past a circular cylinder at a high Reynolds regime.

\subsection{Modelling by dimensionality reduction} \label{sec:strip-theory}

The high computational cost of fully-resolved 3-D simulations of long flexible cylinders has motivated authors to investigate alternative numerical approaches.
Beyond RANS and LES, a dimensionality reduction of the problem can yield even more astonishing gains from the computational cost standpoint.

Flow past a long fixed cylinder is characterised by spanwise cells of synchronised shedding frequency \citep{Noack1991,Kappler2005,Bourguet2011}.
Also, as previously mentioned, the spanwise correlation length of large-scale structures increases for flexible cylinders vibrating in the transverse direction.
Because of this, tackling the problem from a 2-D standpoint becomes reasonable: 2-D planes can be placed at informed positions along the span correctly sampling the spanwise cells, a.k.a. strip theory (see \fref{fig:strips}).
For long cylindrical structures, strip-theory methods allow to simulate the whole physical domain in realistic computational times \citep{Herfjord1999,Willden2001}.
For example, reducing the homogeneous isotropic turbulence case dimensionality from 3-D to 2-D effectively cuts the computational cost with $\mathcal{O}(1.6Re_L^{3/4})$.
This translates to 40 minutes instead of 4 years of computational time using all the Iridis5 supercomputer power.
For multiple 2-D planes, the computational cost is reduced with $\mathcal{O}(1.6Re_L^{3/4}/P)$, where $P$ is the number of 2-D planes.

On the other hand, the dimensionality reduction of strip-theory methods also means that the 3-D physics inherent in the turbulent wake are ignored.
As reviewed, this has severe implications for the forces induced to the cylinder.
For 3-D flows, the vortex-stretching mechanism causes large turbulence scales to break down until the smallest possible scale of turbulence, where the energy is dissipated because of viscous effects.
However, since this mechanism is not present in 2-D systems, vortex-merging and  vortex-thinning processes dominate the wake causing coherent and energetic vortical structures \citep{Boffetta2012,Xiao2009}.
For strip-theory methods, authors include arbitrary dissipative turbulence models, often in the form of LES Smagorinsky models, to reduce the strength of the coherent vortices developing in the 2-D wake \citep{Bao2016}.

The inclusion of 3-D physics into the 2-D planes of strip-theory methods can yield more physical predictions while maintaining the low computational cost of the 2-D approach.
This can improve the simulation of flow past long flexible cylinders arising in different applications such as marine risers and cables, power transmission lines, aircraft wings, wind turbine blades, among other offshore, civil and aerospace engineering projects.

\begin{figure}
\centering
\includegraphics[width=0.9\linewidth]{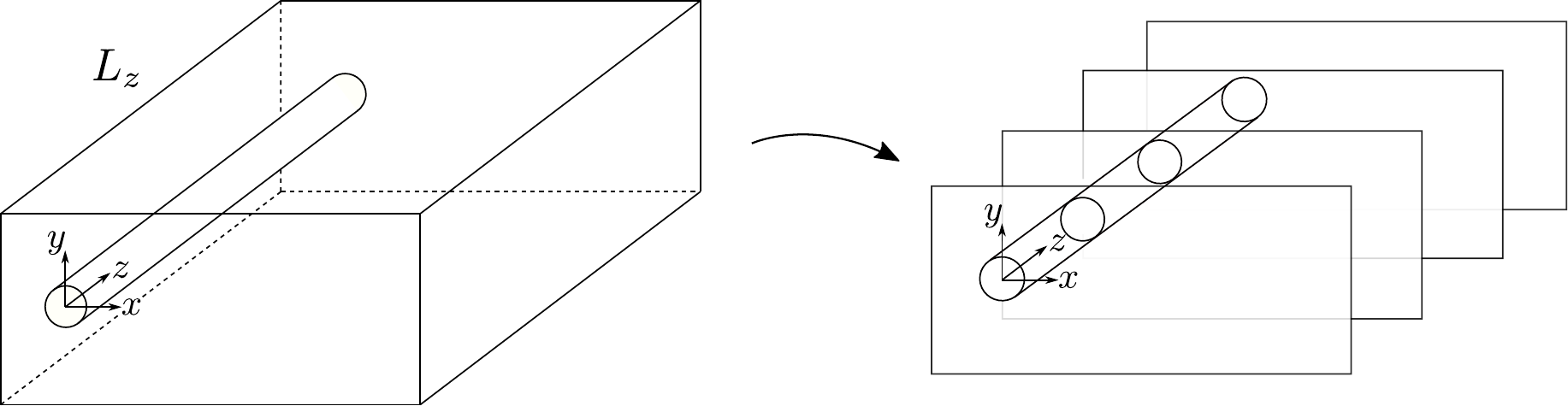}
\caption{Sketch of the standard 2-D strip-theory method.
The original $L_z$ span is decomposed into multiple planes in which the 2-D Navier--Stokes equations are solved.}
\label{fig:stm}
\end{figure}

\section{Research objectives and approach}

As exposed in the previous section, current 2-D strip-theory methods lack the inclusion of 3-D physics which have a significant impact on the prediction of flow past long flexible structures.
This motivates the following fundamental questions which have been investigated in this work:
\begin{enumerate}
	\item Why are standard 2-D systems physically not suited for the simulation of flow past long cylinders at high $Re$?
	\item What role does 3-D turbulence play in the wake of a long cylinder and how does it interact with the 2-D K\'{a}rm\'{a}n dynamics?
	\item How can 3-D effects be incorporated and modelled in a 2-D system?
\end{enumerate}

The following investigations have been conducted to provide novel findings on the stated research questions:

\begin{enumerate}
	\item A parametric study of the span effect on the turbulence dynamics of flow past a circular cylinder at $Re=10^4$.
This provides physical understanding on the implications of using 2-D simulations for 3-D turbulent flows and the cross-over between 2-D and 3-D turbulence.
	\item The derivation of a novel 2-D model based on a spanwise-average decomposition of the flow.
This system accounts for 3-D effects and allows to recover the unsteady spanwise-averaged solution of the flow in a 2-D framework.
	\item The design of a machine-learning (ML) model providing closure to the spanwise-averaged system of equations.
\end{enumerate}

\section{Thesis outline and contributions}

The thesis outline and the author's contributions for each chapter is detailed next.
Note that specific literature review is provided in the introductory part of the main content chapters (\cref{chapter:jfm2019}, \cref{chapter:sans} \& \cref{chapter:zanspy}).
Additionally, chapters also include an opening paragraph summarising their specific motivation, conducted research and main findings.

\textit{Chapter 1} provides an introduction to the current work explaining the background and the justification of the thesis.
The research questions and the adopted investigations are specified as well.

\textit{Chapter 2} poses the incompressible flow system of equations and the numerical approach implemented in the in-house solver used throughout this work.

\textit{Chapter 3} analyses the turbulence dynamics in the wake of a circular cylinder at $Re=10^4$ as the span is constricted.
This chapter provides novel results with respect to the domain geometric anisotropy effect on the wake turbulence dynamics in the presence of a solid wall, in contrast with previous studies only considering obstacle-free flows.
\aref{chapter:appendixA} includes information on the phenomena arising in flow past circular cylinder at different Reynolds regimes.
\aref{chapter:appendixB} contains a validation and verification study of the in-house solver for this particular test case.
The initial stage of this work was published in a conference paper for the \textit{OCEANS 2017} meeting.
The complete study has been published in the \textit{Journal of Fluids Mechanics}:

\begin{itemize}
	\item \cite{Font2017} Analysis of two-dimensional and three-dimensional wakes of long circular cylinders. In \textit{OCEANS 2017 - Aberdeen}. IEEE.
	\item \cite{Font2019} Span effect on the turbulence nature of flow past a circular cylinder. \textit{Journal of Fluid Mechanics} \textbf{878}, 306–323
\end{itemize}

\textit{Chapter 4} introduces the spanwise-averaged Navier--Stokes (SANS) equations, a novel 2-D system of equations including 3-D physics through closure terms in the governing equations.
A SANS-based strip-theory framework accounts for 3-D effects otherwise not considered in standard 2-D strip-theory methods, where unphysical dissipative turbulence models are added in order to mitigate their intrinsic limitations.
Additional derivations related to SANS are provided in \aref{chapter:appendixC}.
The perfect closure is evaluated for the Taylor--Green vortex and the circular cylinder cases.
An a-priori assessment of an standard eddy-viscosity model is performed.
This research has been submitted together with \cref{chapter:zanspy}:

\begin{itemize}
	\item \cite{Font2020b} Deep learning the spanwise-averaged Navier--Stokes equations.
\end{itemize}

\textit{Chapter 5} proposes a ML model for the closure of the novel SANS equations.
A-priori and a-posteriori results are shown for the circular cylinder test case, comprising the second part of \cite{Font2020b}.
A generalisation study of the ML model to flow configuration different to the training case is also presented.
Additionally, a hyper-parametric study of the ML model is provided in \aref{chapter:appendixD}, which constitutes a peer-reviewed paper accepted for the \textit{33rd Symposium on Naval Hydrodynamics},

\begin{itemize}
	\item \cite{Font2020a} Turbulent wake predictions using deep convolutional neural networks. In \textit{33rd Symposium on Naval Hydrodynamics}, Osaka, Japan.
\end{itemize}

\textit{Chapter 6} summarises the main findings and limitations as well as future work recommendations.

This research has been presented by the author in the following international conferences:

\begin{itemize}
	\item Font B., Weymouth, G.D., Nguyen, V.-T. \& Tutty, O.R. \href{https://meetings.aps.org/Meeting/DFD19/Session/L17.5}{2019} Deep learning the spanwise-averaged wake of a circular cylinder. \textit{72nd Meeting of the APS Division of Fluid Dynamics}, Seattle, US.

	\item Font B., Elizalde, I., Weymouth, G.D., Nguyen, V.-T. \& Tutty, O.R. \href{https://etc17.fyper.com/program/show_slot/41}{2019} Turbulence dynamics transition of flow past a circular cylinder and the prediction of vortex-induced forces. \textit{European Turbulence Conference 17}, Torino, Italy.

	\item Font B., Weymouth, G.D. \& Tutty, O.R. \href{https://doi.org/10.1109/OCEANSE.2017.8084904}{2017} Analysis of two-dimensional and three-dimensional wakes of long circular cylinders. \textit{OCEANS 2017}, Aberdeen, UK.

	\item Font B., Weymouth, G.D. \& Tutty, O.R. \href{https://www.imperial.ac.uk/media/imperial-college/faculty-of-engineering/aeronautics/UK-Fluids-Conference-2016-booklet.pdf}{2016} A two-dimensional model for three-dimensional symmetric flows. \textit{UK Fluids Conference}, London, UK.
\end{itemize}
% ---------------------------------------------------------------- 

\chapter{Theoretical background}

\vspace{0.25cm}
This chapter comprises high-level theoretical background in incompressible fluid flow and its numerical solution.
The incompressible flow dynamics is first presented yielding the governing system of equations.
The numerical method providing an approximate solution of the equation system is also explained as implemented in our in-house code, ``Lotus''.
A finite-volume method is employed to spatially discretise the system domain.
The flux-limited quadratic upstream interpolation for convective kinematics (QUICK) scheme numerically approximates the convective term of the momentum equations, which also serves as an implicit model for the subgrid scale (SGS) structures of the flow.
A central-difference scheme is used for the viscous term.
A predictor-corrector scheme integrates the numerical solution in time.
The boundary data immersion method is applied to account for solid boundaries.
Finally, a multigrid method is employed to iteratively solve the pressure-Poisson equation and enforce the continuity condition in the velocity field.

\section{Governing equations of incompressible fluid flow}

Incompressible fluid flow is governed by the conservation of mass and momentum, respectively the continuity equation and the Navier--Stokes momentum equations
\begin{gather}
\nabla\cdot\vect{u}=0,\label{eq:div_u}\\
\pd{\vect{u}}{t}+\pars{\vect{u}\cdot\nabla}\vect{u}=-\nabla p+\nu\nabla^2\vect{u},\label{eq:n-s}
\end{gather}
where $\vect{u}\left(\vect{x}, t\right) = \left(u, v, w\right)$ is the velocity vector field, $p\left(\vect{x}, t\right)$ is the pressure field, $t$ is the time, and $\vect{x}=\left(x,y,z\right)$ is the spatial vector.
The coupling of pressure and velocity arises from an additional relationship obtained by taking the divergence of the momentum equations, namely the pressure-Poisson equation
\begin{gather}
\nabla\cdot\pars{\pd{\vect{u}}{t}+\pars{\vect{u}\cdot\nabla}\vect{u}=-\nabla p+\nu\nabla^2\vect{u}},\\
\nabla^2 p = \nabla\cdot\pars{\vect{h}-\pd{\vect{u}}{t}},\label{eq:ppe}
\end{gather}
where $\vect{h}$ is the force combining convective and viscous terms.

The system of equations is bounded by case-dependant initial and boundary conditions of the system variables.
Importantly, the boundary condition with solid domains is the velocity no-slip condition, i.e. $\vect{u}_b=0$, which yields the flow boundary layer.

The Reynolds number is the non-dimensional parameter describing the ratio between convective and viscous forces,
\begin{equation}
Re=\frac{\rho U L}{\mu},
\end{equation}
where $\rho$ is the fluid density (assumed constant for incompressible flows), $\mu$ is the dynamic fluid viscosity, and $L$ and $U$ are the characteristic length and velocity of the system, respectively.
The Reynolds number defines the scales of motion of the flow.
When the convective forces are much greater than the viscous forces, turbulence develops.
The definition of turbulence is not straight forward, but (loosely) it can be thought as the turning point where the fluid dynamics transitions from ordered to \textit{chaotic} as a result of the nonlinear convective term.
Still, physical laws revealing certain order within the turbulence dynamics have been found in the past.
In particular, it is known that 3-D systems present a constant rate of kinetic energy transfer within a certain range of scales, from large to small structures, a.k.a. direct energy cascade \citep{Richardson1922,Kolmogorov1941}.
Similarly, a constant rate is also found in 2-D systems although energy is inversely transfered from small to large scales, a.k.a. inverse energy cascade \citep{Kraichnan1967,Leith1968,Batchelor1969}.
Such different dynamics will be subject of investigation in the following chapters.

\section{Numerical methods}
\label{sec:lotus}

\subsection{Spatial discretisation}

\begin{figure}[t]
\centering
\includegraphics[width=0.4\linewidth]{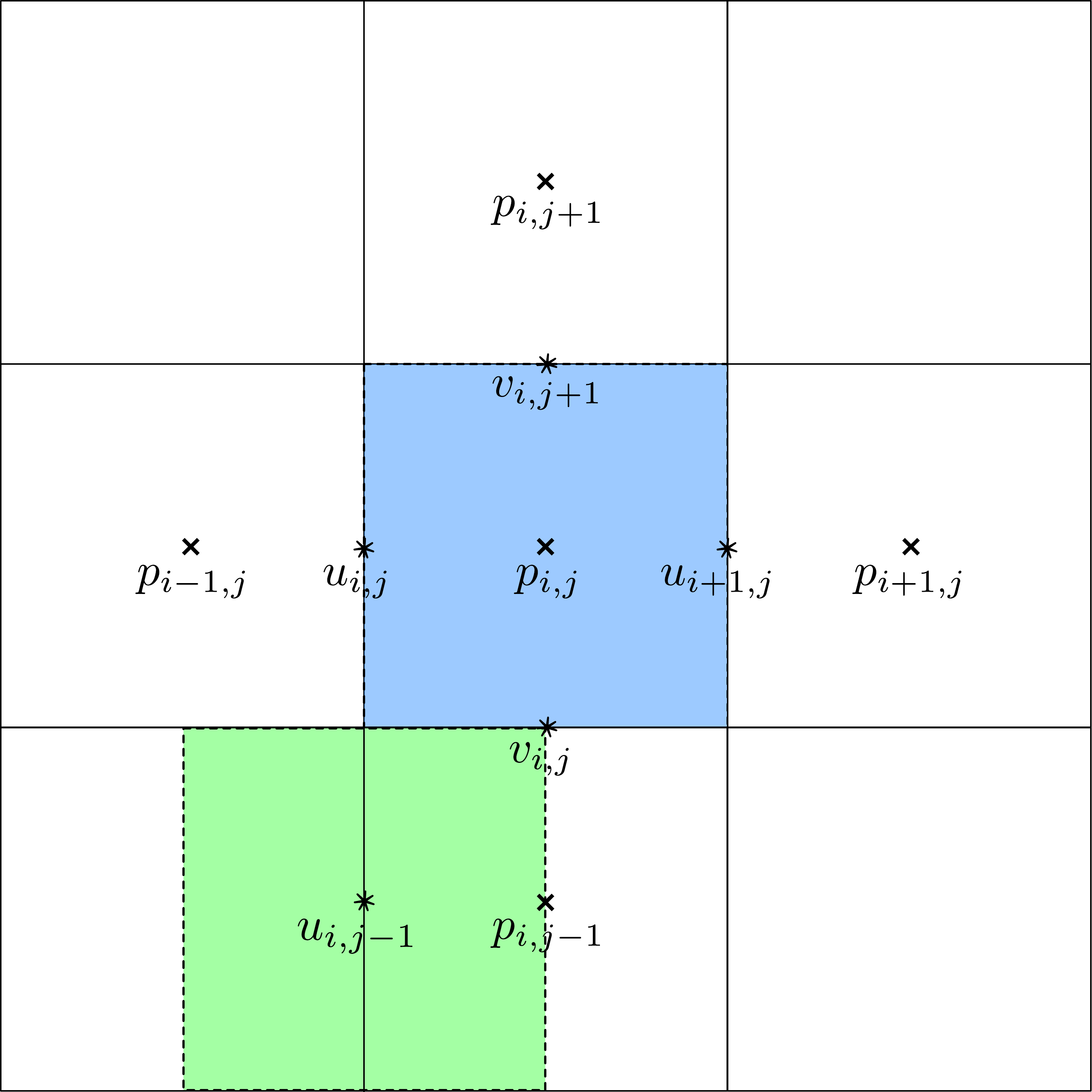}
\caption{Staggered grid.
The control volume for the pressure is defined in blue.
The control volume for the $u$ velocity component is defined in green.}
\label{fig:grid}
\end{figure}

The governing equations comprising \eref{eq:div_u}, \eref{eq:n-s}, and \eref{eq:ppe} are spatially discretised in a structured rectilinear grid using a finite-volume method.
This grid facilitates the implementation of the immersed boundary method accounting for solid boundary conditions, as described in \sref{sec:bdim}.
The flow variables, pressure and velocity, are staggered: the pressure scalar field is stored in the cell centre and the velocity vector field is defined at the cell faces, as displayed in \fref{fig:grid}.

The convective term of the momentum equations is numerically approximated using a flux-limited QUICK scheme.
This scheme interpolates the quantity of interest at the control volume face through a weighted quadratic function fitted with one upstream and two downstream points \citep{Leonard1979}.
The spatial derivative is then approximated using the numerical fluxes given by the cell-face values as defined by the finite-volume approach.

\subsection{Implicit modelling}

As previously exposed, high-Reynolds flows need to be sufficiently discretised in space and time to fully resolve all scales of motion.
The spatial SGS need to be modelled when the grid is not fine enough.
In this sense, we employ an implicit LES (iLES) model, in which the dissipative effect of the SGS is accounted by the numerical dissipation intrinsic in the discretisation scheme of the convective term (the QUICK scheme in the present solver).

Importantly, the instantaneous kinetic energy $(E)$ dissipation rate $(\epsilon)$
\begin{equation}
\dd{E}{t}=-\epsilon
\end{equation}
needs to match the natural dissipation mechanism of the flow.
This balance can be used to compute the iLES effective Reynolds number (or effective kinematic viscosity, $\nu_e$), which can be found through the following evaluation \citep{Domaradzki2003,Aspden2008,Zhou2014}
\begin{equation}
\dd{}{t}\int_\Omega\frac{1}{2}|\vect{u}|^2\,\mathrm{d}\Omega=-2\nu\int_\Omega S_{ij}S_{ij}\,\mathrm{d}\Omega,\label{eq:iles}
\end{equation}
where $S_{ij}$ is the velocity rate-of-strain tensor.
Note that the RHS is the energy dissipation term for incompressible flow arising naturally in the energy transport equation (formally derived from a dot product of the Navier--Stokes momentum equations and the velocity vector field) since \citep[p. 17]{Doering1995}
\begin{equation}
\frac{1}{2}\int_\Omega||\nabla\vect{u}||^2\,\mathrm{d}\Omega=\frac{1}{2}\int_\Omega|\boldsymbol{\omega}|^2\,\mathrm{d}\Omega\equiv Z,
\end{equation}
where $\boldsymbol\omega=\nabla\times\vect{u}$ is the vorticity vector field, $Z$ is the enstrophy, and $||\cdot||$ denotes the tensor Frobenius norm.

Using \eref{eq:iles}, the iLES effective viscosity is computed as
\begin{equation}
\nu_e=\frac{\avg{\epsilon}_{\scaleto{\Omega}{5pt}}}{2\avg{S_{ij}S_{ij}}_{\scaleto{\Omega}{5pt}}},
\end{equation}
where $\avg{\cdot}_{\scaleto{\Omega}{5pt}}$ denotes a volume average.
\cite{Hendrickson2019} verified the present in-house solver by monitoring the grid-scaled effective viscosity $\nu_e\Delta^{-1}$ for the Taylor--Green vortex case.
It was shown that the QUICK scheme linearly scales the effective viscosity with $\Delta$.

\subsection{Temporal discretisation}

The temporal evolution of the system of equations (\eref{eq:div_u}, \eref{eq:n-s} and \eref{eq:ppe}) is discretised using Chorin's projection method \citep{Chorin1967,Chorin1968}.
In short, the Navier--Stokes operator is split in two parts
\begin{gather}
\frac{\vect{u}^*-\vect{u}^n}{\delta t}=\nu\nabla^2\vect{u}^n-\pars{\vect{u}^n\cdot\nabla}\vect{u}^n,\\
\frac{\vect{u}^{n+1}-\vect{u}^*}{\delta t}=-\nabla p^{n+1},
\end{gather}
where the superscript $(\cdot)^n$ refers to the time step level and the intermediate velocity field is noted with the superscript $(\cdot)^*$.
The computation of a non-solenoidal intermediate velocity field allows to decouple the velocity and pressure equations.
In this way, a solenoidal velocity field $\vect{u}^{n+1}$ can be obtained by enforcing the continuity condition into the $p^{n+1}$ pressure-Poisson equation.
Otherwise, using $p^{n}$ to compute $\vect{u}^{n+1}$ would not enforce the velocity field to be divergence-free at the next time step.

The projection method is implemented in a predictor-corrector algorithm.
Next, we note $\vect{h}(\vect{u})=\nu\nabla^2\vect{u}-\pars{\vect{u}\cdot\nabla}\vect{u}$, and $\vect{h}^n=\vect{h}(\vect{u}^n$).
The $(\cdot)^s$ superscript refers to a solenoidal field.
With this, the predictor-corrector algorithm is implemented in two steps as follows

\begin{align}
\mathrm{Pred}&\mathrm{ictor\,\,step:}\nonumber\\
&1.\qquad \vect{u}^* = \vect{u}^n+\vect{h}^n\delta t,\\
&2.\qquad {\delta t}\,\nabla^2p^*=\nabla\cdot\vect{u}^*,\label{eq:ppe1}\\
&3.\qquad \vect{u}^{*,s}=\vect{u}^*-\delta t\nabla p^*.
\nonumber\\
\nonumber\\
\mathrm{Corr}&\mathrm{ector\,\,step:}\nonumber\\
&4.\qquad \vect{u}^{*} = \vect{u}^n+\frac{1}{2}\pars{\vect{h}^n+\vect{h}^{*,s}}\delta t,\\
&5.\qquad {\delta t}\,\nabla^2p^{n+1}=\nabla\cdot\vect{u}^{*},\label{eq:ppe2}\\
&6.\qquad \vect{u}^{n+1}=\vect{u}^{*}-\delta t\nabla p^{n+1}.
\end{align}

Note that the divergence-free constraint for both $\vect{u}^n$ and $\vect{u}^{n+1}$ has been considered.
Also note that the forward Euler time stepping scheme is implemented in the predictor step to compute the intermediate velocity field $\vect{u}^{*}$, finally projected into a solenoidal field by the pressure-Poisson equation yielding $\vect{u}^{*,s}$.
With this, the trapezoidal quadrature can be employed in the corrector step to integrate the solution from time $n$ to $n+1$, as described in step 4.
The solution is again projected into a divergence-free field by solving the pressure-Poisson equation for $p^{n+1}$ (step 5) and correcting the velocity field with the new pressure field (step 6) yielding $\vect{u}^{n+1}$.

The advantage of this method is that it exploits the benefits of both explicit and implicit time-marching schemes: it combines the natural stability of implicit schemes with the low memory requirements of explicit schemes.
It can be shown that the predictor-corrector algorithm is second-order accurate in time.
The reader is referred to \cite{Ferziger2002} for further details.

The time discretisation method is bounded to the limitation of the time step size $(\delta t)$.
For this, the local Courant number $(u\,\delta t/\delta x)$ and Péclet number $(\nu\,\delta t/(\delta x)^2)$ are evaluated at every time step yielding an adaptive step size.
These dimensionless parameters quantify the characteristic convection time and characteristic diffusivity time of the flow, respectively.
A combination of both yields to a time step size limit (for  one-dimensional flow) of
\begin{equation}
\delta t<\left[\frac{2\nu}{\min\left[(\delta x)^2\right]}+\max\pars{\frac{u}{\delta x}}\right]^{-1}.
\end{equation}

\subsection{Solid boundaries} \label{sec:bdim}

The velocity no-slip condition at solid boundaries is implemented using the boundary data immersion method (BDIM) from \cite{Weymouth2011}.
Immersed boundary (IB) methods, such as BDIM, take into account the solid wall effect in a rectilinear non-conforming grid.
The Dirichlet condition on the velocity field is interpolated at the grid nodes from its actual position.
The interpolation method is what differs among IB methods.

A clear advantage of IB methods is the trivial grid-generation process, in contrast to the complicated meshing process regularly encountered for body-conforming grids.
Also in this regard, the body geometry is practically irrelevant for the grid generation process.
Other advantages include the simulation of moving bodies since the grid does not need to be updated at every step, as well as the straightforward implementation of numerical schemes.
On the other hand, accuracy near the solid boundary can be compromised depending on the effective resolution of the non-conforming grid.
Also, the implementation of the solid boundary condition is not as trivial as in body-conforming grids \citep{Mittal2005}.

\begin{figure}[t]
\centering
\includegraphics[width=0.65\linewidth]{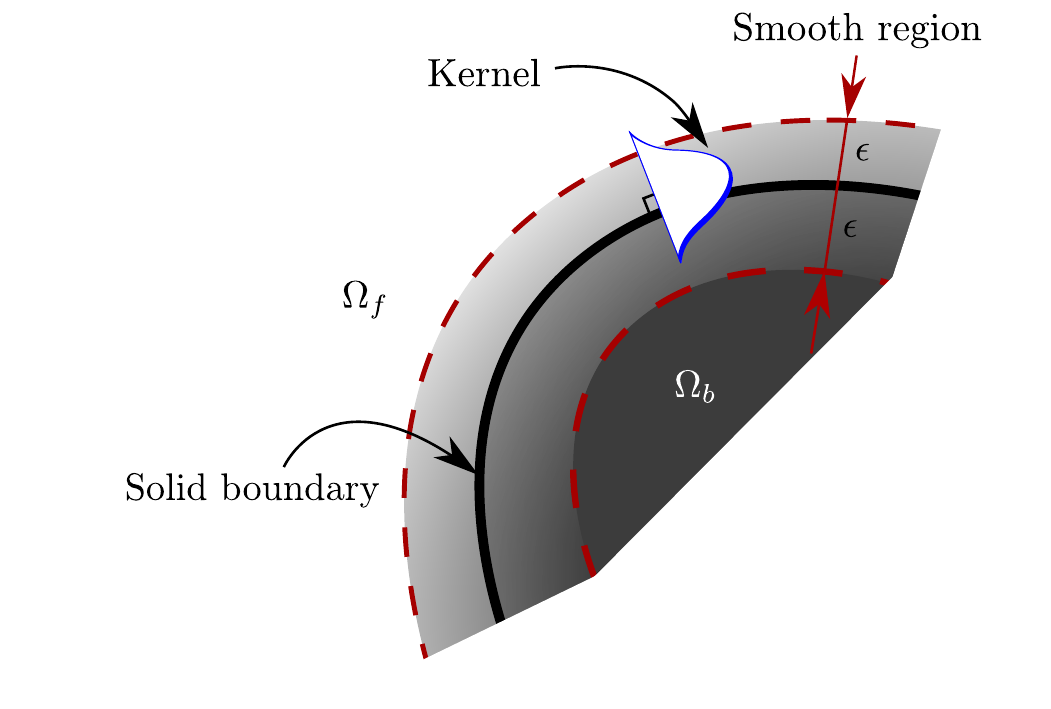}
\caption{BDIM sketch adapted from \cite{Maertens2015}.
A convolution kernel with radius $\epsilon$ smooths the interface between solid $(\Omega_b)$ and fluid $(\Omega_f)$ subdomains.}
\label{fig:bdim}
\end{figure}

The BDIM consists in mapping the fluid governing equation $(\mathcal{F})$ and the solid governing equation $(\mathcal{B})$ into a single meta-equation $(\mathcal{M})$ discretised in a non-conforming rectilinear staggered grid.
A convolution kernel $(K_\epsilon)$ provides a smooth transition between the mediums enforcing the velocity condition at the interface (see \fref{fig:bdim}) while extending the fluid subdomain $(\Omega_f)$ and solid subdomain $(\Omega_b)$ to the full single domain $(\Omega)$.
Imposing the no-slip condition on a static body, this method can be summarised as follows
\begin{align}
&\mathcal{F}(\vect{u},p)=\partial_t\vect{u}-\vect{h}+\nabla p=0,\\
&\mathcal{B}(\vect{u})=\vect{u}=0,\\
&\mathcal{M}(\vect{u},p)=\mathcal{F}(\vect{u},p)\pars{1-\delta_\epsilon}+\mathcal{B}\pars{\vect{u}}\delta_\epsilon=0,
\end{align}
where the convolution between kernel and governing equation has been approximated using (e.g)
\begin{gather}
\mathcal{B}_\epsilon\pars{\vect{x}}=\int_{\Omega_b}\mathcal{B}\pars{\vect{x}_b}K_\epsilon\pars{\vect{x},\vect{x}_b}\,\mathrm{d}\vect{x}_b\approx \mathcal{B}\pars{\vect{x}}\int_{\Omega_b}K_\epsilon\pars{\vect{x},\vect{x}_b}\,\mathrm{d}\vect{x}_b,\\
\mathcal{B}_\epsilon\pars{\vect{x}}\approx\mathcal{B}\pars{\vect{x}}\delta_\epsilon,
\end{gather}
where $\delta_\epsilon$ is the integrated kernel over the subdomain which can take the approximated form
\begin{equation}
\delta_\epsilon(d)=
	\begin{cases}
		\frac{1}{2}[1+\sin(\frac{\pi}{2}\frac{d}{\epsilon})] & \mathrm{for} \left|d\right|<\epsilon\\
		1 & \mathrm{for}\,\,d<-\epsilon\\
		0 & \mathrm{else}
	\end{cases}
\end{equation}
where $d$ is a signed-distance function from a point $\vect{x}$ to the solid interface. This approximation results into a first-order accurate scheme, although high-order terms can be considered to improve the order of accuracy \citep{Maertens2015}.

The BDIM has been tested in multiple applications and the reader is referred to \cite{Maertens2015} for an extended mathematical description of the method, and to \cite{Schulmeister2017} for further validation on bluff body cases.

\subsection{Pressure solver}

A multigrid method is used to solve the pressure-Poisson equation as discretised in the pressure-corrector algorithm (\eref{eq:ppe1} and \eref{eq:ppe2}).
Algebraically, the pressure-Poisson equation can be written as a linear system of equations
\begin{equation}
\matr{A}\vect{x}=\vect{b},
\end{equation}
where $\matr{A}$ is a sparse matrix containing the discretised Laplacian operator, $\vect{x}$ is the pressure field arranged in a column vector, and $\vect{b}$ is the column vector of the intermediate velocity field divergence.

Contrary to direct methods, a multigrid solver is an iterative method in the sense that that a guessed solution $\vect{x}^n$ (at the iteration $n$) is evaluated yielding a residual
\begin{equation}
\vect{r}^n=\matr{A}\vect{x}^n-\vect{b},
\end{equation}
and an error
\begin{equation}
\boldsymbol\epsilon^n=\vect{x}-\vect{x}^n,
\end{equation}
which are related by
\begin{equation}
\matr{A}\boldsymbol\epsilon^n=\vect{r}^n.
\end{equation}
The objective is to iteratively minimise $\vect{r}^n$ so that $\vect{x}^n$ converges to $\vect{x}$.

In multigrid methods, iterations are performed from fine to coarse grids, transferring the residual across grid levels so that iterations become cheaper as the grid is coarsened.
Hence, multigrid methods yield a speed-up of the overall convergence process.
Iterations can be carried with any iterative method, e.g. Gauss--Seidel, Jacobi, conjugate gradient, etc.
In our solver, a single Jacobi iteration to smooth the solution is performed before downsampling the residual to a coarser level.
Downsampling (in one-dimensional form) is performed from
\begin{equation}
\frac{1}{(\delta x)^2}\pars{\epsilon_{i-1}^n-2\epsilon_{i}^n+\epsilon_{i+1}^n}=r_i^n,
\end{equation}
where the subscript $(\cdot)_i$ denotes the fine grid cell index, to
\begin{equation}
\frac{1}{(\delta X)^2}\pars{\epsilon_{I-1}^n-2\epsilon_{I}^n+\epsilon_{I+1}^n}=r_I^n,
\end{equation}
where the subscript $(\cdot)_I$ denotes the coarse grid cell index.
The relationship between both grids might be $\delta X = 2\delta x$, hence the $I$ (coarse grid) control volume is defined as the $i$ (fine grid) control volume plus half of its neighbour control volumes ($i+1$ and $i-1$) \citep{Ferziger2002}.

The residual can be gradually downsampled to an arbitrary coarse grid using linear interpolation for each downsampling step.
Similarly, the residual is upsampled from the coarsest grid to the original fine grid while iterating in the mid level grids to correct the fine grid solution.
We employ a conjugate gradient method after each upsampling step to update the solution.
This downsampling and upsampling process is known as a V-cycle.
Multiple V-cycles can be performed in a single time step to solve the pressure-Poisson equation.
The iterative method is stopped once the convergence tolerance is reached.
The following convergence tolerance criteria is defined in a grid with cell index $i$
\begin{gather}
\frac{1}{\Omega}\sum_i\Big|\oint \vect{u}\cdot\vect{\hat{n}}\,\mathrm{d}S\Big|<10^{-6}\\
\max_i\Big|\oint \vect{u}\cdot\vect{\hat{n}}\,\mathrm{d}S\Big|<10^{-5} \,\,\, \forall i \in \Omega ,
\end{gather}
establishing the allowed average velocity divergence error in $\Omega$ as well as its maximum local error.
% ---------------------------------------------------------------- 

\chapter{Span effect on the turbulence nature of flow past a circular cylinder}

\vspace{0.25cm}
Turbulent flow evolution and energy cascades are significantly different in 2-D and 3-D flows.
Studies have investigated these differences in obstacle-free turbulent flows, but solid boundaries have an important impact on the cross-over from 3-D to 2-D turbulence dynamics.
In this chapter, the span effect on the turbulence nature of flow past a circular cylinder at $Re=10^4$ is investigated.
It is found that even for highly anisotropic geometries, 3-D small-scale structures detach from the walls.
Additionally, the natural large-scale rotation of the K\'{a}rm\'{a}n vortices rapidly two-dimensionalises those structures if the span is 50\% of the diameter or less.
This is linked to the span being shorter than the Mode B instability wavelength.
The conflicting 3-D small-scale structures and 2-D K\'{a}rm\'{a}n vortices result in 2-D and 3-D turbulence dynamics which can coexist at certain locations of the wake depending on the domain geometric anisotropy.

\section{Introduction and literature review}

Incompressible viscous flow past 2-D bluff bodies involves complex physics such as the von-K\'{a}rm\'{a}n street phenomenon as well as 3-D wake dynamics as the Reynolds number is increased \citep{Roshko1954, Williamson1996b, Williamson1996a}.
Due to the two-dimensionality of circular cylinders, some authors have used the 2-D Navier--Stokes equations on multiple planes located along the span of the cylinder as a simplified model of the three-dimensionality without increased computational cost (a.k.a. strip-theory method, see \sref{sec:strip-theory}).
Such strip-theory methods are used in offshore and civil engineering applications to model flow along slender structures where the computational cost of fully resolved simulations is prohibitive, such as marine risers, tow and mooring cable systems, and tall pillars.
However, the physics inherent in 2-D simulations lead to poor 3-D predictions \citep{Bao2016}, and a better understanding of the evolution from 3-D to 2-D turbulent wakes is needed to improve strip-theory methods.

The fluid mechanics of turbulent flows behave quite differently when the spanwise spatial dimension is much more constrained than the others.
Instead of having a direct cascade of the turbulence kinetic energy (TKE) from the integral scales down to the dissipative scales, there is a dual cascade of (direct) enstrophy and (inverse) energy.
This was first suggested by \cite{Kraichnan1967} and further developed by \cite{Leith1968} and \cite{Batchelor1969} (known as the KLB 2-D turbulence theory).
More recently, the dual cascade has been demonstrated both experimentally and computationally \citep[for a comprehensive review see][]{Boffetta2012}.
Studies such as \cite{Xiao2009} have shown that physical processes of vortex thinning and vortex merging dominate the dynamics of 2-D turbulence generating larger and more intense vortical structures the energy of which piles up at the integral scale for bounded domains.

\begin{figure}
\vspace*{0.2cm}
  \centerline{\includegraphics[width=0.4\textwidth]{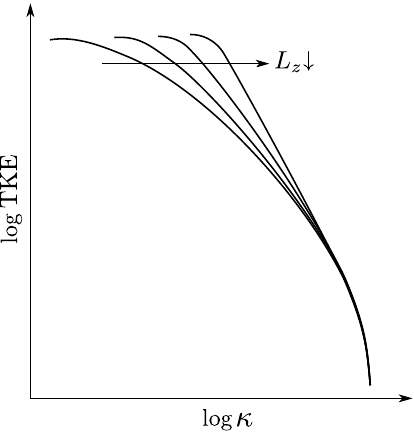}}
  \caption{Sketch of the TKE cascade distribution across the scales as the span is reduced.
By constricting the domain, the integral scale is reduced and the energy-containing scales can feed less energy to the inertial and dissipative scales.}
\label{fig:ek_sketch}
\end{figure}

Previous work has studied the transition between 2-D and 3-D dynamics in obstacle-free turbulent flows in detail.
The effect of the $L_z$ constriction on the TKE distribution across the scales (or wavenumbers $\kappa$) is sketched in \fref{fig:ek_sketch}.
By constricting the domain, the size of the 3-D energy-containing structures (integral-scale structures) is reduced.
Since the energy-containing structures feed the inertial subrange structures down to the 3-D small-scale dissipative structures, smaller integral-scale structures result in less energy fed into the inertial subrange structures and, consequently, to the dissipative structures.
In the 2-D limit, no 3-D dissipative structures are present and, because of the lack of a dissipation mechanism, the turbulent vortical structures can only merge.
This creates larger structures promoting an inverse energy cascade as shown in \cite{Kraichnan1967, Leith1968, Batchelor1969}.

\cite{Smith1996} reviewed the aspect ratio depth effect together with a rotation effect of forced turbulence on a $L_x \times L_y \times L_z$ periodic box.
It was found that the turbulence dimensionality of the flow depended not only on the geometry constriction ($L_z/L_x$), but also on the rotation intensity.
A critical ratio between the span and the turbulence forcing scale was revealed below which two-dimensionalisation occurred for non-rotating cases.
Furthermore, it was found that higher rotation rates induced a more significant 2-D turbulence behaviour and that direct and inverse energy cascades for small and large scales can coexist respectively.
\cite{Celani2010} found a similar splitting of the turbulence cascade for a critical value of the relative forcing scale on a depth-restricted periodic box.

These differences between 2-D and 3-D turbulence dynamics have a significant practical impact.
For example, the forces induced on a cylinder are larger in magnitude and variability in 2-D systems \citep{Mittal1995, Norberg2003}.
In an attempt to dissipate the energised vortical structures and to prevent the vortex-merging dynamics of 2-D turbulence, models based on the eddy-viscosity hypothesis (a.k.a. the Boussinesq hypothesis) are incorporated into 2-D strip-theory methods.
However, these models assume that the anisotropic Reynolds-stress tensor is proportional to the mean rate-of-strain tensor by the scalar eddy viscosity, a relation which has been proven inaccurate even for simple shear flows \citep[p. 94]{Pope2000} and can potentially yield incorrect forces induced to the cylinder.
A further complication for strip theory is that the presence of walls adds a production mechanism of 3-D turbulence which is able to generate very fine 3-D vortical structures even in highly anisotropic geometries.

A ``thick'' strip-theory method proposed by \cite{Bao2016} showed how strips with a certain thickness are able develop to 3-D turbulence when its span is larger than the wavelength of the Mode B instability of circular cylinders, $\lambda_z/D\approx1$ (see \aref{chapter:appendixA} for a summary of phenomena arising in flow past a circular cylinder at different Reynolds regimes).
In fact, this instability creates rib-like streamwise vortical structures along the main K\'{a}rm\'{a}n 2-D vortices \citep{Noack1999}.
Therefore, it can be argued that the two-dimensionalisation of the wake arises from the geometry constriction which prevents the rib-like vortices developing when there is not room enough for its natural wavelength.
However, the connection between wake and wall turbulence and the persistence of 3-D turbulent structures in constricted span flows has not been fully explored.

This work studies the geometry constriction effect on the turbulence nature of a flow past a circular cylinder at $Re=10^4$.
To do this, a series of simulations ranging from $L_z=10$ to pure 2-D planes have been considered.
As discussed above, the inclusion of a body boundary provides an important change to the turbulence production mechanisms compared to previous research as well as novel information on the transition and cross-over between 3-D and 2-D turbulence for very constricted domains in wall-generated turbulent shear flows.
Multiple turbulence statistics are presented for the wide range of constricted wakes, providing new data on the transition from 3-D to 2-D turbulence.

\section{Problem formulation}\label{sec:circular_cylinder_details}

This study considers flow past a circular cylinder with diameter $D$ aligned on the $z$ direction in a 3-D $\left(35\times20\times L_z\right)D$ rectangular domain, where $L_z$ is the non-dimensional span.
To study the span effect on the turbulence nature of the wake, the following cases have been considered: $L_z=10, \pi, 1, 0.5, 0.25, 0.1$ as well as a fully 2-D case.

\begin{figure}
	\vspace*{0.2cm}
  \centerline{\includegraphics[width=0.8\textwidth]{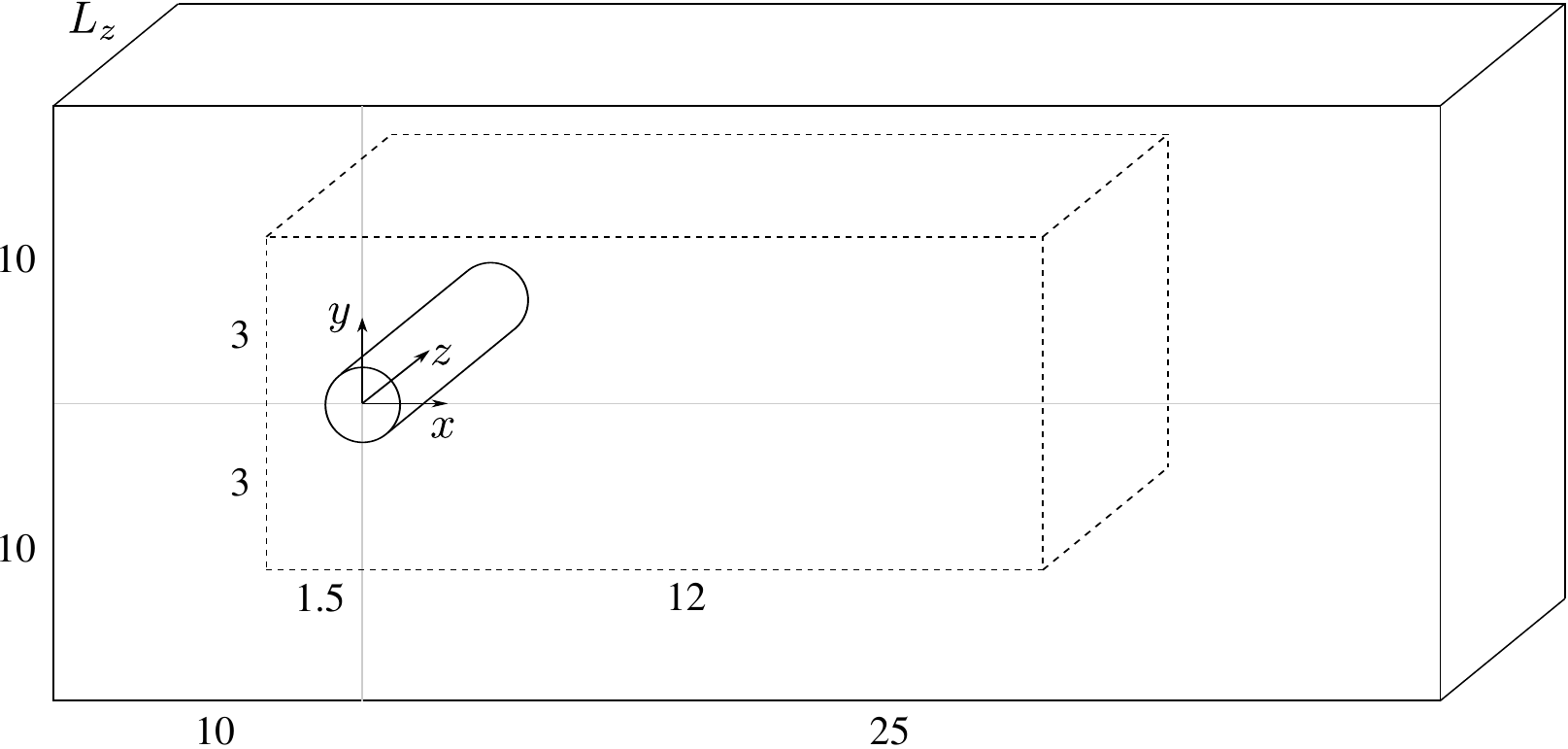}}
  \caption{Computational domain sketch non-dimensionalised with the cylinder diameter $D$.
  A very fine Cartesian grid domain (depicted in discontinuous lines) surrounds the cylinder and the close- and mid-wake regions.
  A stretched rectilinear grid (depicted in solid lines) transitions from the Cartesian grid to the boundaries.}
\label{fig:computational_domain}
\end{figure}

A Reynolds number of $Re=10^4$ is selected.
We define the Reynolds number as
\begin{equation}
Re=\frac{UD}{\nu},
\end{equation}
where $U$ is the scaling velocity and $\nu$ is the kinematic viscosity of the fluid.
Using this Reynolds number we ensure that the flow is well within the turbulence regime encompassing very different spatial and temporal scales in addition to the availability of DNS data \citep{Dong2005} which is used to validate our results in \aref{chapter:appendixB}.

The incompressible viscous fluid motion is described by the continuity equation and the non-dimensional Navier--Stokes momentum equations
\begin{eqnarray}
&\nabla\cdot\vect{u}=0,\label{eq:mass}\\
&\partial_{t}\vect{u}+\pars{\vect{u}\cdot\nabla} \vect{u}=-\nabla p+Re^{-1}\nabla^2\vect{u}.
\end{eqnarray}
Velocity and pressure fields are coupled with the pressure-Poisson equation. 
The initial condition is defined as $\vect{u}\left(\vect{x}, 0\right)=\left(1,0,0\right)$ in the fluid.
The boundary conditions are: a uniform velocity profile on the inlet boundary, a natural convection condition on the outlet boundary, a no-penetration slip condition on the upper and lower boundaries, a periodic condition on the spanwise direction boundaries and a no-slip velocity condition on the cylinder.
Periodic boundary conditions on the constricting planes are used as in previous studies on the cross-over of 2-D and 3-D turbulence for obstacle-free flows such as \cite{Smith1996}, \cite{Celani2010} and \cite{Biancofiore2014}.
This choice is made to avoid the artificial high intensity turbulence enhancements and the deterioration of the 2-D behaviour of the flow when a no-penetration condition is enforced on the constricting planes, as noted on \cite{Biancofiore2012} for obstacle-free turbulent wakes.
Other studies such as \cite{Bao2016} and \cite{Bao2019} also use a periodic spanwise condition for thick strips modelling long circular cylinders.

The numerical methods to solve the governing system of equations have been described in \cref{chapter:theoretical_background}. 
Particularly for the circular cylinder case, the domain is composed of a sufficiently fine Cartesian grid for the close- and mid-wake regions defined as $\left(L_x\times L_y\times L_z\right)D$, where $L_x$ and $L_y$ are the non-dimensional horizontal and vertical lengths respectively.
A stretched grid is considered for the regions far from the cylinder (see \fref{fig:computational_domain}).
A resolution of 90 cells per diameter in all spatial directions is chosen for the Cartesian grid subdomain.
The resolution in all spatial directions is kept constant as the span is reduced.

The 3-D simulations are started from a three-dimensionalised 2-D flow snapshot.
A time length of $t^*_0=200$ $(t^* = tU/D)$ is simulated before starting to record the flow statistics in order to achieve a statistically stationary state of the wake.
The flow statistics are then recorded for a total of $\Delta t^*=t^*-t^*_0=500$ convective time units (around 100 wake cycles).
A verification and validation of the wake turbulence dynamics of the investigated test case is included in \aref{chapter:appendixB}.
Finally, the turbulence statistics of the $L_z=10$ and $L_z=\pi$ cases are very similar as displayed in \fref{fig:TKE}a.
Hence, only the $L_z=\pi$ results are displayed on the other figures for clarity.

\section{Results and discussion}\label{Results and discussion}

\begin{figure}
	\vspace*{0.3cm}
  \centerline{\includegraphics[width=1\textwidth]{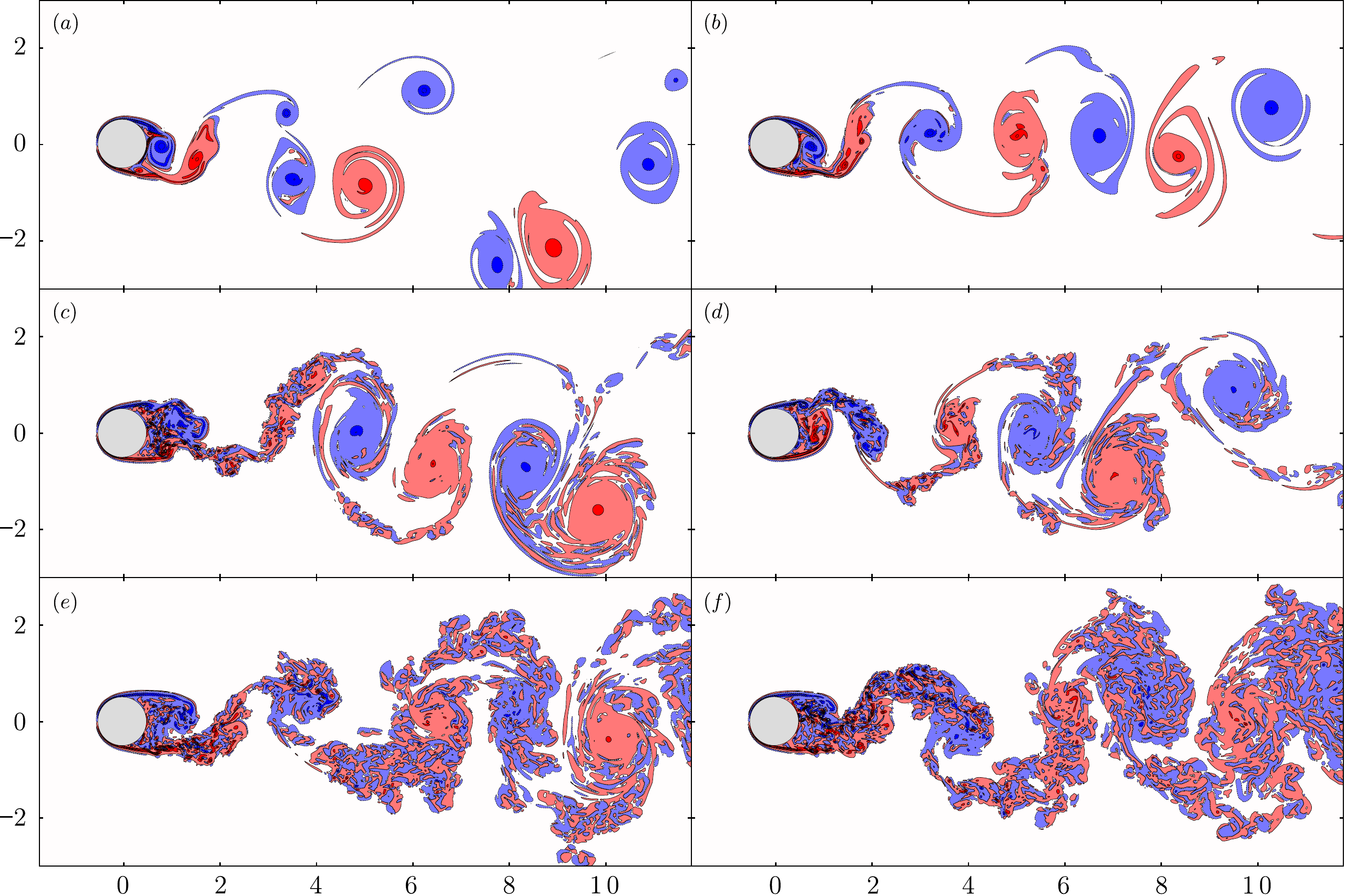}}
  \caption{Instantaneous vorticity $\omega_z$ (red is positive, blue is negative) at the $z=L_z/2$ plane for: $(a)$ $L_z=0$, $(b)$ $L_z=0.1$, $(c)$ $L_z=0.25$, $(d)$ $L_z=0.5$, $(e)$ $L_z=1$, $(f)$ $L_z=\pi$.}
\label{fig:instantaneous_vorticity}
\end{figure}

The flow field is displayed in \fref{fig:instantaneous_vorticity} in terms of the instantaneous vorticity component $\omega_z$ as the span is varied.
The most striking feature is how the coherence of the K\'{a}rm\'{a}n vortices increases as the span is reduced.
However, even in highly-anisotropic geometries such as $L_z=0.25$, small-scale 3-D structures are generated from the cylinder wall.

An important result is that the two-dimensionalisation of these structures is faster (in the sense that it occurs closer to the cylinder) as the domain is constricted because of the geometry constriction and the natural rotation of the K\'{a}rm\'{a}n vortices.
The combination of these two mechanisms as a two-dimensionalisation method is also found in \cite{Smith1996} and \cite{Xia2011}.
For the $L_z\geq1$ cases, the 3-D small-scales structures detaching from the cylinder wall are not two-dimensionalised as rapidly.
In fact, it can be appreciated that only the far wake region of the $L_z=1$ case displays a coherent K\'{a}rm\'{a}n vortex.
This means that less anisotropic geometries promote a direct TKE cascade on the wake so that the 3-D dissipative structures are still sustained far from the cylinder.

\begin{figure}
	\vspace*{0.3cm}
  \centerline{\includegraphics[width=1\textwidth]{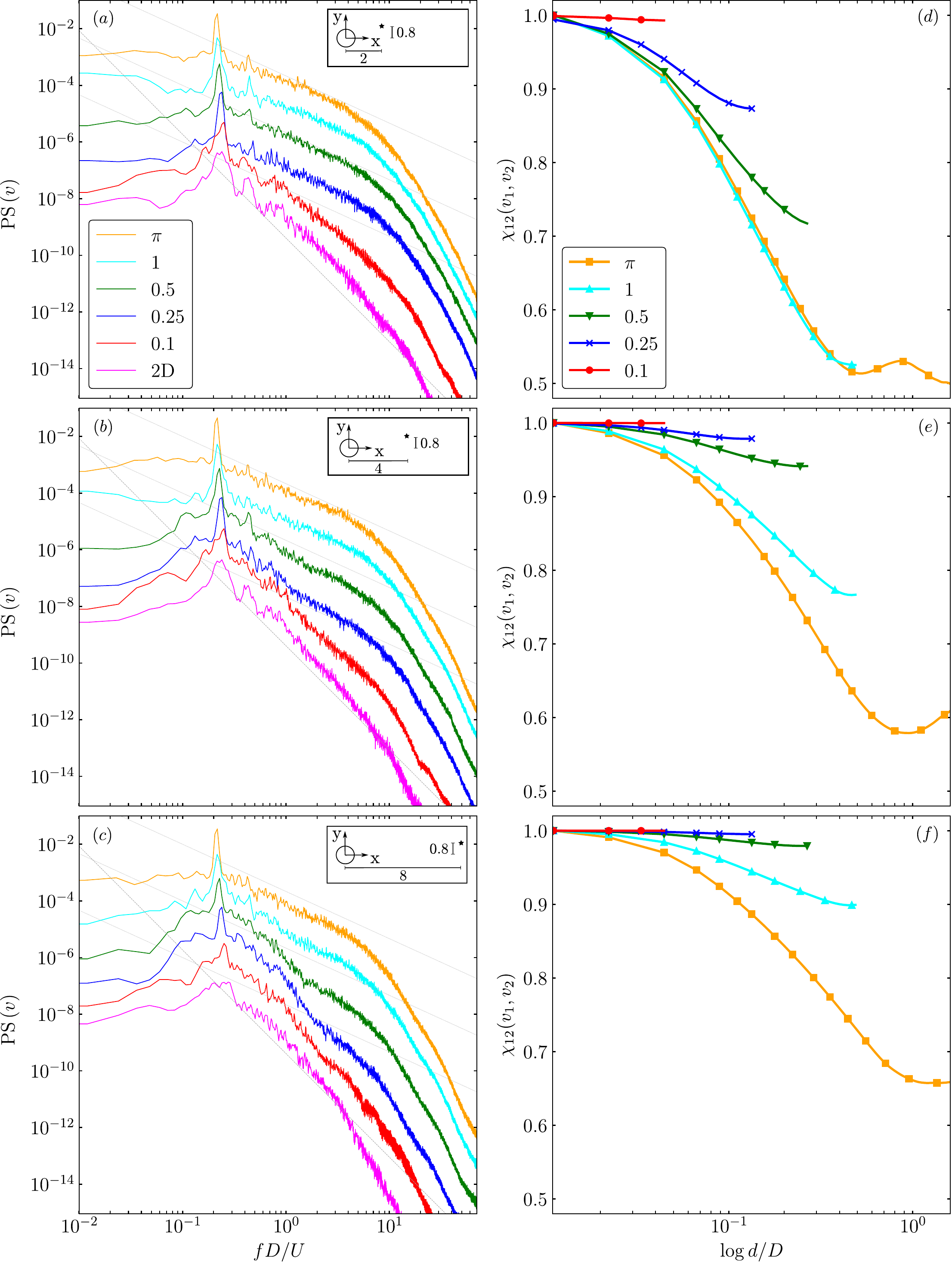}}
  \caption{Left: vertical velocity component temporal PS at different $\left(x,y\right)$ locations on the wake: $\left(a\right)=\left(2,0.8\right)$, $\left(b\right)=\left(4,0.8\right)$, $\left(c\right)=\left(8,0.8\right)$.
The PS lines of each case are shifted a factor of 10 for clarity and the vertical axis ticks correspond to the $L_z=\pi$ case.
The dashed lines have a $-11/3$ slope and the dotted lines have a $-5/3$ slope.
Right: two-point correlations along $z$ at the same $(x,y)$ locations as the left figures respectively.
The correlation value $\chi_{12}$ for a given $d$ corresponds to the averaged value of the multiple correlations of pairs of points separated a distance $d$ along the span.}
\label{fig:velocity_spectras}
\end{figure}

Whether the wake turbulence dynamics is 2-D or 3-D is better captured on the TKE spectra which can be directly compared to classic turbulence theory.
For this, the Taylor's hypothesis is considered and the temporal spectrum at different points of the wake is calculated.
\fref{fig:velocity_spectras} a,b,c shows the temporal power spectrum (PS) of the $v$ velocity component at $(x,y)=(2,0.8),(4,0.8),(8,0.8)$.
The Welch method (using a time signal of 500 units split in 6 parts with 75\% overlapping) has been employed to compute the spectra at 8 different points along the span for each $(x,y)$ point.
These spectra are then averaged resulting in a single spectrum for each case.

First, note that all of the spectra display a peak around the $0.2$ non-dimensional frequency corresponding to the non-dimensional Strouhal number, $St=f_s D/U$, where $f_s$ is the vortex-shedding frequency.
A smaller harmonic peak around $0.4$ non-dimensional frequency is also found for the spectra at $(x,y)=(2,0.8),(4,0.8)$.
Second, the spectrum at the closest analysed point to the cylinder (\fref{fig:velocity_spectras}a) displays a 3-D turbulence behaviour with a $-5/3$ decaying rate with the exception of the pure 2-D and the $L_z=0.1$ cases.
For the latter cases, a decaying slope around $-11/3$ is found.
These 2-D flow spectra are steeper than the $-3$ rate predicted by the classical 2-D turbulence theory \citep{Kraichnan1967} because of the interaction of the coherent large-scale K\'{a}rm\'{a}n vortices, in agreement with the finds of \cite{Dritschel2008} and \cite{Biancofiore2014}.
The filamentary vorticity (filaments of vorticity around the coherent vortices) is likely to be destroyed by the interaction of the large-scale vortices rather than viscous effect (specially for high $Re$ flows), thus limiting its range of scales.
The coherent vortices induce a spiralling effect which limits the range of scales of incoherent filamentary vorticity \citep{Gilbert1988}.
On the other hand, as a $-5/3$ decay rate is captured for the $L_z=0.25$ case, it can be argued that 3-D turbulence is being generated from the cylinder wall even for highly reduced spans.

The $L_z=0.25$ and $L_z=0.5$ cases feature a decay rate that transitions from $-5/3$ to $-11/3$ as the spectra are computed further downstream from the cylinder.
In particular, both cases have $-5/3$ slopes in \fref{fig:velocity_spectras}a and $-11/3$ in \fref{fig:velocity_spectras}c, quantifying the turbulent two-dimensionalisation in the rotating wake.
Also note the coexistence of both 2-D and 3-D turbulence features for the $L_z=0.25,0.5$ and $L_z=1$ cases in \fref{fig:velocity_spectras}b and \fref{fig:velocity_spectras}c respectively.
The low-frequency structures behave mostly two-dimensionally (decay rate between $-3$ and $-11/3$, resulting from the presence of less coherent 2-D structures than the pure 2-D case) up to a certain point where a $-5/3$ rate is briefly recovered.
Hence, high-frequency structures interact in a 3-D fashion while low-frequency structures interact two-dimensionally.
This finding is in agreement with \cite{Smith1996} and \cite{Celani2010}.

Additionally, two-point correlations along the span have been analysed at the same $(x,y)$ locations as the PS plots (\fref{fig:velocity_spectras} d,e,f).
Given a distance $d$ along $z$ ranging from 0 to $L_z/2$, the two-point correlation is calculated with the temporal signals of the vertical velocity component at multiple pairs of points (namely $v_1$ and $v_2$) as follows
\begin{equation}
\chi_{12}\left(v_1(\vect{x},t), v_2(\vect{x}+\vect{r},t) \right) = \frac{\mathrm{cov}(v_1, v_2)}{\sqrt{\mathrm{cov}(v_1, v_1)\mathrm{cov}(v_2, v_2)}},
\end{equation}
where the distance vector is defined as $\vect{r}=(0,0,d)$.
The multiple correlation coefficients for a given $d$ are then averaged corresponding to a data point in the plots.

Very close to the cylinder (\fref{fig:velocity_spectras}d), the correlation coefficient quickly decreases with increasing $d$ for $L_z>0.1$.
This indicates the presence of 3-D structures near the body as also noted on the velocity spectrum plot counterpart.
Also, the decrease is more pronounced as the span increases.
For $L_z=\pi$, it is worth noting a local correlation coefficient maximum around $d \approx 0.9$.
This distance approximately corresponds to the Mode B instability wavelength $(\lambda_z)$.
Since the rib-like vortices associated with the Mode B instability (streamwise and crossflow vorticity) are not very well defined at this $Re$ regime \citep{Chyu1996}, the correlation increase is not as significant as at lower $Re$.
Still, $L_z=\pi$ is the only case displaying such phenomena because of the spanwise boundary conditions periodicity, which only allow the instability to develop if $L_z>\lambda_z$ (the $L_z=1$ case might be too critical to display such phenomenon considering also its intermittent nature).
The correlation coefficient increases when calculated further downstream as shown in \fref{fig:velocity_spectras}e and \ref{fig:velocity_spectras}f, evidencing again the wake two-dimensionalisation.

\begin{figure}
	\vspace*{0.3cm}
  \centerline{\includegraphics[width=1\textwidth]{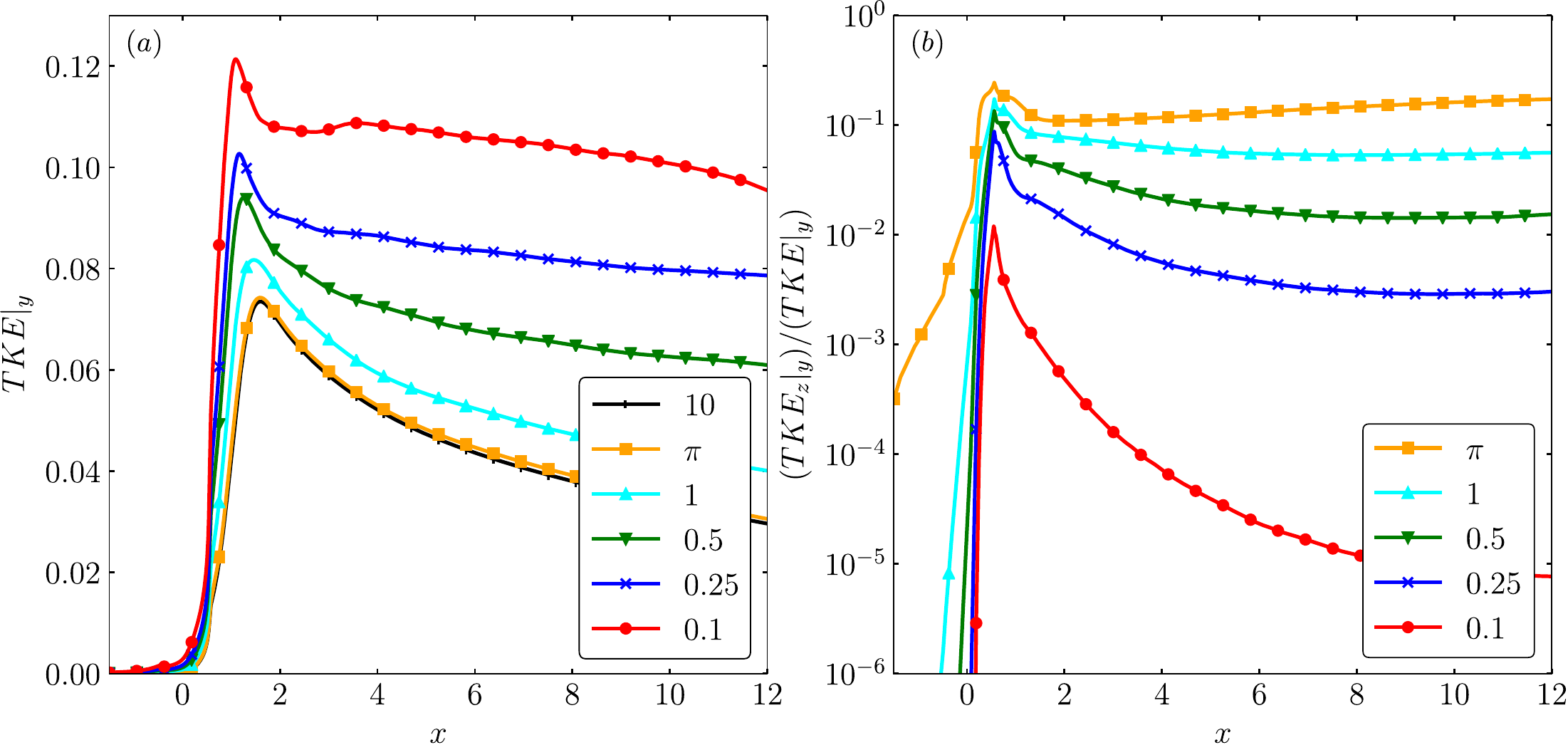}}
  \caption{TKE spatial plots.
The TKE is computed from the normal Reynolds stresses and averaged on the vertical direction: $(a)$ total TKE, $(b)$ ratio of the spanwise component $\overline{w'w'}$ to the total TKE.}
\label{fig:TKE}
\end{figure}

In summary, the transition to a 2-D wake is found at a certain point along the wake on all the cases with $L_z\le1$.
The combination of the large-scale rotation from the K\'{a}rm\'{a}n vortices plus the geometry constriction are mainly responsible for this phenomenon.
The main physical mechanism differing among the compared cases is the ability of the flow to develop Mode B-like 3-D structures in the wake as a result of a sufficiently long span.
The current $Re$ regime is characterised by a transition to turbulent flow at the shear layer (i.e. the TrSL2 regime) as reported in \cite{Bloor1964} and \cite{Kourta1987}.
We argue that when the span is too short for the Mode B instability to develop and thereby sustain these 3-D turbulent structures, the stratification effect of the K\'{a}rm\'{a}n vortices leads to the more coherent and energised wake seen in \fref{fig:velocity_spectras} d,e,f.

Next, the TKE along the $x$ direction, the Lumley triangle of turbulence and the ratio between the vortex-stretching and vortex-advecting terms are examined to further support the observed phenomena.
The TKE is defined as
\begin{equation}
TKE = \frac{1}{2}\left(\overline{u'u'}+\overline{v'v'}+\overline{w'w'}\right),
\end{equation}
where $\overline{\cdot}$ denotes a time average plus a spanwise average and the superscript $(\cdot)'$ denotes a fluctuating quantity such as $a'=a-\overline{a}$.
The six components of the Reynolds-stress tensor $\overline{u_i' u_j'}$ have been computed using the following relation
\begin{equation}
\overline{a'b'} = \overline{ab} - \overline{a}\overline{b}.
\end{equation}

\fref{fig:TKE} shows the streamwise spatial distribution of the TKE averaged along the $y$ direction from $-L_y/2$ to $L_y/2$ (noted as $\cdot|_y$).
From a general point of view, it can be observed that the total TKE (\fref{fig:TKE}a) peaks right after the recirculation region noting that the latter increases slightly with the span.
Also, the total TKE increases as the span is reduced because of the 2-D vortex-merging processes that generate larger and more energised vortical structures.
The contribution of the spanwise normal stress to the total TKE increases with the span as shown in \fref{fig:TKE}b.
Also, a decay of $\overline{w'w'}$ right after being generated from the cylinder wall can be noted and it becomes faster as the span is constricted.

\begin{figure}
 \vspace*{0.3cm}
  \centerline{\includegraphics[width=0.5\textwidth]{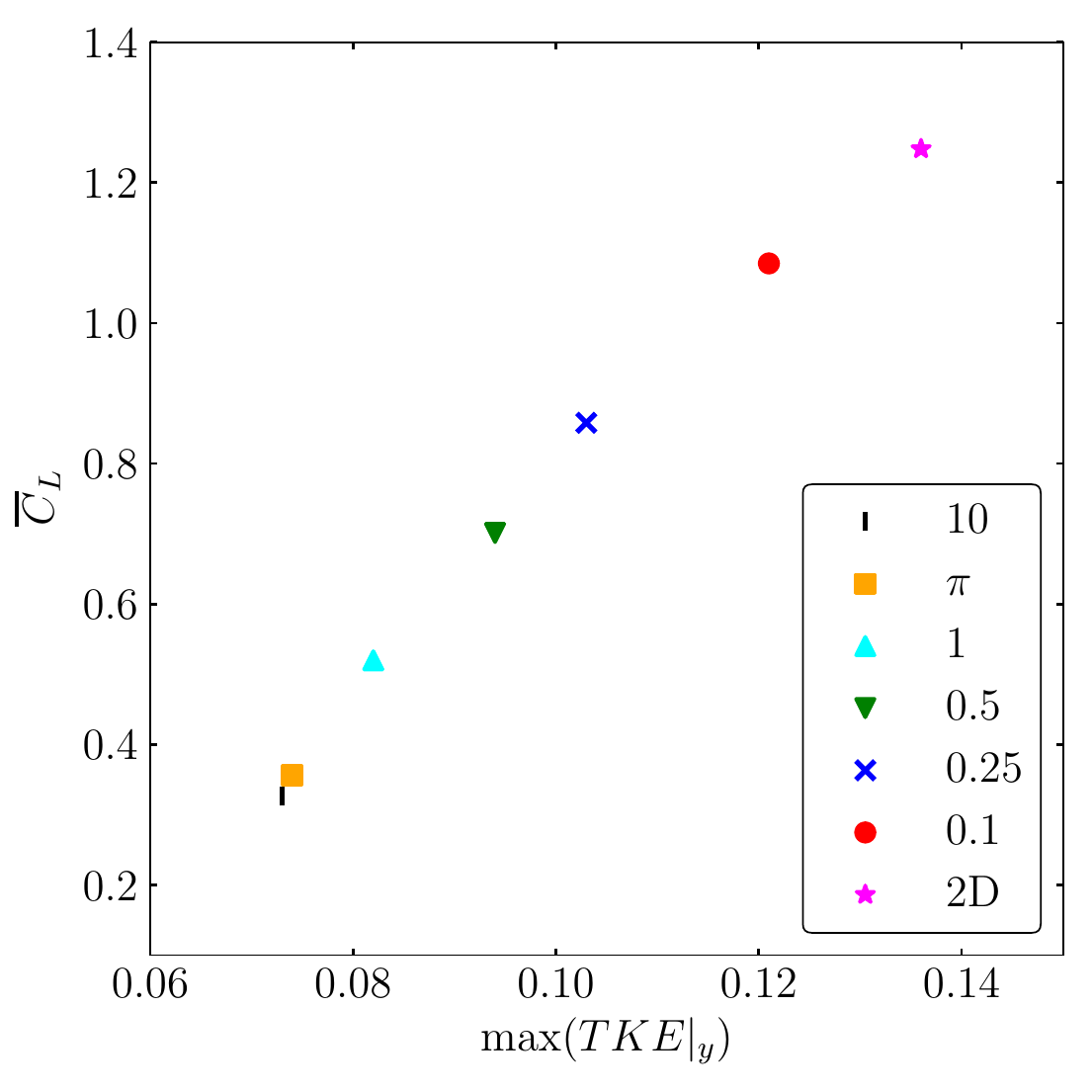}}
  \caption{Effect of the span constriction on the TKE and the $\overline{C}_L$.
The latter is calculated as $\overline{C}_L=|2F_y/(\rho U_\infty^2 D L_z)|_{\mathrm{rms}}$, where $F_y$ is the vertical lift force, $U_\infty$ is the free-stream velocity, and $\rho$ is the constant fluid density.}
\label{fig:TKE_CL}
\end{figure}

The effect of the span on the TKE compared to the lift coefficient root mean square (r.m.s.) value $(\overline{C}_L)$ displays a quasi-linear relation as shown in \fref{fig:TKE_CL}.
Note that the drop on both values is more sensitive to the span constriction at the range where both 2-D and 3-D turbulence dynamics coexist, i.e. $0.25\le L_z\le1$.
On the other hand, a very small change of both values can be appreciated when 3-D turbulence fully dominates the wake, i.e. $L_z\ge\pi$.
Furthermore, as shown in \fref{fig:TKE}a, highly constricted domains yield large values of TKE because of the energised 2-D large-scale vortical structures present at the wake.
Even when small-scale 3-D structures are present in the close wake region and coexist with the large 2-D structures, the values of both the TKE and the $\overline{C}_L$ spike.

\begin{figure}
	\vspace*{0.3cm}
  \centerline{\includegraphics[width=0.9\textwidth]{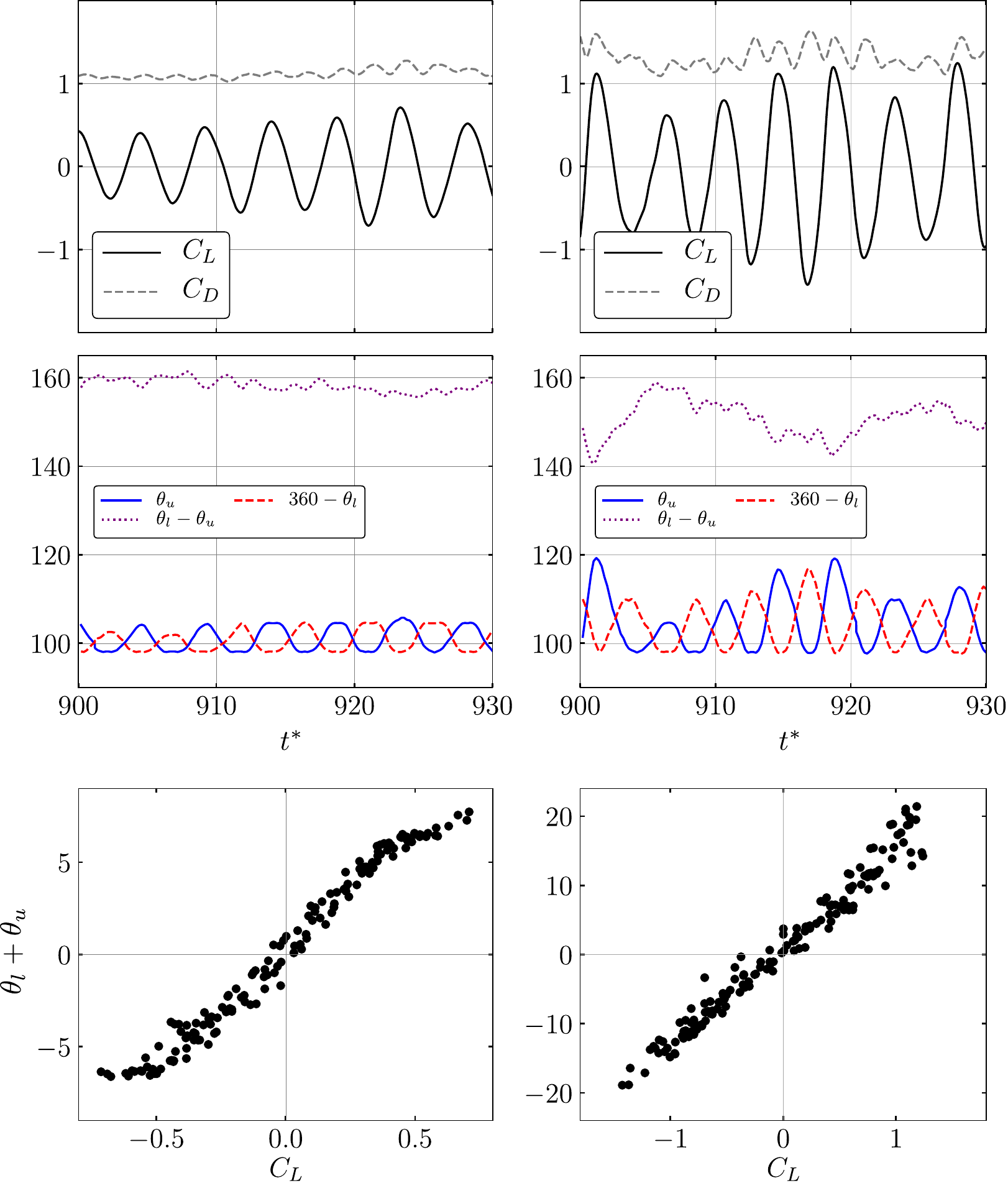}}
  \caption{Top: $C_L$ and $C_D$ temporal signals.
The latter is calculated as $C_D=2F_x/(\rho U_\infty^2 D L_z)$ and the former as detailed in \fref{fig:TKE_CL} (without the r.m.s norm).
Middle: upper $\theta_u$ and lower $\theta_l$ separation angles temporal signal.
The separation angle is calculated from the front stagnation point of the cylinder to the free shear layer separation point (vorticity practically 0 at the wall).
Bottom: correlation between the lift coefficient and the difference between $\theta_u$ and $360-\theta_l$.
Left: $L_z=\pi$.
Right: $L_z=0.5$.}
\label{fig:separation}
\end{figure}

In \fref{fig:separation}, the temporal signals of the lift and drag coefficients as well as the upper and lower separation angles of the free-shear layer are displayed for the $L_z=0.5$ and the $L_z=\pi$ cases.
The oscillation of the separation points increases as the span is constricted inducing larger forces on the cylinder.
A very high correlation of the upper and lower separation points angle with the lift coefficient is shown as well.
Again, the lack of the Mode B instability on the constricted cases yields more coherent 2-D vortical structures in the near wake region.
Therefore, the small-scale 3-D structures which normally dissipate most of the kinetic energy in 3-D turbulence are not present leading to high-intensity vortices.
This can be quantified by the enstrophy ($Z$) of the spanwise-averaged spanwise vorticity in the near wake region, $\Omega$ ($x \in [0.55, 2.1], y \in [-0.8, 0.8]$), which is time averaged for the same temporal length as the body forces signals as follows
\begin{equation}
Z = \int_\Omega\frac{1}{t_2-t_1}\int^{t_2}_{t_1}\frac{1}{L_z}\int^{L_z}_{0} \omega_z^{2} \,\,\mathrm{d}z\,\mathrm{d}t\,\mathrm{d}\Omega.
\label{eq:enstrophy}
\end{equation}
It can be observed in \tref{tab:enstrophy} that the enstrophy increases as the span is reduced, which can be understood as an increase of the rotational energy of the flow.

\begin{table}
  \begin{center}
\def~{\hphantom{0}}
  \begin{tabular}{lrrrrr}
  \toprule
       $L_z$ & $\pi$ & $1$  & $0.5$ & $0.25$ & $0.1$ \\
       \midrule
       $Z/Z_\pi$ & 1 & 1.32 & 1.71  & 2.07   & 2.47  \\
      \bottomrule
  \end{tabular}
  \caption{Near body enstrophy as defined in equation \ref{eq:enstrophy} for the different span cases.
$Z_\pi$ is the enstrophy of the $L_z=\pi$ case.}
  \label{tab:enstrophy}
  \end{center}
\end{table}

The highly energised coherent vortices forming at the near wake region for the constricted cases induce a larger convective force on the free-shear layer.
This translates to the large oscillations observed in \fref{fig:separation} and, ultimately, to the forces induced on the cylinder.
With this, it can be argued that the coherent 2-D structures have a greater impact on the forces induced to the cylinder than the 3-D small-scale structures when both are present.

\cite{Lumley1977} proposed the Lumley triangle of turbulence which provides a way to classify the anisotropic state of turbulence.
The anisotropy property of the Reynolds-stress tensor can be extracted with
\begin{equation}
b_{ij} = \frac{\overline{u_i' u_j'}}{\overline{u_k' u_k'}} - \frac{1}{3}\delta_{ij},
\end{equation}
where $\delta_{ij}$ is the Kronecker delta tensor and $b_{ij}$ is the anisotropic Reynolds-stress tensor (which evidently vanishes for isotropic turbulence).
This dimensionless and traceless tensor has two non-zero invariants, $II=-b_{ij}b_{ji}/2$ and $III=b_{ij}b_{jk}b_{ki}/3$.
These invariants are often rewritten as $\eta^2=-II/3$ and $\xi^3=III/2$ to better appreciate the nonlinear behaviour of the trajectories of return to isotropy of homogeneous turbulence \citep{Choi2001}.
It has been shown that all the possible turbulence states are mapped within the triangle \citep{Lumley1977, Lumley1979}.

The different states of turbulence are classified in the triangle as follows: the top right elbow indicates a one-dimensional (or one component) state with a single non-zero eigenvalue (the eigenvalues can be understood in physical terms as the normal stresses in the principal axes of the anisotropic Reynolds-stress tensor).
The top left elbow indicates a 2-D isotropic state where one eigenvalue vanishes and the two remaining are equal.
The top curve connecting the elbows indicates a 2-D turbulence state where one eigenvalue vanishes and the addition of the two remaining eigenvalues is constant.
The left and right straight lines correspond to a negative or positive $\left( \xi \right)$ axisymmetric state since one eigenvalue is smaller than the other two (which are equal) or greater than the other two (which are equal) respectively.
Finally, the $(0,0)$ point indicates 3-D isotropic turbulence since all of the anisotropic tensor eigenvalues vanish.

\begin{figure}
	\vspace*{0.3cm}
  \centerline{\includegraphics[width=1\textwidth]{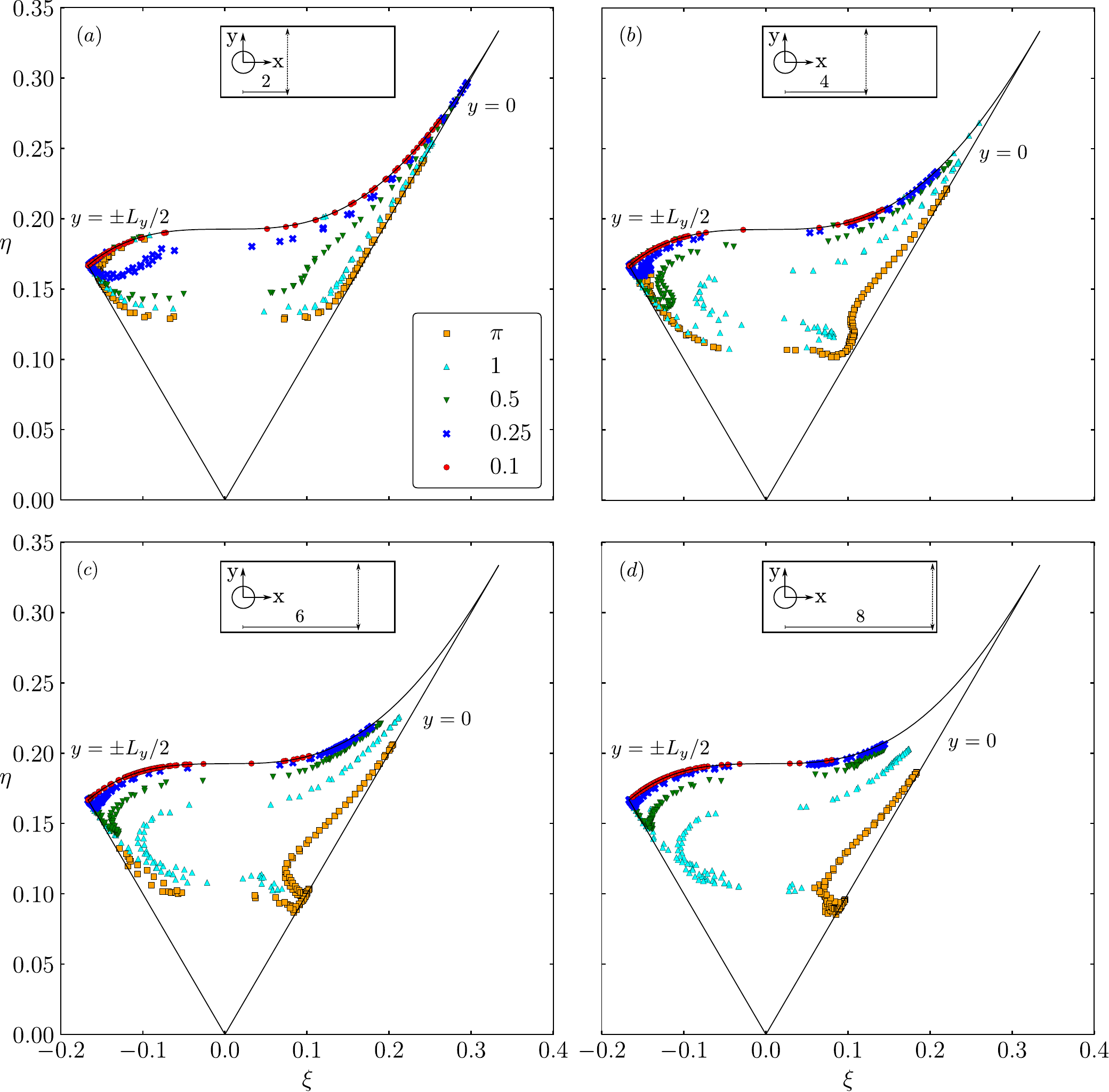}}
  \caption{Lumley triangle constructed across the wake width at: $(a)$ $x=2$, $(b)$ $x=4$, $(c)$ $x=6$, $(d)$ $x=8$.}
\label{fig:lumleys_triangle}
\end{figure}

\fref{fig:lumleys_triangle} displays the Lumley triangle constructed across the wake width at different $x$ locations.
Hence, every point in the triangle corresponds to a point in the domain and the collection of points for each case represents the collection of vertically aligned points in the domain at different $x$ locations on the wake.
It can be appreciated that all of the points are located within the triangle, therefore all of the computed Reynolds stresses are realisable (i.e. have positive and real eigenvalues).
In general, most of the cases transition from one state to another state of turbulence across the wake width.
Only for the almost 2-D case $\left(L_z=0.1\right)$, the state of turbulence is always 2-D (or two-component), since $\overline{w'w'}$ is effectively negligible everywhere.

As the triangle is constructed further downstream on the wake, the trajectories (or collection of points) of the $L_z=0.25$ and $L_z=0.5$ cases move closer to the 2-D turbulence state location (upper curve).
This emphasises once again the wake two-dimensionalisation caused by the large-scale vortical structures on cases with critical span which contain a cross-over of 2-D and 3-D turbulence, as shown in previous results.
In contrast, the $L_z=1$ and $L_z=\pi$ cases remain approximately at the same $\eta$ region showing that the two-dimensionalisation is not as effective.

Additionally, the trajectories present a negative axisymmetric almost two-component state for the locations far from the wake centreline, i.e. the location close to $y=\pm L_y/2$, since one of the normal turbulent stresses $(\overline{w'w'})$ is smaller than the others.
On the centreline, $\overline{v'v'}$ is larger than the other stresses causing a shift to the positive axisymmetric state.

A comparison of the different cases at the same $x$ location also shows a more noticeable 2-D turbulence state as the span is constricted.
This difference is less noticeable close to the cylinder (\fref{fig:lumleys_triangle}a), where even the $L_z=0.5$ case trajectory resembles the $L_z =\pi$ case.
Again, this evidences that 3-D turbulence is present close to the wall even in considerably constricted cases.

Consider now the vorticity transport equation (VTE) which can be written as
\begin{equation}
\partial_{t}\boldsymbol\omega+\pars{\vect{u}\cdot\nabla} \boldsymbol\omega=\pars{\boldsymbol\omega\cdot\nabla}\vect{u}+Re^{-1}\nabla^2\boldsymbol\omega,
\end{equation}
where $\boldsymbol\omega\left(\vect{x}, t\right) = \left(\omega_x, \omega_y, \omega_z\right)$ is the vorticity vector field defined as $\boldsymbol\omega=\nabla\times\vect{u}$.
The vortex-stretching term, $\pars{\boldsymbol\omega\cdot\nabla}\vect{u}$, is often pointed to as the term responsible for the direct energy cascade of the TKE.
The stretching of a vortex tube causes a reduction on its diameter while increasing the rotation speed of the vortex by conservation of angular momentum.
This term vanishes in the 2-D formulation of the VTE since the stretch of the vortex tube is perpendicular to the plane of rotation.
From this mathematical and physical difference, different turbulence dynamics is captured in 2-D or 3-D computations and, therefore, 3-D turbulence can be directly linked to this term.

\begin{figure}
	\vspace*{0.3cm}
  \centerline{\includegraphics[width=1\textwidth]{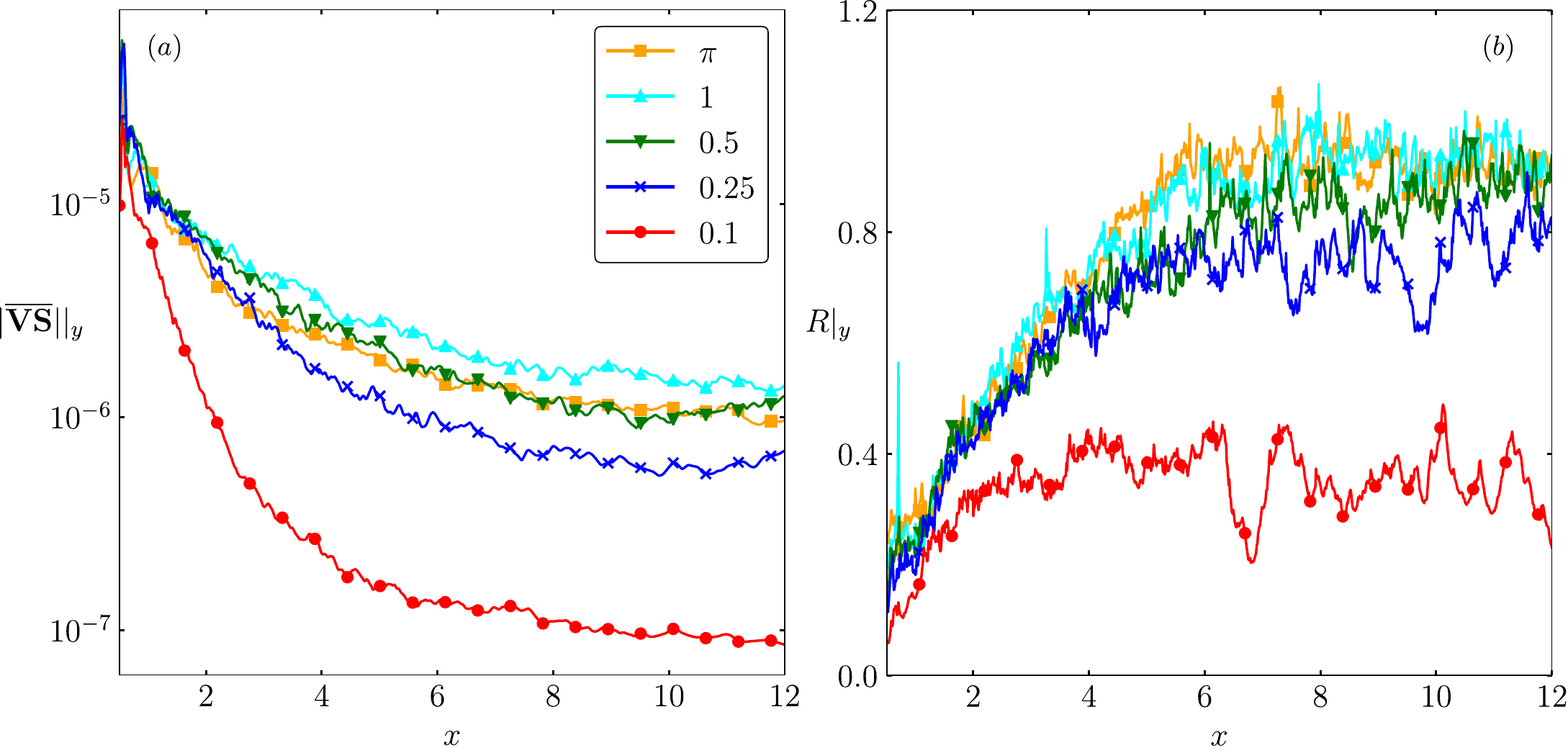}}
  \caption{$(a)$ Modulus of the mean vortex-stretching term averaged along the vertical direction.
$(b)$ Ratio of the modulus of the mean vortex-stretching term to the modulus of the mean vortex-advecting term averaged along the vertical direction.}
\label{fig:vortex_stretching}
\end{figure}

\fref{fig:vortex_stretching} displays the modulus of the mean vortex-stretching term and its ratio $R$ to the mean vortex-advecting term.
These quantities are averaged in the vertical direction.
On \fref{fig:vortex_stretching}a, it can be observed that the vortex-stretching term decays faster while moving downstream from the cylinder for cases with shorter span.
The vortex-stretching decay demonstrates again the two-dimensionalisation of the flow by the large-scale vortices.
The ratio $R$ in \fref{fig:vortex_stretching}b shows that the vortex-advecting term decays faster (because of the wake momentum deficit) than the vortex-stretching term along the streamwise direction up to $x=6$ where the ratio is kept constant.
It can also be observed that the vortex-stretching term becomes as important as the vortex-advecting term with increasing span.

\section{Conclusion}\label{Conclusion}

The span effect on the turbulence dynamics of a flow past a circular cylinder at $Re=10^4$ has been investigated using spectra  and two-point correlations at different locations in the domain, the TKE along the wake, the separation points, the Lumley triangle of turbulence and the mean vortex-stretching term of the VTE.
It has been shown that 3-D turbulence is present even for highly constricted cases (for example $L_z=0.25$) which is generated by the cylinder wall (\fref{fig:velocity_spectras}a).
However, the small-scale structures rapidly get two-dimensionalised by the large-scale K\'{a}rm\'{a}n vortices when the span is 50\% of the diameter or less.
This is linked to the Mode B instability wavelength being longer than the periodic span, in agreement with \cite{Bao2016}.
Since the Mode B instability helps sustaining the turbulent structures advected from the shear layer, the lack of it prevents large-scale 3-D structures to be created and less dissipative structures can be sustained.
In this scenario, 2-D turbulence takes over and dominates the wake dynamics creating larger, stronger and more coherent vortices.
Ultimately, the coherent and energised vortical structures induce a larger convective force on the free-shear layer.
This translates to larger oscillations and, finally, higher forces on the cylinder.

The flow turbulence transition from 3-D to 2-D caused by a geometry constriction found in this work is in agreement with the physical mechanisms described in the obstacle-free turbulence work of \cite{Smith1996}, \cite{Celani2010} and \cite{Biancofiore2014}.
In the present study with solid boundaries, the main difference is found in the presence of small-scale 3-D turbulence even in highly constricted geometries $\left(L_z=0.25\right)$ which leads to a coexistence of 2-D and 3-D turbulence close to the cylinder wall.
This is observed not only in the $L_z=0.25$ case, but also for the $L_z=0.5$ and $L_z=1$ cases as shown in \fref{fig:velocity_spectras}b and \fref{fig:velocity_spectras}c respectively.
Note that the cross-over between 2-D and 3-D turbulence dynamics arises at different points in the spatial domain depending on the span length.
The shorter the span, the closer to the cylinder it takes place evidencing that the wake two-dimensionalisation transitions at different locations in function of the domain geometric anisotropy.

On the other hand, a very rapid two-dimensionalisation is found in the present cases because of the natural large-scale rotation motion of the K\'{a}rm\'{a}n vortices.
A large-scale rotation as a mechanism of two-dimensionalisation has been also found in other works such as \cite{Smith1996} and \cite{Xia2011}.
These two mechanisms combined yield to a rapid transition from the 3-D to 2-D turbulence dynamics when the span is shorter than the Mode B instability wavelength of the wake.
% ---------------------------------------------------------------- 

\chapter{The spanwise-averaged Navier--Stokes equations}

\vspace{0.25cm}
Simulations of turbulent fluid flow around long cylindrical structures are computationally expensive because of the vast range of length scales, requiring simplifications such as dimensional reduction.
Current dimensionality reduction techniques such as strip-theory and depth-averaged methods do not take into account the natural flow dissipation mechanism inherent in the small-scale 3-D vortical structures.
In this chapter, a novel flow decomposition based on a local spanwise average of the flow is proposed, yielding the spanwise-averaged Navier--Stokes (SANS) equations.
The SANS equations include closure terms accounting for the 3-D effects otherwise not considered in 2-D formulations.
The SSR tensor modifies the flow dynamics from standard 2-D Navier--Stokes to spanwise-averaged dynamics.
It is shown that the SANS perfect closure recovers the unsteady spanwise-averaged solution of the Taylor--Green vortex and flow past a circular cylinder at $Re=10^4$.
Finally, it is demonstrated that eddy-viscosity models (EVMs) are not suited for the prediction of the spanwise stresses, motivating the need of a different closure as proposed in the following chapter.

\section{Introduction and literature review}

The vast range of spatial and temporal scales inherent in turbulence renders direct numerical simulations of most engineering and environmental flows impossible.
This is because the required resolution is governed by the balance of the flow convective and viscous forces, quantified by the Reynolds number.
For illustration, turbulent engineering or atmospheric flows have $Re\smallgtrsim 10^6-10^{10}$ and, as reviewed in \sref{sec:turbulence_cost}, directly simulating homogeneous isotropic turbulence in an $L^3$ box for one convection cycle would require $O(Re^3)$ operations \citep[p. 346]{Coleman2010,Pope2000}.
To avoid resolving the smallest time and length scales, turbulence models such as RANS, LES, and others have been developed to account for the unresolved scales of motion on coarse discretisations.

However, in many cases the scale of the flow is so large that turbulence models do not sufficiently reduce the computational cost and further simplifications are required, such as reducing the flow from three spatial dimensions down to two.
In the case of very long slender bodies, strip theory can be used to place 2-D strips along the structure span, reducing cost by orders of magnitude.
In offshore engineering, strip-theory methods have been used to simulate the vortex-induced vibrations of marine risers \citep{Herfjord1999}. \cite{Willden2001} considered a marine riser at a low Reynolds regime where the strips were linked by an inviscid unsteady 3-D vortex lattice.
Further investigations on marine risers multi-modal VIV phenomena were performed by the previous authors considering a more realistic Reynolds number ($\mathcal{O}(10^5)$) and including a LES turbulence model for the 2-D strips \citep{Willden2004}.
Other authors such as \cite{Yamamoto2004, Meneghini2004, Sun2012} employed a strip-theory discrete-vortex method to study the VIV phenomena on marine risers with $L_z/D=400$ and $Re=10^5$, modelling the flow viscous effects through an exponentially decaying function.
Other slender-body problems tackled with strip theory are found in naval engineering to simulate ship hydrodynamics \citep{Newman1979}, or in aerospace engineering for wing fluttering prediction \citep{Yates1966}.
A different dimensionality reduction technique is the depth-averaged method, typically used in the simulation of shallow open-channel flow such as rivers \citep{Molls1995}, where the dimensions of the problem are simplified by averaging along the shear direction of the flow.

The main shortcoming of the dimensionality reduction methods above is that the fluid dynamics of 2-D systems are intrinsically different from their 3-D counterparts.
In particular, 3-D turbulence cannot develop when the spanwise dimension is neglected.
Instead, 2-D turbulence promotes coherent and high-intensity vortices via vortex-thinning and vortex-merging events \citep{Xiao2009} ultimately inducing larger forces to the cylinder \citep{Mittal1995, Norberg2003}.
Because of this, these methods require additional dissipation mechanisms, often in the form of LES Smagorinsky models.
These turbulence models applied to 2-D systems decrease the vortices intensity thus inducing weaker forces on solid structures which can resemble those found in 3-D cases.
However, the dissipation mechanism is arbitrary and we show that it cannot be used to simulate 3-D turbulence dynamics in a 2-D simulation.

To address this issue, Bao et al. \citep{Bao2016, Bao2019} proposed a \textit{thick} strip-theory method, as hinted in \cite{Herfjord1999}.
Providing spanwise thickness to the strips allows the flow to locally develop 3-D turbulence, thus recovering the natural dissipation mechanism.
However, the minimum strip thickness at which 3-D turbulence dominates the flow is in the order of the structure diameter \citep{Bao2016,Font2019}, as discussed in \cref{chapter:jfm2019}. Hence, the inclusion of 3-D effects comes at the expense of increased computational cost, which can limit the simulation of realistic spans or Reynolds regimes. 

\begin{figure}[t]
\centering
\includegraphics[width=0.9\linewidth]{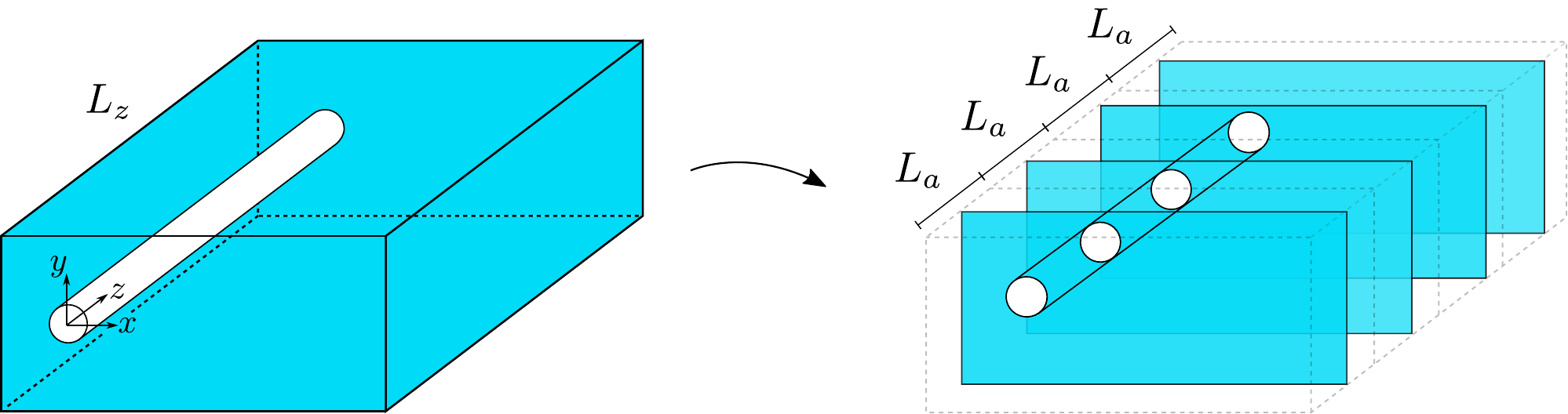}
\caption{
Sketch of the spanwise-averaged strip-theory method.
The original $L_z$ span is decomposed into multiple $L_a$ spanwise segments for which the SANS equations are simultaneously solved, hence reducing the computational domain (coloured region).
Solving the SANS equations for every spanwise-averaged strip allows to recover the local integrated forces induced to the cylinder.
The 2-D strips can be linked with a structural dynamic model similarly to \cite{Bao2016, Bao2019}.}
\label{fig:strips}
\end{figure}

In this work, we propose a novel flow decomposition which separates the spanwise-averaged and the spanwise-fluctuating parts of the flow.
Instead of directly resolving the 3-D spatial scales, these are taken into account with an additional spanwise-fluctuating term inserted into the 2-D governing equations.
We call the resulting system of equations the spanwise-averaged Navier--Stokes equations (SANS).
In contrast to RANS or LES, the closure terms account for the 3-D turbulence effects enabling the simulation of turbulent wake flows in two dimensions.
Similarities can be drawn with depth-averaged methods.
Uniquely, the SANS equations do not assume hydrostatic pressure or approximately uniform velocity, instead being developed specifically to deal with fully 3-D wake turbulence.
Unlike RANS simulations, which can only (at best) reproduce the time-averaged forces, the SANS system is capable of reproducing the instantaneous spanwise-integrated forces on a locally cylindrical body. 
As such, a SANS strip theory is ideal for predicting the 3-D fluid-structure interactions with widely spaced 2-D simulation planes (see \fref{fig:strips}).

\section{The spanwise-averaged Navier--Stokes equations}\label{sec:SANS_eqs}

When a flow is essentially homogeneous in one direction, this can be exploited to simplify modelling in a number of ways;
the turbulent flow statistics are invariant to translation along the homogeneous direction, there are no mean flow gradients in the homogeneous direction, and the component of mean velocity in the homogeneous direction can always be set 0 in an appropriate reference frame \citep[p. 76]{Pope2000}.
Therefore, a single homogeneous direction can be used to reduce a 3-D flow to a 2-D system of equations, potentially reducing the cost of simulation by orders of magnitude.

\subsection{Mathematical description} \label{sec:sans_math}

Starting from the conservation form of the non-dimensional 3-D Navier--Stokes momentum equations
\begin{equation}
\partial_t\vect{u}+\nabla \cdot \pars{\vect{u}\otimes\vect{u}}=-\nabla p+Re^{-1}\nabla^2\vect{u},\label{eq:N-S}
\end{equation}
the velocity vector field $\vect{u}\pars{\vect{x},t}=\pars{u,v,w}$ and the pressure scalar field $p\pars{\vect{x},t}$ are decomposed into a spanwise-averaged and a spanwise-fluctuating part as follows
\begin{equation}
q(\vect{x},t) = \avg{q}(x,y,t)+q\p(x,y,z,t),
\end{equation}
where $\avg{\cdot}$ and $(\cdot)\p$ denote the averaged and the fluctuating parts of the decomposition, respectively.
The average, also noted as $\avg{q}=Q$, can be defined on a $z\in\left[a,b\right]$ interval as
\begin{equation}
Q=\avg{q}\pars{x,y,t} = \frac{1}{b-a}\int^b_a q\pars{x,y,z,t}\,\mathrm{d}z.
\label{eq:spanwise_average}
\end{equation}
Such decomposition applied to the $\vect{e}_x$ component of \eref{eq:N-S} yields
\begin{gather}
\partial_t\pars{U+u\p}+\partial_x\pars{UU+2Uu\p+u\p u\p}+\partial_y\pars{VU+Vu\p+v\p U+v\p u\p}+\nonumber\\
+\partial_z\pars{WU+Wu\p+w\p U+w\p u\p}=\nonumber\\
=-\partial_x\pars{P+p\p}+Re^{-1}\left[\partial_{xx}\pars{U+u\p}+\partial_{yy}\pars{U+u\p}+\partial_{zz}\pars{U+u\p}\right].
\label{eq:N-S_decomp}
\end{gather}

Next, the following relations arising from \eref{eq:spanwise_average} are considered,
\begin{alignat}{3}
&\,\,q=Q+q\p,                      &&\qquad \,\,s=S+s\p,                       &&\qquad \avg{q+s}=Q+S,              \label{eq:rule1}\\
&\avg{Q}=Q,                  &&\qquad \avg{q\p}=0,                             &&\qquad \avg{Qs\p}=Q\avg{s\p}=0,      \label{eq:rule2} \\
&\avg{q\p q\p} \neq0,                    &&\qquad \avg{\partial_z q}\neq\partial_zQ, &&\qquad \,\,\partial_zQ=0,                \\
&\avg{\partial_t q} =\partial_t Q, &&\qquad \avg{\partial_x q} =\partial_x Q,  &&\qquad \avg{\partial_y q} =\partial_y Q.\label{eq:rule3}
\end{alignat}
Note that the spanwise average does not commute with the spanwise derivative, i.e. $\avg{\partial_z q}=\avg{\partial_z q\p}\neq\partial_zQ$, since the average and the derivative operate over the same variable $z$.
However, terms such as $\avg{\partial_z^n q}$ vanish under infinitely-long (periodic) averaging intervals, as discussed in \sref{sec:spanwise-periodic_interval}.

Applying the averaging operator to the whole \eref{eq:N-S_decomp} and considering the relations from \eref{eq:rule1} to \eref{eq:rule1} yields the governing equation of $U$,
\begin{gather}
\partial_tU+\partial_x\pars{UU+\avg{u\p u\p}}+\partial_y\pars{VU+\avg{v\p u\p}}+\avg{\partial_z\pars{Wu\p+w\p U+w\p u\p}}=\nonumber \\
=-\partial_xP+Re^{-1}\pars{\partial_{xx}U+\partial_{yy}U+\avg{\partial_{zz} u\p}}, \\
\partial_tU+\partial_x\pars{UU}+\partial_y\pars{VU}+U\avg{\partial_z w\p}=-\partial_xP+Re^{-1}\pars{\partial_{xx}U+\partial_{yy}U}-\nonumber\\
-\partial_x\avg{u\p u\p}-\partial_y\avg{v\p u\p}-W\avg{\partial_z u\p}-\avg{\partial_z\pars{w\p u\p}}+Re^{-1}\avg{\partial_{zz} u\p}. \label{eq:N-S_decom_avg}
\end{gather}

In order to rewrite \eref{eq:N-S_decom_avg} in its non-conservation form, the continuity equation is considered
\begin{equation}
\nabla\cdot\vect{u}=0, \label{eq:cont}
\end{equation}
and applying the same decomposing and averaging procedure yields
\begin{equation}
\partial_xU+\partial_yV+\avg{\partial_z w\p}=0.\label{eq:cont_decomp_avg}
\end{equation}
By substituting \eref{eq:cont_decomp_avg} into \eref{eq:N-S_decom_avg}, the non-conservation form is obtained and the $U\avg{\partial_z w\p}$ term vanishes yielding
\begin{gather}
\partial_tU+U\partial_xU+V\partial_yU=-\partial_xP+Re^{-1}\pars{\partial_{xx}U+\partial_{yy}U}-\partial_x\avg{u\p u\p}-\partial_y\avg{v\p u\p}-\nonumber\\
-W\avg{\partial_z u\p}-\avg{\partial_z\pars{w\p u\p}}+Re^{-1}\avg{\partial_{zz} u\p}.\label{eq:z-avg_X}
\end{gather}

Analogously to the spanwise decomposition for the $\vect{e}_x$ component of the momentum equations, the same procedure can be applied to the $\vect{e}_y$ and $\vect{e}_z$ components which respectively leads to
\begin{gather}
\partial_tV+U\partial_xV+V\partial_yV=-\partial_yP+Re^{-1}\pars{\partial_{xx}V+\partial_{yy}V}-\partial_x\avg{u\p v\p}-\partial_y\avg{v\p v\p}-\nonumber\\
-W\avg{\partial_z v\p}-\avg{\partial_z\pars{w\p v\p}}+Re^{-1}\avg{\partial_{zz} v\p},\\
\partial_tW+U\partial_xW+V\partial_yW=-\avg{\partial_z p\p}+Re^{-1}\pars{\partial_{xx}W+\partial_{yy}W}-\partial_x\avg{u\p w\p}-\partial_y\avg{v\p w\p}-\nonumber\\
-W\avg{\partial_z w\p}-\avg{\partial_z\pars{w\p w\p}}+Re^{-1}\avg{\partial_{zz} w\p}.\label{eq:z-avg_Z}
\end{gather}
Note that Eqs. \ref{eq:z-avg_X} to \ref{eq:z-avg_Z} are the 2-D incompressible Navier--Stokes equations with additional spanwise-fluctuating closure terms (first line, also referred in future mentions as ``spanwise stresses'') and terms arising from the finite spanwise-averaging length (second line).

With this, the complete 2-D SANS momentum equations can be written in a more compact form as
\begin{gather}
\partial_t\vect{U}+\vect{U}\cdot\nabla\vect{U}=-\nabla P+Re^{-1}\nabla^2\vect{U}-\nabla\cdot\boldsymbol\tau_{ij}^R-\nonumber\\
-W\avg{\partial_z \vect{u}\p}-\avg{\partial_z\pars{w\p \vect{u}\p}}+Re^{-1}\avg{\partial_{zz}\vect{u}\p},\label{eq:sans_full}
\end{gather}
where $\vect{U}=\pars{U,V}$, and the spanwise-stress residual (SSR) tensor $\boldsymbol\tau_{ij}^R$ is defined as
\begin{equation}
\boldsymbol\tau_{ij}^R= \avg{\vect{u}\otimes\vect{u}}-\avg{\vect{u}}\otimes\avg{\vect{u}} =
  \begin{pmatrix}
    \avg{u\p u\p} & \avg{u\p v\p} \\
    \avg{u\p v\p} & \avg{v\p v\p} 
  \end{pmatrix}.
\end{equation}

For the sake of completeness, the analogous spanwise-averaged VTE derivation is included in \aref{sec:savte}, where the relation between the velocity-velocity and vorticity-velocity spanwise stresses is derived as well.

\section{Columnar vortex between two infinite parallel plates}

\begin{figure}[!t]
    \centering
    \includegraphics[width=0.45\textwidth]{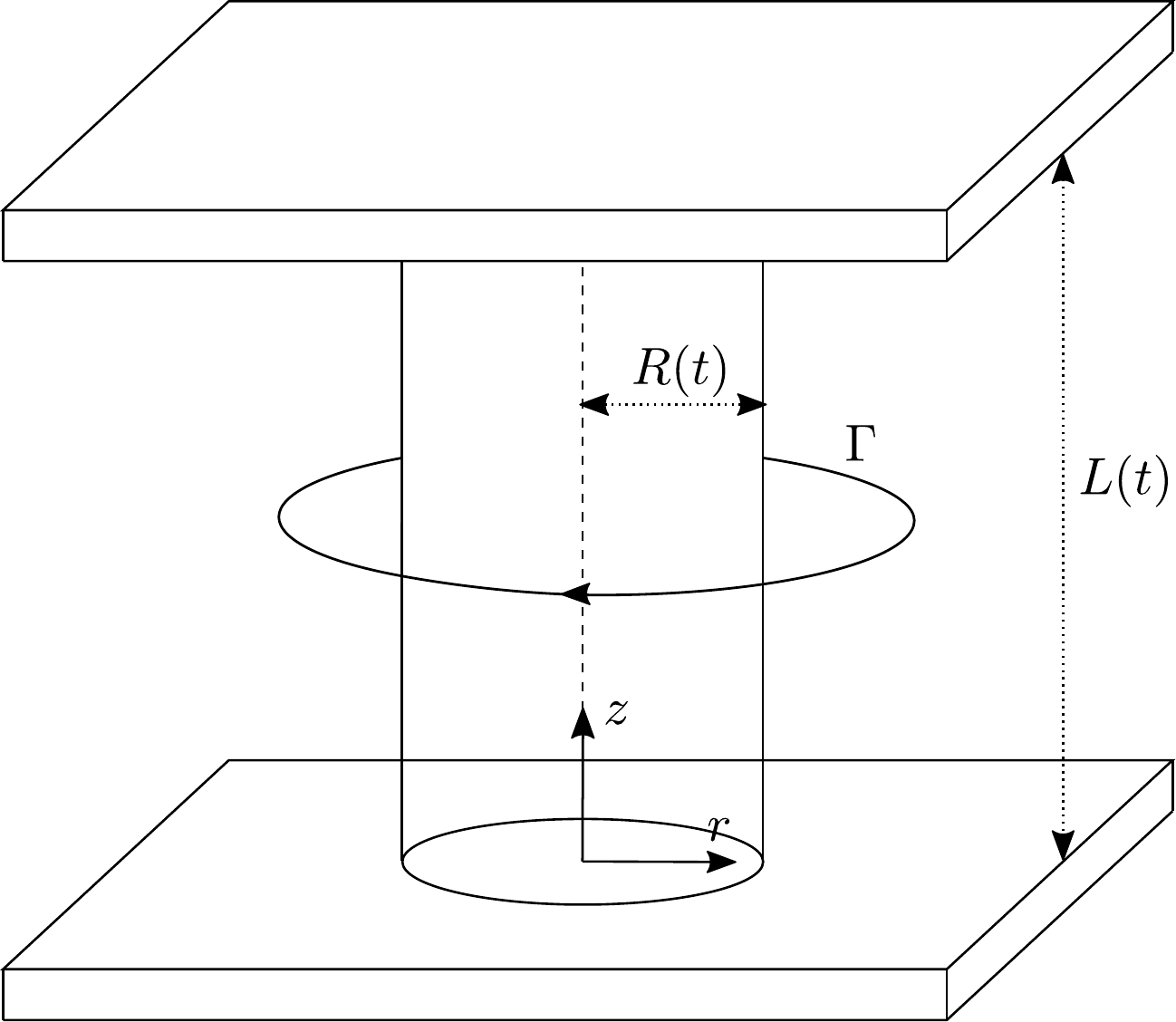}
    \caption{Columnar vortex of radius $R$ between two infinite parallel plates.
Adapted from \cite{Marshall2001}.}
    \label{fig:columnar_vortex}
\end{figure}

The SANS equations dynamics is explored for an analytical test case where a known solution exists.
For this, a columnar vortex between two infinite parallel plates is considered \citep{Marshall2001}.
Marshall's analytical test case shows the stretching of a vortex tube by the superposition of an external straining field (\fref{fig:columnar_vortex}).

In a cylindrical coordinate system, we consider unit vectors $\vect{e}_r,\vect{e}_\theta,\vect{e}_z$ and define the vector fields $\vect{u}=\left(u_r,u_\theta,u_z\right),\,\boldsymbol\omega=\left(\omega_r,\omega_\theta,\omega_z\right)$ in a $\vect{x}=\left(r,\theta,z\right)$ cylindrical space. In this coordinate system, a columnar vortex is defined as
\begin{equation}
{\boldsymbol\omega_v}\pars{\vect{x},t}=\begin{cases}
\begin{array}{c}
\omega\left(t\right)\vect{e}_3\\
0
\end{array} & \begin{array}{c}
r\leq R\left(t\right),\\
r>R\left(t\right),
\end{array}\end{cases}
\end{equation}
where uniform vorticity within the vortex core is considered for the $\vect{e}_3$ direction.
Also note that the subscript $(\cdot)_v$ refers to the columnar vortex field.
The velocity field associated to the columnar vortex is
\begin{equation}
{\vect{u}_v}\pars{\vect{x},t}=\left(0,v\left(r,t\right),0\right).
\end{equation}

Next, we demonstrate that the superposition of an axisymmetric straining field to the vortex tube increases the vorticity magnitude (for incompressible flow) as a result of the vortex-stretching mechanism.
Let us consider the following straining field
\begin{equation}
\vect{u}_s=\pars{-\frac{\gamma r}{2},0,\gamma z},\label{eq:strain_field}
\end{equation}
where $\gamma$ is the rate of strain $(dL/dt=\gamma L_0)$, and the subscript $(\cdot)_s$ refers to the straining field.
Note that the straining field is irrotational, $\nabla\times\vect{u}_s=0$, so no vorticity is added.
With this, the superposition of the columnar vortex and the straining field results into the following vorticity and velocity fields, respectively
\begin{gather}
\boldsymbol\omega=\boldsymbol\omega_v=\pars{0,0,\omega(t)}\label{eq:vorticity_field}\\
\vect{u}=\vect{u}_v+\vect{u}_s=\pars{-\frac{\gamma r}{2},v(r,t),\gamma z}.\label{eq:velocity_field}
\end{gather}

Substituting \eref{eq:vorticity_field} and \eref{eq:velocity_field} into the $\vect{e}_3$ component of the non-dimensional VTE, written in cylindrical coordinates as
\begin{gather}
\frac{\partial\omega_{z}}{\partial t}+u_r\frac{\partial \omega_{z}}{\partial r}+\frac{u_{\theta}}{r}\frac{\partial \omega_{z}}{\partial\theta}+u_z\frac{\partial \omega_{z}}{\partial z}=\omega_r\frac{\partial u_z}{\partial r}+\frac{\omega_{\theta}}{r}\frac{\partial u_z}{\partial\theta}+\omega_{z}\frac{\partial u_z}{\partial z}+\nonumber\\
+\frac{1}{Re}\left(\frac{\partial^{2}\omega_{z}}{\partial r^{2}}+\frac{1}{r}\frac{\partial\omega_{z}}{\partial r}+\frac{1}{r^{2}}\frac{\partial^{2}\omega_{z}}{\partial\theta^{2}}+\frac{\partial^{2}\omega_{z}}{\partial z^{2}}\right),\label{eq:vte_cyl_non_conserv}
\end{gather}
yields
\begin{gather}
\pd{\omega}{t}+0+0+0=0+0+\omega\frac{\partial}{\partial z}\pars{\gamma z}+0+...+0,\\
\dd{\omega}{t}=\gamma\omega.
\end{gather}
Therefore, it is shown that an increase in the straining field $\left(\gamma>0\right)$ induces an increase of the vorticity magnitude (vortex-stretching), and a decrease in the straining field $\left(\gamma<0\right)$ induces a decrease of the vorticity magnitude (vortex-compression).
Also note that the $\gamma\omega$ term arises from  $\omega_{z}\left(\partial u_{z}/\partial z\right)$, indeed part of the vortex-stretching term.

Next, the Marshall's vortex stretching analytical test case is considered using the spanwise-averaged VTE.
In cylindrical coordinates, the non-dimensional spanwise-averaged VTE can be written as (see derivation in \sref{sec:cyl_coords})
\begin{gather}
\pd{\Omega_z}{t}+U_r\pd{\Omega_z}{r}+\frac{U_\theta}{r}\pd{\Omega_z}{\theta}=\Omega_r\pd{U_z}{r}+\frac{\Omega_\theta}{r}\pd{U_z}{\theta}+\frac{1}{Re}\pars{\ddn{\Omega_z}{r}{2}+\frac{1}{r}\pd{\Omega_z}{r}+\frac{1}{r^2}\ddn{\Omega_z}{\theta}{2}}\nonumber\\
-\frac{1}{r}\bracs{\pd{}{r}\pars{\avg{r u_r\p\omega_z\p}-\avg{r \omega_r\p u_z\p}}+
\pd{}{\theta}\pars{\avg{u_\theta\p\omega_z\p}-\avg{\omega_\theta\p u_z\p}}}+\nonumber\\
+\Omega_z\avg{\pd{u_z\p}{z}}-U_z\avg{\pd{\omega_z\p}{z}}+\frac{1}{Re}\avg{\ddn{\omega_z\p}{z}{2}}.\label{eq:vte_cyl_avg}
\end{gather}

Applying the averaging procedure to the velocity field resulting from the superposition of the columnar vortex and the straining field (\eref{eq:velocity_field}) yields
\begin{gather}
\vect{u}=\left(-\frac{\gamma r}{2},0,\gamma z\right),\\
U_r =\frac{1}{b-a}\int^b_a -\frac{\gamma r}{2}\,dz=-\frac{\gamma r}{2},\quad u_r\p=0,\\
U_z =\frac{1}{b-a}\int^b_a \gamma z\,dz=\gamma\frac{a+b}{2},\quad u_z\p=u_z-U_z=\gamma\pars{z-\frac{a+b}{2}},\\
\vect{U}=\pars{-\frac{\gamma r}{2},0,\gamma\frac{a+b}{2}},
\end{gather}	
while the vorticity field stays as $\vect{\Omega}=\pars{0,0,\omega(t)}$.
Substituting the averaged and fluctuating parts of the velocity and the vorticity fields into \eref{eq:vte_cyl_avg} yields
\begin{gather}
\pd{\omega}{t}+0+0=0+0+...+\omega\frac{\partial}{\partial z}\pars{\gamma z}+0+...+0,\\
\dd{\omega}{t}=\gamma\omega.
\end{gather}
This shows that the vortex-stretching mechanism is still present in the spanwise-averaged VTE.
However, note that $\gamma\omega$ arises from $\Omega_z \avg{\partial_z u_z\p}$, a term that vanishes when an infinite (spanwise-periodic) averaging interval is considered, as explained next.

\section{Averaging over a spanwise-periodic interval} \label{sec:spanwise-periodic_interval}

The SANS equations can be greatly simplified when the averaging is performed over a periodic interval since
\begin{equation}
\avg{\partial_z q} = \avg{\partial_z q\p}  =\frac{1}{b-a}\int^b_a \frac{\partial q\p}{\partial z}\,dz=\frac{q\p\left(b\right)-q\p\left(a\right)}{b-a}=0
\label{eq:periodic_assumption}
\end{equation}
because $q\p(a)=q\p(b)$.
This yields a simplified version of SANS which can be written as
\begin{gather}
\partial_tU+U\partial_xU+V\partial_yU=-\partial_xP+Re^{-1}\pars{\partial_{xx}U+\partial_{yy}U}-\partial_x\avg{u\p u\p}-\partial_y\avg{v\p u\p},\\
\partial_tV+U\partial_xV+V\partial_yV=-\partial_yP+Re^{-1}\pars{\partial_{xx}V+\partial_{yy}V}-\partial_x\avg{u\p v\p}-\partial_y\avg{v\p v\p},\\
\partial_tW+U\partial_xW+V\partial_yW=Re^{-1}\pars{\partial_{xx}W+\partial_{yy}W}-\partial_x\avg{u\p w\p}-\partial_y\avg{v\p w\p}.
\end{gather}
Note that the spatial dependency on the spanwise direction is lost during the process (there are no $z$ derivatives present in the final form), and also the momentum equations for $U$ and $V$ are decoupled from $W$.
The momentum equation for $W$ becomes a convective-diffusive equation (with additional forcing terms) since the pressure term vanishes under the assumption of spanwise periodicity.
In a more compact form and dropping the $\vect{e}_z$ momentum equation, the simplified SANS equations are written as
\begin{equation}
\partial_t\vect{U}+\vect{U}\cdot\nabla\vect{U}=-\nabla P+Re^{-1}\nabla^2\vect{U}-\nabla\cdot\boldsymbol\tau_{ij}^R.\label{eq:sans_simple}
\end{equation}

Under this assumption, the vortex-stretching term appearing in \eref{eq:vte_cyl_avg} vanishes and the SANS equations would fail to capture the previous analytical test case mechanism.
On the other hand, the SANS equations are still valid in a $z$-periodic domain, and the only difference with a standard 2-D Navier--Stokes system resides in the SSR term, $\nabla\cdot\boldsymbol\tau_{ij}^R$. 
From here, the concept ``SANS'' assumes spanwise periodicity and refers to the simplified SANS equations (\eref{eq:sans_simple}).

\section{Closure methods}

\begin{figure}[t]
\centering
\includegraphics[width=0.42\linewidth]{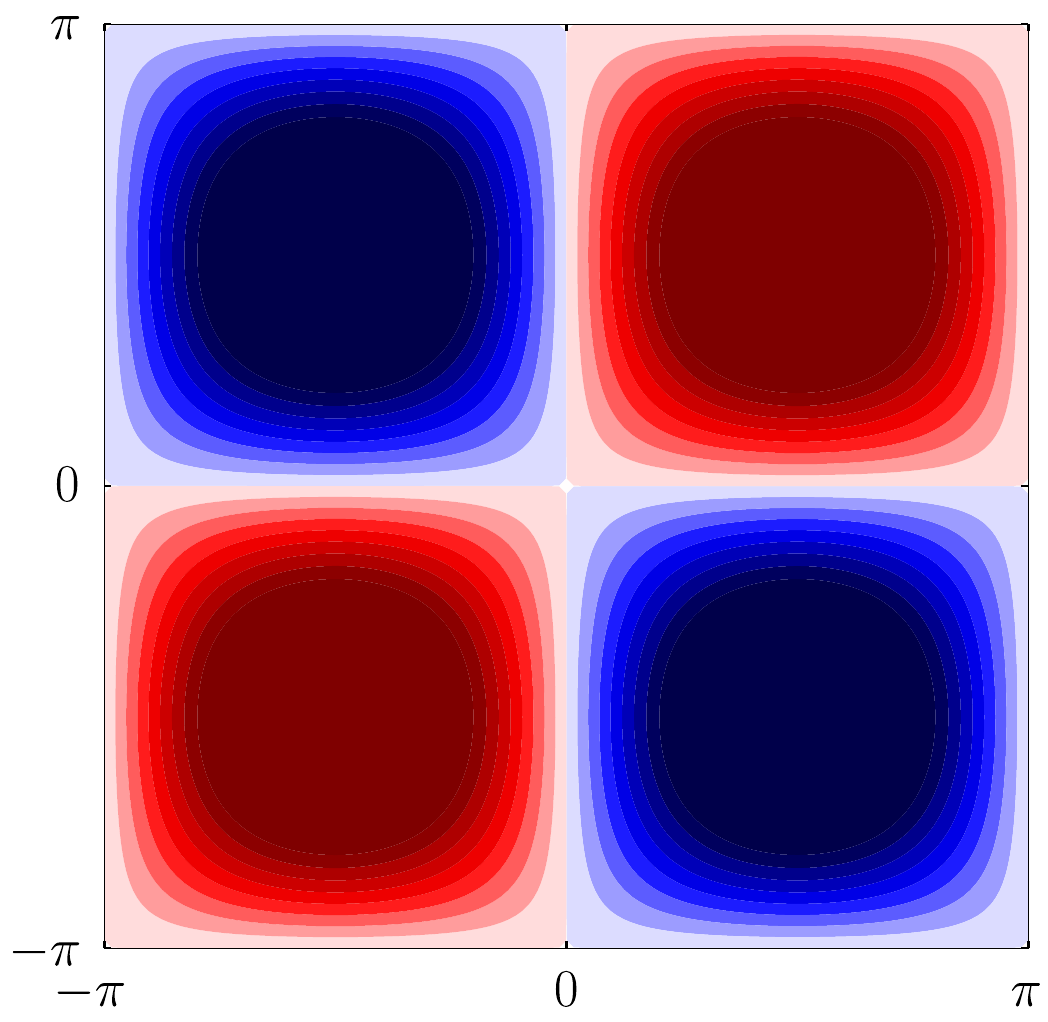}
\caption{Taylor--Green vortex initial $\omega_z$ vorticity field at $z=\pi/4$.}
\label{fig:t-g_0}
\end{figure}

The challenging part of the SANS equations resides in the closure term which drives the fluid dynamics from its natural 2-D evolution to a 3-D-like state.
The importance of the SSR term is illustrated using a fully turbulent 3-D Taylor--Green vortex flow.
The Taylor--Green vortex \citep{Taylor1937} is broadly used in the computational fluid dynamics literature for the validation of numerical methods and turbulence models.
It consists of an energy decaying flow defined in a triple periodic box $(L\times L\times L)$ with initial flow field $\vect{u}_0$
\begin{gather}
u_0(x,y,z) = U_0\sin(\kappa x)\cos(\kappa y)\cos(\kappa z), \nonumber \\
v_0(x,y,z) = -U_0\cos(\kappa x)\sin(\kappa y)\cos(\kappa z), \nonumber \\
w_0(x,y,z) = 0,
\end{gather}
where $\kappa=2\pi/L$, and we select $L=2\pi$ and $U_0=1$.
The initial flow field (\fref{fig:t-g_0}) breaks down into smaller structures because of the vortex-stretching mechanism, yielding a phase of global enstrophy growth.
Once most of the large-scale structure have been destroyed, the inertial subrange and the dissipative structures cannot be sustained, which leads to an enstrophy decay because of viscous effects.

\begin{figure*}
\centering
\setlength{\columnsep}{-0.35cm}
\begin{multicols}{3}
    \includegraphics[width=0.9\linewidth]{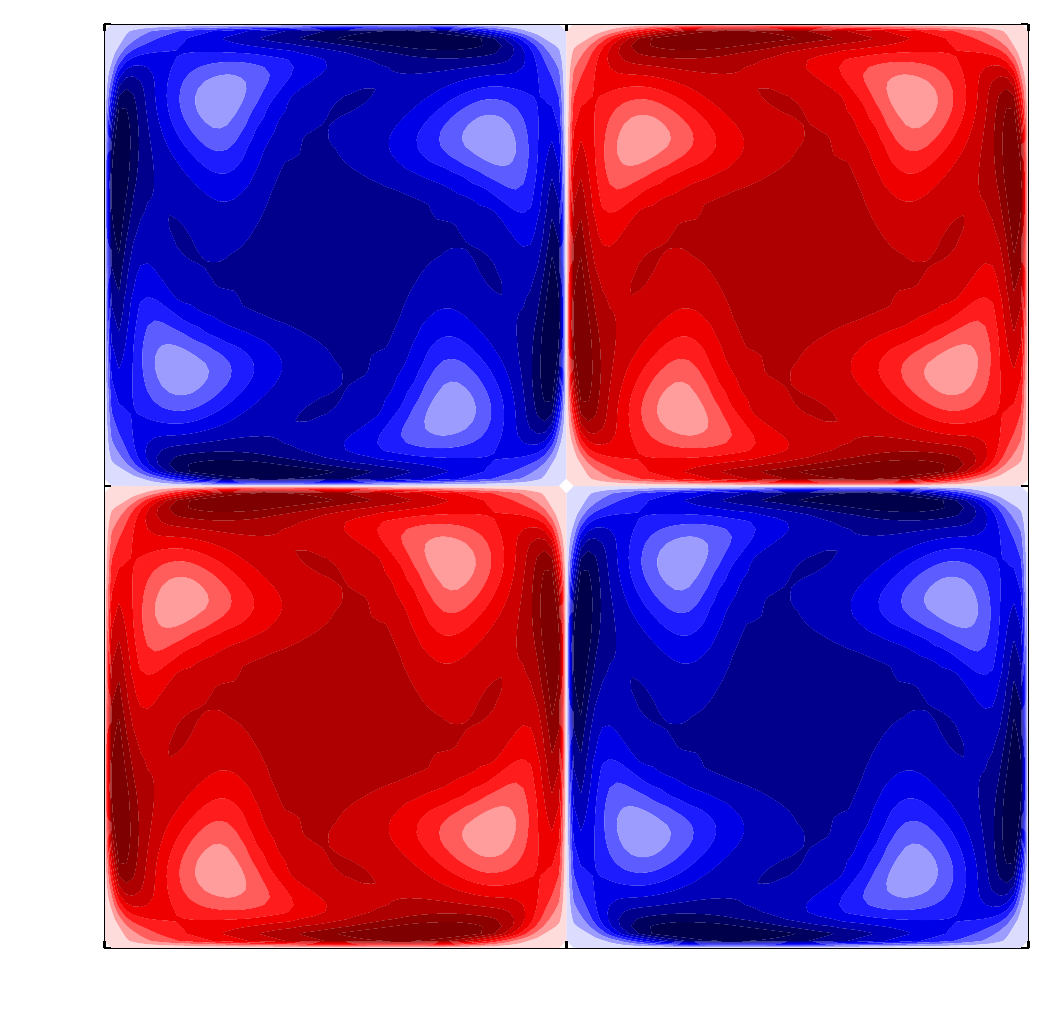}\par
    \includegraphics[width=0.9\linewidth]{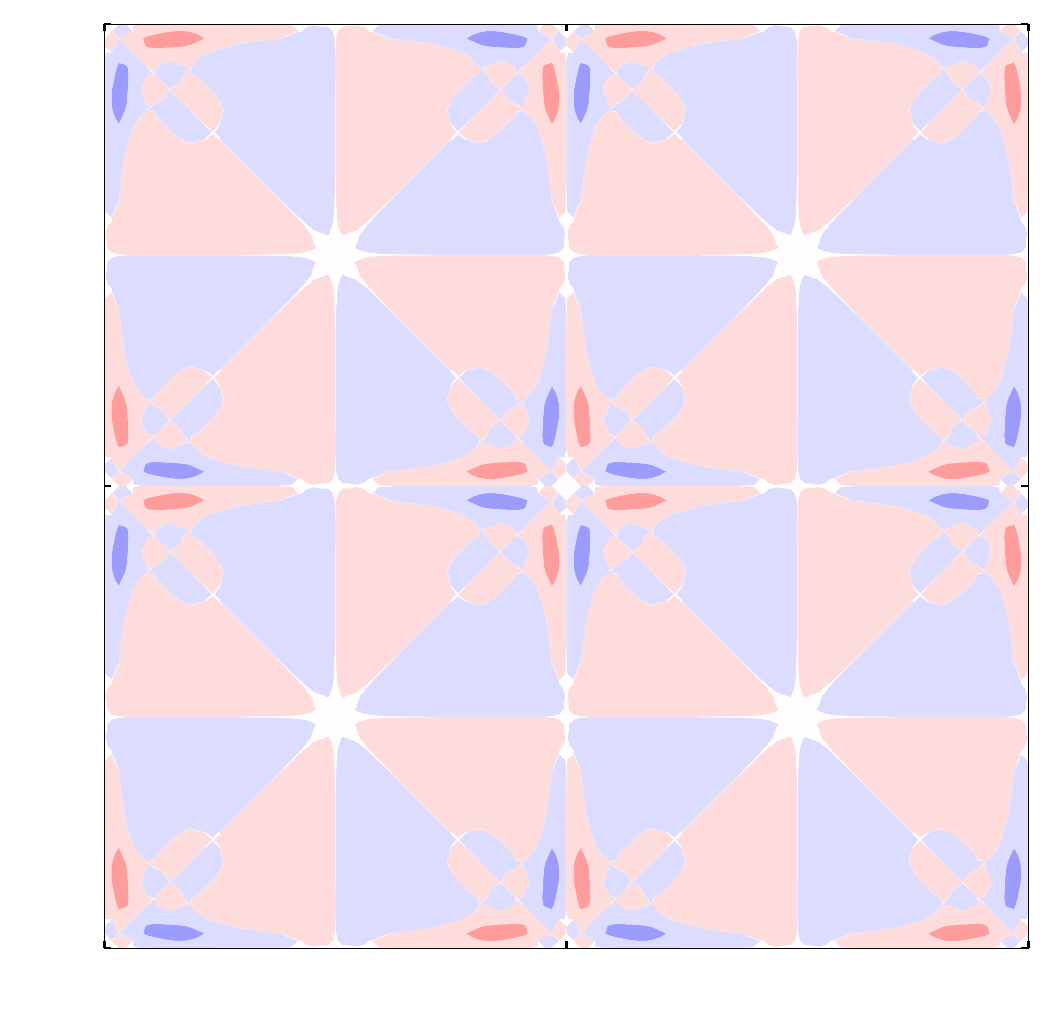}\par
    \includegraphics[width=0.9\linewidth]{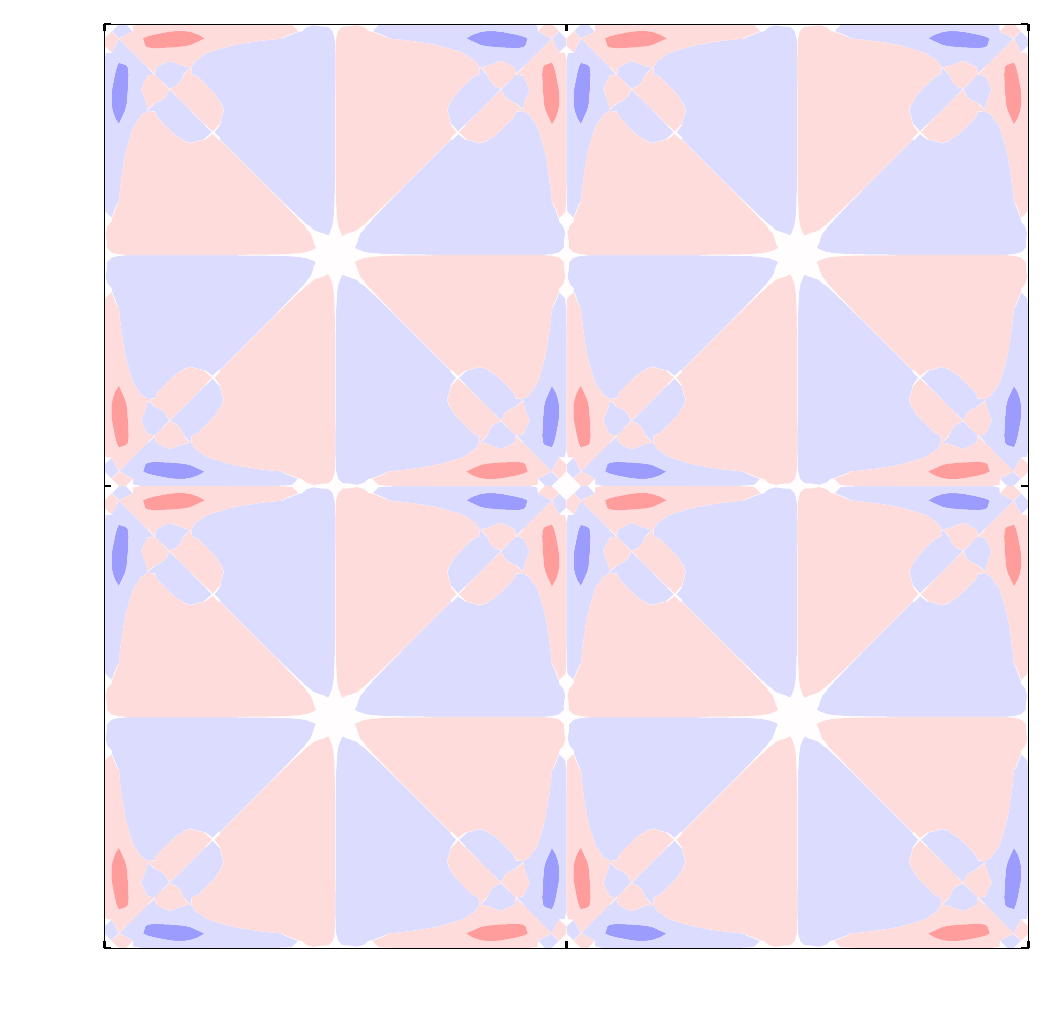}\par
\end{multicols}
\vspace{-0.7cm}
\begin{multicols}{3}
    \includegraphics[width=0.9\linewidth]{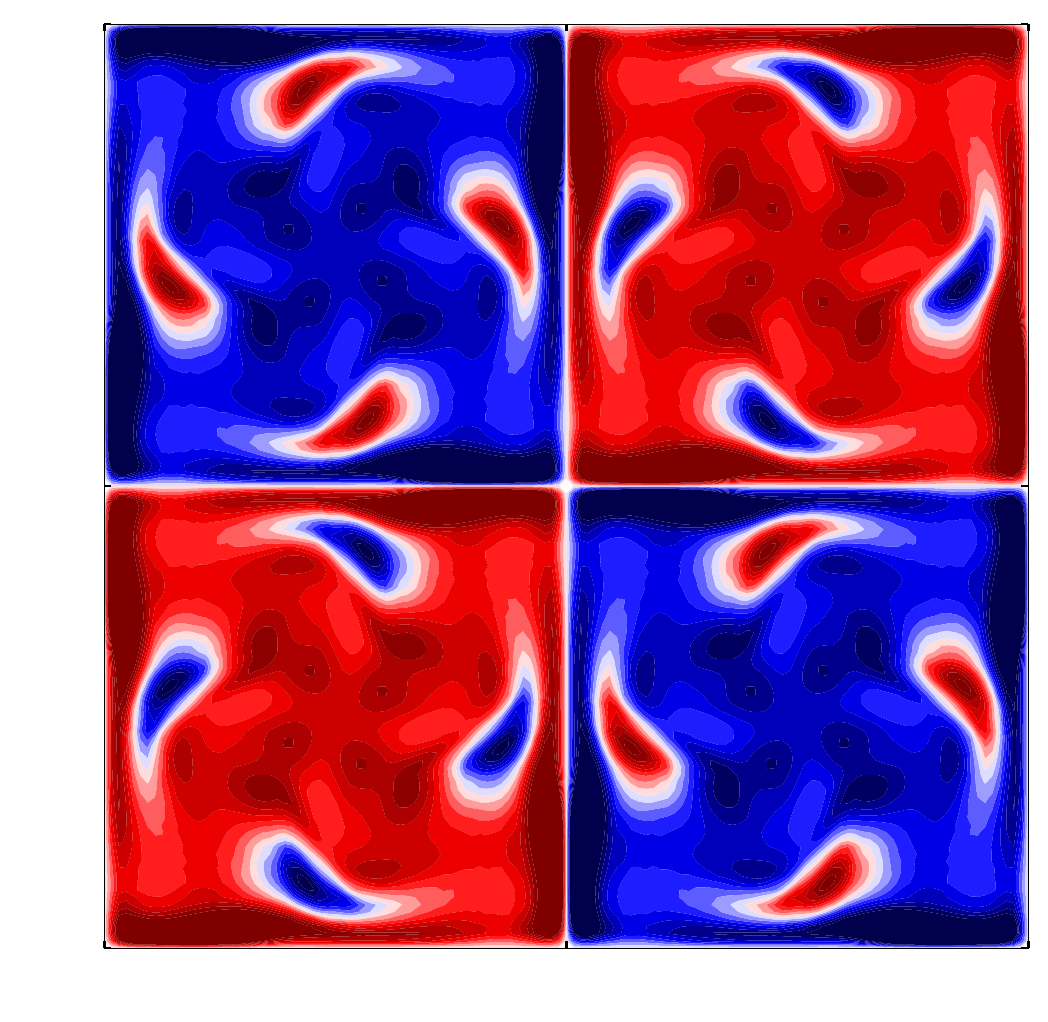}\par
    \includegraphics[width=0.9\linewidth]{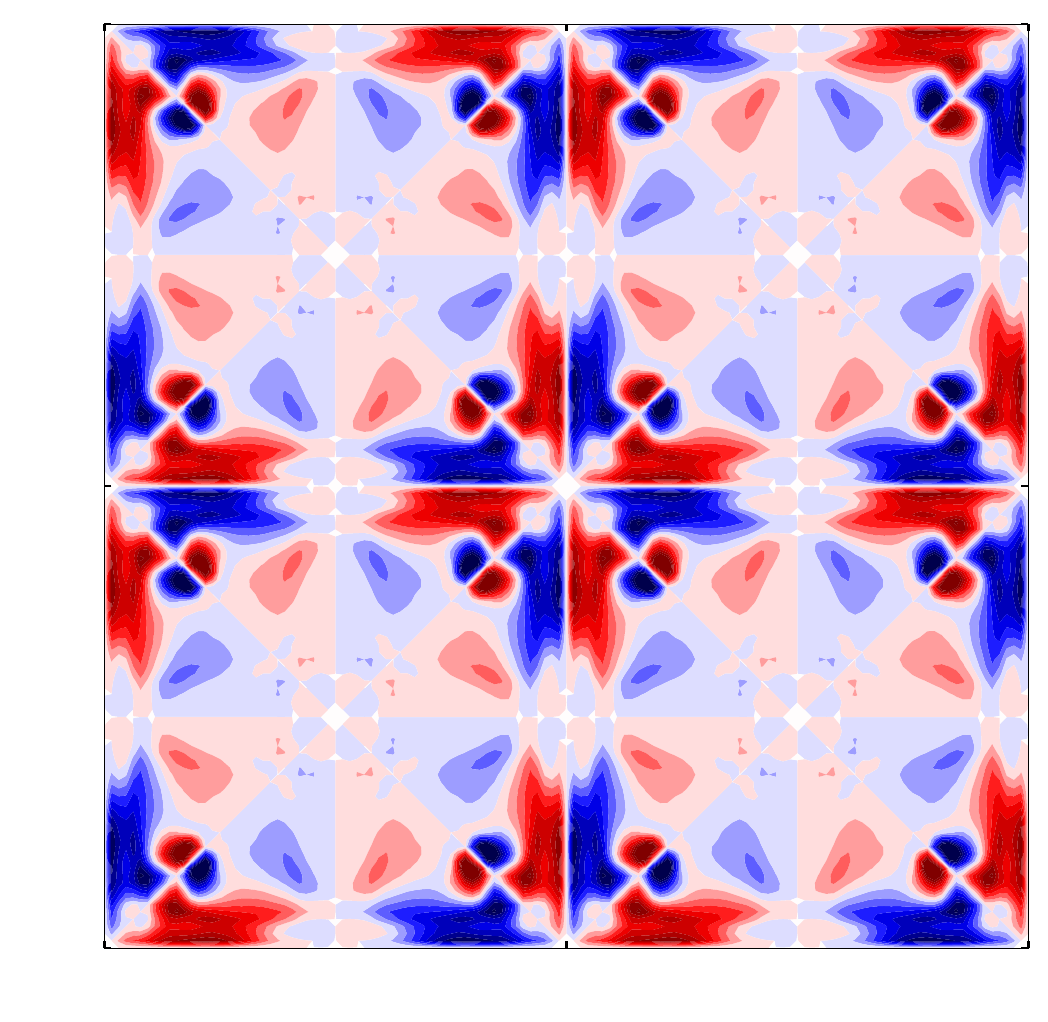}\par
    \includegraphics[width=0.9\linewidth]{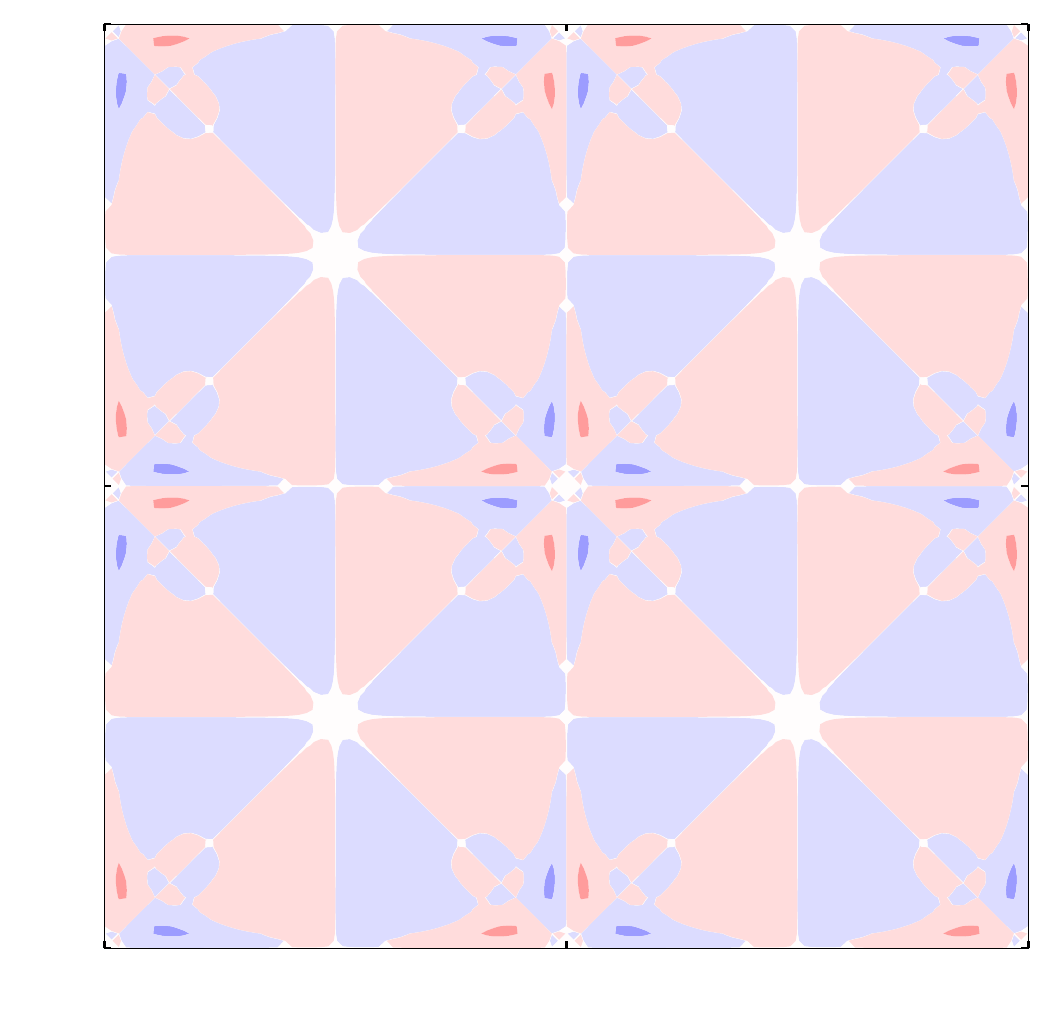}\par
\end{multicols}
\vspace{-0.7cm}
\begin{multicols}{3}
    \includegraphics[width=0.9\linewidth]{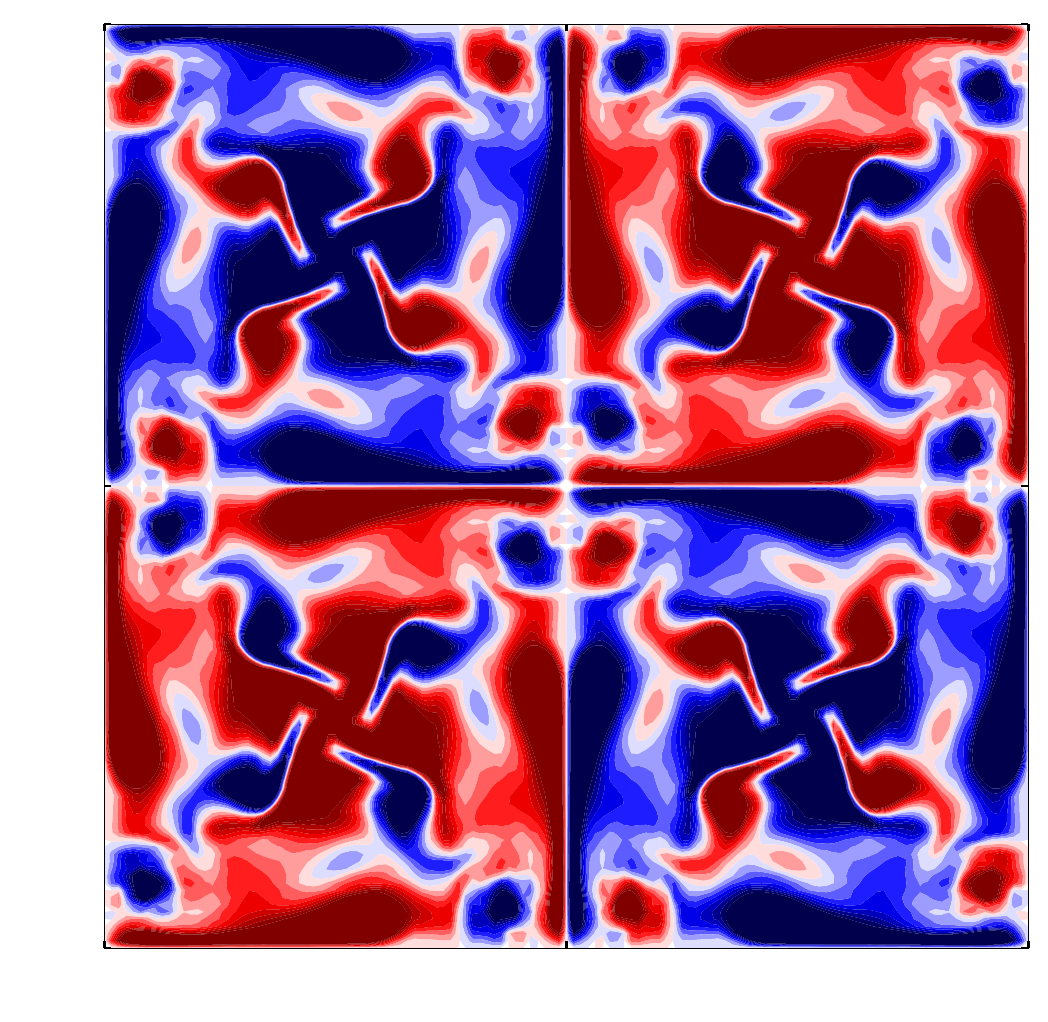}\par
    \includegraphics[width=0.9\linewidth]{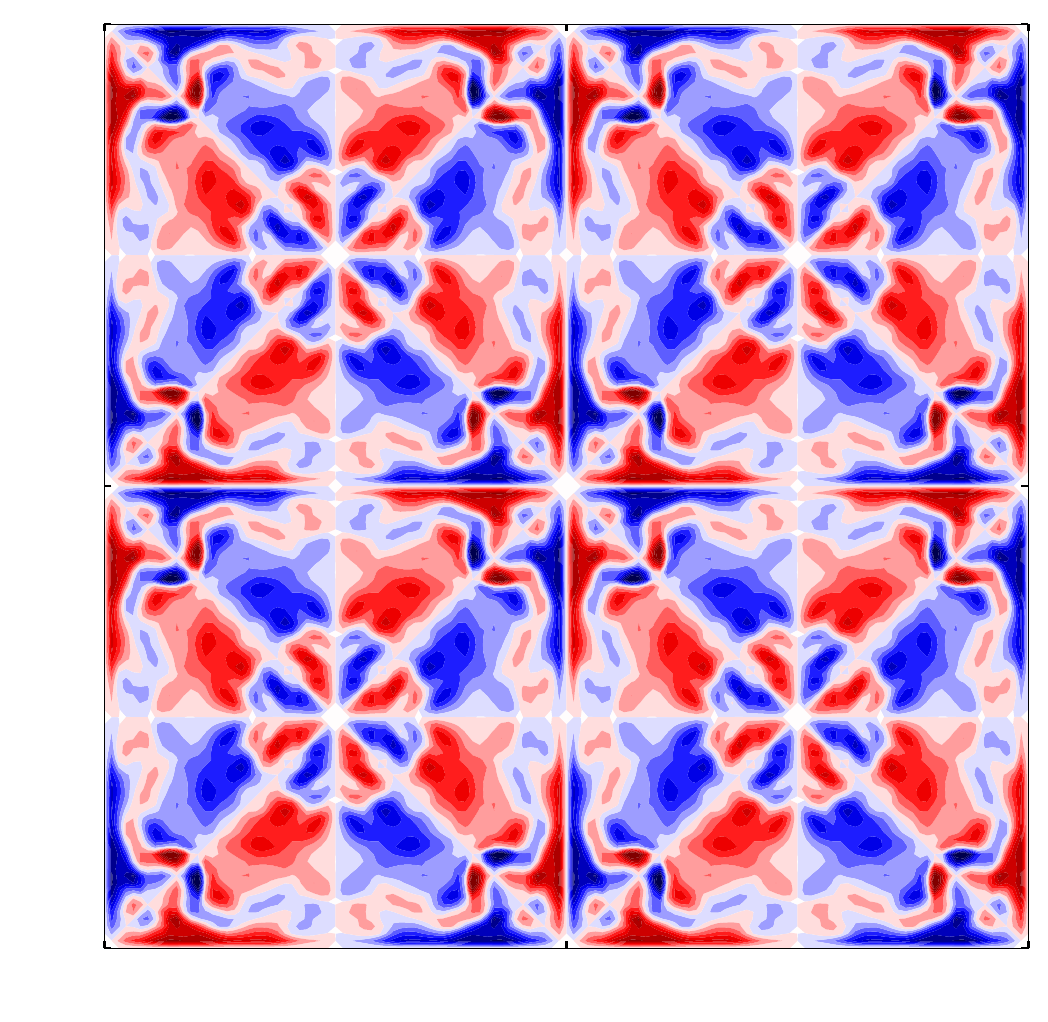}\par
    \includegraphics[width=0.9\linewidth]{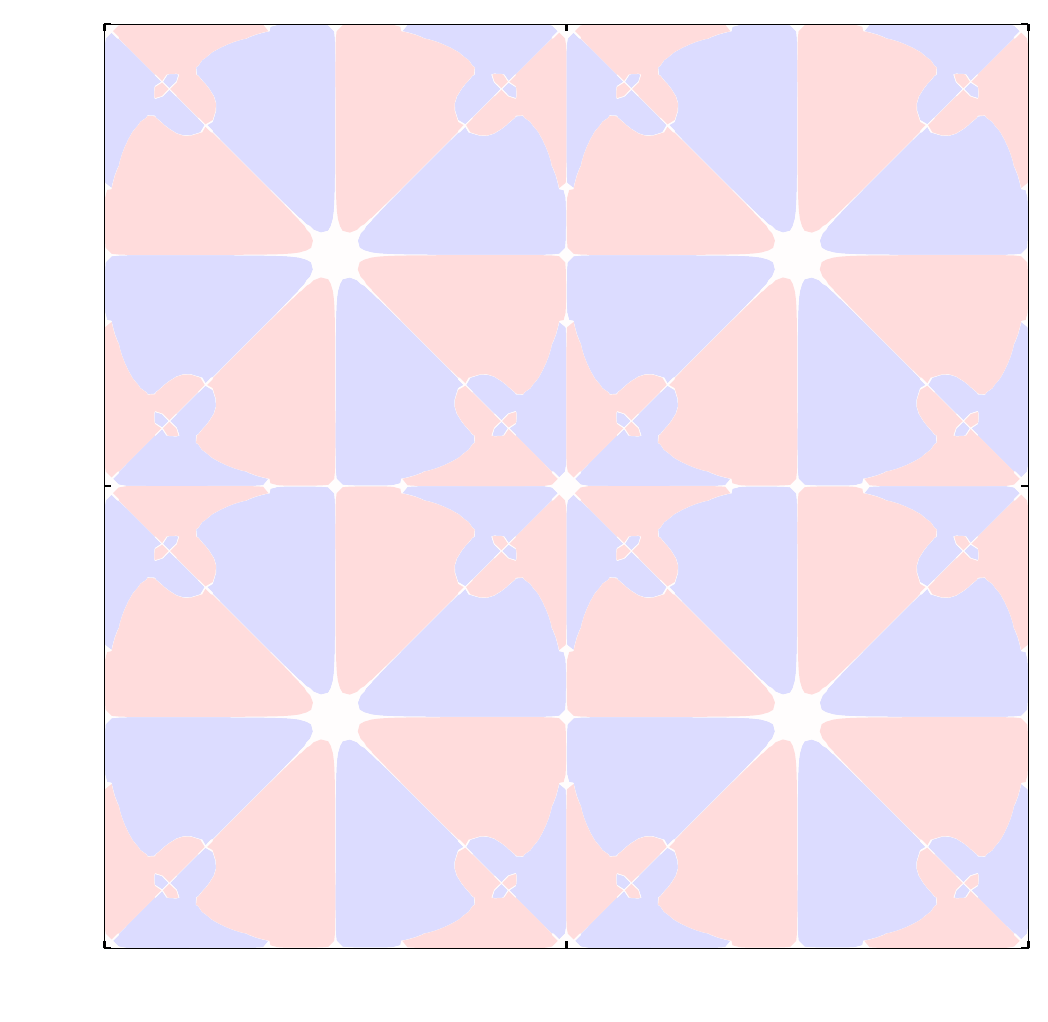}\par
\end{multicols}
\vspace{-0.7cm}
\begin{multicols}{3}
    \includegraphics[width=0.9\linewidth]{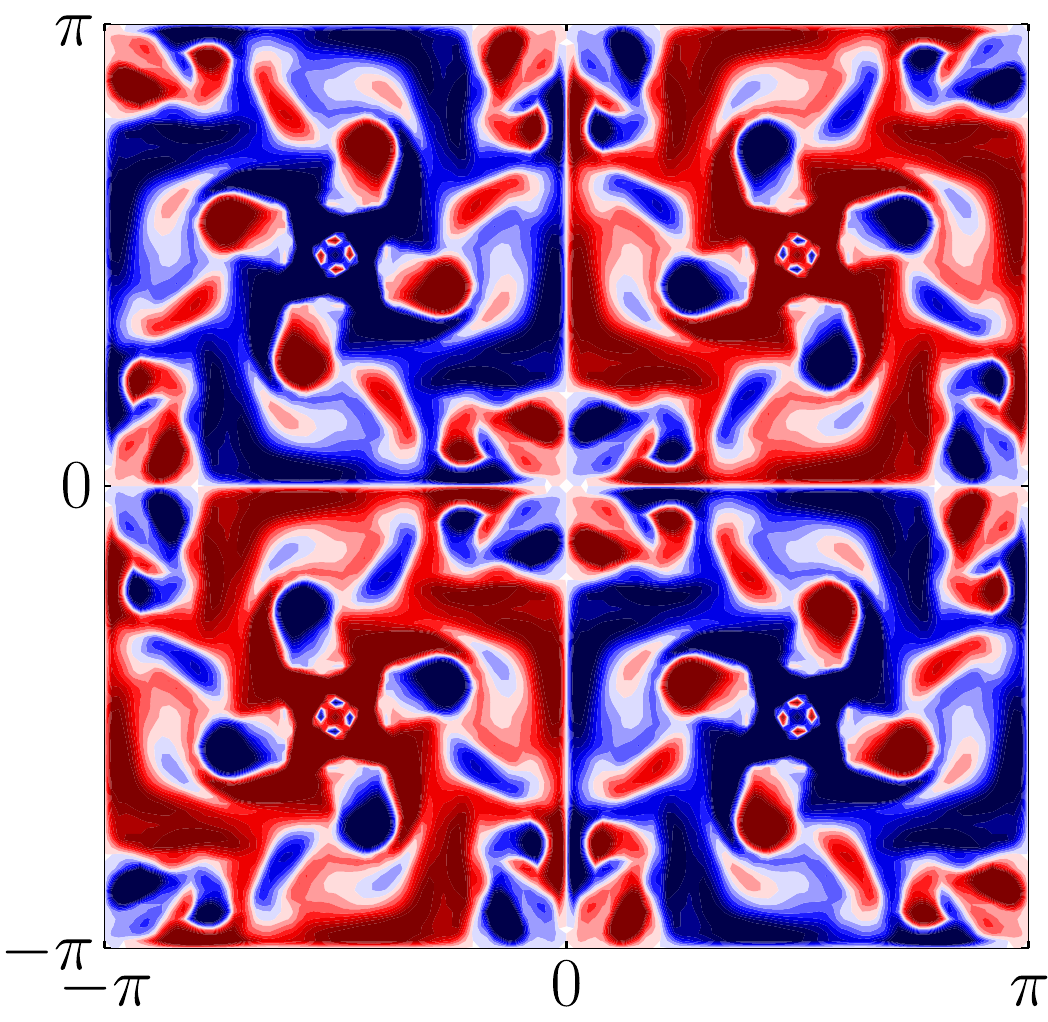}\par
    \includegraphics[width=0.9\linewidth]{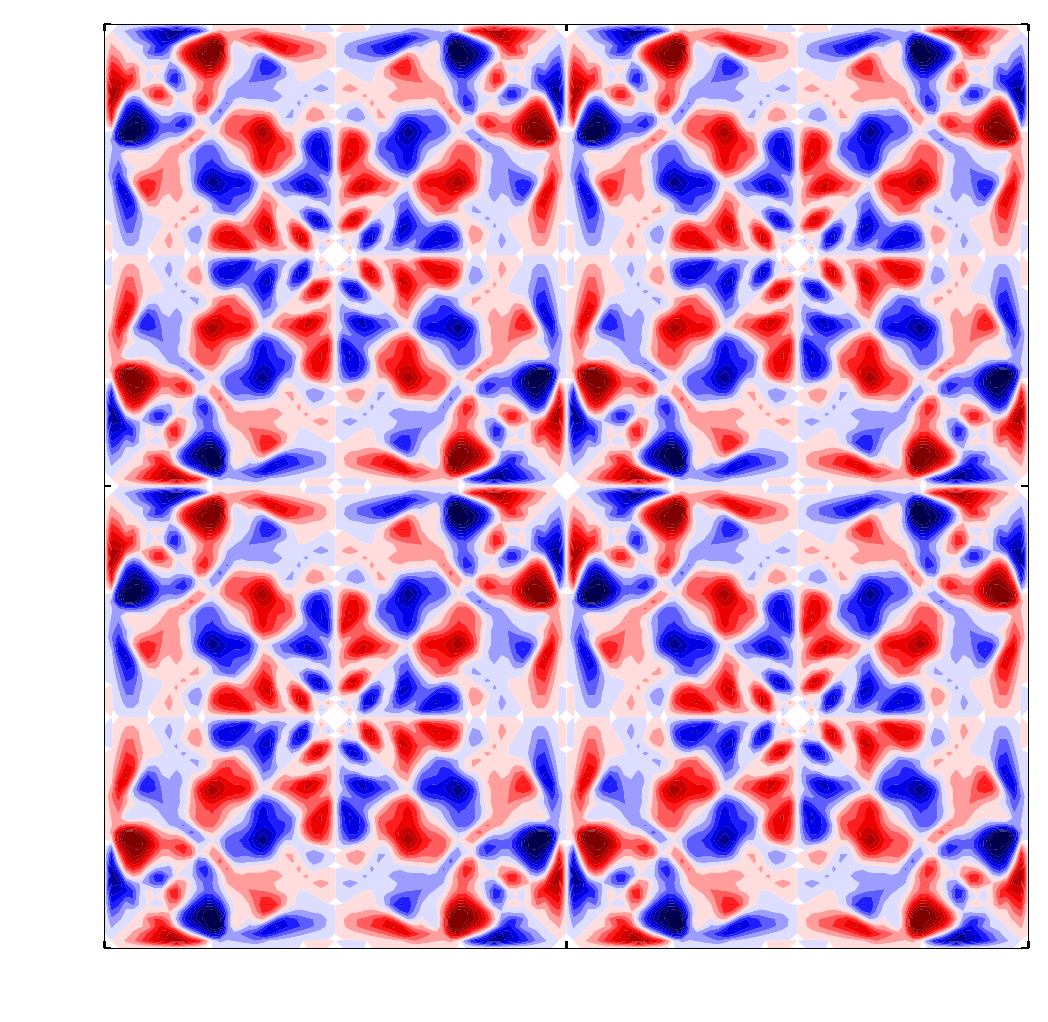}\par
    \includegraphics[width=0.9\linewidth]{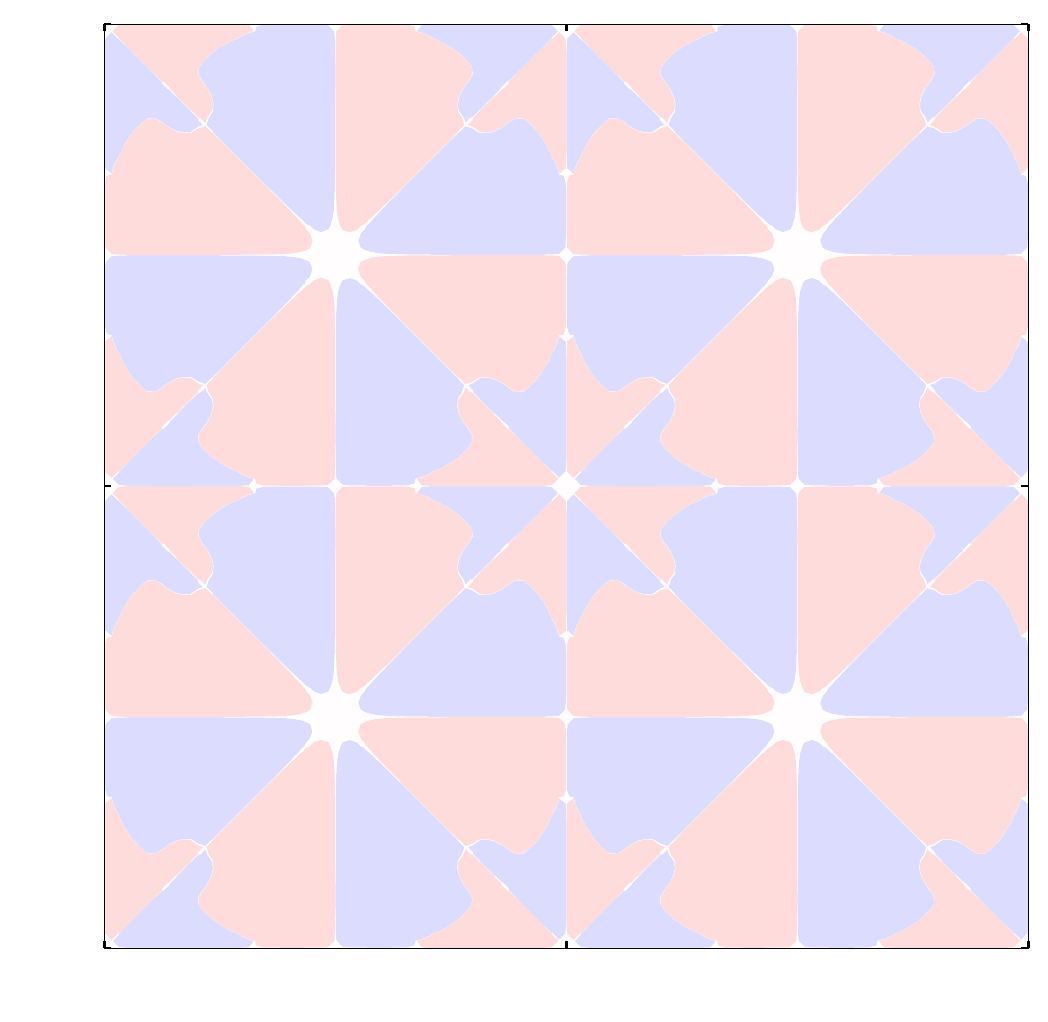}\par
\end{multicols}

\caption{Spanwise vorticity evolution of the Taylor--Green vortex at $Re=1600$ in a $2\pi$ periodic box.
From top to bottom: $t^*=4,6,8,10$.
Left: 3-D slice at $z=\pi/4$.
Middle: 3-D spanwise-averaged.
Right: 2-D started from the 3-D spanwise-averaged snapshot at $t^*=4$, where transition to turbulence is expected in the 3-D system.}
\label{fig:t-g_flow_field}
\end{figure*}

\fref{fig:t-g_flow_field} displays how the spanwise-averaged Taylor--Green vortex evolution, governed by the SANS equations, is completely different from the purely 2-D or 3-D turbulence dynamics.
The inclusion of the spanwise stresses (which can be computed in a 3-D simulation) into the 2-D Navier--Stokes solver allows to directly simulate spanwise-averaged dynamics, as demonstrated in \sref{sec:perfect_closure}.
Mathematically, the kinetic energy $\pars{E=1/2\int \vect{u}^2\,\mathrm{d}\Omega}$ and enstrophy $\pars{Z=1/2\int \boldsymbol{\omega}^2\,\mathrm{d}\Omega}$ of 2-D incompressible flows are invariant in the absence of viscosity \citep{Kraichnan1967}, i.e. $\mathrm{D}E/\mathrm{D}t=\mathrm{D}Z/\mathrm{D}t=0$.
This is different in a 3-D system, where the vortex-stretching term $\pars{\boldsymbol\omega\cdot\nabla\vect{u}}$ transfers energy from large scales to small scales of motion inducing local changes in the vorticity field.
In this scenario, enstrophy is no longer conserved and kinetic energy together with helicity $\pars{H=1/2\int \vect{u}\cdot\boldsymbol{\omega}\,\mathrm{d}\Omega}$ are the non-zero quadratic invariants.
Similarly to the vortex-stretching mechanism in 3-D flows, the inclusion of spanwise stresses into the 2-D equations locally adds or removes rotational energy into the system; hence enstrophy and kinetic energy are no longer invariant in the absence of viscosity.

\subsection{Perfect closure} \label{sec:perfect_closure}

The SANS equations are numerically verified by including the \textit{perfect closure} of the spanwise stresses in a 2-D simulation.
The perfect closure $(\mathcal{S}^R)$ is defined as the difference between the 2-D Navier--Stokes spatial operator $\tilde{\mathcal{S}}$ and the spanwise-averaged 3-D Navier--Stokes spatial operator $\avg{\mathcal{S}}$.
The inclusion of the perfect closure into the 2-D system allows to calculate the unsteady spanwise-averaged solution of the flow at every time step.
It can be written as
\begin{equation}
\mathcal{S}^R\pars{\vect{u}\p}\equiv \tilde{\mathcal{S}}-\avg{\mathcal{S}},
\label{eq:sans_perfect}
\end{equation}
where
\begin{align}
\mathcal{S}\pars{\vect{u},p}= \vect{u}\cdot\nabla \vect{u} + \nabla p - \nu \nabla^2 \vect{u}, \qquad &\vect{u}=\pars{u,v,w},\,\,\, \vect{x}=\pars{x,y,z},\\
 \tilde{\mathcal{S}}\pars{\vect{U},P}= \vect{U}\cdot\nabla \vect{U} + \nabla P - \nu \nabla^2 \vect{U}, \qquad &\vect{U} = \pars{U,V},\,\,\, \vect{x}=\pars{x,y}.
\end{align}

With the previous definitions, the $\nabla\cdot\boldsymbol\tau_{ij}^R$ closure term in \eref{eq:sans_simple} is redefined as the residual between the operators
\begin{gather}
\partial_t\vect{U}+\tilde{\mathcal{S}}=\tilde{\mathcal{S}}-\avg{\mathcal{S}}\equiv\mathcal{S}^R\pars{\vect{u}\p},\label{eq:perfect_SANS}\\
\partial_t\vect{U}+\tilde{\mathcal{S}}=\mathcal{S}^R.
\end{gather}

Note that, differently from  $\nabla\cdot\boldsymbol\tau_{ij}^R$, the $\mathcal{S}^R$ residual term includes the spatial discretisation error of the resolved spanwise-averaged quantities since $\tilde{\mathcal{S}}$ cancels out in \eref{eq:perfect_SANS} (see \cite{Beck2019} for an analogous explanation for LES closure terms).
Hence, the perfect spanwise-averaged solution can be recovered from the original definition of the SANS equations, i.e. $\avg{\partial_t\vect{u}+\mathcal{S}}=0$.

Beyond the spatial discretisation error inherent in $\nabla\cdot\boldsymbol\tau_{ij}^R$, the quadrature error itself could also affect the perfect recovery of the spanwise-averaged solution.
In this sense, we show in \aref{sec:sans_quadrature_errors1} that the quadrature error for the full SANS equations (\eref{eq:sans_full}) cancels out for the initial state of the Taylor--Green vortex case.
On the other hand, \aref{sec:sans_quadrature_errors2} shows that the quadrature error balance is broken when assuming spanwise periodicity simplifications (\eref{eq:sans_simple}).
Still, the midpoint and trapezoidal quadratures, normally converging with $\mathcal{O}(h^2)$, can achieve exponential convergence rates for periodic functions \citep{Weideman2002,Trefethen2014}.
Hence, the quadrature error should not have a practical impact given it decays faster than the discretisation error of the governing equations.

\subsubsection*{Taylor--Green vortex}

\begin{figure}[!t]
\centering
\includegraphics[width=0.48\linewidth]{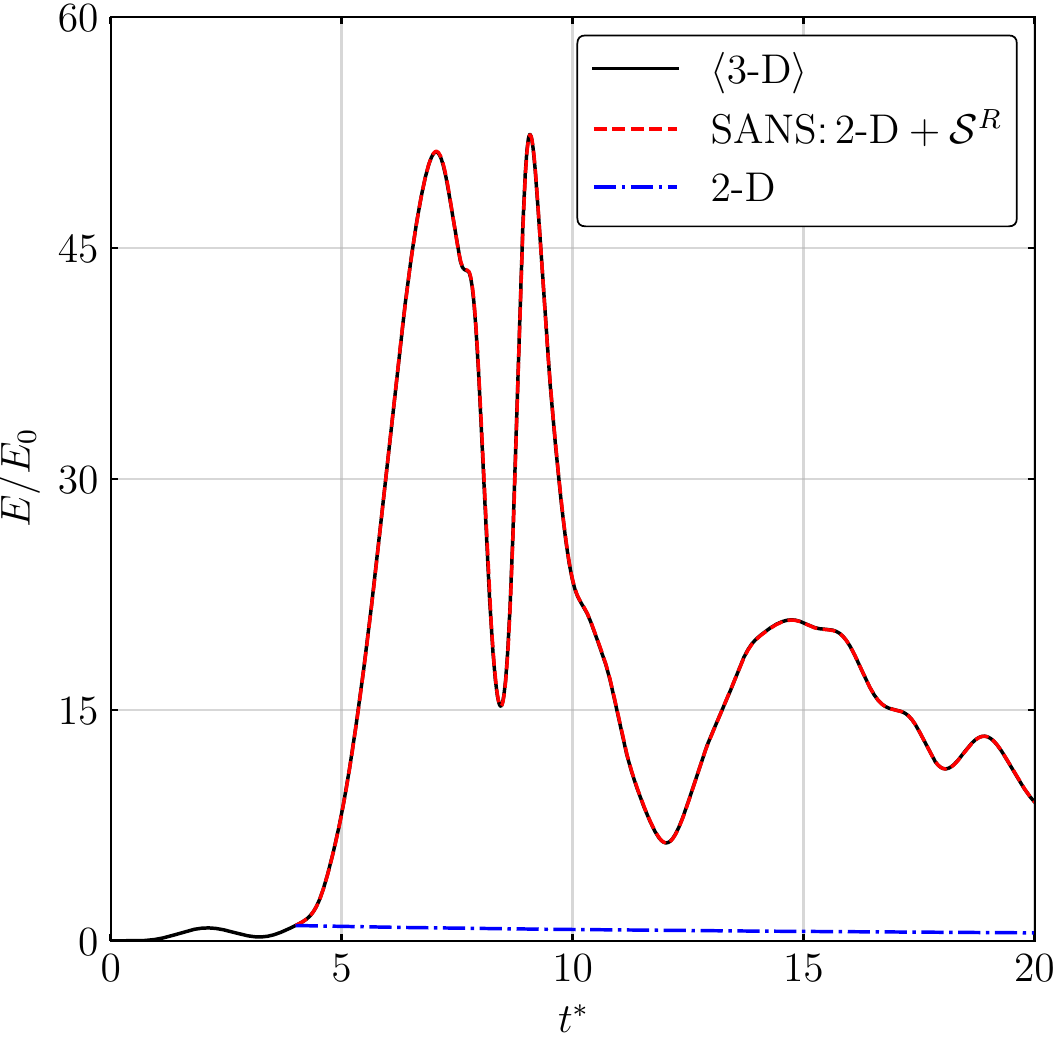}
\includegraphics[width=0.48\linewidth]{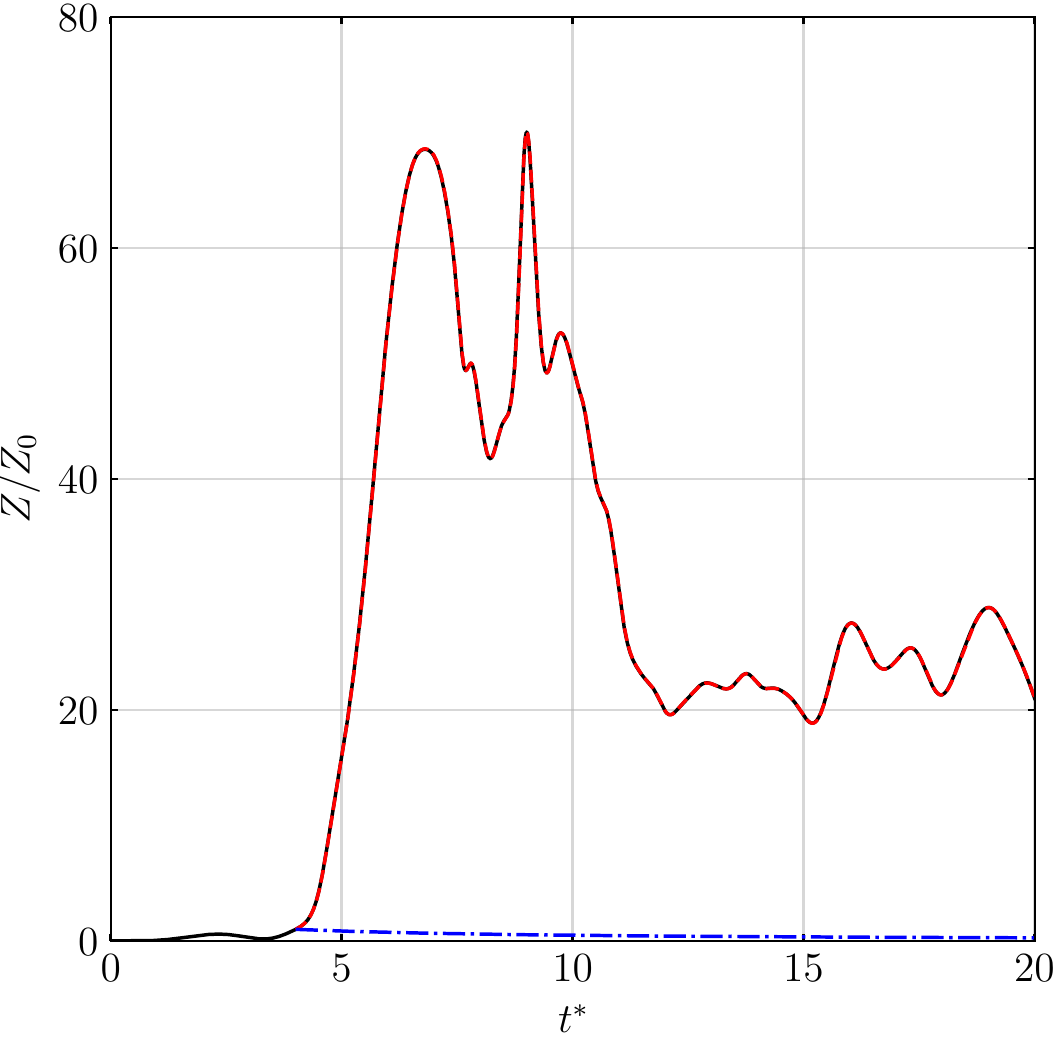}
\caption{Kinetic energy ($E$, left) and enstrophy ($Z$, right) of the Taylor--Green vortex case.  
$E_0$ and $Z_0$ correspond to the energy and enstrophy at $t^*=4$. 
Energy is computed as $E=1/(2\Omega)\int \vect{U}^2\,\mathrm{d}\Omega$.
Enstrophy is computed as $Z=1/(2\Omega)\int\Omega_z^2\,\mathrm{d}\Omega$, where $\Omega_z=\nabla\times\vect{U}$.}\label{fig:perfect_t-g}
\end{figure}

The Taylor--Green vortex flow is used to validate and verify the perfect closure of the SANS equations by computing the closure terms in a 3-D simulation, where their exact solution is known, and then adding them in the 2-D solver at every time step.
We select a uniform resolution of 128 cells in all directions and set $Re=1600$, noting that the purpose of this exercise is to validate the correct derivation of the SANS equations rather than fully resolving the temporal and spatial scales of the problem.
The global kinetic energy and enstrophy are recorded for the spanwise-averaged 3-D,  SANS, and 2-D systems, as displayed in \fref{fig:perfect_t-g}. 
The SANS and 2-D simulations are started from a spanwise-averaged snapshot of the 3-D system at $t^*=4$, where $t^*=t\kappa U_0$.
The evolution of the perfect closure matches the spanwise-averaged 3-D case to machine accuracy.
It can also be observed that energy and enstrophy are not conserved.
Again, this is due to the inclusion of the perfect closure term $(\mathcal{S}^R)$, which changes the system dynamics from standard 2-D Navier--Stokes to SANS.
On the contrary, the 2-D case  almost conserves energy and enstrophy, and the decrease of the metrics is due to viscous and numerical effects.
 
\begin{figure}[!t]
\centering
\includegraphics[width=0.50\linewidth]{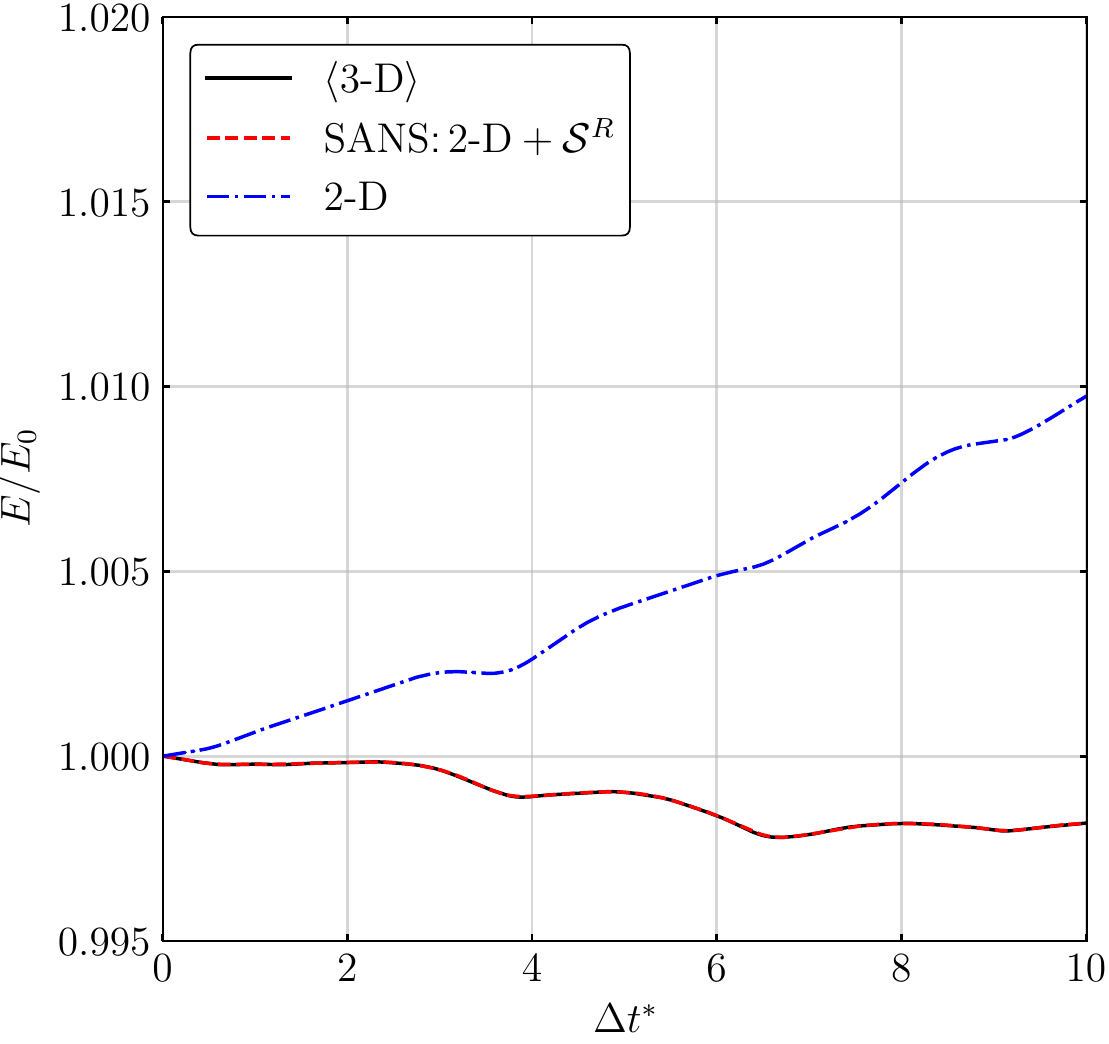}
\includegraphics[width=0.48\linewidth]{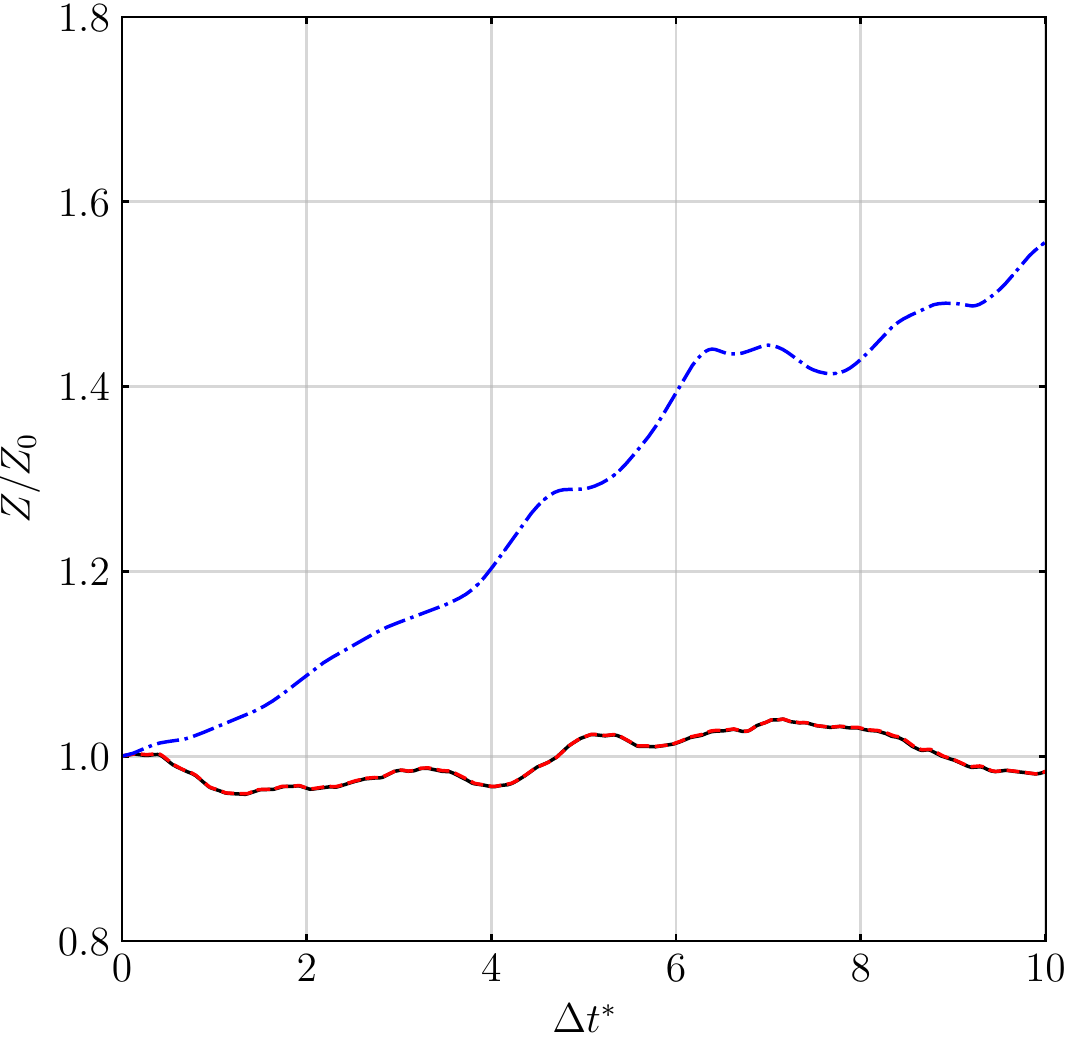}
\caption{Kinetic energy (left) and enstrophy (right) evolution of flow past a circular cylinder at $Re=10^4$.
$E_0$ and $Z_0$ correspond to the energy and enstrophy at $t_0^*$.
Energy and enstrophy are computed as detailed in \fref{fig:perfect_t-g} caption.}
\label{fig:perfect_cc_EZ}
\end{figure}

\subsubsection*{Flow past a circular cylinder}\label{sec:SANS_circular_cylinder}

Similarly to the Taylor--Green vortex case, the SANS equations are validated with the perfect closure for flow past a circular cylinder at $Re=10^4$.
We employ the same case set-up as \sref{sec:circular_cylinder_details} using a fixed cylinder of $L_z=1$ (span non-dimensionalised with the cylinder diameter).

\begin{figure}[!t]
\centering
\includegraphics[width=0.48\linewidth]{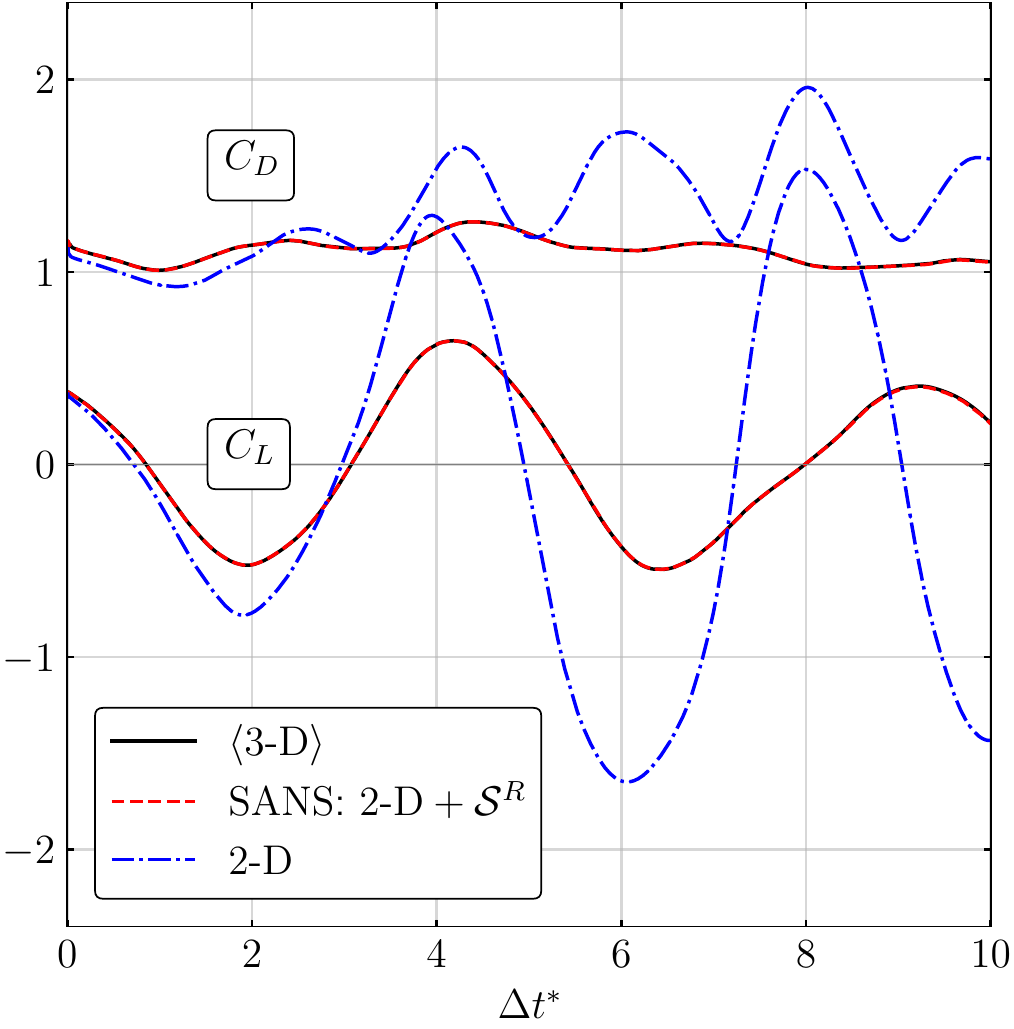}
\caption{Lift (bottom) and drag (top) force coefficients of flow past a circular cylinder at $Re=10^4$.
The force coefficients in the 3-D system have been calculated as $C_{L,D}=2F_{y,x}/(\rho U_\infty^2 D L_z)$, where $F_y$ and $F_x$ are respectively the vertical and horizontal pressure force, $\rho$ is the constant density, $U_\infty$ is the free-stream velocity, and $D$ is the cylinder diameter.
The force coefficients in the 2-D system are calculated analogously without factoring by $1/L_z$.}
\label{fig:perfect_cc_forces}
\end{figure}

Starting from a spanwise-averaged 3-D snapshot, the inclusion of the perfect closure in the 2-D solver at every time step allows to recover the unsteady spanwise-averaged flow.
Quantitatively, the difference between the SANS and the 2-D dynamics is emphasised in the kinetic energy and enstrophy evolution (\fref{fig:perfect_cc_EZ}), and the forces induced to the cylinder (\fref{fig:perfect_cc_forces}).
In the 2-D system, both kinetic energy and enstrophy rapidly increase as a result of its natural 2-D system attractor.
On the other hand, the spanwise-averaged dynamics dictate the evolution of the flow when the perfect closure is inserted into the 2-D solution.

\begin{figure}[!t]
    \centering
    \begin{subfigure}[t]{0.7\linewidth}        
        \centering
        \includegraphics[width=\linewidth]{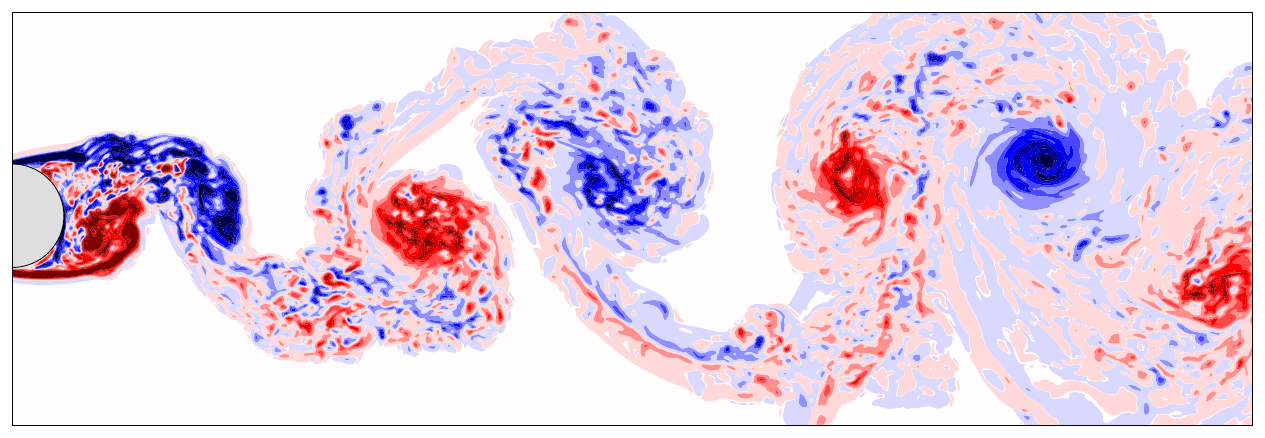}
        \caption{$\avg{3\text{-}\mathrm{D}}$, $t^*_0$.}\vspace{0.4cm}
    \end{subfigure}
    \begin{subfigure}[t]{0.7\linewidth}
        \centering
        \includegraphics[width=\linewidth]{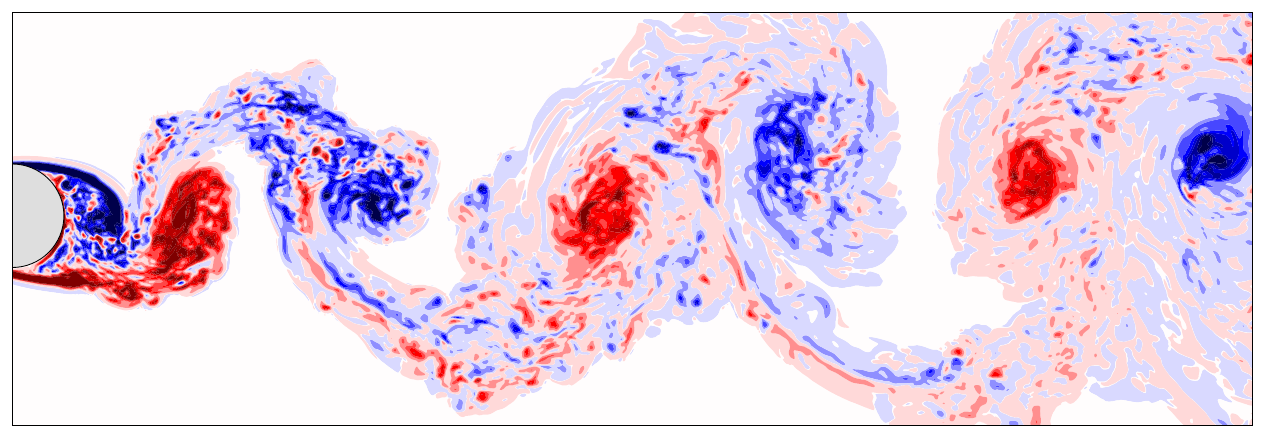}
        \caption{SANS: $2\text{-}\mathrm{D}+\mathcal{S}^{R}$, $\Delta t^*=2$.}\vspace{0.4cm}
    \end{subfigure}
    \begin{subfigure}[t]{0.7\linewidth}
        \centering
        \includegraphics[width=\linewidth]{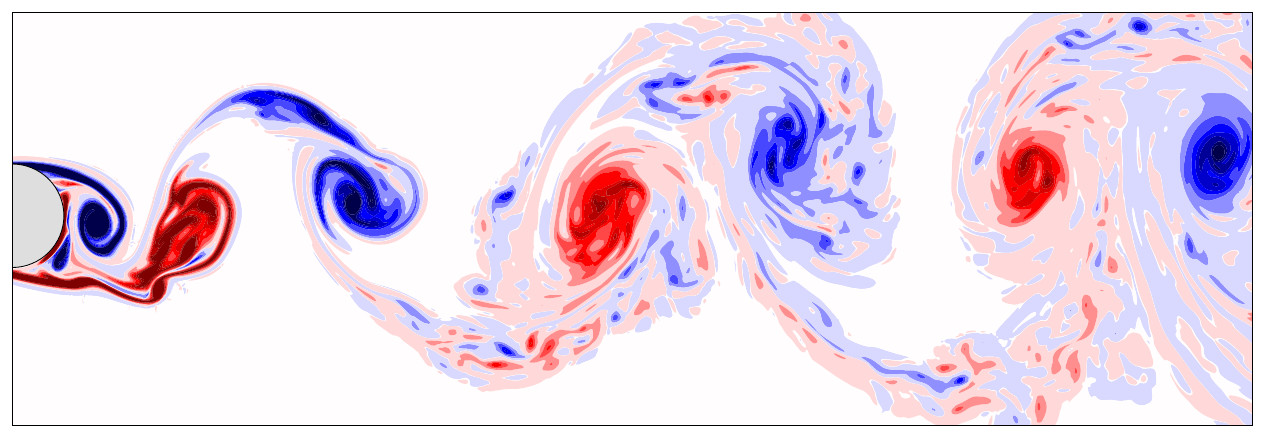}
        \caption{$2\text{-}\mathrm{D}$, $\Delta t^*=2$.}
    \end{subfigure}
    \captionsetup[subfigure]{width=\linewidth}
    \caption{Top: Vorticity of the spanwise-averaged 3-D flow used as initial condition for the SANS and the 2-D simulations.
Middle: Vorticity obtained in the SANS system after 2 convective time units.
Bottom: Vorticity obtained in the 2-D system after 2 convective time units.}
\label{fig:perfect_cc_vort}
\end{figure}

Qualitatively, it can be observed in \fref{fig:perfect_cc_vort} that a rapid two-dimensionalisation of the wake develops when the perfect closure is not included in the 2-D solver.
In this scenario, vortical structures are highly coherent after just two convective time units (i.e. $\Delta t^*=t^*-t_0^*=2$, where $t^*=tU/D$).
This is reflected in the over-prediction of the forces induced to the cylinder as a result of the two-dimensionalisation mechanism yielding energised vortices in the near wake region.

The over-prediction of the hydrodynamic forces is the main short-coming of standard 2-D strip-theory methods, which arbitrarily include additional dissipation mechanisms such as 2-D turbulence models to tackle this issue.
Here we show that correctly modelling the SSR terms can bypass this problem by explicitly including 3-D effects into the governing equations while keeping the low computational cost of 2-D strip-theory methods.

\subsection{Eddy-viscosity model} \label{sec:evm}

The most common model for the closure terms arising from time-averaged (RANS) or spatially-filtered (LES) Navier--Stokes equations is based on the eddy-viscosity (Boussinesq) hypothesis.
Under this hypothesis, it is assumed that the unresolved scales of motion can be linked to an effective (positive scalar) viscosity additional to the molecular viscosity of the fluid.
This implies that the closure terms can only have a dissipative effect in the governing equations.
Analogous to the definition of the stress tensor for Newtonian fluids, the eddy-viscosity hypothesis is defined as
\begin{equation}
\boldsymbol\tau^r_{ij}\equiv\avg{\vect{u}\p\otimes\vect{u}\p}-\frac{2}{3}k\delta_{ij}=-2\nu_t \avg{S_{ij}},
\label{eq:Boussinesq}
\end{equation}
where $\boldsymbol\tau^r_{ij}$ is the anisotropic SSR tensor, $\avg{S_{ij}}=(\nabla\otimes\avg{\vect{u}}+\avg{\vect{u}}\otimes\nabla)/2$ is the mean rate-of-strain tensor, $\nu_t$ is the eddy viscosity, $k=\mathrm{tr}\avg{\vect{u\p}\otimes\vect{u\p}}/2$ is the turbulence kinetic energy, and $\delta_{ij}$ is the Kronecker delta tensor.
The following relations arise for 2-D incompressible flow
\begin{gather}
\avg{u\p u\p}-2k/3=-2\nu_t\partial_x U,\label{eq:evm1}\\
\avg{v\p v\p}-2k/3=-2\nu_t\partial_yV,\\
\avg{u\p v\p}=-\nu_t\pars{\partial_yU+\partial_xV}.\label{eq:evm3}
\end{gather}

The eddy-viscosity field in \eref{eq:Boussinesq} has been a subject of research for decades and many different eddy-viscosity models (EVMs) have been developed for RANS and LES flow descriptions.
One of the most successful is the LES Smagorinsky model \citep{Smagorinsky1963}
\begin{equation}
\nu_t=\pars{C_s \Delta}^2\sqrt{2\avg{S_{ij}}\avg{S_{ij}}},
\end{equation}
where $\Delta$ is the averaging filter width, and $C_s$ is the Smagorinsky constant which tunes the amount of SGS kinetic energy to be removed.

\subsubsection*{A-priori results} \label{sec:EVM_a-priori}

To test the applicability of EVMs for SANS flow, we apply the Smagorinsky model to the circular cylinder flow field, manually selecting the value of $C_s$ to provide the best magnitude fit with the anisotropic part of the residual tensor.
As this a-priori assessment assumes knowledge of the best $C_s$ and does not require any modelling of the isotropic part of the stress tensor (i.e. the SSR kinetic energy, $k$), it represents the best possible results for the EVM\footnotemark.
In the context of this work and as often found in the literature, \textit{a-priori} analysis refers to the analysis of the model performed before deploying it in a live simulation.
Such analysis uses reference data previously generated with the full system of equations, i.e. the 3-D Navier--Stokes equations.
On the other hand, \textit{a-posteriori} analysis refers to the deployment of the trained model in a live 2-D simulation, where metrics such as the unsteady hydrodynamic forces can be used as a performance indicator of the model.

\footnotetext{In an a-posteriori framework, turbulence modelling requires the full residual stress tensor to solve the averaged governing equations. Hence, EVMs usually incorporate the isotropic part $\pars{2k/3}$ in the total pressure defining a new pressure field, $p^*=p+2k/3$, on which a divergence-free velocity field is projected.
Alternatively, a transport equation for the TKE can also be considered.}

\begin{figure}[!t]
\centering
    \includegraphics[width=0.45\linewidth]{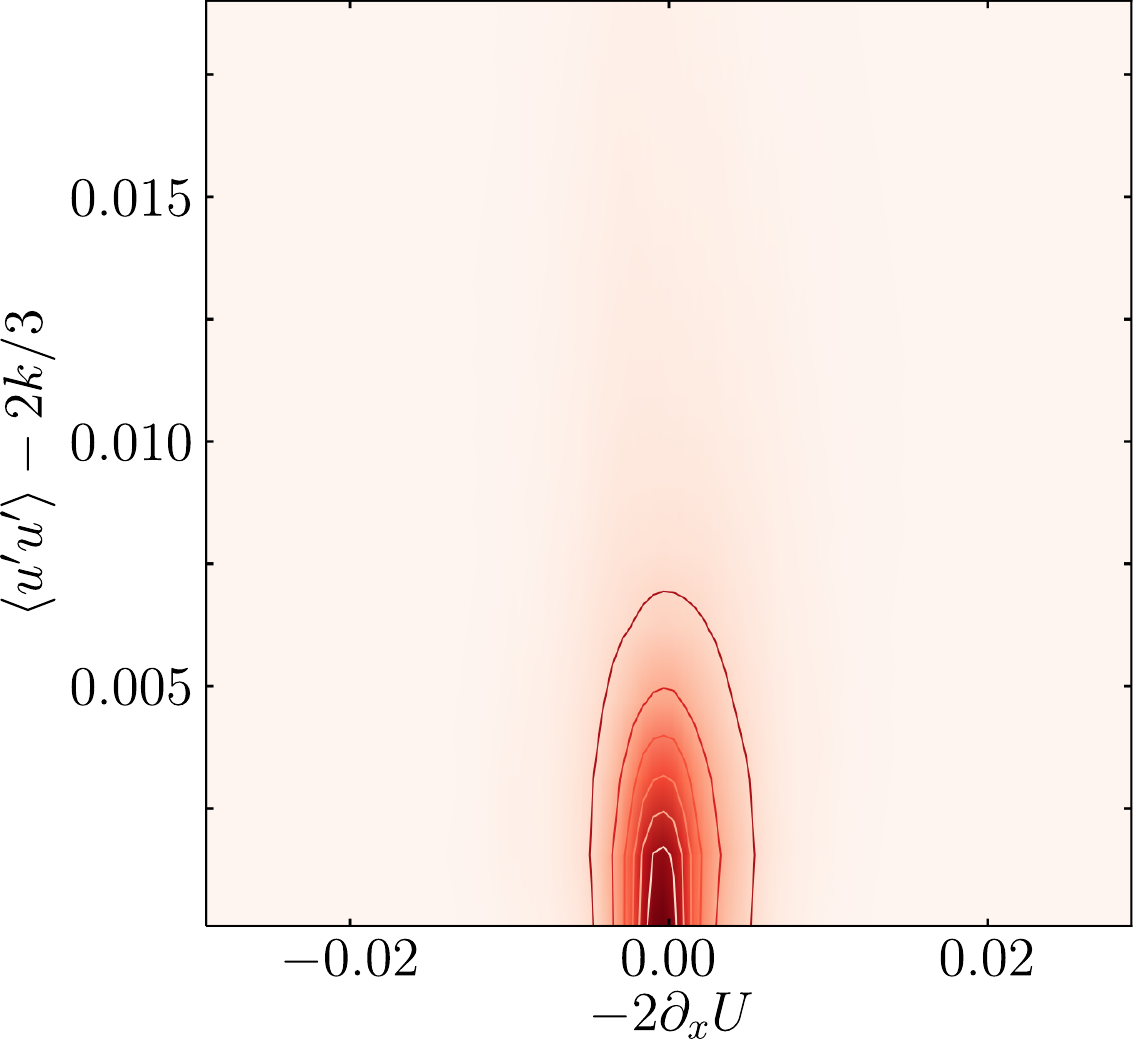}\hspace{1em}
    \includegraphics[width=0.45\linewidth]{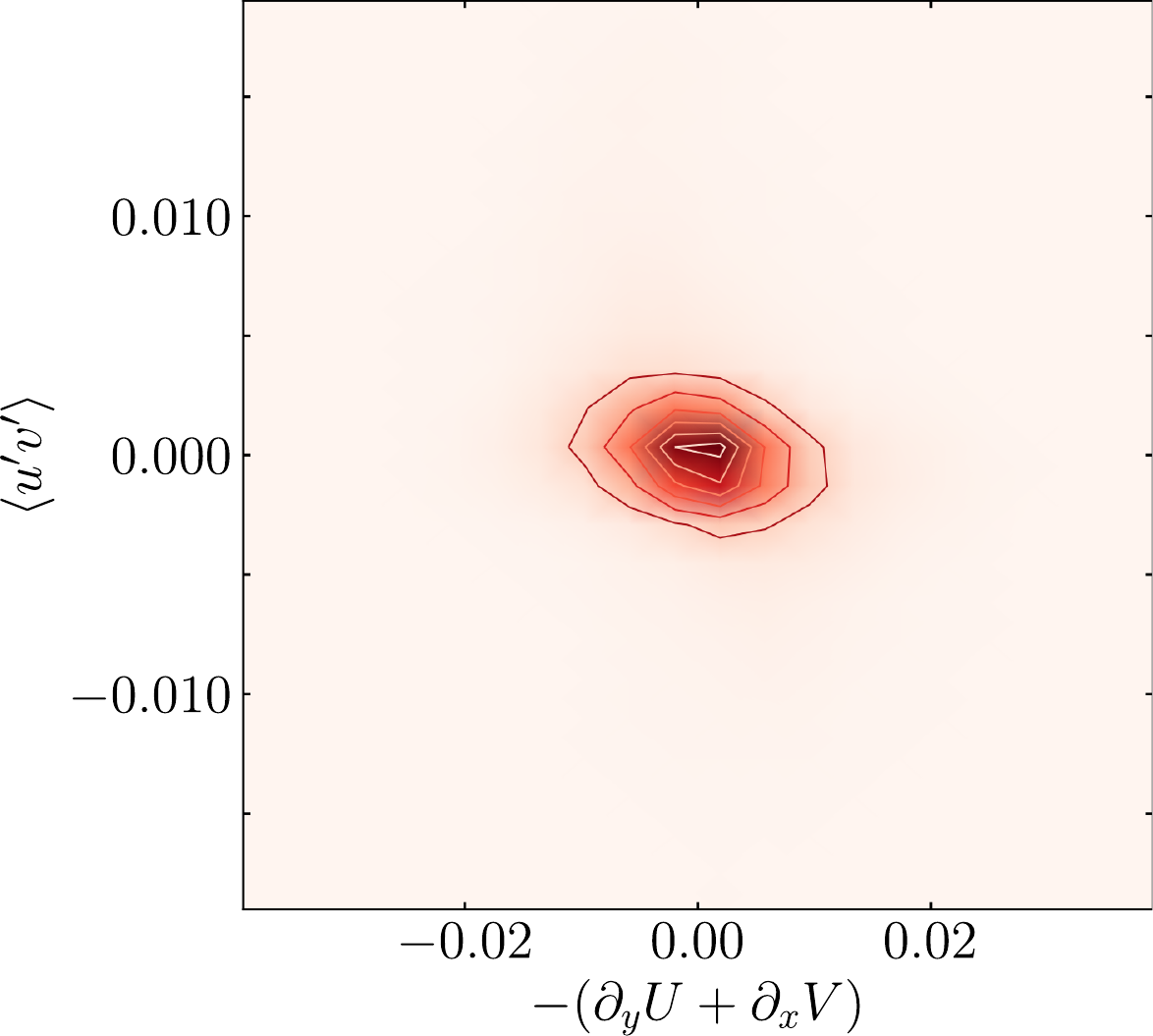}
\caption{Density plots of the Boussinesq relationship between resolved and fluctuating quantities computed within the $x \in [0, 12], y \in [-2, 2]$ domain of the cylinder wake.}
\label{fig:evm_density_plots}
\end{figure}

The a-priori correlation of the EVM as defined by the Boussinesq hypothesis (Eqs. \ref{eq:evm1} to \ref{eq:evm3}) is displayed in \fref{fig:evm_density_plots} in terms of density plots, where the eddy-viscosity field is not accounted yet.
The observed density distributions hint that a closure based on the eddy-viscosity hypothesis will most likely fail to correctly capture the SSR anisotropic tensor components.
This arises from the fact that no eddy-viscosity field (scalar) will be able map the resolved quantities to the different components of the closure tensor, as the density plot is significantly different between normal and shear components.

\begin{figure}[!ht]
\centering
\begin{subfigure}[t]{0.6\linewidth}
    \includegraphics[width=\linewidth]{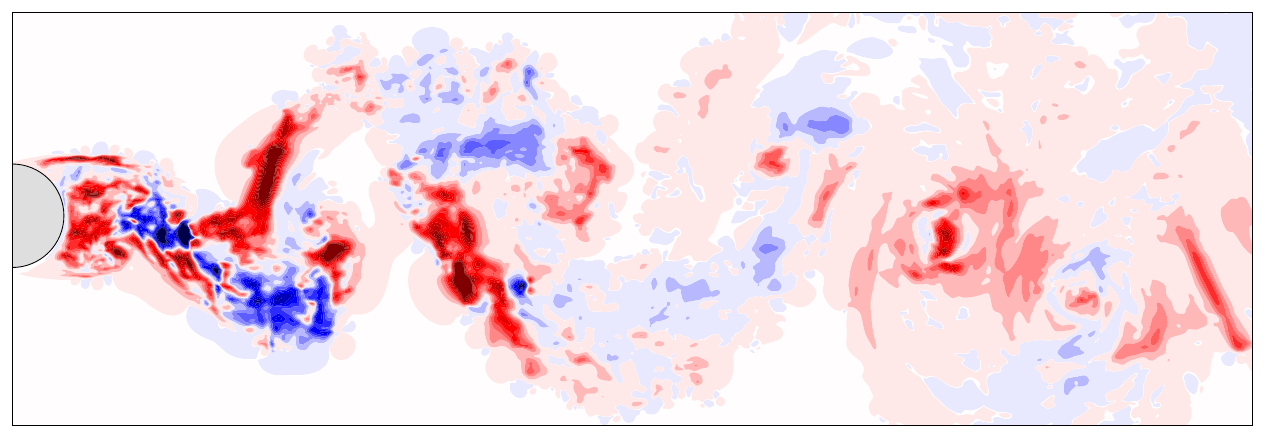}
    \caption{$\boldsymbol\tau^r_{11}$} 
\end{subfigure}
\begin{subfigure}[t]{0.6\linewidth}
    \includegraphics[width=\linewidth]{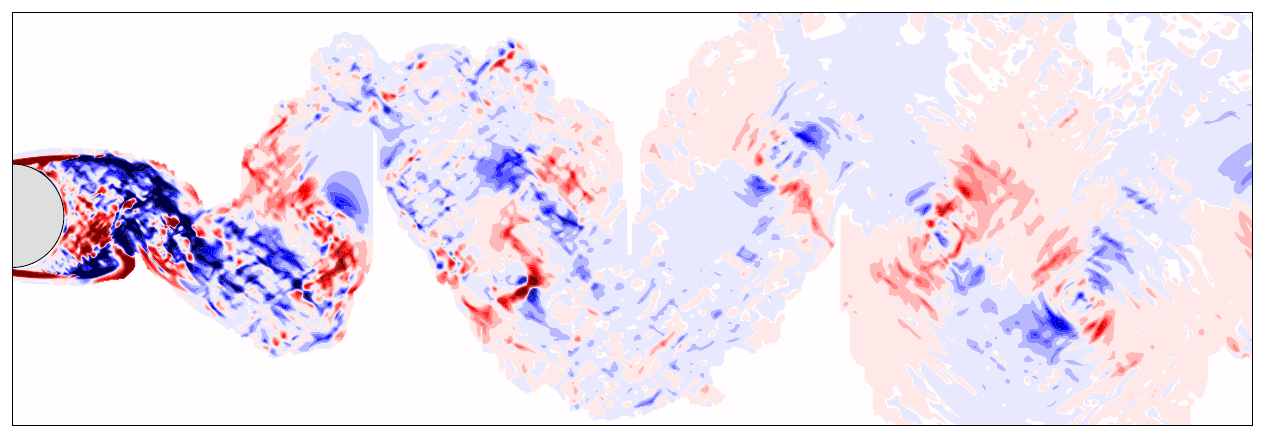}
    \caption{$\boldsymbol\tau^{r,\,\mathrm{EVM}}_{11}$}
\end{subfigure}
\begin{subfigure}[t]{0.6\linewidth}
    \includegraphics[width=\linewidth]{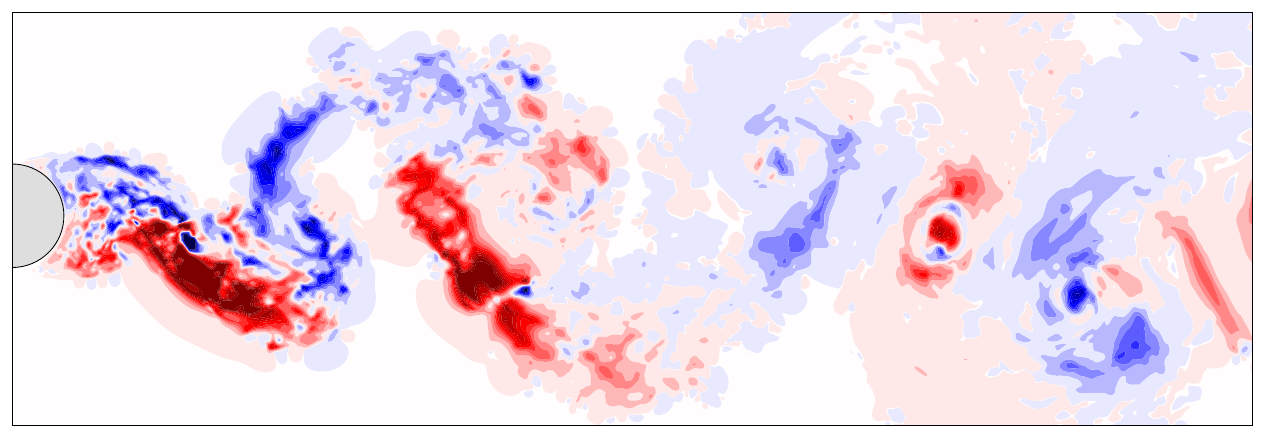}
    \caption{$\boldsymbol\tau^r_{12}$}
\end{subfigure}
\begin{subfigure}[t]{0.6\linewidth}
    \includegraphics[width=\linewidth]{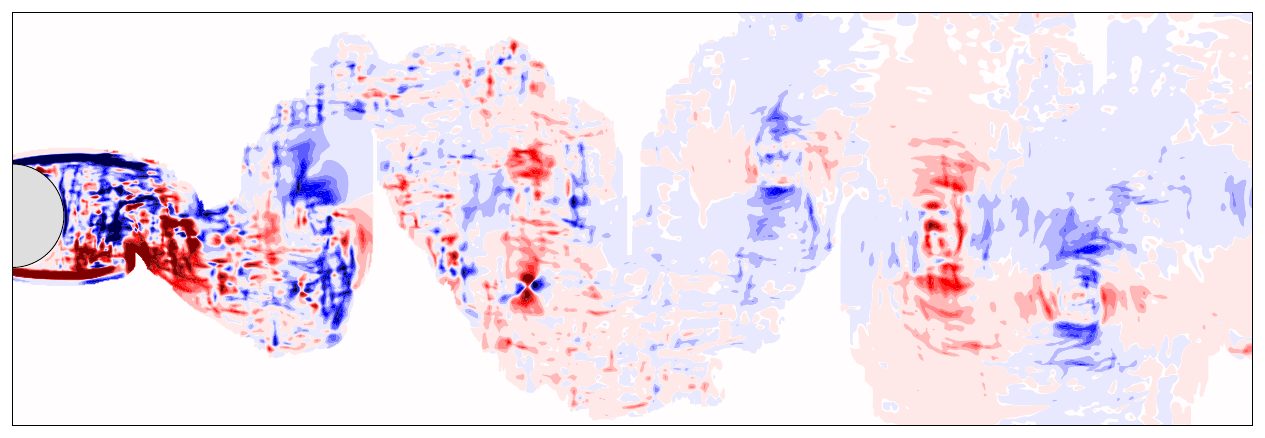}
    \caption{$\boldsymbol\tau^{r,\,\mathrm{EVM}}_{12}$}
\end{subfigure}
\begin{subfigure}[t]{0.6\linewidth}
    \includegraphics[width=\linewidth]{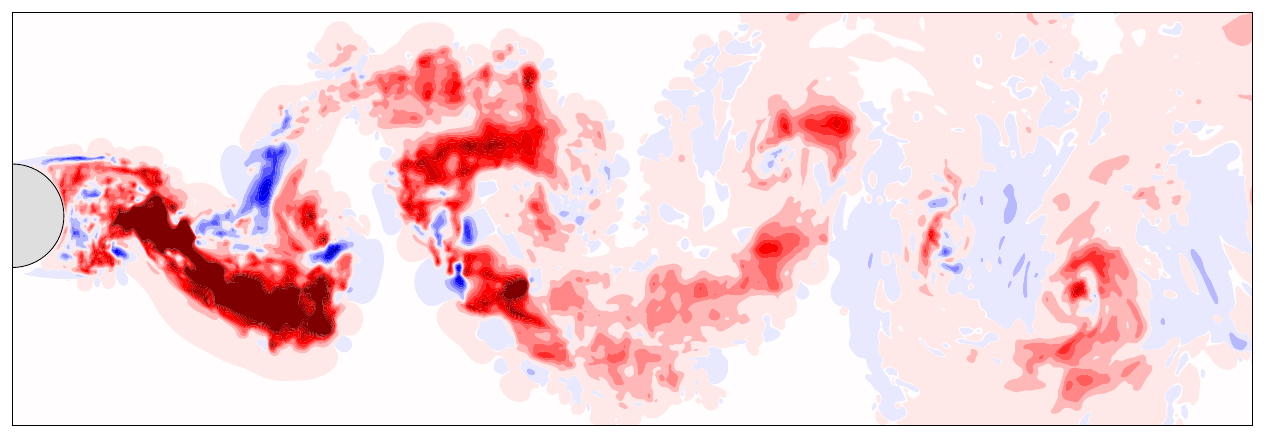}
    \caption{$\boldsymbol\tau^r_{22}$}
\end{subfigure}
\begin{subfigure}[t]{0.6\linewidth}
    \includegraphics[width=\linewidth]{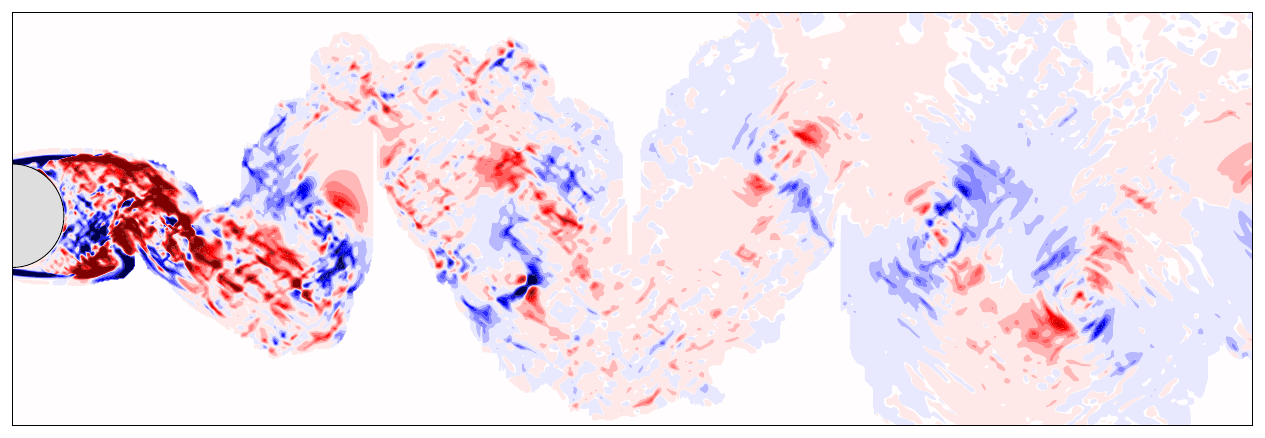}
    \caption{$\boldsymbol\tau^{r,\,\mathrm{EVM}}_{22}$}
\end{subfigure}
\caption{EVM predictions of the anisotropic SSR tensor components $(\boldsymbol\tau^{r,\,\mathrm{EVM}}_{ij})$ compared to reference data $(\boldsymbol\tau^{r}_{ij})$.}
\label{fig:EVM_a-priori}
\end{figure}

The a-priori results of the EVM with the Smagorinsky closure are shown in \fref{fig:EVM_a-priori}.
Almost no qualitative similarity with the target components of the SSR anisotropic tensor can be observed, thus evidencing the poor performance of the EVM on capturing the spanwise fluctuations.
This poor performance is quantified in \tref{tab:a-priori_EVM} using the Pearson correlation coefficient
\begin{equation}
\mathcal{CC}\pars{\boldsymbol\tau^r_{ij},\boldsymbol\tau^{r,\,\mathrm{EVM}}_{ij}}=\frac{\mathrm{cov}\pars{\boldsymbol\tau^r_{ij},\boldsymbol\tau^{r,\,\mathrm{EVM}}_{ij}}}{\sqrt{\mathrm{cov}\Big(\boldsymbol\tau^r_{ij},\boldsymbol\tau^r_{ij}\Big)\mathrm{cov}\pars{\boldsymbol\tau^{r,\,\mathrm{EVM}}_{ij},\boldsymbol\tau^{r,\,\mathrm{EVM}}_{ij}}}},
\end{equation}
where the superscript $(\cdot)^{\mathrm{EVM}}$ denotes the EVM prediction.
The low correlation values provide clear evidence that the EVM is not suited for the prediction of the spanwise stresses.

These poor results are ultimately expected since EVMs only account for the dissipative effects of turbulent fluctuations.
In the spanwise-averaged system, the spanwise stresses can have both a dissipative and energising physical effect, and this is a fundamental difference which no EVM can reproduce.

\begin{table}[t]
\centering
\caption{Correlation coefficients between components of the target anisotropic SSR tensor and predictions provided by the EVM.}
\begin{tabular}{lrrr}
\toprule
$\mathcal{CC}$ & $\boldsymbol\tau^r_{11}$ & $\boldsymbol\tau^r_{12}$ & $\boldsymbol\tau^r_{22}$ \\
\midrule
EVM &  0.06& 0.06 & 0.14\\
\bottomrule
\label{tab:a-priori_EVM}
\end{tabular}

{\footnotesize The correlation coefficients are calculated for 500 snapshots and the average values are provided. \par}
\end{table}

\clearpage
\section{Conclusion}

A flow decomposition based on the local spanwise average has been proposed  aiming to reduce the system dimensionality of flows presenting an homogeneous direction.
The SANS equations yield the spanwise-averaged solution of a 3-D flow in a 2-D system.
In practical terms, the SANS equations are equivalent to the incompressible 2-D Navier--Stokes equations with additional forcing terms representing the spanwise fluctuating part of the flow.

First, we examine the SANS equations in an analytical test case consisting in the superposition of a vortex tube and a straining field.
It is demonstrated that the SANS equations still include the vortex-stretching mechanism.
On the other hand, the SANS equations are greatly simplified when the flow is assumed to be spanwise periodic.
The vortex-stretching term plus other terms arising from averaging over a finite non-periodic interval vanish as a result of this constraint.

In spanwise-periodic conditions, the SANS equations only contain the divergence of the SSR tensor as a closure term.
In this sense, the perfect closure is presented as the difference between the 2-D Navier--Stokes spatial operator and the spanwise-averaged 3-D Navier--Stokes spatial operator.
The SANS equations are validated by incorporating the perfect closure term computed in a 3-D system into a 2-D simulation.
This allows to recover the unsteady spanwise-averaged solution of a Taylor--Green vortex and flow past a circular cylinder at $Re=10^4$.
The inclusion of the perfect closure radically changes the flow dynamics, where energy and enstrophy are no longer conserved in the inviscid limit.

Finally, an EVM has been investigated to provide closure to the SANS equations for the cylinder flow test case.
The a-priori analysis has shown that the Boussinesq hypothesis completely fails to capture the nature of the closure terms.
This underlines the need for a different SANS model bearing in mind that, differently from RANS or LES, general information on the physical laws governing the closure terms is still limited.
The possibility of directly computing the closure terms in 3-D systems can be exploited with the use of data-driven models such as machine-learning algorithms, as reviewed in the following chapter.
% ---------------------------------------------------------------- 

\chapter{A machine-learning model for SANS}

The modelling of the SANS closure terms is critical for the prediction of 3-D spanwise-averaged dynamics in 2-D systems.
An standard EVM, such as the Smagorinsky model, has failed to correctly predict the SANS closure terms because of if intrinsic dissipative and linear behaviour.
While other physical models allowing energy backscatter (energy transfer from small to large scales) such as the dynamic Smagorinsky model or a transport equation model could be explored in the future, in this chapter a data-driven model is proposed.
The limited knowledge on the spanwise stresses dynamics motivates this choice.
In particular, a supervised machine-learning (ML) model based on a deep convolutional neural network has been designed for this purpose.
A-priori results show up to 92\% correlation between target and predicted closure terms; more than an order of magnitude better than the EVM correlation.
Additionally, the ML model performance is assessed for a different flow configuration (Reynolds regime and body shape) to the training case and high correlation values are still captured.
The new SANS equations and ML closure model are also used for a-posteriori prediction.
Even though evidence of known stability issues with long-time ML predictions for dynamical systems is found for the present data-driven model, the closed SANS simulations are still capable of predicting wake metrics and induced forces with errors from 1-10\%.
This results in approximately an order of magnitude improvement over standard 2-D simulations while reducing the computational cost of 3-D simulations by 99.5\%.

\section{Introduction and literature review}

In recent years, machine learning, i.e. algorithms that reveal information by processing raw data inputs, have made an impact across numerous disciplines: from image recognition, to improved web searches, content filtering, natural language processing, among others \citep{LeCun2015}.
The fluid mechanics community has also adopted ML techniques both in experimental and computational work, specifically with the use of deep-learning methods \citep{Kutz2017}, where multiple layers of trainable parameters are stacked sequentially.
As data-driven models learn from observations, the vast amounts of data generated through field measurements and high-fidelity simulations has accelerated the application of ML models to fluid flow problems \citep{Brunton2020}. 

Supervised learning, or model optimisation based on labelled examples, has yielded the exploration of novel flow classification, control, and regression techniques.
Within classification and uncertainty quantification, detection of low RANS accuracy regions in the flow has been accomplished with the use of different ML strategies such as support vector machines and random trees \citep{Ling2015}.
Deep artificial neural networks (ANNs) have been employed for uncertainty quantification in RANS \citep{Geneva2019}, pointwise classification and blending of appropriate turbulence closures in LES \citep{Maulik2019}, and discovery of active flow control strategies such as drag reduction in turbulent channel flow \citep{Lee1997}.

Different control strategies have been developed using reinforcement learning which, in contrast to supervised learning, does not rely in labelled examples and the algorithm objective is to maximise a long-term reward signal based on self-proposed actions \citep{SuttonBarto2018}. 
Reinforcement learning has been used for (e.g.) active wake flow control \citep{Rabault2019a,Rabault2019b} and efficient control of collective swimmers \citep{Verma2018}.
Other unsupervised learning models such as generative adversarial networks \citep{Goodfellow2014} have been used for unsteady flow field generation \citep{Lee2019}.

The use of ML models for regression problems in fluid flows has also bloomed in the last few years. ANNs have been used for the reconstruction of turbulent spatio-temporal sampled data \citep{Deng2019}, physics-based discovery of governing equations parameters such as Reynolds number \citep{Raissi2019}, quantitative prediction of flow fields from scalar concentration snapshots \citep{Raissi2020}, and RANS modelling \citep{Duraisamy2015,Ling2016,Lapeyre2019}, among others.
Convolutional neural networks (CNNs) have also become a popular tool to solve regression problems.
CNNs have been used for turbulent flow field super-resolution \citep{Fukami2019,Liu2020}, modelling of surface pressure distribution from sparse wake flow data \citep{Ye2020}, nonlinear mode decomposition \citep{Murata2019}, and prediction of turbulent fields \citep{Lapeyre2019,Kim2020b,Kim2020a}, among others. 

Specifically within turbulence closures, CNNs have been applied in RANS \citep{Weatheritt2016} and LES \citep{Gamahara2017,Beck2019} frameworks, establishing its utility for revealing hidden correlations between system variables for which physical laws are, a-priori, unknown \citep{Brenner2019}.
In particular, CNNs have been developed to efficiently learn spatial structures in data arrays by exploiting the translational invariance inherent in convolutions \citep{LeCun1995}.
This makes them a very attractive tool for modelling novel and spatially-complex fields such as the new SANS closures.

With respect to the strip-based SANS framework, the general methodology of the data-driven ML model is based on the idea that training a SANS closure on a single local spanwise segment (see \fref{fig:strips}) automatically generalises to all the spanwise-averaged strips because of the system homogeneity in that direction.
Hence, the database required to train the model can be generated in a single 3-D simulation with a span equivalent to the local spanwise-averaging length.
For example, a structure of aspect ratio 100 can be reduced to 100 strips, or 2-D simulations, resolving the (1-diameter) SANS equations.
The required SANS closure model can be previously trained using data of a 3-D high-fidelity simulation of a 1-diameter-long spanwise-periodic cylinder, noting that the 3-D simulation is the most computationally expensive part of the method as detailed in \aref{chapter:appendixE}.
In this chapter, the ML SANS closure model is assessed for a single spanwise-averaged strip of 1 diameter.

\section{Machine-learning model} \label{sec:ML_model}

Data-driven closures are particularly helpful when physical information of the system to be modelled is not known a-priori, and this is certainly the case for the novel spanwise stresses developed in this work.
Hence, in this section we present a ML model for the spanwise fluctuations thus providing closure to the SANS equations.

\begin{figure}[t]
\centering
\includegraphics[width=0.95\linewidth]{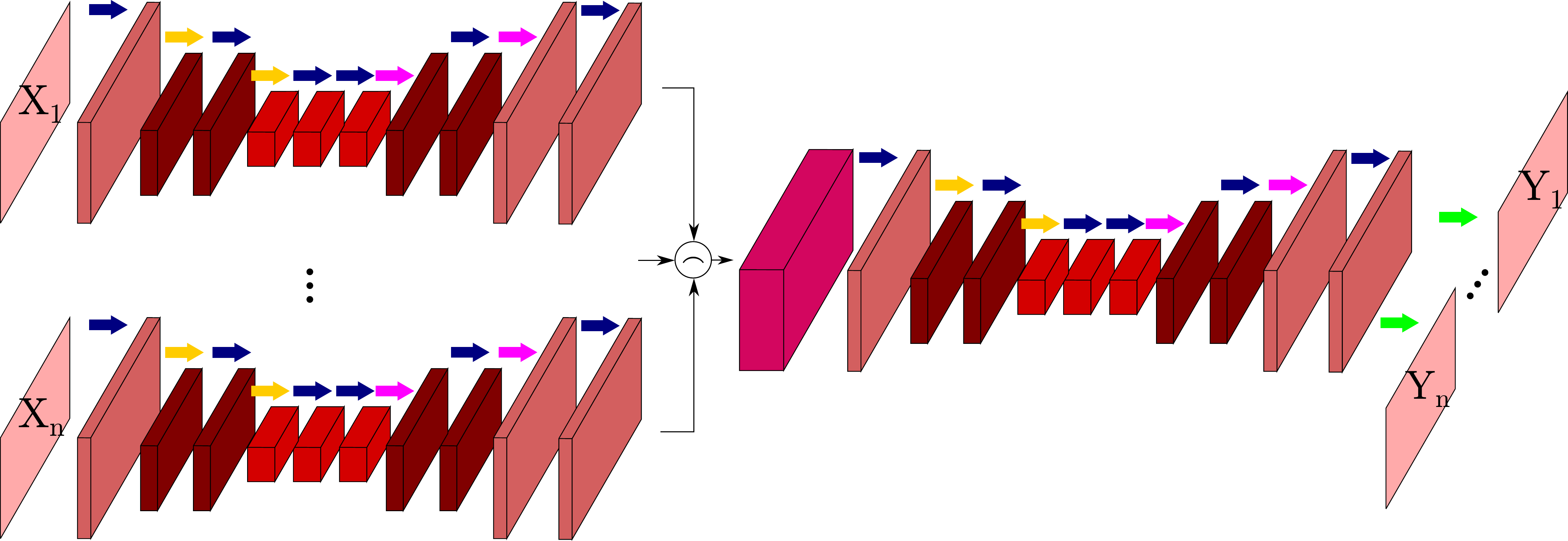}\vspace{-0.1cm}
\caption{CNN architecture.
An encoding-decoding branch is considered for each input field.
The input branches are then concatenated $(\frown)$ and another encoding-decoding branch follows.
The closure terms are separately extracted on the last layer.
Arrows legend: Dark blue: Conv2D $5\times5$ $+$ batch normalisation $+$ activation function.
Yellow: MaxPooling $2\times2$.
Magenta: UpSampling $2\times2$.
Green: Conv2D $1\times1$ $+$ wake mask $+$ Gaussian filter.
Block thickness indicates the number of filters, which double/halve at each convolutional operation in the encoding/decoding parts.}
\label{fig:cnn2}
\end{figure}

A CNN is proposed to map resolved quantities, $\mathrm{X}_n=\left\lbrace U,V,P\right\rbrace$, to closure terms, $\mathrm{Y}_n$.
The CNN architecture is depicted in \fref{fig:cnn2}.
The closure terms of the a-priori analysis are the SSR tensor components (similarly to \sref{sec:evm}), $\mathrm{Y}_n=\left\lbrace \avg{u\p u\p},\avg{u\p v\p},\avg{v\p v\p}\right\rbrace$.
On the other hand, the perfect closure is targeted as the output of the a-posteriori analysis, $\mathrm{Y}_n=\lbrace \mathcal{S}^R_x,\mathcal{S}^R_y\rbrace$, in order to improve the model stability as will be discussed in \sref{sec:ML_a-post}.

The ML model architecture is based on a multiple-input multiple-output CNN (MIMO-CNN). 
The need of a high-dimensional tunable function that can reveal hidden correlations in spatial data motivates the CNN choice \citep{LeCun1989}.
The same model architecture is used for both a-priori and a-posteriori analysis, where only the number of outputs changes from three to two, respectively.
Each input branch is composed of an encoding part and a decoding part.
The number of filters doubles at every convolutional layer in the encoding part, reaching a maximum of 64 filters, and halves in the decoding part.
Each convolutional layer consists of a 2-D convolution operation with a $5\times5$ filter (convolution kernel), a batch normalisation operation, and a rectified linear unit (ReLU) activation function.
The total number of trainable parameters is 944,739.
The batch normalisation helps to speed up the neural network learning rate and improves its stability with respect to the random weights initialisation \citep{Ioffe2015}. 
The ReLU function allows higher learning rates and handles vanishing-gradient problems better than other standard activation functions such as the Sigmoid or the $\tanh$ functions \citep{Nair2010}.
All input branches are then concatenated into a single branch, for which another encoding-decoding branch follows.
The last layer consists of a wake mask function and a Gaussian filter.

\begin{figure}[!ht]
\centering
\begin{subfigure}{0.7\linewidth}
 	\includegraphics[width=\linewidth]{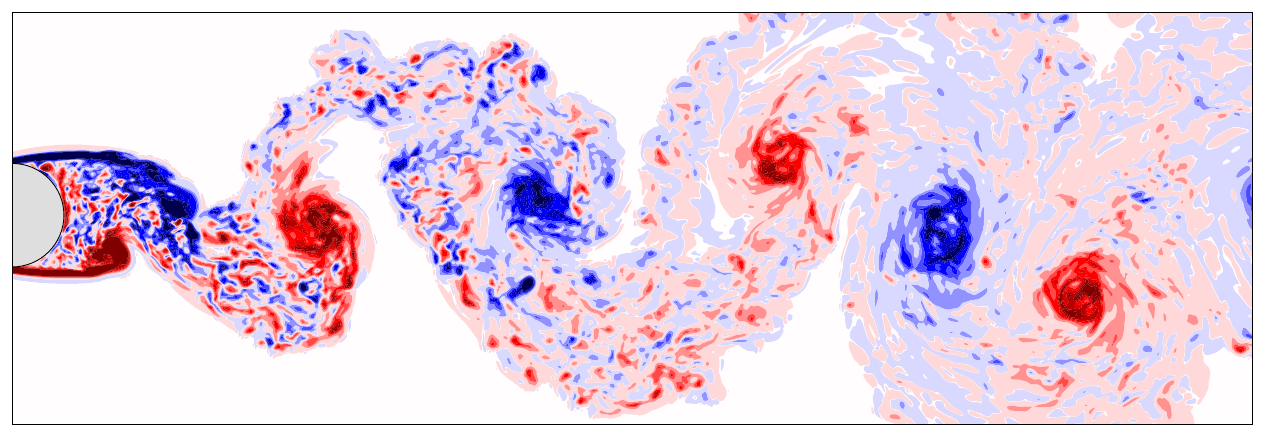}
 	\caption{$\Omega_z$} 
\end{subfigure}
\begin{subfigure}{0.7\linewidth}
    \includegraphics[width=\linewidth]{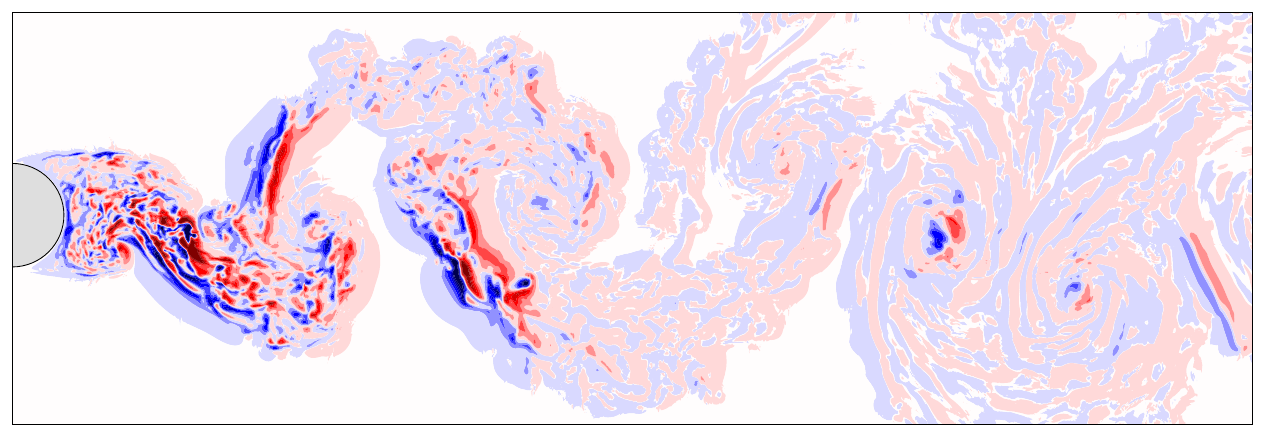}
	\caption{$\mathcal{S}^R_x$}
\end{subfigure}
\begin{subfigure}{0.7\linewidth}
	\includegraphics[width=\linewidth]{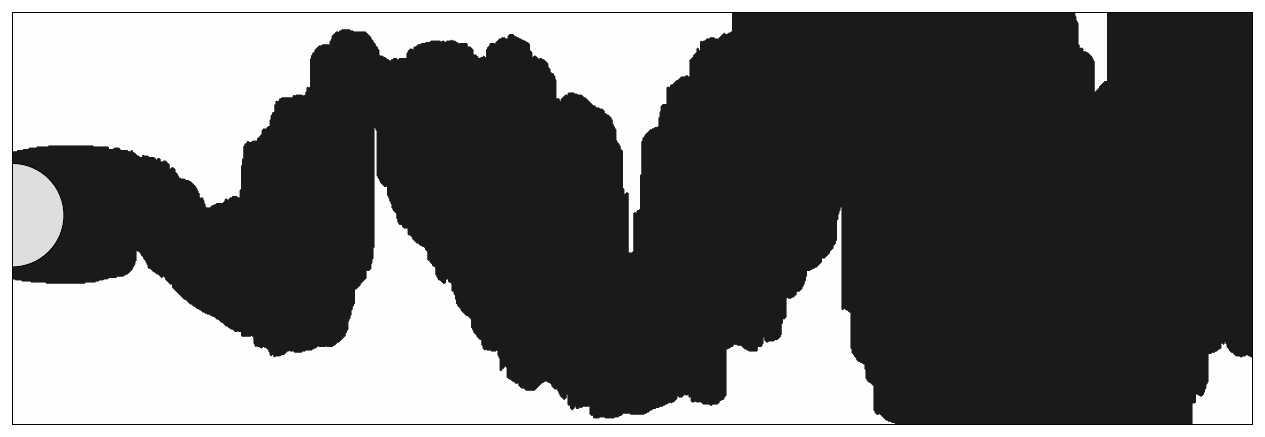}
	\caption{Wake mask used in the a-priori analysis.}\label{fig:wake1}
\end{subfigure}
\begin{subfigure}{0.7\linewidth}
	\includegraphics[width=\linewidth]{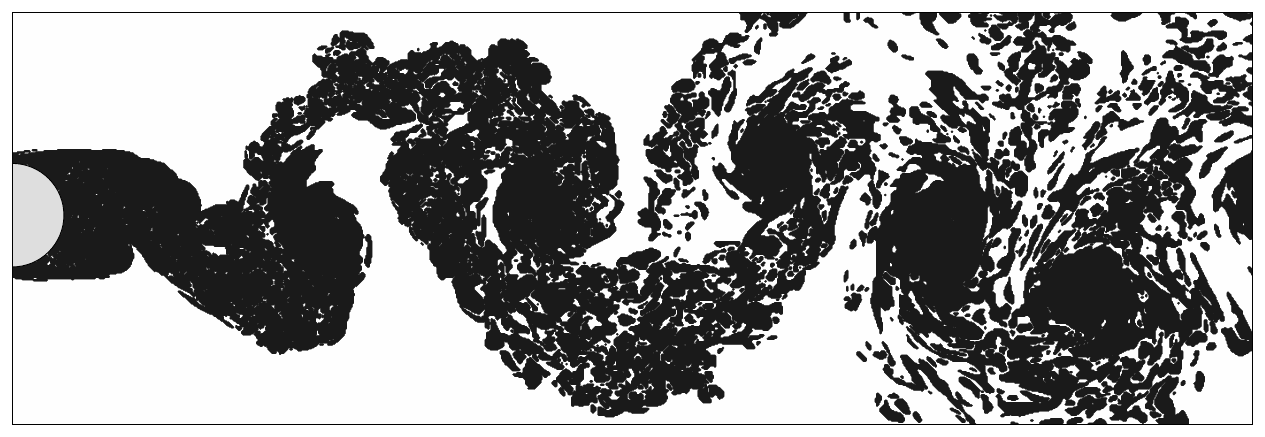} 
	\caption{Wake mask used in the a-posteriori analysis.}\label{fig:wake2}
\end{subfigure}
\caption{A wake mask from the vorticity field is generated to focus the ML model outputs on the non-trivial wake region alone. The aim is to match the region where the target output $(\mathcal{S}^R_x)$ is activated. In the a-posteriori analysis, the far-wake region is not fully activated in order to aid the closure stability.}
\end{figure}

The wake mask helps providing a clean output in the regions of the domain where the closure terms are negligible.
The wake mask is based in the vorticity field for which the condition $|\Omega_z|>\epsilon$ is applied.
Hence, a field of 1s (and 0s) is generated when the condition is met (or not).
The wake mask for the a-priori stage considers all the points within the wake limit.
The wake limit is activated using the $\epsilon=1\cdot10^{-3}$ threshold (\fref{fig:wake1}).
This helps the model focusing on the non-trivial region during optimisation while still providing some areas within the wake width where the closure term is not active.
The wake mask for the a-posteriori stage applies the criterion in a pointwise manner, hence points within the wake region might not be activated, particularity in the far wake region (\fref{fig:wake2}).
Also, the criterion is more strict by setting $\epsilon=3.5\cdot10^{-3}$.
This is enforced to improve the stability of the a-posteriori analysis so longer temporal dynamics can be evaluated.
Finally, a Gaussian filter of $\sigma=0.5$ (resulting in a $5\times5$ kernel matching the feature maps size) is applied to the output of the wake detection layer.

A detailed hyper-parametric study of the ML model has been conducted for the circular cylinder case (similarly to \sref{sec:SANS_circular_cylinder}) in \cite{Font2020a}, and is also included in \aref{chapter:appendixD}.
In short, the present model architecture was compared against two different MIMO-CNNs, and this architecture offered the best accuracy among the tested models.
Also, it was found that the larger the CNN (in terms of trainable parameters), the faster high correlation coefficients were obtained for a similar number of training epochs (where epoch refers to an optimisation iteration on the training dataset).
The number of trainable parameters was modified by changing the number of filters per layer.
The sum of the squared error as a loss function provided the best performance when compared to the sum of the absolute error.
The use of primitive variables (velocity and pressure) did not yield significant differences when compared to the velocity gradient tensor components set. 
Both these input sets showed a better performance than a third input set containing the vorticity field and the distance function to the cylinder.

Unsteady turbulent flow past a circular cylinder with a 1 diameter span $(L_z=1)$ at $Re=10^4$ (as considered in \sref{sec:SANS_circular_cylinder}) has been used to generate a dataset containing a total of 7000 samples of $1216\times540$ size during $\Delta t^*=1650$ and using a sampling rate of $\delta t^*=0.25$.
The 1 diameter span is sufficiently long to sustain 3-D turbulence on a meaningful wake region and it is also a sensible choice of the minimum structural mode wavelength that would need to be resolved in a SANS-based strip-theory framework.
The input fields are normalised with the standard score, i.e. $\hat{x} = (x-\overline{x})/\sigma_x$.
As in \cite{Beck2019, Kim2020a}, a single $Re$ regime is used in the dataset generation, which is deemed sufficient because of the turbulent nature and unsteadiness of the flow.
The constant energy transfer rate across different scales in the inertial subrange of turbulence implies that closures trained on one $Re$ should generalise reasonably well to unseen $Re$ as long as the mean flow characteristics are similar.
Such study is conducted in \sref{sec:generalisation}.

The dataset is split with 6000 samples for optimisation (training), 500 samples for validation during training, and 500 for evaluation (testing) purposes.
The dataset split is chosen as often encountered in the machine learning literature, probably adopted from power law (or Pareto) distributions \citep{Newman2005}.
To ensure the closure's ability to generalise and avoid overfitting, the time windows for testing, optimisation and validation do not overlap, i.e. the closure is assessed on a continuous window of previously unseen flow.
A mini-batch stochastic gradient descent method based on the Adam optimiser \citep{Kingma2014} is used to update the network weights during training.
The sum of the squared error function is used for optimisation
\begin{equation}
\mathrm{Loss}=\sum^m_i\sum^n_j \pars{\mathrm{Y}_{i,j}-\mathrm{Y}_{i,j}^{\mathrm{ML}}}^2,
\end{equation}
where $m$ is the size (number of samples) of the mini-batch, $n$ is the output index, and the superscript $(\cdot)^{\mathrm{ML}}$ denotes the ML model prediction.
When the validation error does not decrease for 5 epochs or a total of 60 epochs is reached, training is stopped (a.k.a. early-stopping).

\begin{figure}[!t]
\setlength{\columnsep}{-1pt} 
\begin{multicols}{2}
\centering
\includegraphics[width=1.02\linewidth]{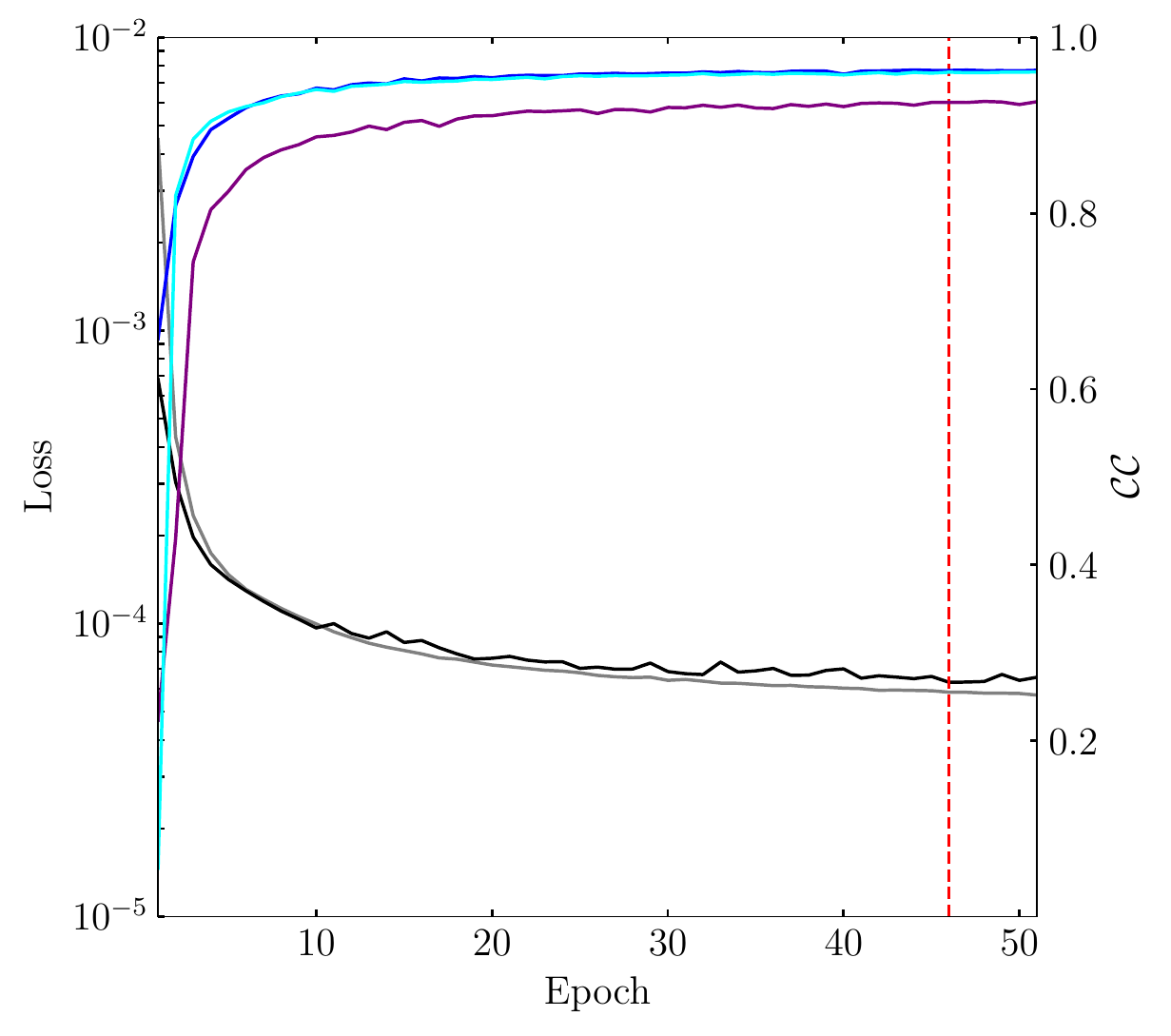}
\caption{Training of the $\mathrm{Y}_n=\big\lbrace\langle u' u'\rangle,\langle u' v'\rangle,\langle v' v'\rangle\big\rbrace$ output set.
The vertical red dashed line indicates the early stop.
Grey: combined training loss.
Black: combined validation loss.
Blue: $\langle u' u'\rangle$ correlation coefficient.
Purple: $\langle u' v'\rangle$ correlation coefficient.
Cyan: $\langle v' v'\rangle$ correlation coefficient.}\label{fig:ab_history}

\includegraphics[width=1.02\linewidth]{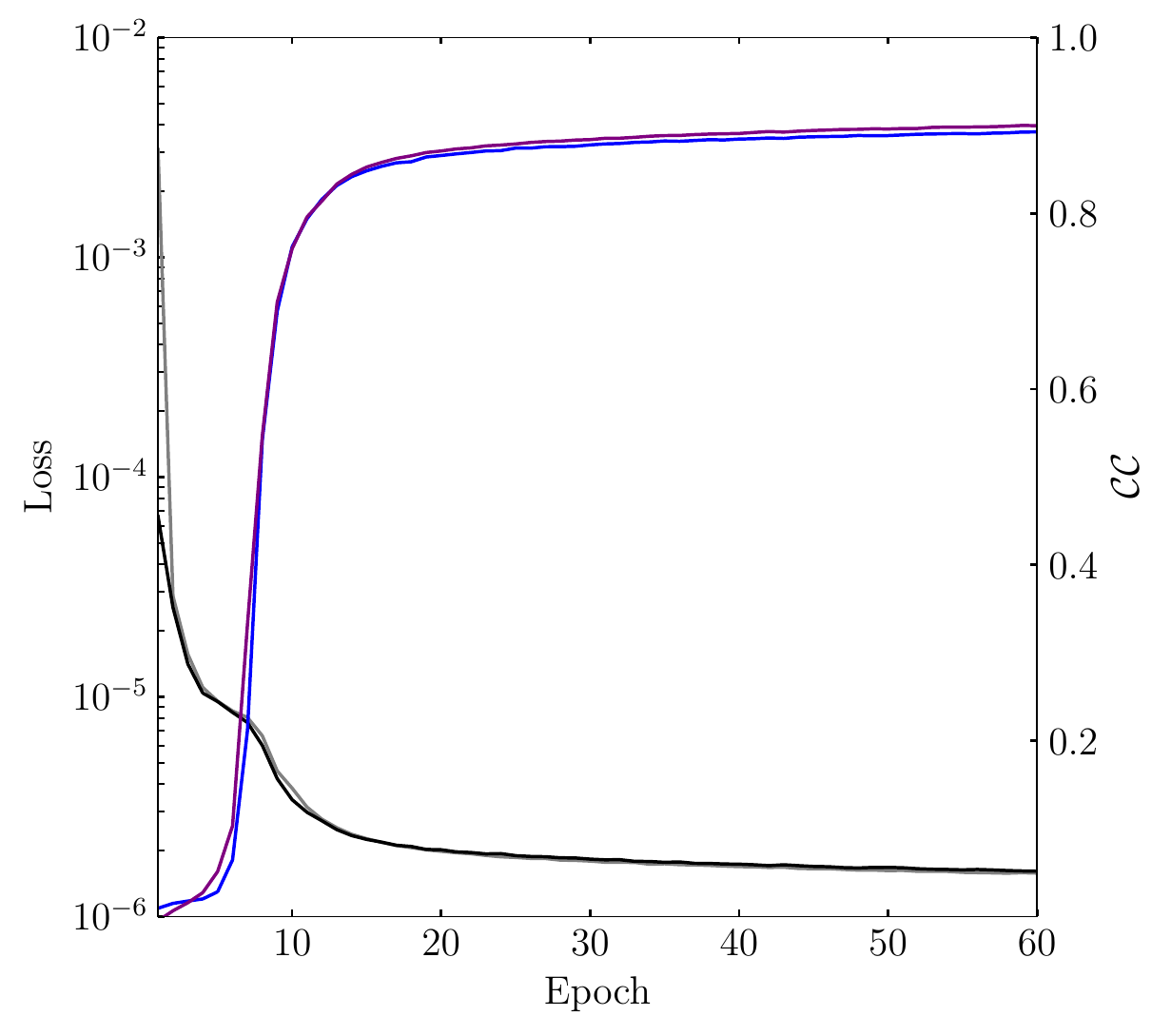}
\caption{Training of the $\mathrm{Y}_n=\big\lbrace \mathcal{S}^R_x, \mathcal{S}^R_y \big\rbrace$ output set.
Grey: combined training loss.
Black: combined validation loss.
Blue: $\mathcal{S}^R_x$ correlation coefficient.
Purple: $\mathcal{S}^R_y$ correlation coefficient.}
\label{fig:SR_history}
\end{multicols}
\end{figure}

The ML model training history of the SSR tensor components is plotted in \fref{fig:ab_history}, where a rapid convergence cap be appreciated.
The early-stopping criterion is triggered on the 46th epoch since it provides the lowest validation error not improved for the next 5 epochs.
On the other hand, the training history of the perfect closure components is plotted in \fref{fig:SR_history}.
In this case, the early-stopping criterion is not triggered and optimisation is performed during 60 epochs (the maximum defined).

\section{Results}

\subsection{A-priori analysis}

To allow a comparison with the EVM, the anisotropic part of the SSR tensor is recovered from the ML model predictions of the full SSR tensor.
The fact that the ML model yields the full SSR tensor already provides an advantage with respect to the EVM since no further modifications of the governing equations are required, nor an additional transport equation for the SSR kinetic energy.

\begin{figure}[!ht]
\centering
\begin{subfigure}[t]{0.6\linewidth}
    \includegraphics[width=\linewidth]{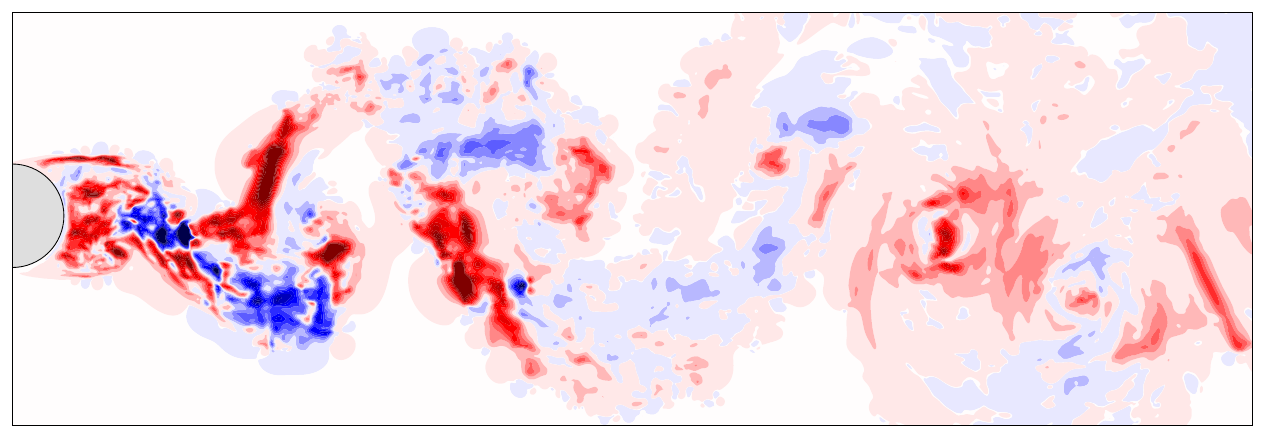}
    \caption{$\boldsymbol\tau^r_{11}$} 
\end{subfigure}
\begin{subfigure}[t]{0.6\linewidth}
    \includegraphics[width=\linewidth]{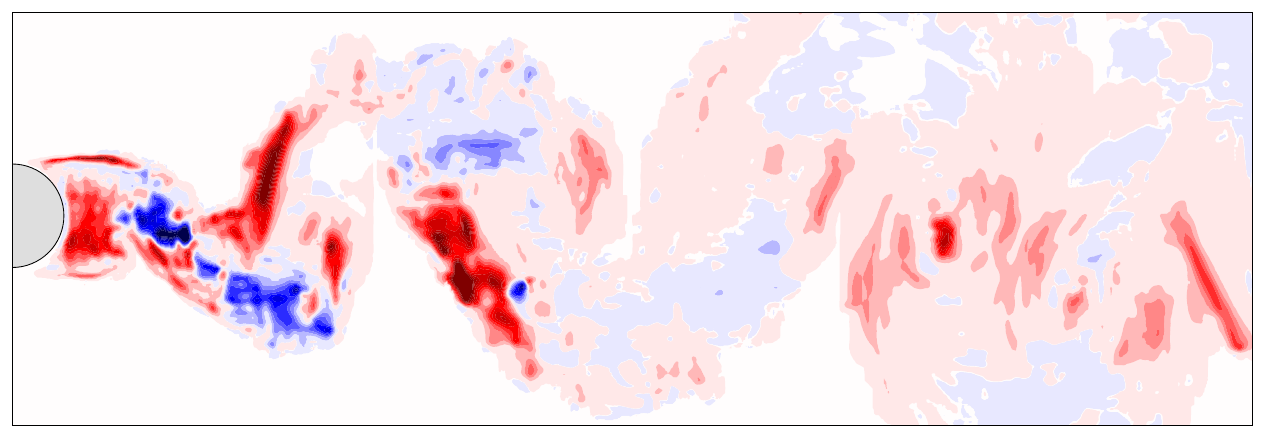}
    \caption{$\boldsymbol\tau^{r,\,\mathrm{ML}}_{11}$}
\end{subfigure}
\begin{subfigure}[t]{0.6\linewidth}
    \includegraphics[width=\linewidth]{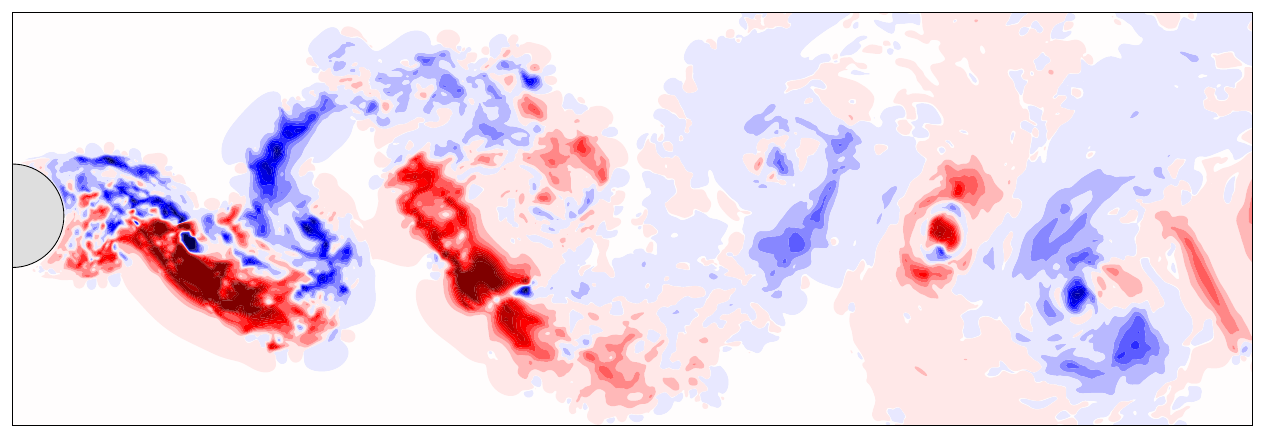}
    \caption{$\boldsymbol\tau^r_{12}$}
\end{subfigure}
\begin{subfigure}[t]{0.6\linewidth}
    \includegraphics[width=\linewidth]{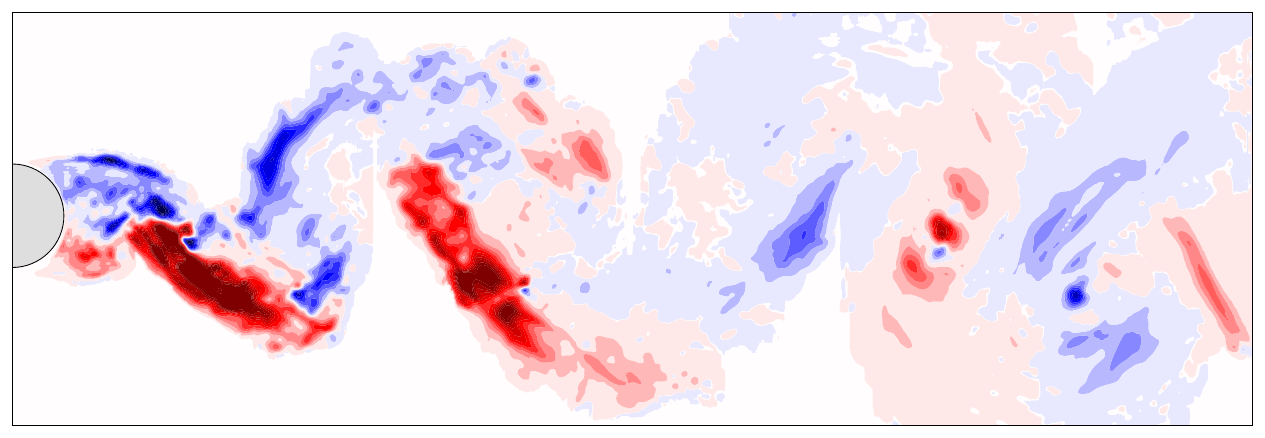}
    \caption{$\boldsymbol\tau^{r,\,\mathrm{ML}}_{12}$}
\end{subfigure}
\begin{subfigure}[t]{0.6\linewidth}
    \includegraphics[width=\linewidth]{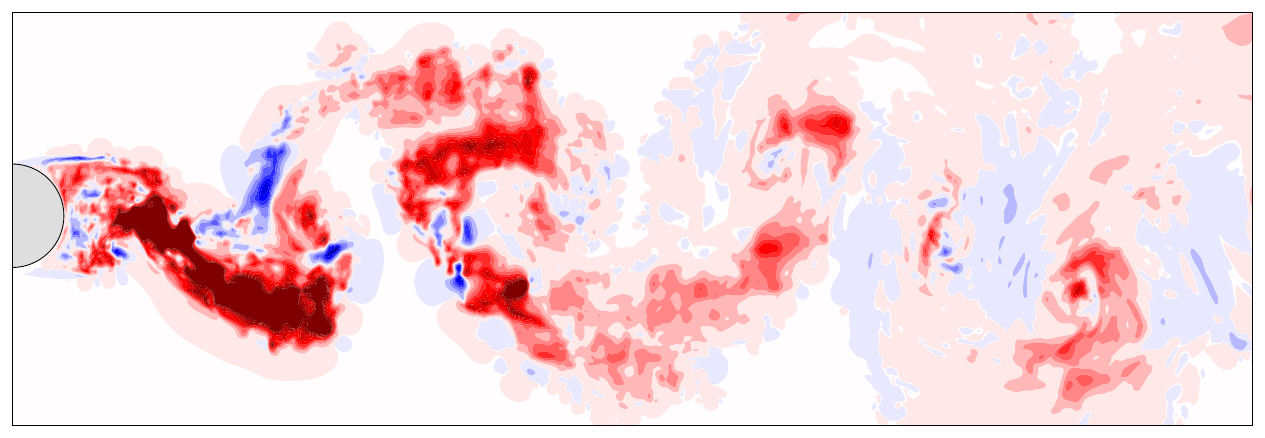}
    \caption{$\boldsymbol\tau^r_{22}$}
\end{subfigure}
\begin{subfigure}[t]{0.6\linewidth}
    \includegraphics[width=\linewidth]{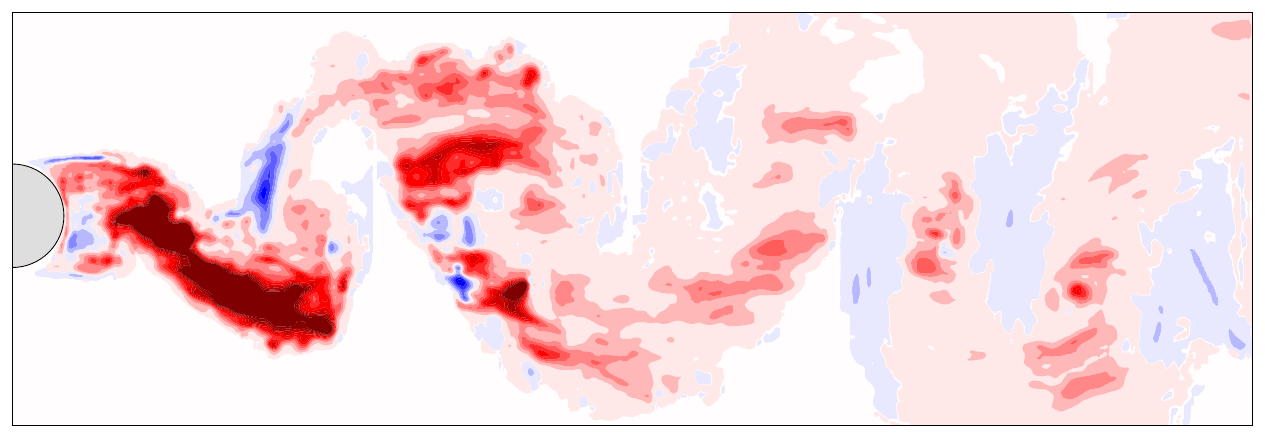}
    \caption{$\boldsymbol\tau^{r,\,\mathrm{ML}}_{22}$}
\end{subfigure}
\caption{ML model prediction of the anisotropic SSR tensor components compared to reference data.}
\label{fig:ML_a-priori}
\end{figure}

\begin{figure}[!ht]
\centering
\begin{subfigure}[t]{0.6\linewidth}
    \includegraphics[width=\linewidth]{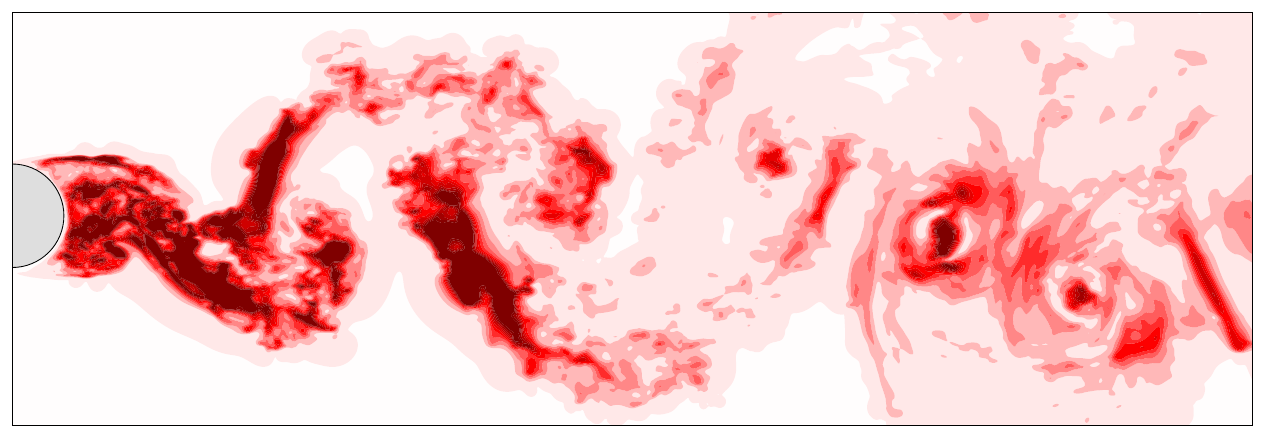}
    \caption{$\avg{u\p u\p}$} 
\end{subfigure}
\begin{subfigure}[t]{0.6\linewidth}
    \includegraphics[width=\linewidth]{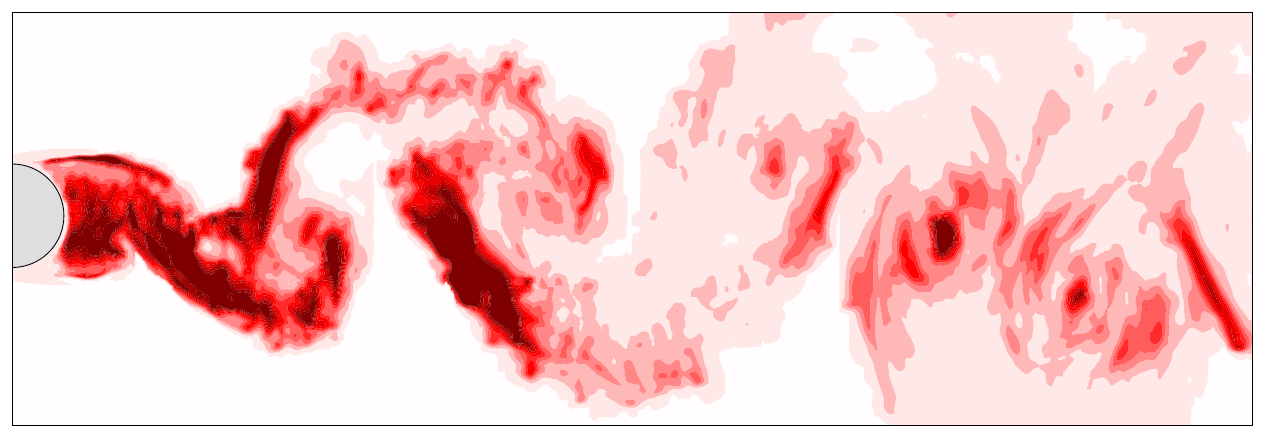}
    \caption{$\avg{u\p u\p}^{\mathrm{ML}}$}
\end{subfigure}
\begin{subfigure}[t]{0.6\linewidth}
    \includegraphics[width=\linewidth]{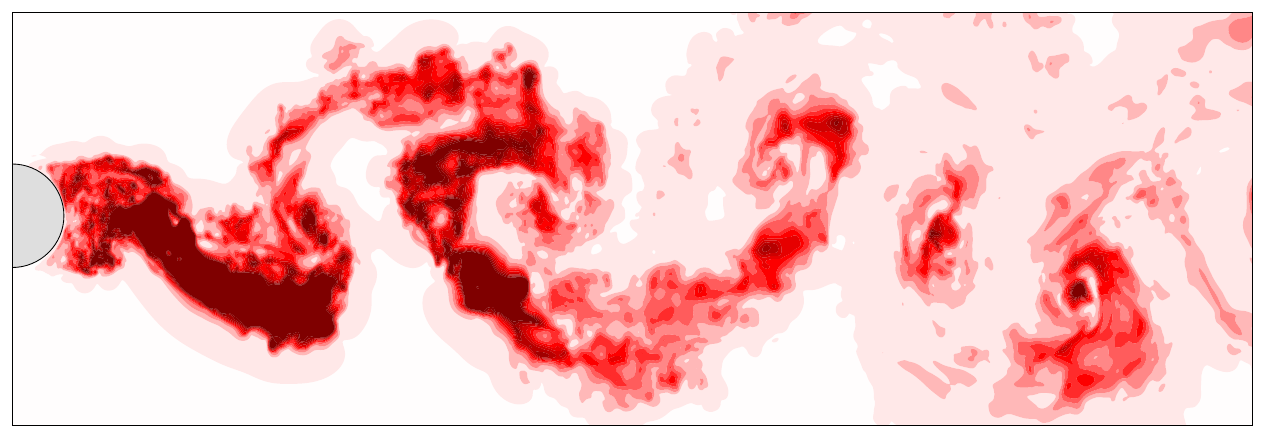}
    \caption{$\avg{v\p v\p}$}
\end{subfigure}
\begin{subfigure}[t]{0.6\linewidth}
    \includegraphics[width=\linewidth]{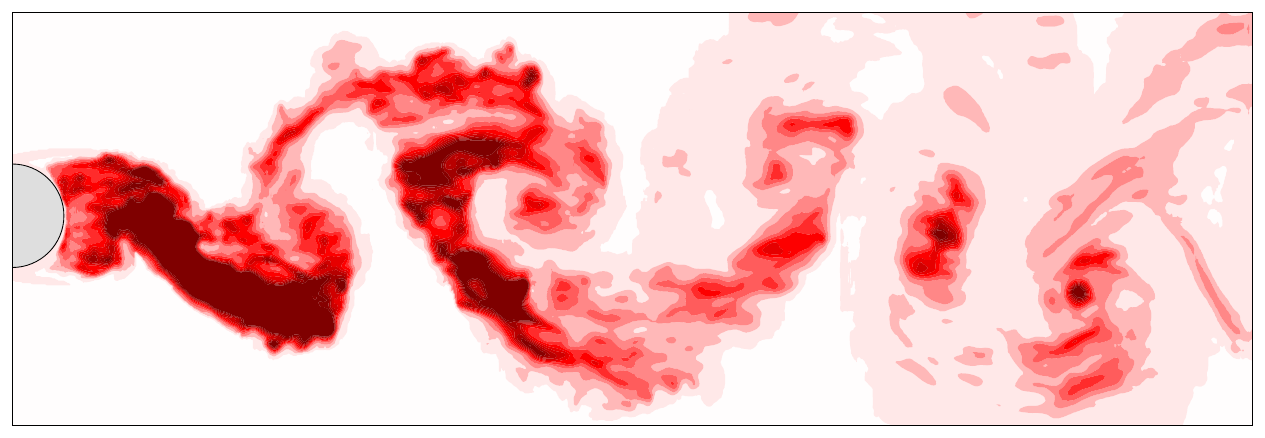}
    \caption{$\avg{v\p v\p}^{\mathrm{ML}}$}
\end{subfigure}
\caption{ML model prediction of the full SSR tensor components compared to reference data.}
\label{fig:ML_ab_a-priori}
\end{figure}

The predicted anisotropic SSR tensor components $(\boldsymbol\tau^{r,\,\mathrm{ML}}_{11}, \boldsymbol\tau^{r,\,\mathrm{ML}}_{12}, \boldsymbol\tau^{r,\,\mathrm{ML}}_{22})$ are depicted in \fref{fig:ML_a-priori}, and the predicted components of the full SSR tensor $(\avg{u\p u\p}^{\mathrm{ML}}, \avg{v\p v\p}^{\mathrm{ML}})$ are depicted in \fref{fig:ML_ab_a-priori}.
It can be appreciated that large-scale structures are overall correctly captured.
On the other hand, small-scale structures, such as those encountered in near-wake region, are more difficult to predict.
Similar issues have been found for other CNN-based data-driven models \citep{Lee2019}.
A Gaussian filtering operation together with a wake detection feature aids the CNN to provide less noisy predictions focused on the wake region alone.
This helps mitigating error propagation in an a-posteriori framework (see next section), however, it can have an impact on the small-scale structures as well.

A significant improvement is qualitatively observed when comparing the predictions against the EVM, where the latter fails to capture almost all flow structures except for the largest ones (\fref{fig:EVM_a-priori}).
Furthermore, correlation values above $90\%$ have been achieved for all the anisotropic SSR tensor components (\tref{tab:ML_a-priori}), as well as the full SSR tensor components (\tref{tab:ML_ab_a-priori}).
It can be concluded that the ML model provides a significantly better prediction of the closure terms compared to the EVM, which is critical for SANS to correctly model turbulence wake dynamics.

\begin{table}[t]
\centering
\caption{Correlation coefficients between components of the anisotropic SSR tensor and predictions provided by the EVM and the ML model.}
\begin{tabular}{lrrr}
\toprule
$\mathcal{CC}$ & $\boldsymbol\tau^r_{11}$ & $\boldsymbol\tau^r_{12}$ & $\boldsymbol\tau^r_{22}$ \\
\midrule
EVM &  0.06& 0.06 & 0.14\\
ML &  0.90   & 0.92 & 0.92\\
\bottomrule \label{tab:ML_a-priori}
\end{tabular}
\end{table}

\begin{table}[t]
\centering
\caption{Correlation coefficients between target SSR tensor components and ML model predictions.}
\begin{tabular}{lrrr}
\toprule
$\mathcal{CC}$ & $\avg{u\p u\p}$ & $\avg{u\p v\p}$ &$\avg{v\p v\p}$  \\
\midrule
ML & 0.95 & 0.92 & 0.96\\
\bottomrule \label{tab:ML_ab_a-priori}
\end{tabular}

{\footnotesize The correlation coefficients are calculated for the 500 snapshots of the test dataset and the average values are provided. \par}
\end{table}

\clearpage

\subsection{Generalisation across flow regimes and geometries}\label{sec:generalisation}

So far, it has been demonstrated that the trained ML model can correctly predict the SANS closure terms of previously unseen snapshots of the cylinder case, i.e. the model has correctly generalised in time.
Here, the generalisation of the ML model to other flow configurations such as body shapes and Reynolds regimes is assessed.
This analysis is important because, first, it provides information on whether training the ML model in a fixed set-up can still be useful for the prediction of the SANS stresses in a different case.
Second, it is a good test to check that the ML model has not over-fitted the training data even with the use of the early-stopping criterion.

Four different cases are studied:
Flow past a circular cylinder at $Re=1000$.
Flow past a circular cylinder at $Re=3900$.
Flow past an ellipse with axis $(x,y)=(2D,D)$ (ellipse 1) at $Re=10^4$.
Flow past an ellipse with axis $(x,y)=(0.5D,D)$ (ellipse 2) at $Re=10^4$.
Similarly to the circular cylinder case at $Re=10^4$, a test dataset of 500 snapshots has been generated during $\Delta t^*=125$ time units after reaching a statistically stationary state of the wake for all cases.
Both output target sets, i.e. the full SSR tensor components and the perfect closure components, are analysed.

\begin{figure}[t]
\centering
\includegraphics[width=0.75\linewidth]{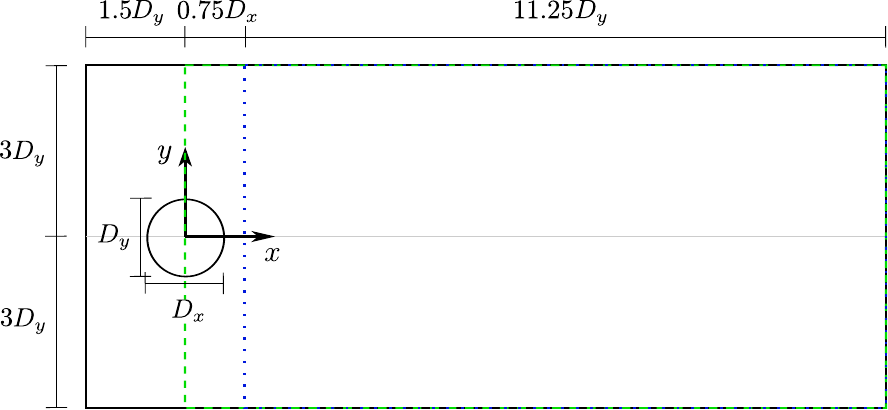}
\caption{Regions used for the ML model analysis.
Black: Input and output region of the ML model. 
Green dashed: Region used to compute the correlation coefficient between ML predictions and target data (full wake).
Blue dotted: Region used to compute the correlation coefficient between ML predictions and target data omitting the near-body domain (downstream wake).}
\label{fig:cc_regions}
\end{figure}

\begin{figure*}[!ht]
\begin{multicols}{2}
\begin{subfigure}{\linewidth}
    \includegraphics[width=\linewidth]{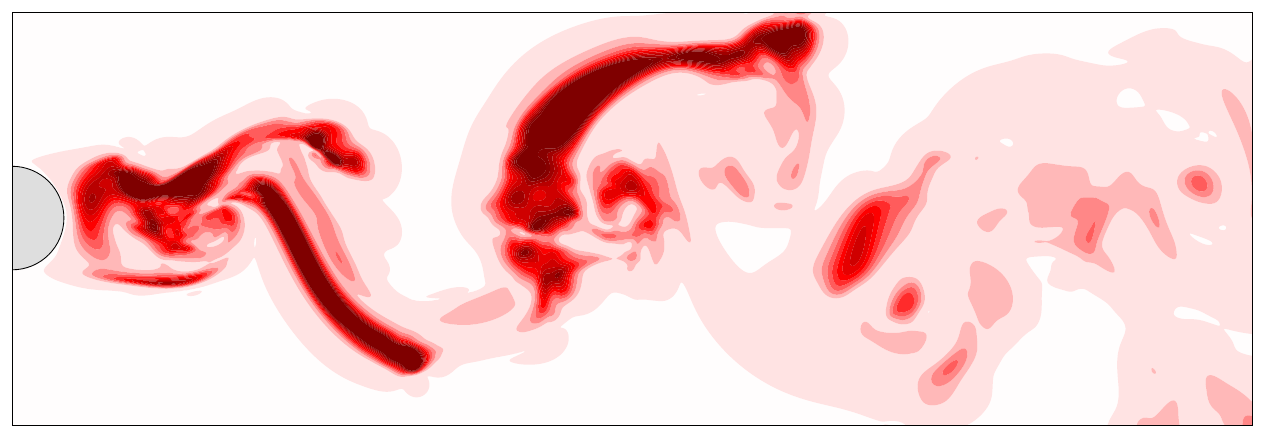}
    \caption{$\avg{u\p u\p}$} 
\end{subfigure}
\begin{subfigure}{\linewidth}
    \includegraphics[width=\linewidth]{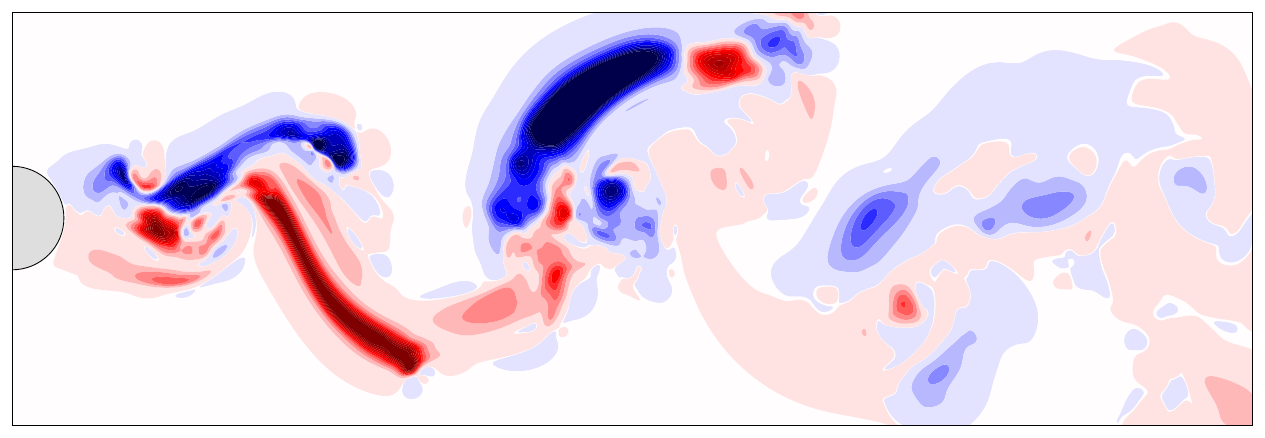}
    \caption{$\avg{u\p v\p}$}
\end{subfigure}
\begin{subfigure}{\linewidth}
    \includegraphics[width=\linewidth]{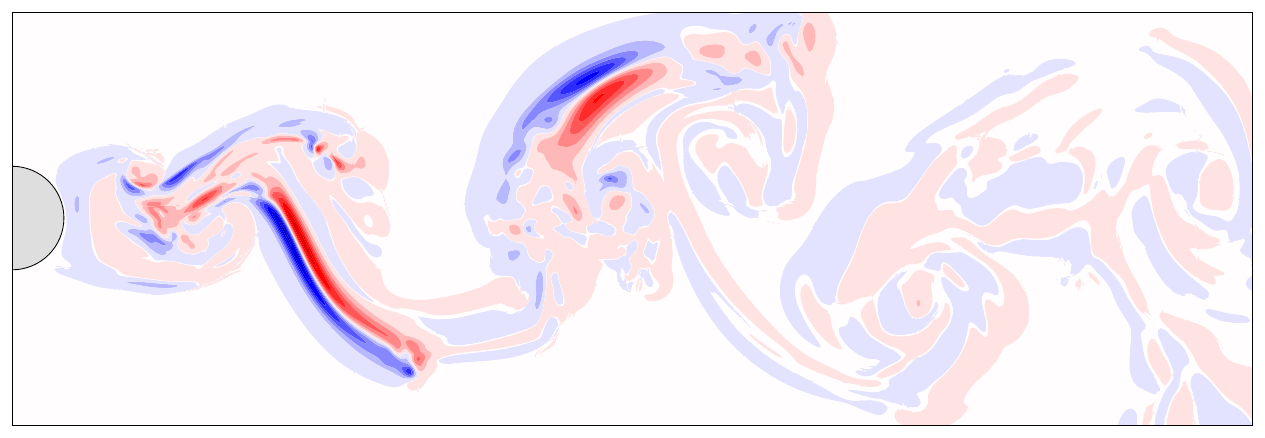}
    \caption{$\mathcal{S}^R_x$}
\end{subfigure}
\begin{subfigure}{\linewidth}
    \includegraphics[width=\linewidth]{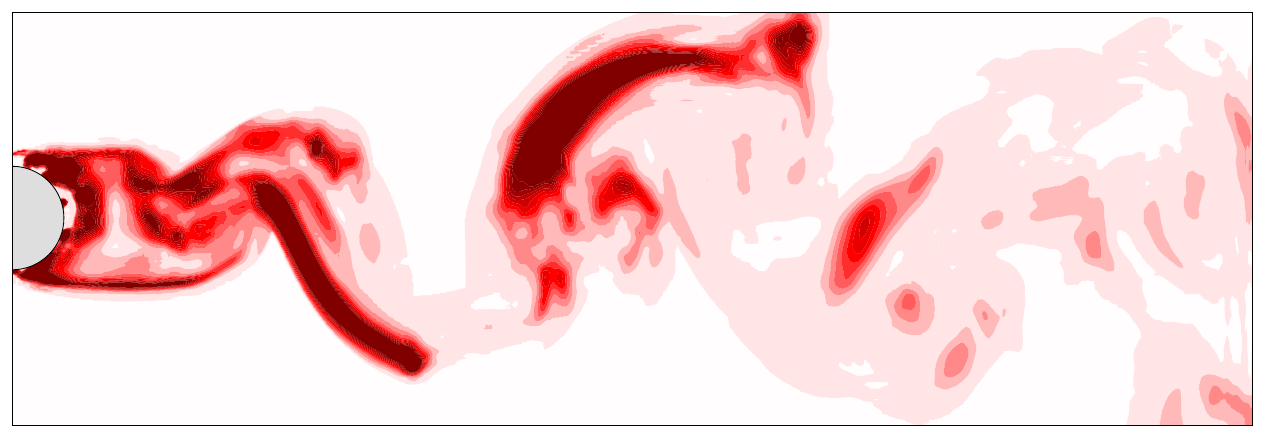}
    \caption{$\avg{u\p u\p}^{\mathrm{ML}}$} 
\end{subfigure}
\begin{subfigure}{\linewidth}
    \includegraphics[width=\linewidth]{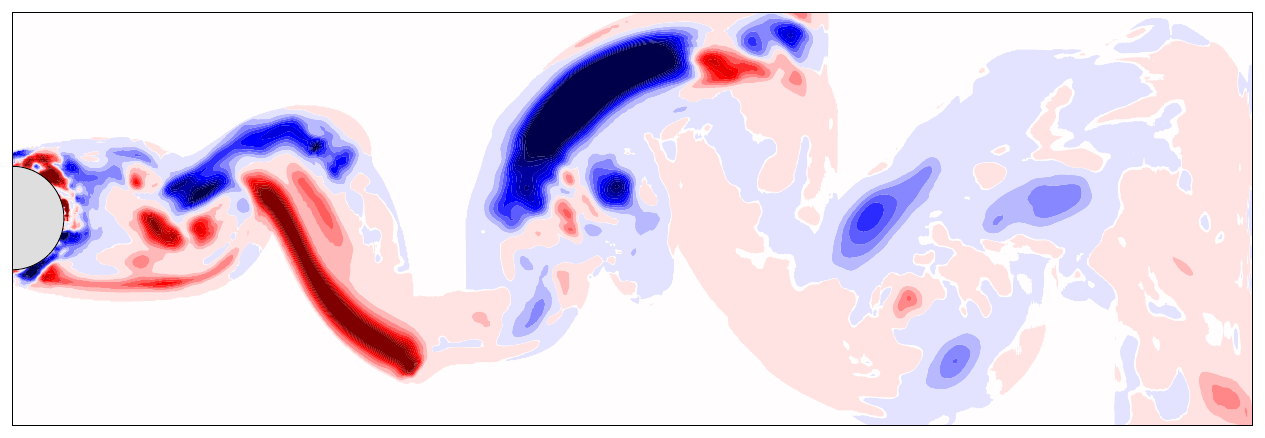}
    \caption{$\avg{u\p v\p}^{\mathrm{ML}}$}
\end{subfigure}
\begin{subfigure}{\linewidth}
    \includegraphics[width=\linewidth]{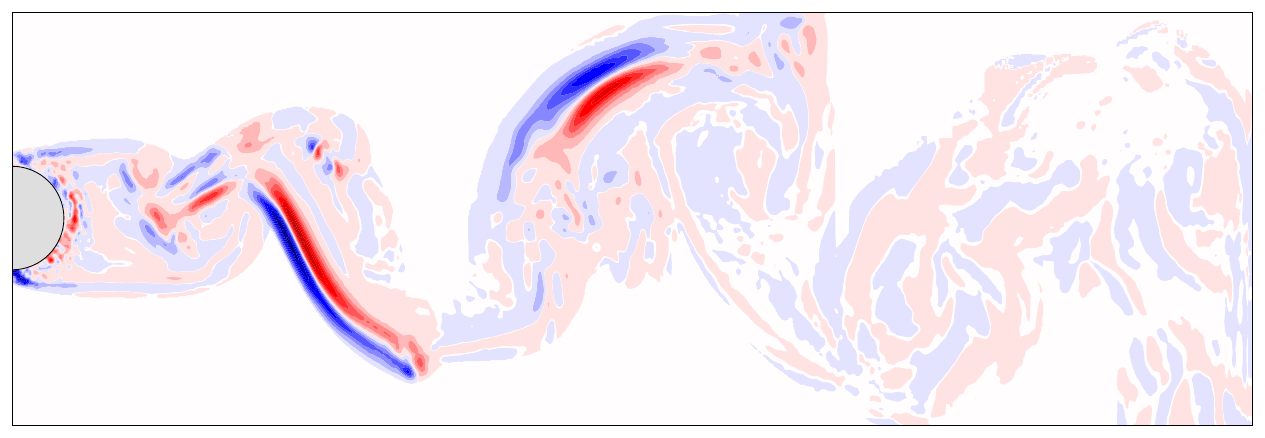}
    \caption{$\mathcal{S}_x^{R,\,\mathrm{ML}}$}
\end{subfigure}
\end{multicols}
\caption{Cylinder, $Re=1000$ case: ML model predictions of components of the SSR tensor and the perfect closure compared to reference data.}
\label{fig:ML_generalisation_R1}
\end{figure*}

\begin{figure*}[!ht]
\begin{multicols}{2}
\begin{subfigure}{\linewidth}
    \includegraphics[width=\linewidth]{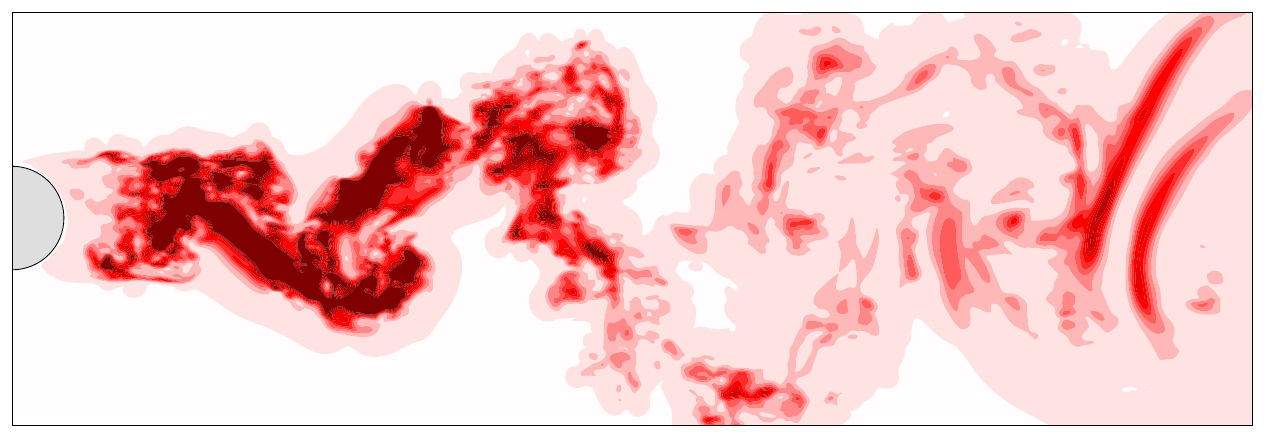}
    \caption{$\avg{u\p u\p}$} 
\end{subfigure}
\begin{subfigure}{\linewidth}
    \includegraphics[width=\linewidth]{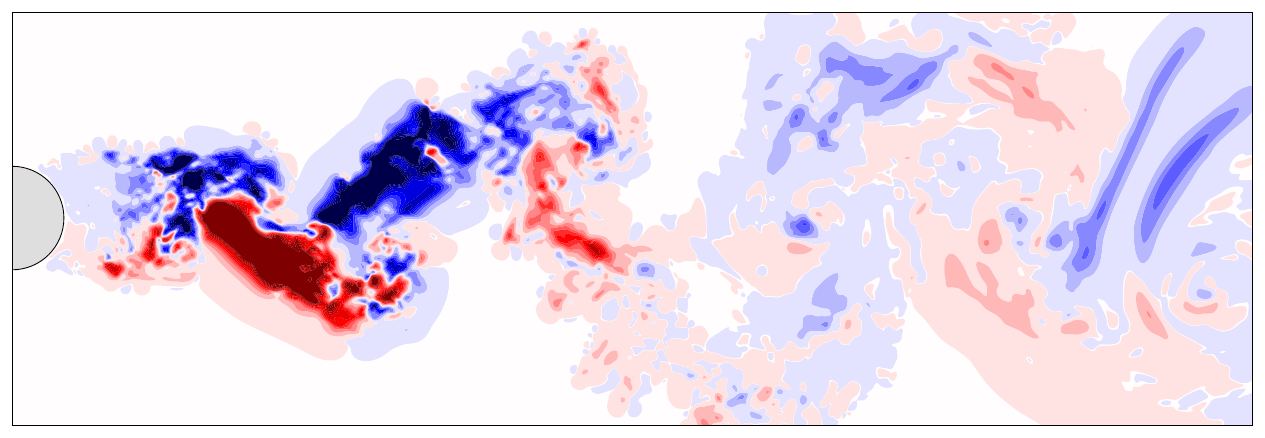}
    \caption{$\avg{u\p v\p}$}
\end{subfigure}
\begin{subfigure}{\linewidth}
    \includegraphics[width=\linewidth]{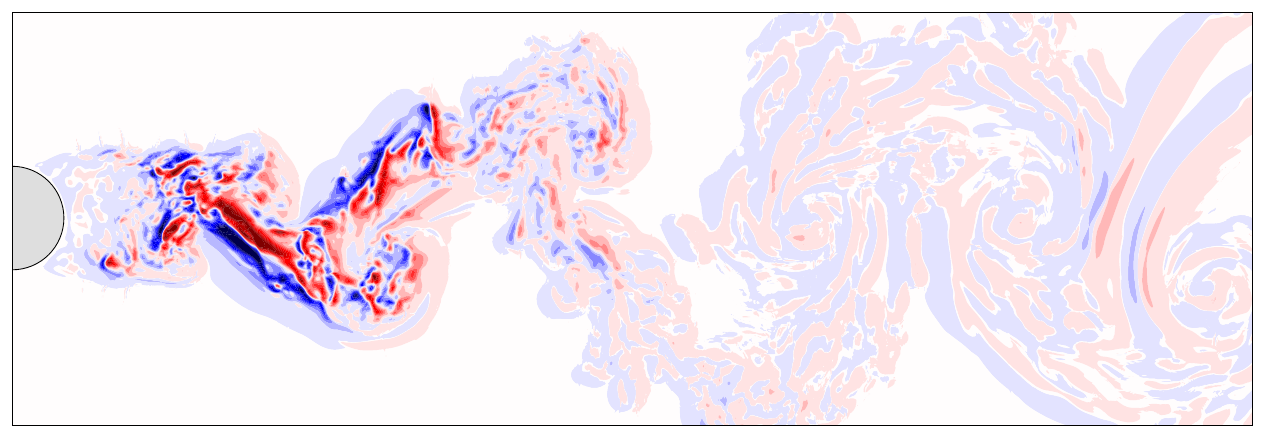}
    \caption{$\mathcal{S}^R_x$}
\end{subfigure}
\begin{subfigure}{\linewidth}
    \includegraphics[width=\linewidth]{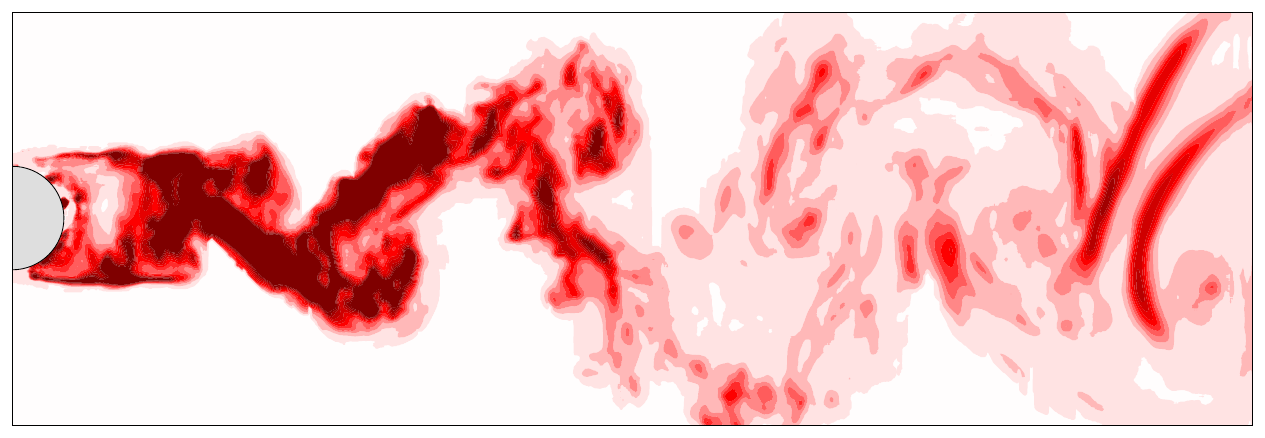}
    \caption{$\avg{u\p u\p}^{\mathrm{ML}}$} 
\end{subfigure}
\begin{subfigure}{\linewidth}
    \includegraphics[width=\linewidth]{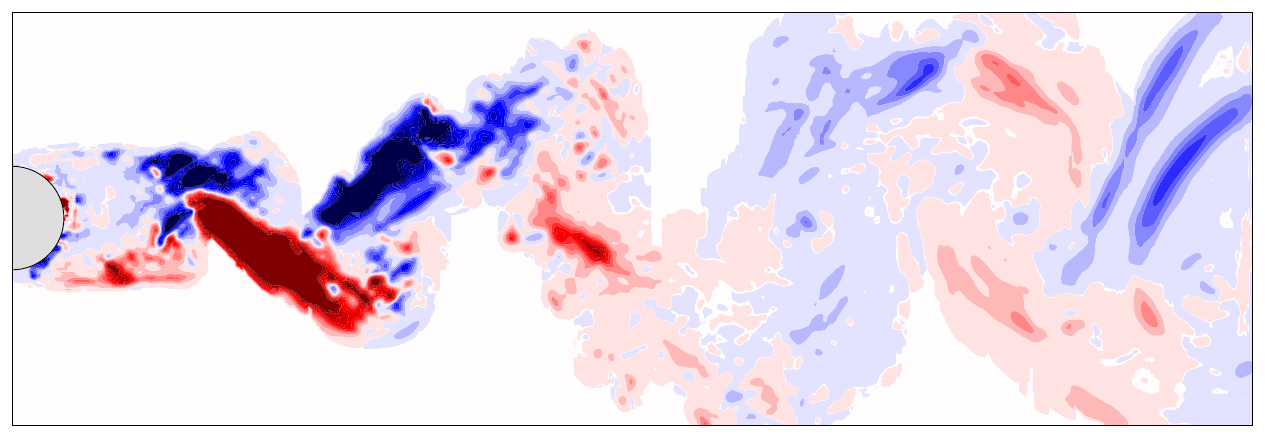}
    \caption{$\avg{u\p v\p}^{\mathrm{ML}}$}
\end{subfigure}
\begin{subfigure}{\linewidth}
    \includegraphics[width=\linewidth]{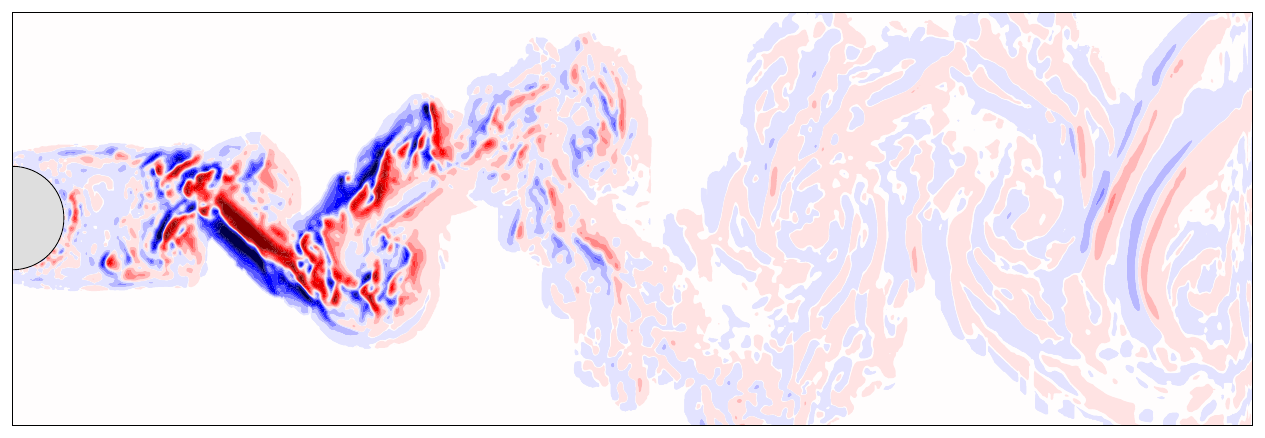}
    \caption{$\mathcal{S}_x^{R,\,\mathrm{ML}}$}
\end{subfigure}
\end{multicols}
\caption{Cylinder, $Re=3900$ case: ML model predictions of components of the SSR tensor and the perfect closure compared to reference data.}
\label{fig:ML_generalisation_R2}
\end{figure*}

\tref{tab:ML_generalisation} quantifies the accuracy of the ML model in terms of mean correlation coefficient to target data. 
Note that correlation values are provided for two wake regions, and these can be visualized in \fref{fig:cc_regions}:
one involving the non-trivial wake region (same region as used in previous results, noted in green dashed lines) and one discarding the region surrounding the body (noted in blue dashed lines), i.e. $x\in[0.75D_x,12D_y]$ where $D_x$ and $D_y$ are the streamwise and crossflow body lengths, respectively.
With respect to the model generalisation in different Reynolds regimes, its prediction ability on the $Re=3900$ case is similar to the training case ($Re=10^4$) since the correlation values provided in both cases are very close.
For the $Re=1000$ case, similar results are also obtained to the training case, specially when the near-body region is cropped out of the correlation analysis.
Qualitatively, the ML model prediction of the spanwise stresses is displayed in \fref{fig:ML_generalisation_R1} and \fref{fig:ML_generalisation_R2} for the $Re=1000$ and $Re=3900$ cases, respectively.
Again, a similar performance to the training case is observed;
mid- and large-scale structures are correctly captured whereas small-scale structures are not reconstructed as in the target data.
For the $Re=1000$ case, it can also be observed how the ML model struggles to predict the spanwise stresses in the region near to the body.

\begin{figure*}[!ht]
\begin{multicols}{2}
\begin{subfigure}{\linewidth}
    \includegraphics[width=\linewidth]{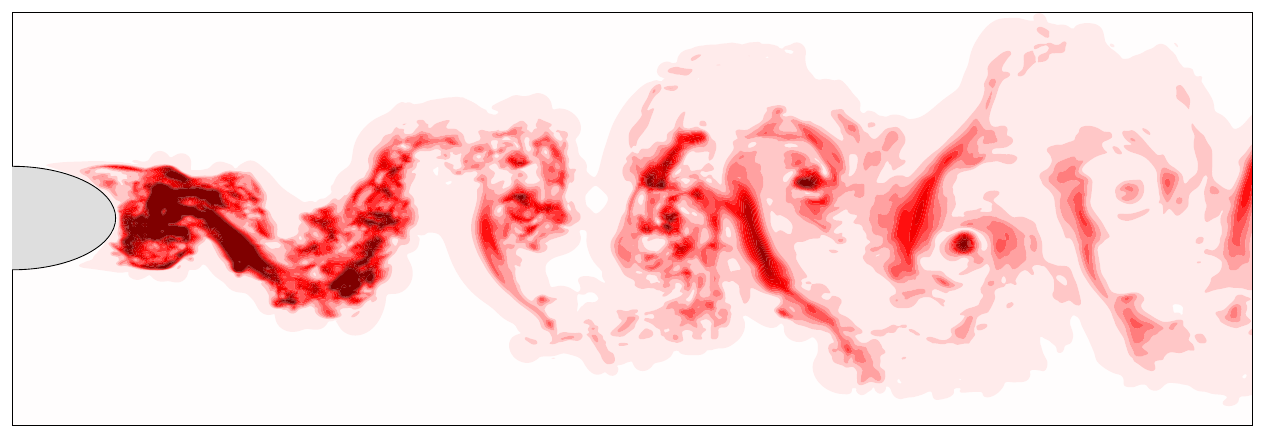}
    \caption{$\avg{u\p u\p}$} 
\end{subfigure}
\begin{subfigure}{\linewidth}
    \includegraphics[width=\linewidth]{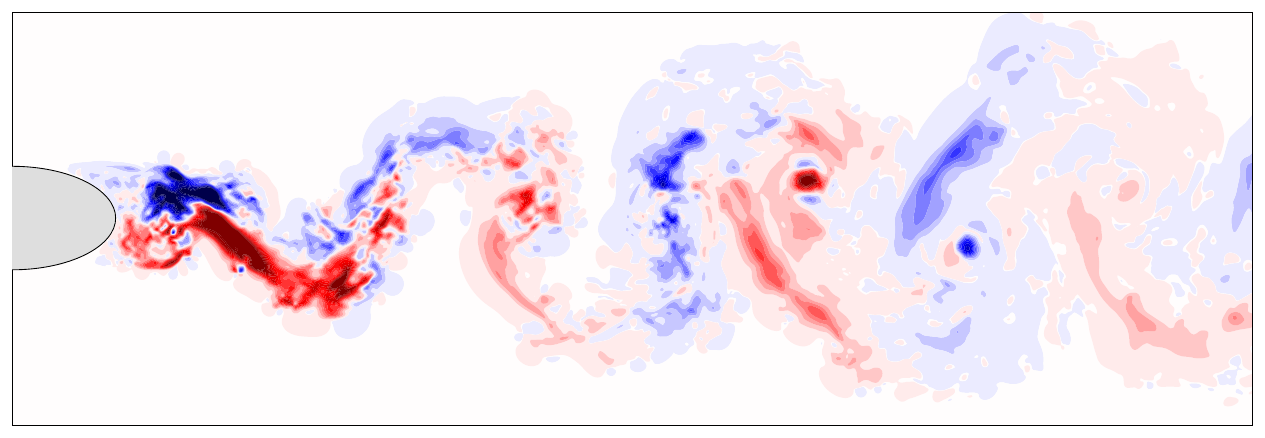}
    \caption{$\avg{u\p v\p}$}
\end{subfigure}
\begin{subfigure}{\linewidth}
    \includegraphics[width=\linewidth]{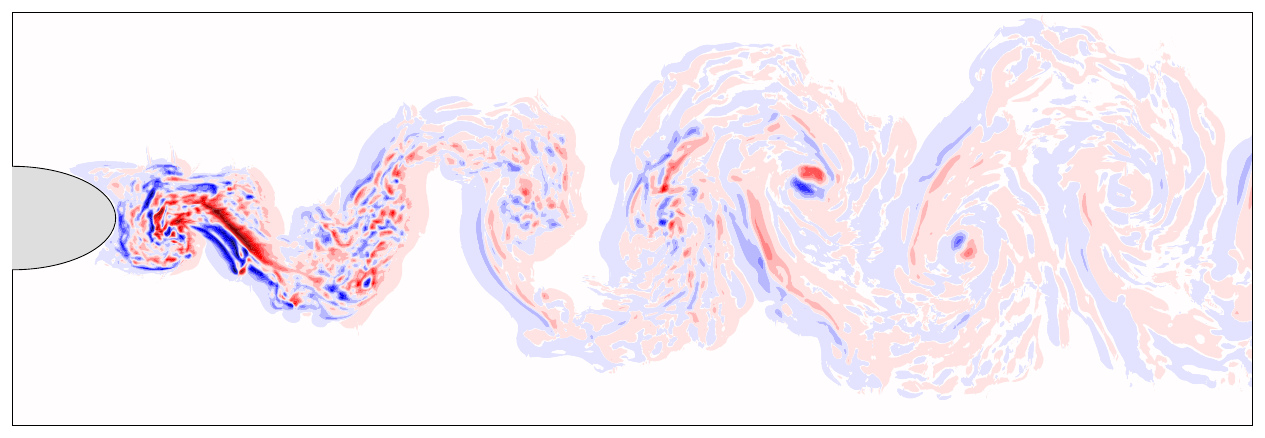}
    \caption{$\mathcal{S}^R_x$}
\end{subfigure}
\begin{subfigure}{\linewidth}
    \includegraphics[width=\linewidth]{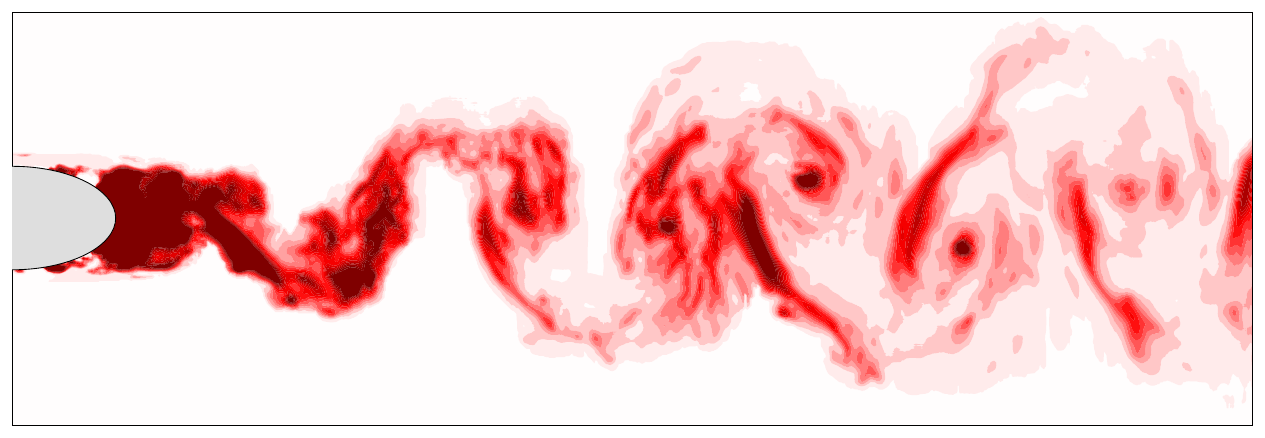}
    \caption{$\avg{u\p u\p}^{\mathrm{ML}}$} 
\end{subfigure}
\begin{subfigure}{\linewidth}
    \includegraphics[width=\linewidth]{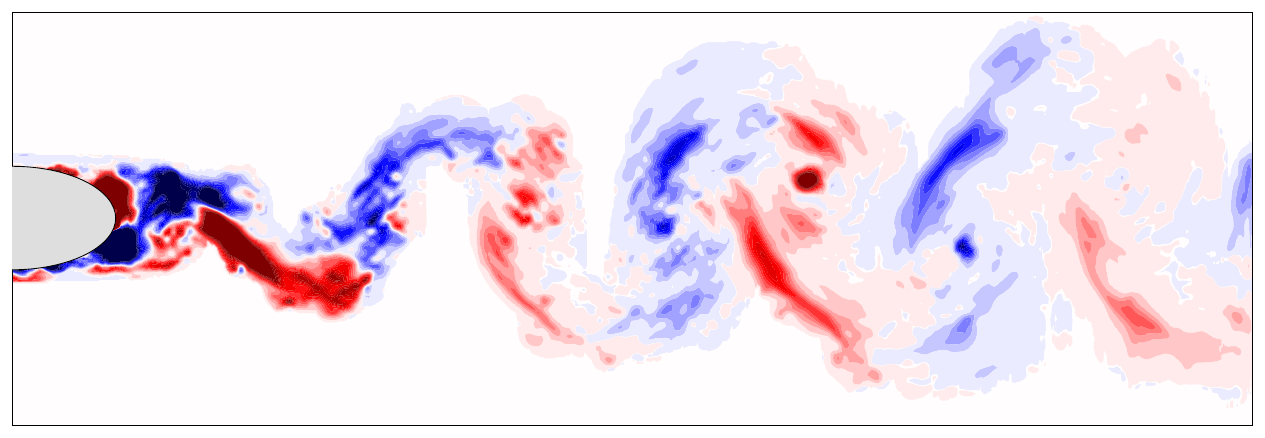}
    \caption{$\avg{u\p v\p}^{\mathrm{ML}}$} \label{fig:ML_generalisation_E1_uv}
\end{subfigure}
\begin{subfigure}{\linewidth}
    \includegraphics[width=\linewidth]{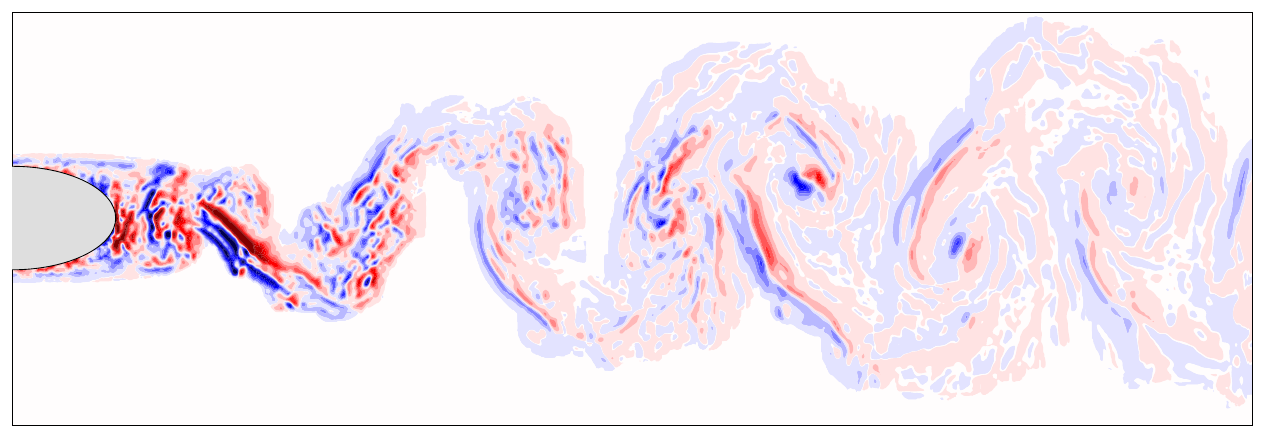}
    \caption{$\mathcal{S}_x^{R,\,\mathrm{ML}}$}
\end{subfigure}
\end{multicols}
\caption{Ellipse 1, $Re=10^4$ case: ML model predictions of components of the SSR tensor and the perfect closure compared to reference data.}
\label{fig:ML_generalisation_E1}
\end{figure*}

\begin{figure*}[!ht]
\begin{multicols}{2}
\begin{subfigure}{\linewidth}
    \includegraphics[width=\linewidth]{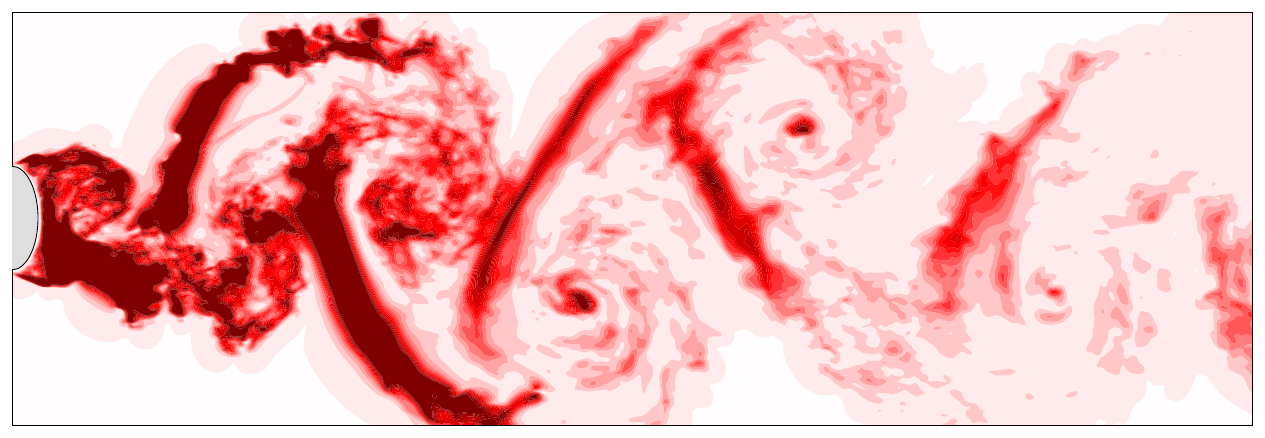}
    \caption{$\avg{u\p u\p}$} 
\end{subfigure}
\begin{subfigure}{\linewidth}
    \includegraphics[width=\linewidth]{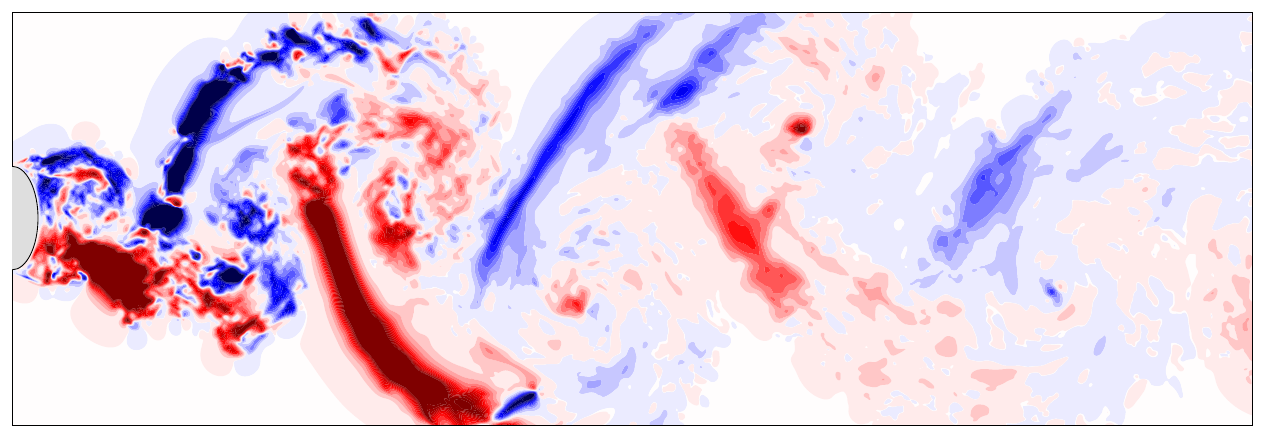}
    \caption{$\avg{u\p v\p}$}
\end{subfigure}
\begin{subfigure}{\linewidth}
    \includegraphics[width=\linewidth]{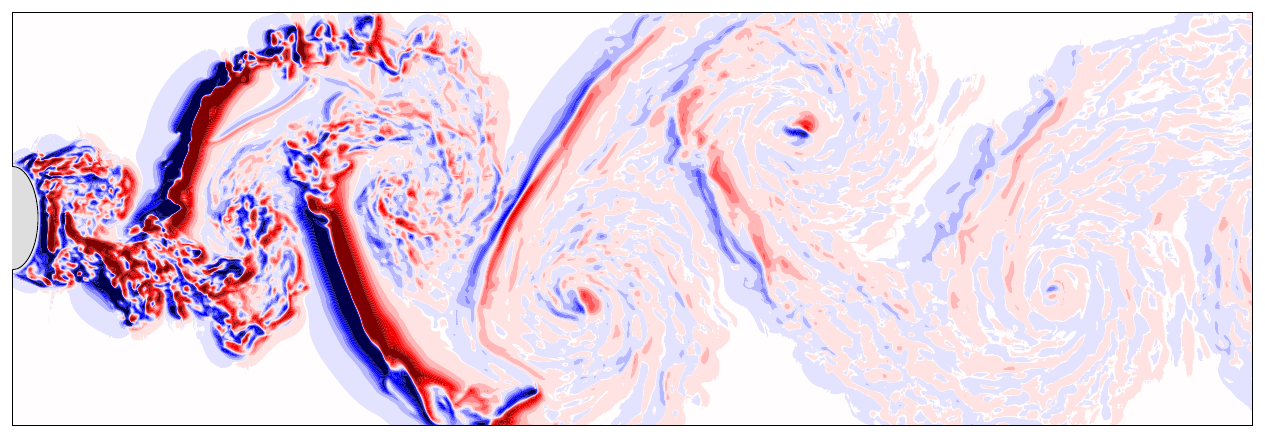}
    \caption{$\mathcal{S}^R_x$}
\end{subfigure}
\begin{subfigure}{\linewidth}
    \includegraphics[width=\linewidth]{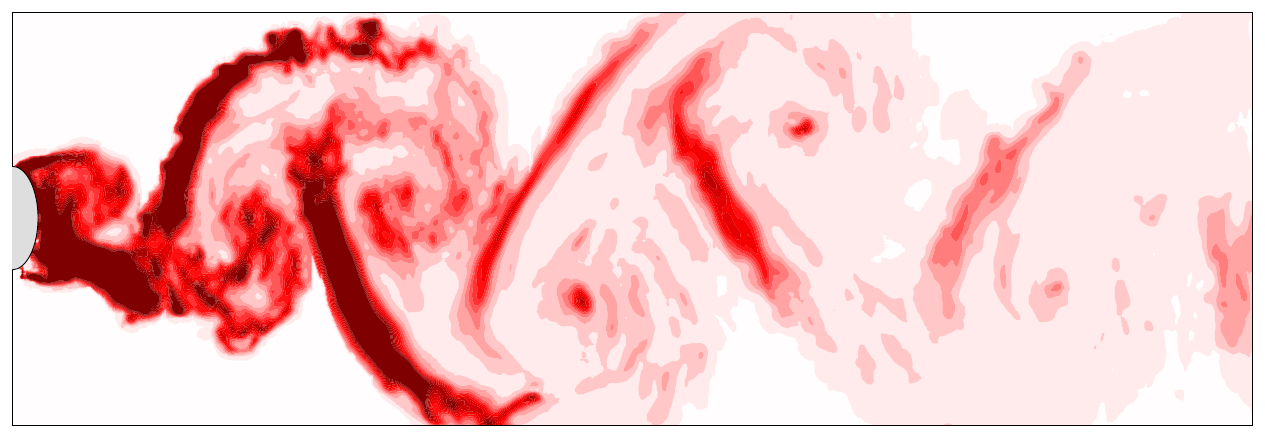}
    \caption{$\avg{u\p u\p}^{\mathrm{ML}}$} 
\end{subfigure}
\begin{subfigure}{\linewidth}
    \includegraphics[width=\linewidth]{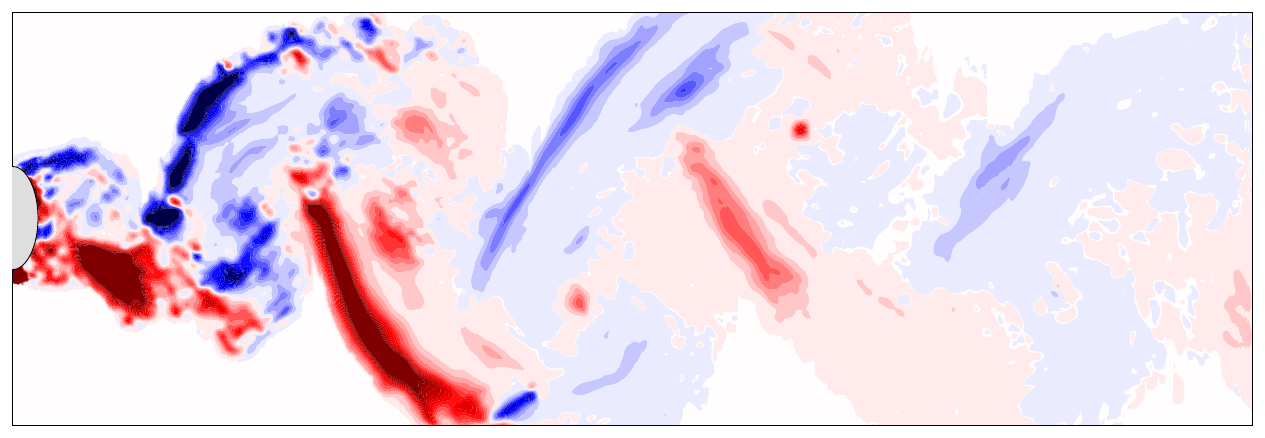}
    \caption{$\avg{u\p v\p}^{\mathrm{ML}}$}
\end{subfigure}
\begin{subfigure}{\linewidth}
    \includegraphics[width=\linewidth]{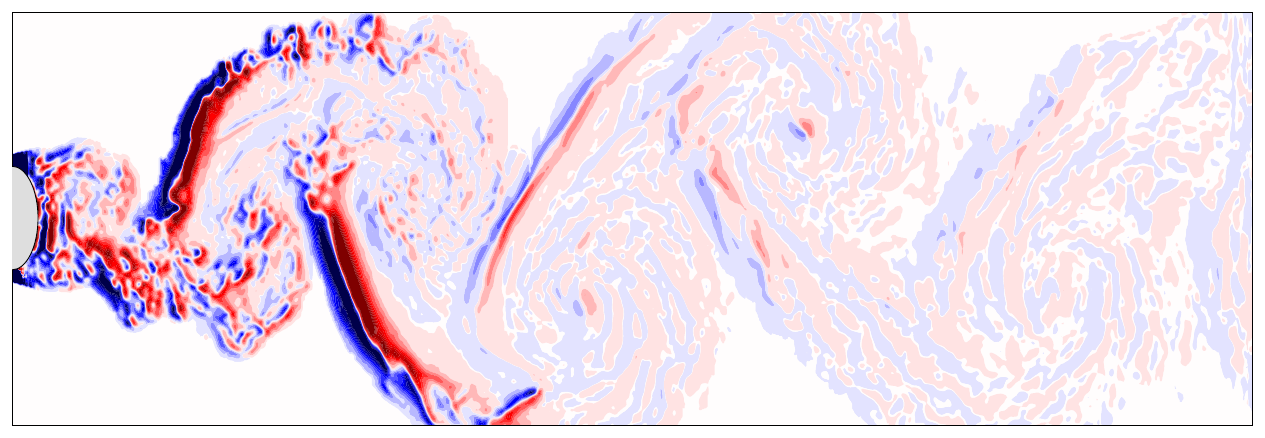}
    \caption{$\mathcal{S}_x^{R,\,\mathrm{ML}}$}
\end{subfigure}
\end{multicols}
\caption{Ellipse 2, $Re=10^4$ case: ML model predictions of components of the SSR tensor and the perfect closure compared to reference data.}
\label{fig:ML_generalisation_E2}
\end{figure*}

Regarding the ellipse cases at $Re=10^4$, high correlation values are observed when the near-body region is cropped out of the correlation analysis as displayed \tref{tab:ML_generalisation}.
On the other hand, the ML model fails to correctly predict the spanwise stresses in the shear-layer region.
This behaviour can be appreciated in \fref{fig:ML_generalisation_E1} and \fref{fig:ML_generalisation_E2}, where patches of high-intensity spanwise stresses arise at the vicinity of the body.
To a lesser extent, such discrepancies can also be observed for the $Re=1000$ and $Re=3900$ circular cylinder cases.
It indicates that training the ML model in a single flow configuration generalises well for lower Reynolds regimes and body shape across the wake except for the the shear-layer region, which performs better when it resembles the training data.
It is important to stress that the shear layer information contained in the training database is limited, so the spatial patterns learnt by the CNN related to this region cannot be recombined for previously unseen cases.
This limitation has also been reported in \citet{Lee2019}, where flow past a circular cylinder at $Re=300$ and $Re=500$ is used as training data and the model is then tested at $Re=150,400,3900$.
Similarly, their results show a better agreement for the test cases that resemble the flow within the training Reynolds regime.
Another issue arising in our generalisation results is the over- or under-prediction of the spanwise stresses intensity (not reflected in the correlation coefficients), noting that the ML model predictions and the reference data are plotted using the same scale. 
This can be observed in \fref{fig:ML_generalisation_E1_uv}.
The current input data normalization method, based in each snapshot standard score, is a factor that might result in this difference.

Overall, the ML model has shown an excellent performance on the generalisation across lower $Re$ and different body shapes providing correlation values similar to the training case.
Additionally, this indicates that the training data is not over-fitted.
Two limitations on the generalisation have been observed: the prediction of the spanwise stresses in the shear-layer region and the over- or under-prediction of the spanwise stresses intensity in some cases.

\clearpage

\begin{table}[t]
\centering
\caption{Correlation coefficients between target data and ML model predictions for the different generalisation cases. FW and DW refers to the correlation coefficient for the green-dashed region and blue-dotted region of \fref{fig:cc_regions}, respectively.}
        \begin{tabular}{ccccccc}
            \toprule
            \multicolumn{1}{c}{Case}& $\mathcal{CC}$ & $\avg{u\p u\p}$ & $\avg{u\p v\p}$ &$\avg{v\p v\p}$ & $\mathcal{S}^R_x$ & $\mathcal{S}^R_y$ \\
            \midrule
            \multirow{2}{*}{Cylinder, $Re=1000$} & \multicolumn{1}{c}{FW} & \multicolumn{1}{c}{0.86} & \multicolumn{1}{c}{0.90} & \multicolumn{1}{c}{0.92} & \multicolumn{1}{c}{0.86} & \multicolumn{1}{c}{0.91} \\
                                & \multicolumn{1}{c}{DW} & \multicolumn{1}{c}{0.94} & \multicolumn{1}{c}{0.94} & \multicolumn{1}{c}{0.96} & \multicolumn{1}{c}{0.93} & \multicolumn{1}{c}{0.95} \\ 
            \midrule
            \multirow{2}{*}{Cylinder, $Re=3900$} & \multicolumn{1}{c}{FW} & \multicolumn{1}{c}{0.94} & \multicolumn{1}{c}{0.91} & \multicolumn{1}{c}{0.95} & \multicolumn{1}{c}{0.89} & \multicolumn{1}{c}{0.92} \\
                                & \multicolumn{1}{c}{DW} & \multicolumn{1}{c}{0.96} & \multicolumn{1}{c}{0.93} & \multicolumn{1}{c}{0.96} & \multicolumn{1}{c}{0.91} & \multicolumn{1}{c}{0.93} \\ 
            \midrule
            \multirow{2}{*}{Ellipse 1, $Re=10^4$} & \multicolumn{1}{c}{FW} & \multicolumn{1}{c}{0.61} & \multicolumn{1}{c}{0.60} & \multicolumn{1}{c}{0.79} & \multicolumn{1}{c}{0.67} & \multicolumn{1}{c}{0.69} \\
                                & \multicolumn{1}{c}{DW} & \multicolumn{1}{c}{0.93} & \multicolumn{1}{c}{0.90} & \multicolumn{1}{c}{0.95} & \multicolumn{1}{c}{0.86} & \multicolumn{1}{c}{0.91} \\ 
            \midrule
            \multirow{2}{*}{Ellipse 2, $Re=10^4$} & \multicolumn{1}{c}{FW} & \multicolumn{1}{c}{0.91} & \multicolumn{1}{c}{0.21} & \multicolumn{1}{c}{0.92} & \multicolumn{1}{c}{0.07} & \multicolumn{1}{c}{0.06} \\
                                & \multicolumn{1}{c}{DW} & \multicolumn{1}{c}{0.94} & \multicolumn{1}{c}{0.91} & \multicolumn{1}{c}{0.94} & \multicolumn{1}{c}{0.89} & 0.89 \\ 
            \bottomrule
        \end{tabular}\label{tab:ML_generalisation}

\vspace{0.5cm}
{\footnotesize The correlation coefficients are calculated for the 500 snapshots of the test dataset and the average values are provided. \par}
\end{table}

\newpage

\newpage

\subsection{A-posteriori analysis} \label{sec:ML_a-post}

The use of the ML model in the a-posteriori framework is assessed next, i.e. the trained closure is embedded directly into the simulation and no knowledge of the correct flow is assumed.
As before, the goal is to recover the unsteady spanwise-averaged solution of the original 3-D flow using only a 2-D system.
Starting from a spanwise-averaged initial condition, the ML model is incorporated into a 2-D solver providing closure to the SANS equations at every time step.

The most important aspect of the a-posteriori analysis revolves around the system stability.
Ideally, a stable closure would be provided at every time step even when noisy inputs are given to the ML model.
For the system to be stable, the spanwise-average attractor should dictate the temporal evolution of the flow and prevent the 2-D flow dynamics dominating as well as prevent the growth of non-physical effects.

For our test case, it has been found that directly incorporating the raw output of the SSR tensor ML model prediction $(\boldsymbol\tau_{ij}^{R,\,\mathrm{ML}})$ into the fluid solver makes it to rapidly diverge from the physical solution as the flow evolves in time.
The noise in the predicted fields is amplified by the divergence operator, which in turn significantly contaminates the resolved quantities at every time step in a feedback loop.
To mitigate this issue, the ML model has been trained to predict the perfect closure $(\mathcal{S}^{R})$ instead of the SSR tensor.
Hence, the new target outputs of the CNN are $\mathrm{Y}_n=\lbrace \mathcal{S}^R_x,\mathcal{S}^R_y\rbrace$.
Issues related to data-driven closure stability are discussed in detail in \cite{Cruz2019} under a RANS formulation.
It is suggested that learning the divergence of the residual tensor instead of its components helps reduce the error accumulation on the resolved quantities.

\begin{table}[t]
\centering
\caption{Correlation coefficients between the perfect closure and the ML model predictions.}
\begin{tabular}{lcc}
\toprule
$\mathcal{CC}$ & $\mathcal{S}^R_x$ & $\mathcal{S}^R_y$ \\
\midrule
ML & 0.89 & 0.90\\
\bottomrule \label{tab:ML_SR_a-priori}
\end{tabular}

{\footnotesize The correlation coefficients are calculated for the 500 snapshots of the test dataset and the average values are provided. \par}
\end{table}
\begin{table}[t]
\centering
\caption{Comparison of different metrics recorded during $\Delta t^*=10$ time units for the 3-D, SANS, and 2-D systems.}
\begin{tabular}{lrrr}
\toprule
 &$\left\langle 3\text{-}\mathrm{D} \right\rangle$ & SANS & $2\text{-}\mathrm{D}$\\
\midrule
$\overline{C}_D$ & 1.17 & 1.14 (-2.6\%)& 1.35 (+15.4\%)\\
$\overline{C}_L$ & 0.46 & 0.51 (+10.9\%) & 0.91 (+97.8\%)\\
$\overline{E}$ & 374.34 & 374.84 (+0.1\%) & 376.44 (+0.6\%) \\
$\overline{Z}$ & 0.0150 & 0.0151 (+0.6\%) & 0.0189 (26.0\%) \\
Cost [h] & 80.4 & 0.43 (-99.5\%) & 0.21 (-99.7\%) \\
\bottomrule
\label{tab:ML_a-posteriori}
\end{tabular}

{\footnotesize The overline indicates a temporal average except for the lift coefficient, for which the r.m.s. value is provided.
Energy $(E)$ and enstrophy $(Z)$ of the 3-D system are calculated using the streamwise and crossflow velocity components alone.
The relative error to the 3-D case is expressed in parenthesis.
The computational cost is expressed in hours and normalised by number of processes and clock speed.
The same constant time step is used in all systems.}
\end{table}

\begin{figure}[t]
\centering
\begin{subfigure}[t]{0.6\linewidth}
    \includegraphics[width=\linewidth]{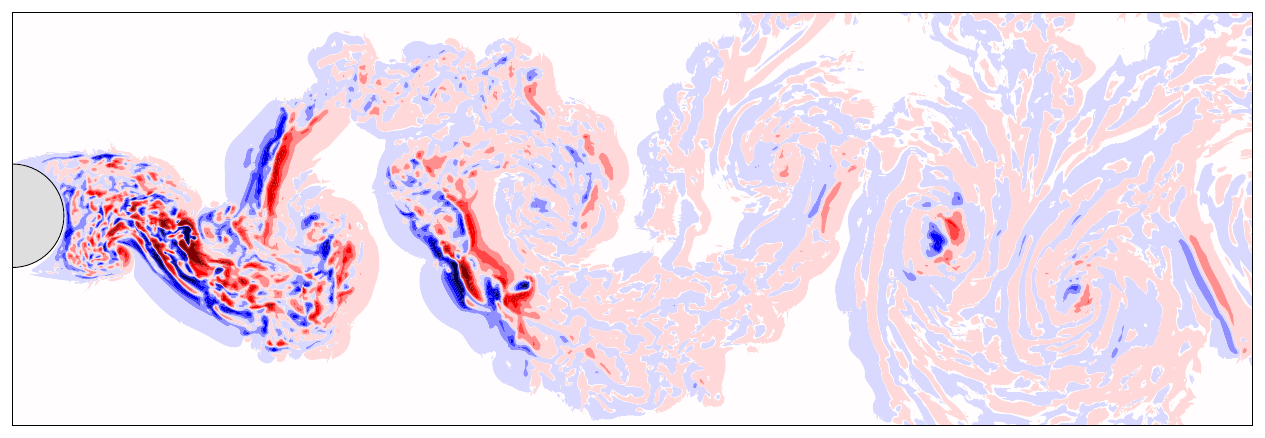}
    \caption{$\mathcal{S}^R_x$} 
\end{subfigure}
\begin{subfigure}[t]{0.6\linewidth}
    \includegraphics[width=\linewidth]{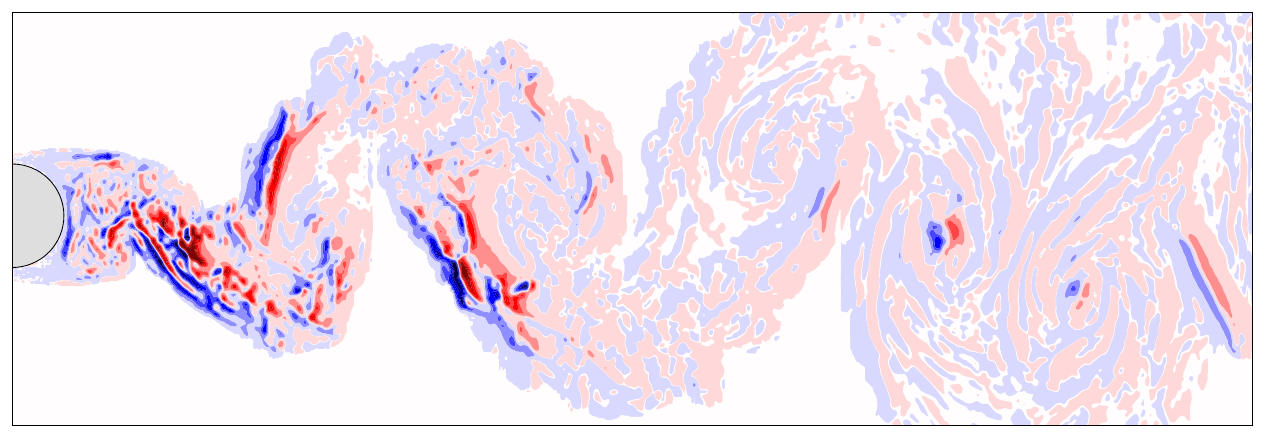}
    \caption{$\mathcal{S}^{R,\,\mathrm{ML}}_x$} 
\end{subfigure}
\begin{subfigure}[t]{0.6\linewidth}
    \includegraphics[width=\linewidth]{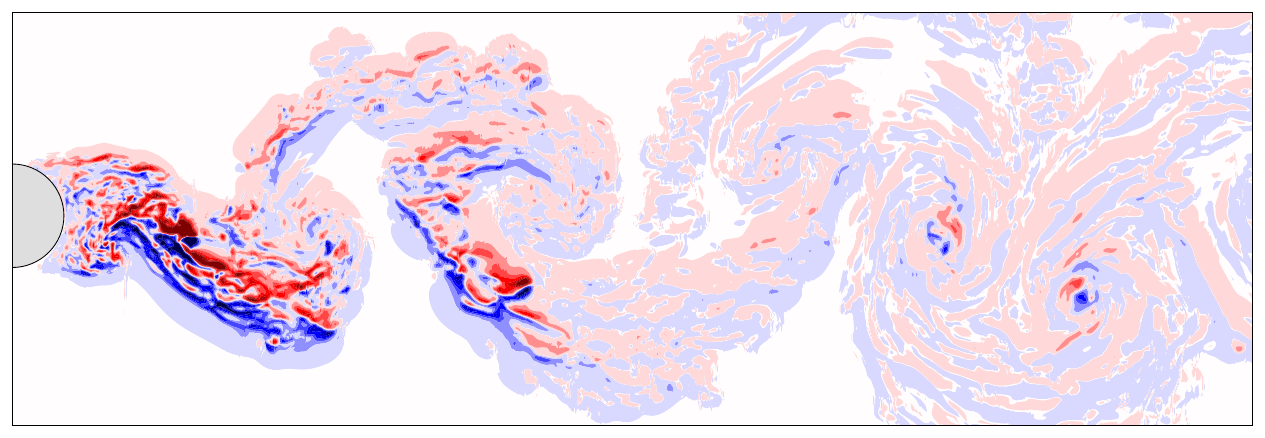}
    \caption{$\mathcal{S}^R_y$} 
\end{subfigure}
\begin{subfigure}[t]{0.6\linewidth}
    \includegraphics[width=\linewidth]{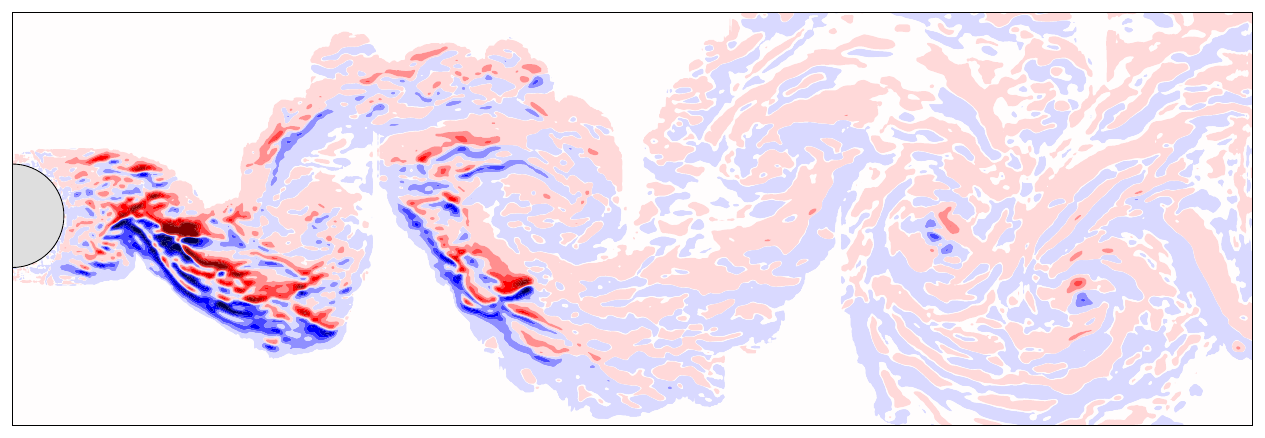}
    \caption{$\mathcal{S}^{R,\,\mathrm{ML}}_y$} 
\end{subfigure}
\caption{ML model prediction of the perfect closure components.}
\label{fig:ML_SR_a-priori}
\end{figure}

Before inserting the ML model prediction of the perfect closure into the 2-D system, an a-priori analysis of the trained model is performed (the training history is displayed in \fref{fig:SR_history}).
High correlation values are found for both perfect closure components, as presented in \tref{tab:ML_SR_a-priori}.
Qualitatively, the ML model prediction of both components is displayed in \fref{fig:ML_SR_a-priori}.
Again, mid- and large-scale structures are correctly captured while small-scale structures present a weaker correlation.

\begin{figure}[t]
\centering
\includegraphics[width=0.50\linewidth]{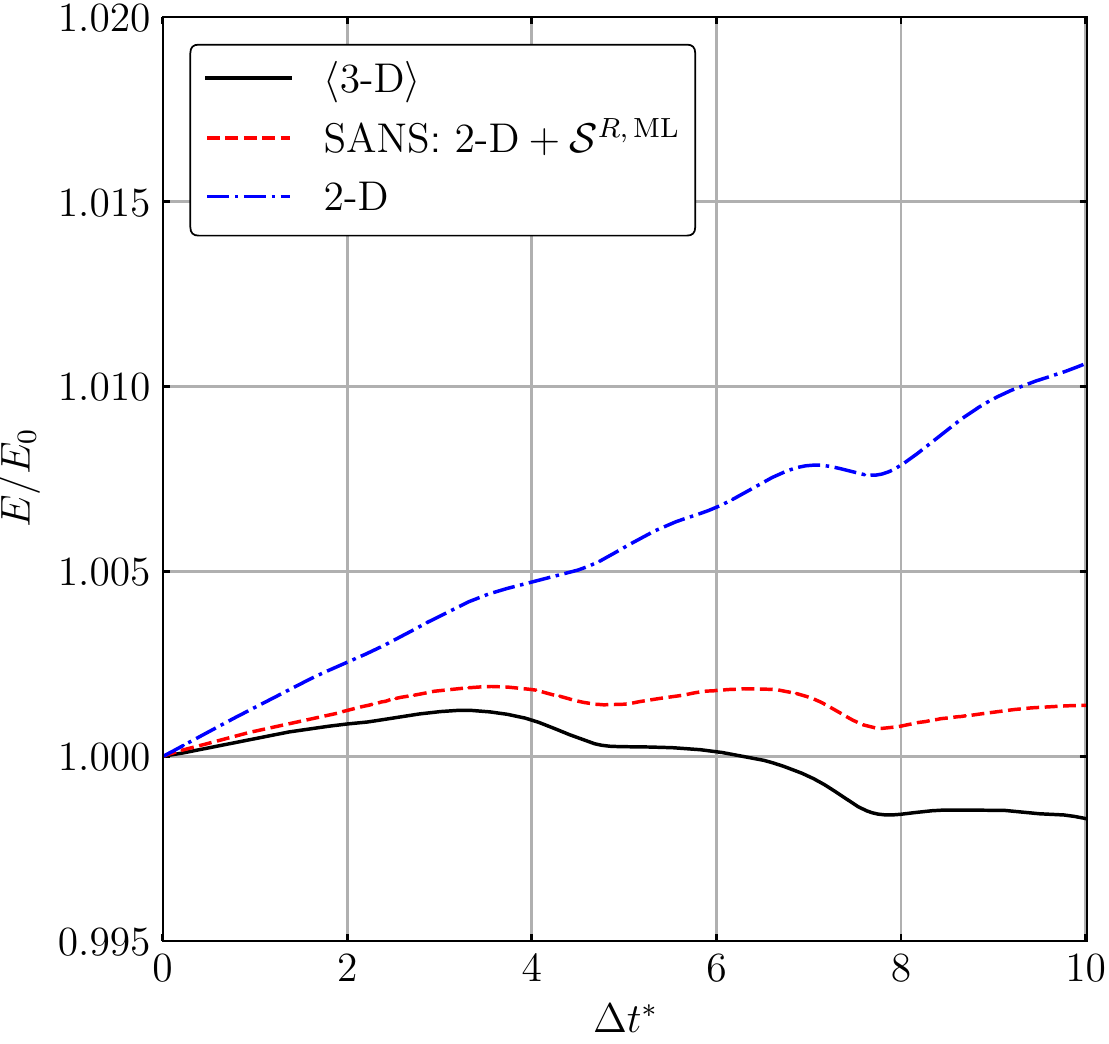}
\includegraphics[width=0.48\linewidth]{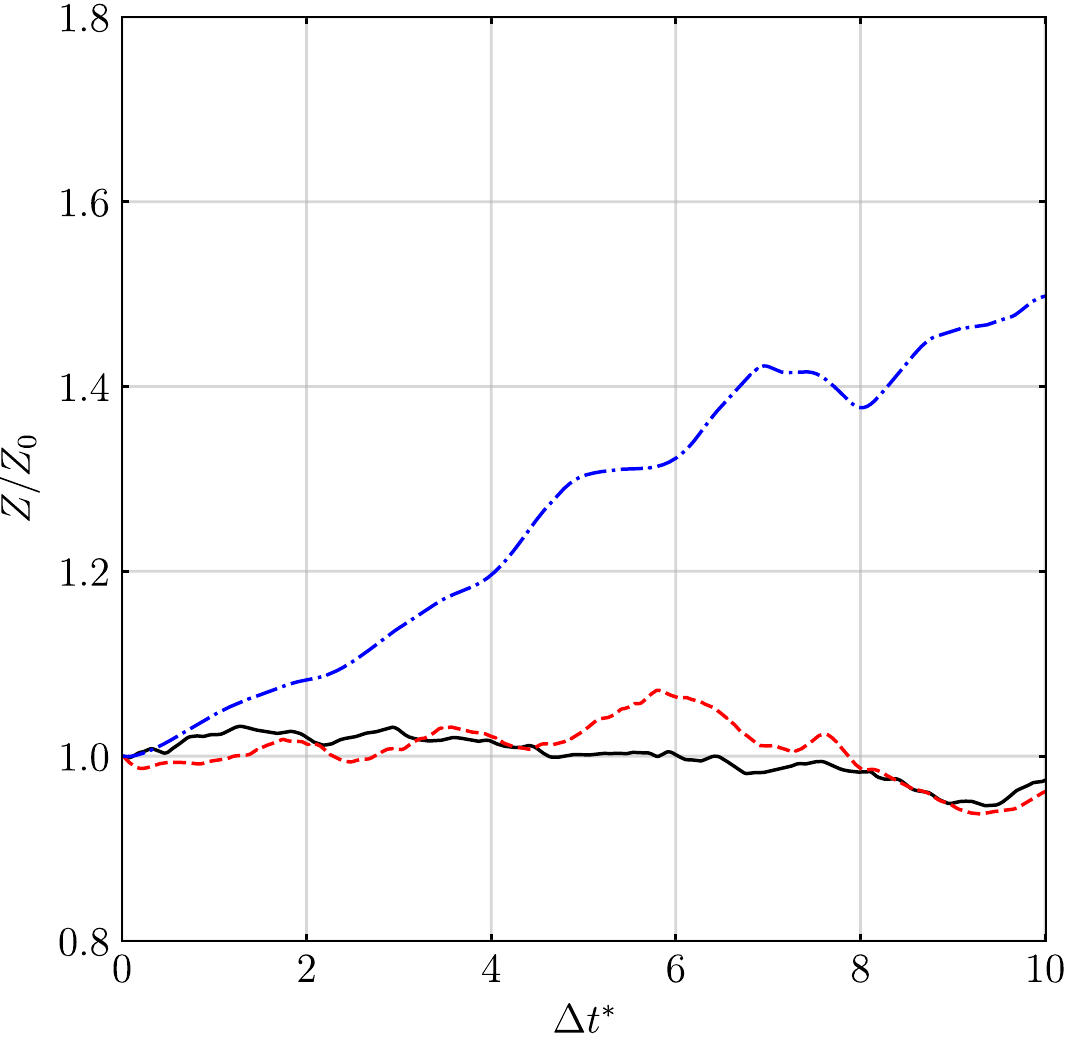}
\caption{Kinetic energy (left) and enstrophy (right) evolution during $\Delta t^*=10$ for the 3-D spanwise-averaged (reference data), SANS, and 2-D systems.
Simulations are started from a 3-D spanwise-averaged snapshot at $t^*_0$, corresponding to the scaling energy $(E_0)$ and enstrophy $(Z_0)$.
Energy and enstrophy are computed as detailed in \fref{fig:perfect_t-g} caption.}
\label{fig:E_Z_a-posteriori}
\end{figure}

\begin{figure}[t]
\centering
\includegraphics[width=0.48\linewidth]{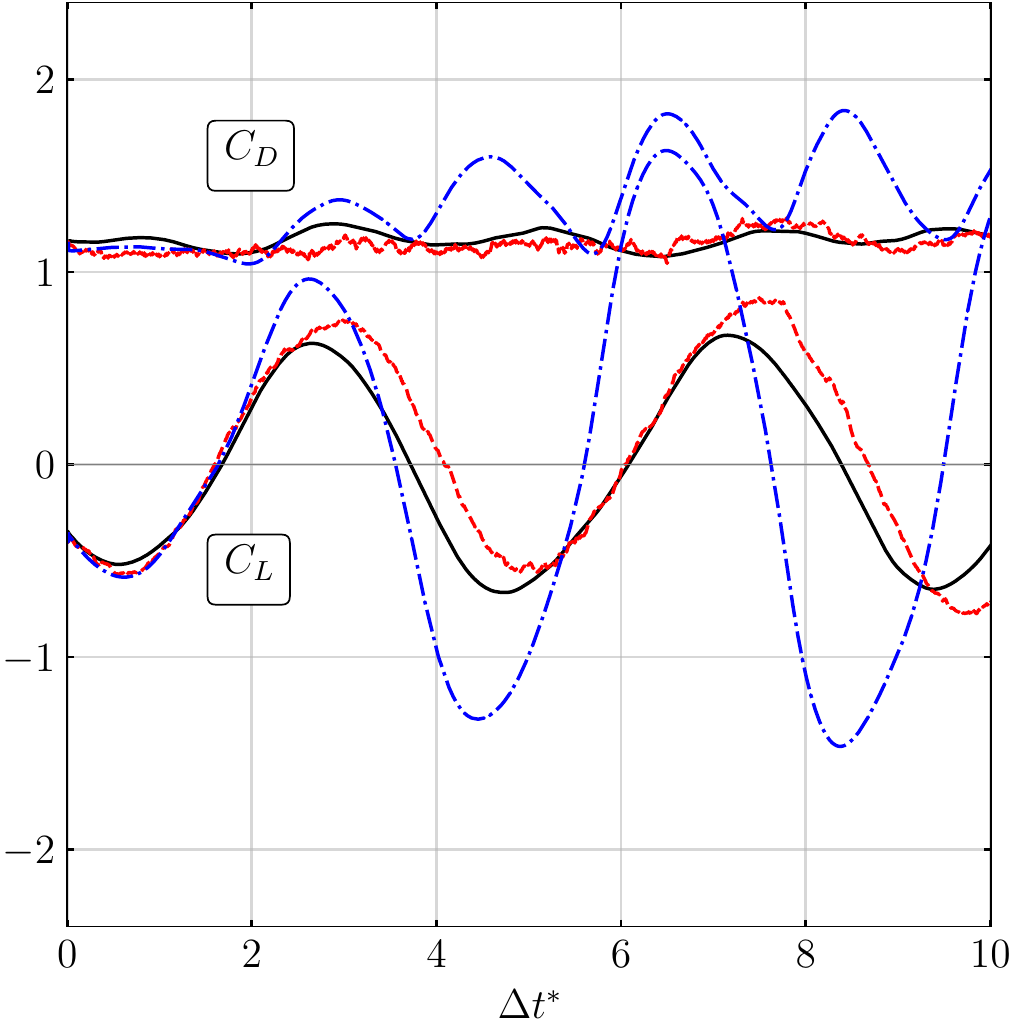}
\caption{Lift (bottom) and drag (top) forces induced to the cylinder (computed as described in \fref{fig:perfect_cc_forces}).
The lines legend is equivalent to \fref{fig:E_Z_a-posteriori}.}
\label{fig:forces_a-posteriori}
\end{figure}

Learning the perfect closure improves the system stability and delays the transition to 2-D dynamics.
Starting from a spanwise-averaged snapshot of the 3-D flow, a-posteriori statistics are recorded as the flow evolves over $\Delta t^*=10$ time units, and these are summarised in \tref{tab:ML_a-posteriori}.
As shown in \fref{fig:E_Z_a-posteriori}, the global enstrophy deviates from the spanwise-averaged reference data obtained in the 3-D system as a result of the ML model error accumulation.
This is also reflected in the kinetic energy evolution, which increases with respect to the reference data.
On the other hand, both metrics still provide a notable improvement compared to the pure 2-D prediction.
The forces induced to the cylinder (\fref{fig:forces_a-posteriori}) also appear to be in better agreement with the reference data when including perfect closure ML predictions.
It becomes clear from \tref{tab:ML_a-posteriori} that the SANS equations yield results closer to a 3-D system while offering the same computational cost order of 2-D simulations.
The computational cost overhead between the SANS and 2-D systems is related to the data transfer between the fluid solver and the ML model, plus the ML model prediction time.


\begin{figure}[t]
\centering
\begin{subfigure}[t]{0.7\linewidth}
 	\includegraphics[width=\linewidth]{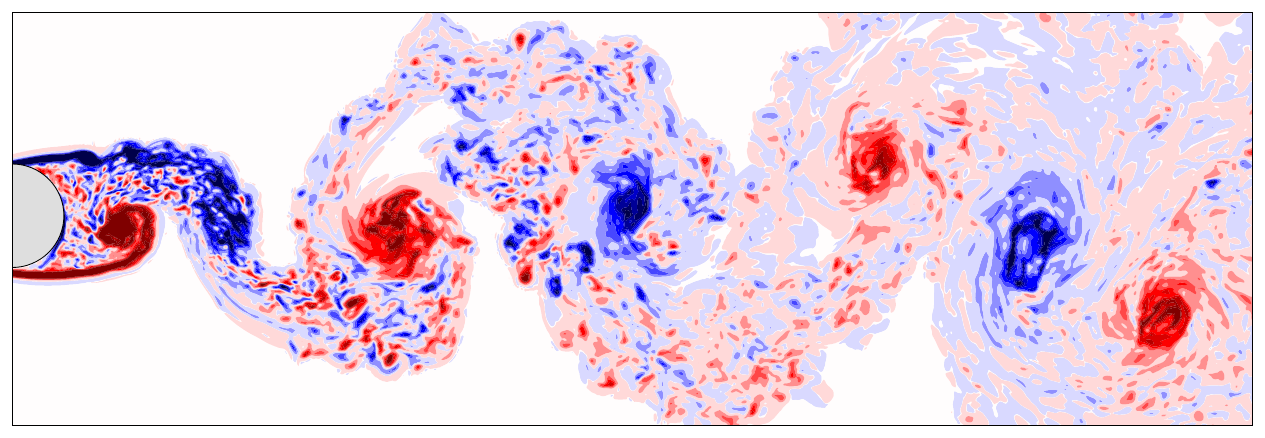}
 	\caption{$\left\langle 3\text{-}\mathrm{D} \right\rangle$}
\end{subfigure}
\begin{subfigure}[t]{0.7\linewidth}
	\includegraphics[width=\linewidth]{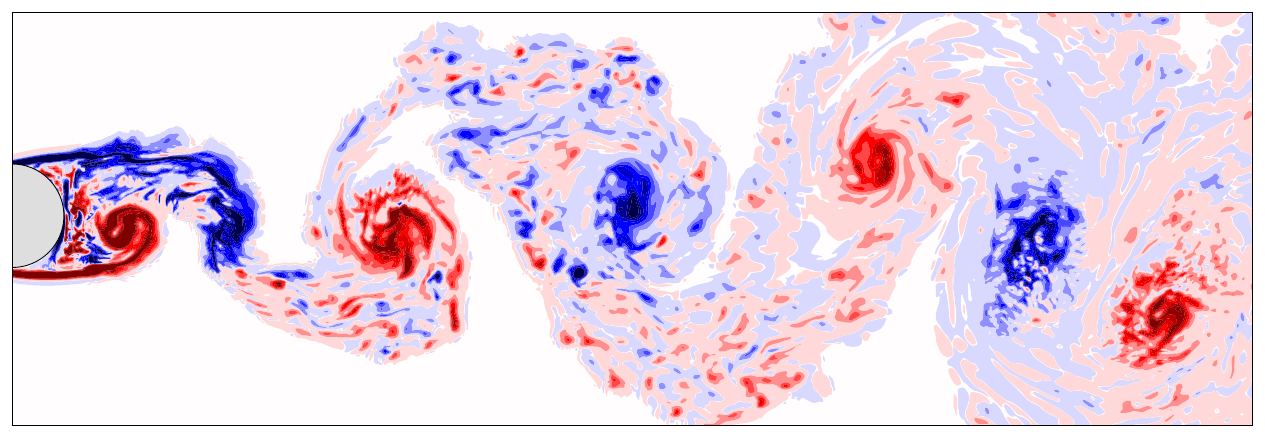}
	\caption{SANS}
\end{subfigure}
\begin{subfigure}[t]{0.7\linewidth}
	\includegraphics[width=\linewidth]{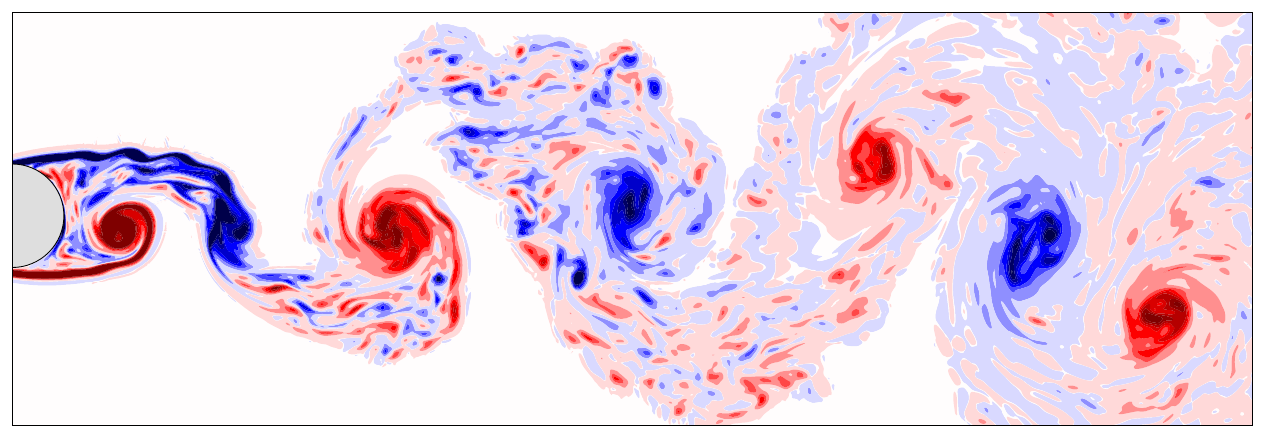}
	\caption{$2\text{-}\mathrm{D}$}
\end{subfigure}
\caption{Spanwise vorticity snapshots for: (a) Spanwise-averaged 3-D flow (reference data).
(b) SANS with $\mathcal{S}^{R,\,\mathrm{ML}}$.
(c) 2-D simulation (no-model).
The SANS and the 2-D systems are started from a spanwise-averaged snapshot of the 3-D simulation at $t^*_0$.
The snapshots above correspond to $\Delta t^*=1$.}
\label{fig:ML_a-posteriori}
\end{figure}

Additionally, the instantaneous spanwise vorticity is depicted in \fref{fig:ML_a-posteriori}.
It is worth noting that the shear layer roll-up region of the SANS snapshot is not as coherent as the 2-D system, where structures have already reorganised into large scales.
In particular, the bottom detaching vortex of the SANS simulation is not as energised as the pure 2-D simulation one, and small-scale structures are still found in the SANS snapshot.
The close-wake region dictates the forces induced to the cylinder, explaining the significant difference in lift and drag forces shown in \tref{tab:ML_a-posteriori}.
In general, vortices found in the SANS simulation do not appear as coherent as in the 2-D case, where the inverse energy cascade promoted via vortex-merging and vortex-thinning events rapidly two-dimensionalises the wake structures.
A closer look on the SANS small-scale structures also shows that these are not correlated to the reference data.
This can also be appreciated at the top shear layer roll-up, where the error starts to propagate to the far field.
The ML closure error is responsible for such phenomenon, and ultimately yields the divergence of the system dynamics.

\section{Discussion and conclusion}

This chapter introduces a ML model for the SANS equations, enabling 3-D turbulent flows with a homogeneous flow direction to be modelled using a 2-D system of equations with an observed speed up of 99.5\%.
The SSR tensor is critical to the success of this description, driving the 2-D flow dynamics to match the 3-D statistics and flow forces.
While EVMs are completely inadequate to model the SSR tensor, we develop a data-driven closure based on a deep-learning network with tested correlation statistics above 90\%.
The trained ML model also performs well in different Reynolds regimes and body geometries to the training case, despite some discrepancies in the near-body region.

The success of ML models for turbulent flows in an a-posteriori set-up is still limited, and the error propagation into the resolved quantities remains an open issue \citep{Wang2017, Beck2019}.
Stability has been previously achieved by projecting the ML model output into a known stable closure such as EVMs for time-averaged (RANS) or filtered (LES) systems \citep{Maulik2018, Beck2019, Cruz2019}.
Also, highly-resolved simulations help to mitigate the closure weight resulting in more stable computations \citep{Gamahara2017}.
Unfortunately, these solutions cannot be implemented in SANS because, as previously exposed, the EVMs hypothesis is intrinsically different from the nature of the SSR tensor.

Employing recurrent neural networks (RNN) for the prediction of SANS closure terms is a research direction worth exploring in the future.
RNNs incorporate the system temporal dynamics so they can offer a more stable a-posteriori solution targeting statistically stationary metrics \citep{Vlachas2018, Kim2020b}.
In this sense, the use of long-short term neural networks (a type of RNN) similarly to \cite{Vlachas2018}, together with CNNs could help overcome the issues exposed in the SANS closure.
Additionally, a physics-informed loss function can also improve the model stability as shown in \cite{Lee2019}.
Finally, while the current model is initialised from a 3-D averaged simulation, a more stable closure should be capable of developing 3-D turbulence dynamics even when initialised with a pure 2-D flow snapshot.

Overall, the new SANS flow description combined with advances in data-driven modelling has the potential to vastly improve the speed and accuracy of turbulence simulations for a prevalent class of engineering and environmental applications.
% ----------------------------------------------------------------

\chapter{Conclusions and future work}

\section{Summary of contributions}

This work proposes to drastically speed up turbulent flow simulation by averaging in a homogeneous direction, such as along the span of a marine riser, a slender building, or an aircraft wing.
Differently from standard strip-theory methods, 3-D turbulence effects are directly incorporated in the 2-D system as closure terms, thus significantly improving current 2-D models.
This has been accomplished with the following main contributions,
\begin{itemize}
	\item Improved physical understanding of the role of small-scale 3-D structures in high-Reynolds incompressible wake flow and the forces induced to the cylinder.
	\item A new 2-D model arising from a dimensionality reduction of the 3-D flow by local spanwise averaging.
	\item The design of a ML closure for the spanwise-averaged equations system based on a CNN.
\end{itemize}

First, the 3-D effects inherent in turbulent flows which present an homogeneous direction have been investigated by a systematic reduction of the span of a circular cylinder at $Re=10^4$.
Multiple span lengths ranging from a pure 2-D system to 10 diameters have been analysed using highly-resolved simulations.
Different turbulence metrics of the wake have revealed that small-scale 3-D structures can be observed even when the span is less than half a diameter long.
The small-scale structures generated in the shear layer are rapidly two-dimensionalised by the natural large-scale rotation of the K\'{a}rm\'{a}n vortices, a mechanism also encountered in obstacle-free turbulent flows \citep{Smith1996,Xia2011}.
It has been found that the critical span at which 3-D turbulence dynamics dominate the wake is the Mode B instability wavelength (approximately one diameter), in agreement with \cite{Bao2016}.
Since the Mode B instability helps sustaining the turbulent structures advected from the shear layer, the lack of it prevents large-scale 3-D structures to be created and less dissipative structures can be sustained.
Additionally, it has been also shown that 2-D and 3-D turbulence dynamics can coexist at certain positions of the wake depending on the domain geometric anisotropy.

In order to reduce the computational cost of simulating turbulent wake flow, a model based on the flow spatial average in its homogeneous direction is proposed, thus reducing the problem dimensionality from 3-D to 2-D.
This 2-D model, namely the spanwise-averaged Navier--Stokes (SANS) equations, incorporates 3-D effects in the form of SSR closure terms, which change the flow dynamics from standard 2-D Navier--Stokes to spanwise-averaged dynamics.
The verification of the SANS equations has been conducted for the Taylor--Green vortex and the cylinder wake flow cases using the perfect closure model.
The perfect closure is the residual between the spanwise-averaged 3-D spatial operator and the 2-D spatial operator of the momentum equations.
For both test cases, extracting the perfect closure from a 3-D simulation and inserting it into the 2-D solver has allowed to recover the unsteady spanwise-averaged flow, as shown in the temporal evolution of global energy and enstrophy. 
For the circular cylinder test case, including the perfect closure in the SANS simulation yields forces induced to the cylinder as obtained in the high-fidelity 3-D simulation at a fraction of its computational cost.

Since the perfect closure is not available when deploying SANS in a live simulation, alternative models have been investigated.
A standard EVM employing the Smagorinsky closure has been tested in the a-priori framework.
The EVM has provided very low accuracy for the modelling of the SANS closure terms.
This is related to the intrinsic formulation of the eddy-viscosity hypothesis, which assumes that turbulent fluctuations only have a dissipative effect.
In the spanwise-averaged system, the spanwise stresses can be both dissipative and energising, and this is a fundamental difference which no EVM can reproduce.

Because of the EVM poor performance, a data-driven approach has been designed to improve the SANS closure terms prediction.
The use of data-driven model for a SANS strip-theory method is advantageous because the training dataset can be obtained in a 3-D high-fidelity simulation only comprising the spanwise-averaging length, instead of the full span (as in \fref{fig:strips}, where $L_a\ll L_z$).
This is justified by the flow spanwise homogeneity, since an equivalent dataset would be obtained for a 3-D simulation in a spanwise-shifted location.
The trained ML model could then be used for all the 2-D strips in a SANS strip-theory framework.
A ML model based in a CNN has been considered for the modelling of the spanwise stresses, and a dataset of the closure terms has been generated for a circular cylinder of one diameter span.

In the a-priori analysis, the ML model prediction of the anisotropic SSR tensor components has provided correlations of 90-92\%, thus greatly improving the EVM performance.
While large-scale structures were overall correctly captured, small-scale structures were still hard to predict, as also found in other CNN-based data-driven models for turbulent flows \citep{Lee2019}.
Additionally, the ML model trained with a single test case (circular cylinder at $Re=10^4$) has been assessed for other Reynolds regimes and body geometries.
It has been observed that the ML model is still capable of correctly predicting the SANS closure terms in the downstream wake region with high correlation values ($86\%-96\%$).
On the other hand, the prediction of the near-body region has shown a poorer performance because of the case-dependent shear layer that different flow regimes and body geometries develop.

In the a-posteriori analysis, it has been found that applying the divergence operator to the raw output of the SSR tensor ML model prediction amplifies the modelling error significantly, hence yielding an unstable closure as also observed in \cite{Cruz2019}.
Because of this, the perfect closure has been targeted as the new ML model output, instead of the SSR tensor components, so that no derivatives are directly applied to the ML predictions.
The ML model a-priori performance on the perfect closure terms showed correlations of 89-90\% with the target quantities.
In the a-posteriori analysis, the ML model predicts the perfect closure at every time step based only on the resolved flow quantities $(\vect{U},P)$.
While we found evidence of known stability issues with long-time ML model predictions for dynamical systems, the closed SANS simulations are still capable of predicting wake metrics and induced forces with errors from 1-10\%.
This results in approximately an order of magnitude improvement over standard 2-D simulations while reducing the computational cost of 3-D simulations by 99.5\%.

\section{Technical achievements}

The fluid solver used throughout this thesis is the in-house 3-D iLES immersed-boundary code ``Lotus'', described in \cref{chapter:theoretical_background} and developed by Gabriel D. Weymouth as an extension of the 2-D real-time fluid solver ``Lily Pad'' \citep{Weymouth2015-LilyPad}.
Contributions to ``Lotus'' have been made to accommodate the solver for the needs of this work (or not) such as an extension for internal flows, addition of a passive scalar equation, adaptation of the software package to be run in HPC facilities (supercomputers), and an interface to allow data transfer between the Fortran fluid solver and the Python ML model \citep{Font2019-f2py}.
More details of the Fortran-Python interface can be found in \aref{chapter:appendixE} together with specifications of the hardware employed.

As any computational investigation, a significant part of the workload has been spent in post-processing data.
All the developed post-processing tools have been gathered in \cite{Font2018-postproc}, such as the analysis of turbulence statistics, tracking flow separation points on the cylinder surface, data plotting and video rendering, among others.
On the other hand, the implemented ML model providing closure to the SANS equations can be found in \cite{Font2019-sanspy}. 
It has been written using Keras \citep{Chollet2015}, a high-level deep-learning Python library employing TensorFlow \citep{Tensorflow} as backend.

\section{Limitations}

The SANS equations have been proposed as an improvement of current 2-D strip-theory methods, which do not take into account the 3-D turbulence effects of the flow and instead use arbitrarily dissipative turbulence models.
However, some intrinsic limitations of standard strip-theory methods would also apply to a SANS-based strip-theory method. 
For example, strip-theory methods cannot directly capture variations of the flow out-of-plane angle of attack or twisting of the cylinder.
In this scenario, the flow would no longer be homogeneous in the spanwise direction and the SANS system would struggle to provide a physical answer.
Regarding the strips spanwise resolution, the minimum number of 2-D planes still depends on the highest mode of the structural response, and three 2-D planes per half wavelength would be required \citep{Willden2004}.

With respect to the SANS equations closure, known limitations of ML models have also been faced.
A certain degree of randomness of the ML model has been observed both in a-priori and a-posteriori frameworks.
This is derived from the fact that the CNN weights are initialised in a random manner and that stochastic backpropagation is used for the CNN weights optimisation.
This has translated in failed trainings of the CNN as well as converged optimisations which yield poor results in the a-posteriori analysis.
In these cases, the training has been repeated until a better local minima was reached.

The generalisation of the ML model to flow configurations different from the training case (in terms of Reynolds number and body geometry) has also presented difficulties in the prediction of the SANS closure terms near the body.
This arises from fact that the shear-layer region is strongly case-dependant and such information was only provided by the single training case.
Adding other flow configurations to the database could help mitigate this issue.

Furthermore, the ML model error accumulation in the resolved quantities during an a-posteriori run is an important limitation of the current approach.
Although successful for short temporal dynamics, the implemented ML model diverges from the spanwise-averaged attractor at every time step.
Eventually, it yields a non-physical solution and the simulation has to be stopped.
Still, the current model works as a proof-of-concept which simultaneously serve the purpose of motivating other authors to develop new closures of the SANS equations.

\section{Recommendations for future work}

All the exposed limitations of the current work offer an opportunity for improvement.
The main impediment encountered is, as exposed, the limited closure stability of the current ML model for the SANS equations.
The development of a general and stable closure for SANS could represent a turning point in the simulation of slender bodies as a result of new SANS strip-theory methods.
In this regard, the following ideas could be worth exploring:
\begin{itemize}
	\item A CNN model embedded in a RNN architecture to capture both the spatial patterns and the temporal dynamics of the SANS closure terms.
	In convolutional RNNs, a series of consecutive snapshots are provided as input data so that the model learns how spatial structures evolve in time.
	RNNs have demonstrated to be successful for the prediction of statistically-stationary metrics in chaotic systems \citep{Vlachas2018,Kim2020a}.

	\item Adding a physical constraint in the loss function (a.k.a physics-based machine learning, or PBML) has also been useful for improving the stability of ML-based closures, as shown in \cite{Lee2019}.
	At the same time, the dataset size required for optimisation can be significantly reduced when using PBML models \citep{Weymouth2019}.

	\item Projection of the ML model predictions into a stable closure. This has been done for a ML model of SGS stresses in a LES framework, where the ML model output is fit into an EVM \citep{Beck2019}.
	This approach would not work for the SANS equations since it has been shown that EVMs are not suited for the modelling of spanwise stresses.
	Still, a stable closure accounting for production and dissipation of the closure terms similarly to Reynolds-stress models in RANS could be investigated.

	\item Learning how to reverse the 2-D dynamics domination over the spanwise-averaged dynamics. 
	This could be explored by feeding the ML model a spanwise-averaged snapshot followed by snapshots evolved without the inclusion of the SSR terms.
	The model would be informed that this temporal dynamics are to be corrected by providing the actual spanwise-averaged snapshot of the next time step in the loss function.
	In this way, the ML model weights can be optimise to prevent or reverse 2-D dynamics back to spanwise-averaged flow.

	\item Use of a generative adversarial neural network \citep{Goodfellow2014} able to discern when the ML model prediction is not within the spanwise-averaged system attractor.
	This can also help identifying the 2-D dynamics dominating the flow and trigger a corrective action.
	Such corrective action could be the projection of the 2-D dominated flow field into a realistic spanwise-averaged snapshot according to examples previously shown during training.
\end{itemize}
% ---------------------------------------------------------------- 

\appendix

\chapter{Reynolds regimes of flow past a circular cylinder}

The classification of flow past a circular cylinder according to its Reynolds regime is summarised in \tref{tab:Re_regimes}.
The flow remains laminar, steady and 2-D along the span up to $Re<47$ and, from this point onwards, vortex shedding starts to develop.
A supercritical Hopf bifurcation is responsible for the onset of the laminar vortex shedding \citep{Williamson1996a}.
An oblique shedding with respect to the axis of the cylinder might also be observed when perturbations are introduced into the end-boundaries of the cylinder (often arising in experiments rather than numerical simulations).
Otherwise, the wake remains 2-D up to $Re=160-180$, where the first 3-D instability at the wake (Mode A) comes into the picture.
This mode arises naturally and is characterised by streamwise vortices with a spanwise wavelength ($\lambda_z$) of approximately 4 diameters.
The shedding frequency ($f_s$) experiences a rapid decrease at the onset of this instability.
There is no general agreement on the physical mechanism driving this instability even though some authors suggest a core instability, braid instability, elliptical vortex instability or centrifugal instability \citep{Noack1999}.

\begin{table}[!htpb]
\centering
\captionsetup{width=0.82\textwidth}
  \caption{Reynolds regimes of incompressible viscous flow past a circular cylinder. Note that some Reynolds numbers might differ among authors.}
  \begin{tabular}{lAAA}
    \toprule
    \multicolumn{1}{l}{Flow phenomenon}&\multicolumn{2}{c}{Reynolds regime}&\\ 
    \midrule
    Laminar 2-D steady & &\phantom{<} \hspace{0.2cm} Re<47\\
    Laminar 2-D vortex shedding & 47 &<Re<80\\
    Mode A 3-D wake transition ($\lambda_z/D\approx4$) & 180&<Re<230\\
    Mode B 3-D wake transition ($\lambda_z/D\approx1$) & 200&<Re<10^4-10^5\\
    Mode C 3-D wake transition ($\lambda_z/D\approx2$) & 170&<Re<270\\ 
    Turbulent wake transition & 260&<Re<300\\   
    Subcritical regime & 250&<Re<2\cdot 10^5\\
    Drag crisis & 2\cdot 10^5&<Re<4\cdot 10^5\\
    Supercritical regime & 4\cdot 10^5&<Re\\        
    \bottomrule
  \end{tabular}  
  \label{tab:Re_regimes}
\end{table}

Around $Re=200-230$, Mode B takes place and Mode A gradually decreases.
The Mode B instability presents a spanwise wavelength of roughly $1D$.
Rib-like streamwise vortices are generated and the mode becomes more dominant as $Re$ increases, eventually replacing all other transition modes at $Re=260$.
During this process, the wake spanwise correlation length increases as well.
With this, it can be noted that the natural wake transition scenario is successively: 2-D vortex shedding, Mode A, Mode B.
The Mode C instability is not a natural mode since it requires direct excitation from added perturbations or specific boundary conditions.
It is characterised by a $2T_s$ vortex-shedding period, a spanwise wavelength in the order of 2 diameters, and it has been observed typically within $Re<300$.
For $Re>270$, the Mode C instability periodicity becomes less apparent as the wake transitions to a turbulent state \citep{Jiang2020}.
Further information including diagrams and flow visualisations of 3-D instability mechanisms can be found in \cite{Williamson1996b, Williamson1996a}.

It is difficult to accurately define the actual turbulence transition point on the wake and this is normally indicated through temporal averages.
The transition to turbulence is triggered by the interaction of the different wake instabilities.
With increasing Reynolds number, the subcritical regime takes place.
A sharp decrease of the lift force on the cylinder (increase on base suction) is found as well as a decrease in the Strouhal number.
Up to $Re=5\cdot 10^3$, the transition appears to be caused by rib-like vortices of Mode B-type and the transition point is in a fixed streamwise position with respect to the cylinder.
As the Reynolds number increases, the transition is driven by 3-D structures generated through the interaction of the K\'{a}rm\'{a}n vortices with the separated shear layer, a.k.a. shear-layer transition regime.
This results into an upstream displacement of the transition point \citep{Bloor1964,Williamson1996a,Norberg2001}.

The drag crisis is observed at $Re=2\cdot 10^5$ and arises because of the cylinder boundary layer becoming turbulent.
This causes the separated shear layer to narrow as the separation points are moved towards the rear part of the cylinder.
A sharp reduction in drag and base suction can be appreciated.
Laminar separation bubbles can also be observed on the cylinder surface.
Finally, above $Re=4\cdot 10^5$, the supercritical (and transcritical for $Re>10^6$) regime is defined.
The wake is further narrowed and the shedding frequency increases.
Asymmetrical (laminar/transitional) states of separation bubbles at both upper and lower sides of the cylinder can be observed \citep{Noack1999b}.
% ---------------------------------------------------------------- 

\chapter{Validation and verification of the in-house solver}

A grid convergence study has been conducted in order verify the correct implementation of the governing equations and to show that the selected grid resolution and averaging time length is suitable for a proper analysis.
Three grids with different resolution $(\Delta)$ designed as shown in \fref{fig:computational_domain} have been investigated for the $L_z=\pi$ case (see details in \tref{tab:verification_grids}).
The refinement ratio is $\sqrt{2}$.

\begin{figure}
  \centerline{\includegraphics[width=0.94\textwidth]{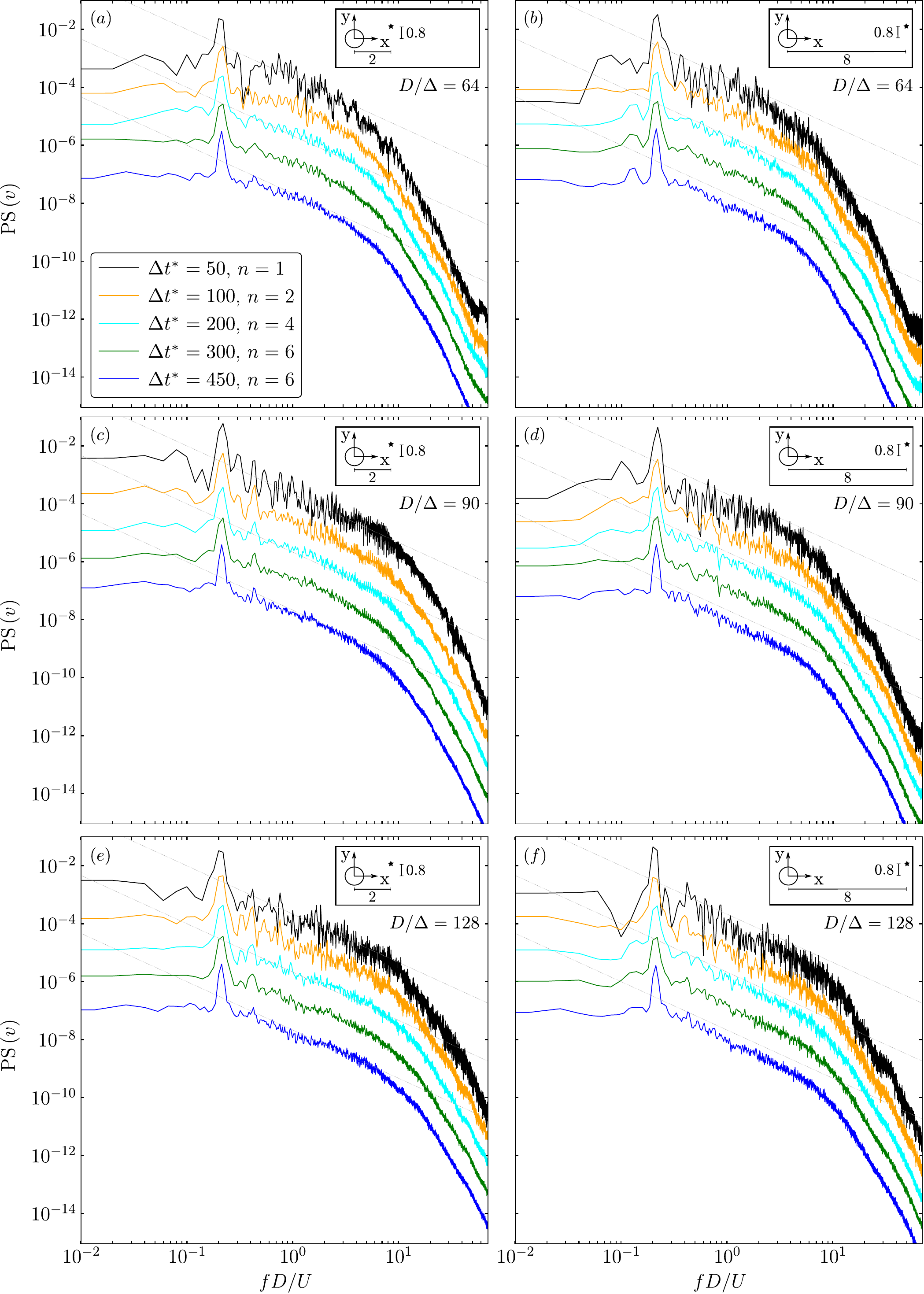}}
  \caption{Vertical velocity component temporal PS at the close (left) and far (right) wake regions.
The coarse, medium and fine grid results are displayed at the first, second and third row respectively.
The spectra are shifted a factor of 10 and the vertical axis ticks correspond to the $\Delta t^*=50$ case.
The number of the time signal splits selected for the Welch method ($n$) ensures a minimum of $\Delta t^*=50$ per split (at least 8 times larger than the lowest frequency of interest, i.e. the shedding frequency which is approximately $\Delta t^*=5$), and a maximum of 6 splits per total time signal.
The dotted line corresponds to a $-5/3$ slope.}
\label{fig:velocity_spectras_GC}
\end{figure}

\begin{figure}
  \centerline{\includegraphics[width=1\textwidth]{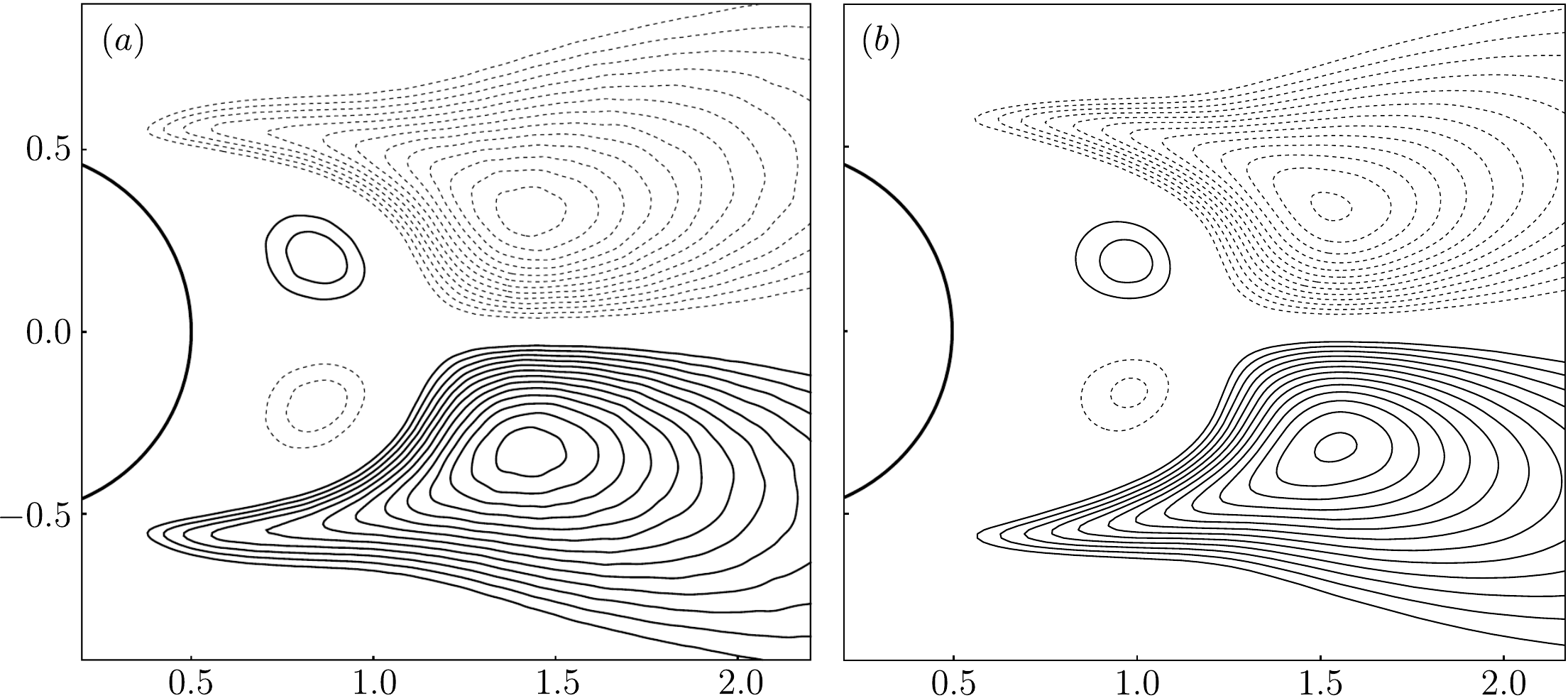}}
  \caption{Comparison of the Reynolds shear stress $\overline{u'v'}$ between: (a) DNS data from \cite{Dong2005} and (b) present solver.
The contour levels criteria is the same as the reference data: ${|\overline{u'v'}|}_{\mathrm{min}}=0.03$ and $\Delta\overline{u'v'}=0.01$.
Positive and negative levels are noted with continuous and dashed lines respectively.}
\label{fig:uv_validation}
\end{figure}

The quantities of interest of the present work are the turbulence statistics arising from time and spatial averages.
Because of this, we need to ensure that the presented results are statistically stationary and have converged in spatial resolution terms.
The velocity temporal power spectrum constructed analogously to \fref{fig:velocity_spectras} at the $(x,y)=\left(2,0.8\right),\left(8,0.8\right)$ locations is used with this purpose.
The spectrum produced by the investigated grids is shown for different time signal lengths in \fref{fig:velocity_spectras_GC}.

\begin{table}
  \begin{center}
\def~{\hphantom{0}}
  \begin{tabular}{rrr}
  	   \toprule
       $D/\Delta$ & $N_{CG}$ (M) & $N_{T}$ (M) \\[3pt]
       \midrule
       64  & 65.3 & 109.2\\
       90  & 189.5 & 311.4 \\
       128 & 510.6 & 855.6\\
       \bottomrule
  \end{tabular}
  \caption{Details of the three different grids ranging from coarse ($D/\Delta=64$, i.e. 64 cells per diameter) to fine ($D/\Delta=128)$ for $L_z=\pi$.
$N_{GC}$ refers to the total number of cells (in millions) for the Cartesian grid subdomain.
$N_T$ refers to the total number of cells (in millions).}
  \label{tab:verification_grids}
  \end{center}
\end{table}

At the near wake region, the coarse grid shows a converged spectrum ($-5/3$ inertial subrange decay) for time signals with $\Delta t^*>200$.
A similar behaviour is captured with the medium grid, which shows a converged spectrum for time signals with $\Delta t^*>100$ and the resulting inertial subrange accommodates more frequencies.
At the far wake region, the coarse grid presents slightly steeper spectra than at the close wake region.
This can be caused by an insufficient resolution in span, which would induce a two-dimensionalisation effect.
On the other hand, the medium grid still presents converged spectra for time signals with $\Delta t^*>100$.
The fine grid leads to results very similar to the medium grid.
Hence, it is shown that using the medium grid with simulation times over $\Delta t^*>100$ provides statistically stationary results both in the close and far wake regions ($\Delta t^*=500$ is used in the results presented in \cref{chapter:jfm2019}).

To validate the wake turbulence statistics, the Reynolds shear stress $\overline{u'v'}$ has also been qualitatively analysed.
It can be observed in \fref{fig:uv_validation} that the shear stress predicted by the present solver is in very good agreement with the DNS data from \cite{Dong2005}, with just a slight shift of the field structures on the streamwise direction.
All positive and negative regions are correctly captured which display an antisymmetric pattern with respect to the centreline of the wake.
% ---------------------------------------------------------------- 

\chapter{Additional SANS derivations}

\section{The spanwise-averaged vorticity transport equation}
\label{sec:savte}
The non-dimensional vorticity transport equation (VTE) can be written in its conservation form as
\begin{equation}
\partial_t\boldsymbol\omega+\nabla \cdot \pars{\vect{u}\otimes\boldsymbol{\omega}}=\nabla \cdot \pars{\boldsymbol{\omega}\otimes\vect{u}}+Re^{-1}\nabla^2\boldsymbol{\omega}.\label{eq:VTE}
\end{equation}
Decomposing and averaging similarly to \sref{sec:sans_math}, the following spanwise-averaged VTE can be obtained for the $\vect{e}_z$ component of \eref{eq:VTE},
\begin{gather}
\partial_t {\Omega}_z + U\partial_x{\Omega}_z + V\partial_y{\Omega}_z = Re^{-1}\pars{\partial_{xx}{\Omega}_z + \partial_{yy}{\Omega}_z}-\nonumber \\
-\partial_x\pars{\avg{u\p\omega_z\p}-\avg{\omega_x\p w\p}}-\partial_y\pars{\avg{v\p\omega_z\p}-\avg{\omega_y\p w\p}}+\nonumber \\
+{\Omega}_x\partial_xW+{\Omega}_y\partial_yW+{\Omega}_z\avg{\partial_z w\p}-W\avg{\partial_z\omega_z\p}+Re^{-1}\avg{\partial_{zz}\omega_z\p}, \label{eq:vort_e3_avg}
\end{gather}
where the spanwise-averaged divergence of the velocity and vorticity vector fields, respectively
\begin{gather}
\partial_x{U}+\partial_y{V}=-\avg{\partial_z{w\p}},\\
\partial_x{{\Omega}_x}+\partial_y{{\Omega}_y}=-\avg{\partial_z{\omega_z\p}},
\end{gather}
has been introduced to obtain the non-conservation form.

When spanwise periodicity is assumed, the third line in \eref{eq:vort_e3_avg} vanishes by its last three terms cancelling by definition (\eref{eq:periodic_assumption}), and the first two terms cancelling each other as follows,
\begin{gather}
{\Omega}_x\partial_xW+{\Omega}_y\partial_yW=\avg{\partial_y{w}-\partial_z{v}}\partial_x{W}+\avg{\partial_z{u}-\partial_z{w}}\partial_y{W}=\nonumber\\
=\avg{\partial_y{\pars{W+w\p}}-\partial_z{\pars{V+v\p}}}\partial_x{W}+\avg{\partial_z{\pars{U+u\p}}-\partial_z{\pars{W+v\p}}}\partial_y{W}=\nonumber\\
=\pars{\partial_y{W}-\cancelto{0}{\avg{\partial_z{v\p}}}}\partial_x{W}+\pars{\cancelto{0}{\avg{\partial_z{u\p}}}-\partial_x{W}}\partial_y{W}=0.
\end{gather}

The spanwise-averaged VTE is then simplified to
\begin{equation}
\partial_t {\Omega}_z + \vect{U}\cdot\nabla{\Omega}_z = Re^{-1}\nabla^2{\Omega}_z-\nabla\cdot\boldsymbol\zeta^R,
\end{equation}
where $\vect{U}=\pars{U,V}$ and $\boldsymbol \zeta^R = \pars{\avg{u\p\omega_z\p}-\avg{\omega_x\p w\p}, \avg{v\p\omega_z\p}-\avg{\omega_y\p w\p}}$.

\subsection{Cylindrical coordinates} \label{sec:cyl_coords}

In a cylindrical coordinates system, using $(\vect{e}_r,\vect{e}_\theta,\vect{e}_z)$ and
\begin{equation}
\nabla = \pars{\frac{1}{r}\pd{\pars{r\cdot}}{r}, \frac{1}{r}\pd{\pars{\cdot}}{\theta}, \pd{\pars{\cdot}}{z}},
\end{equation}
the $\vect{e}_z$ component of the non-dimensional VTE (\eref{eq:VTE}) can be written as
\begin{gather}
\pd{\omega_z}{t}+\frac{1}{r}\pd{\pars{r u_r\omega_z}}{r}+\frac{1}{r}\pd{\pars{u_\theta\omega_z}}{\theta}+\cancel{\pd{\pars{u_r\omega_z}}{z}}=\nonumber\\
=\frac{1}{r}\pd{\pars{r \omega_r u_z}}{r}+\frac{1}{r}\pd{\pars{\omega_\theta u_z}}{\theta}+\cancel{\pd{\pars{\omega_z u_r}}{z}}+\frac{1}{Re}\pars{\ddn{\omega_z}{r}{2}+\frac{1}{r}\pd{\omega_z}{r}+\frac{1}{r^2}\ddn{\omega_z}{\theta}{2}+\ddn{\omega_z}{z}{2}}.
\label{eq:VTE_cyl}
\end{gather}

A decomposition and average similarly to \sref{sec:sans_math} yields
\begin{gather}
\pd{{\Omega}_z}{t}+\frac{1}{r}\pd{\pars{r U_r{\Omega}_z+\avg{r u_r\p\omega_z\p}}}{r}+\frac{1}{r}\pd{\pars{U_\theta{\Omega}_z+\avg{u_\theta\p\omega_z\p}}}{\theta}=\nonumber\\
=\frac{1}{r}\pd{\pars{r {\Omega}_r U_z + \avg{r\omega_r\p u_z\p}}}{r}+\frac{1}{r}\pd{\pars{{\Omega}_\theta U_z+\avg{\omega_\theta\p u_z\p}}}{\theta}+\nonumber\\
+\frac{1}{Re}\pars{\ddn{{\Omega}_z}{r}{2}+\frac{1}{r}\pd{{\Omega}_z}{r}+\frac{1}{r^2}\ddn{{\Omega}_z}{\theta}{2}+\avg{\ddn{\omega_z\p}{z}{2}}}.
\label{eq:VTE_cyl_avg}
\end{gather}

The spanwise-averaged divergence of the velocity and vorticity fields can be respectively written in cylindrical coordinates as
\begin{gather}
\frac{1}{r}\pd{\pars{rU_r}}{r}+\frac{1}{r}\pd{U_\theta}{\theta}=-\avg{\pd{u_z\p}{z}},\label{eq:cont_avg_cyl}\\
\frac{1}{r}\pd{\pars{r{\Omega}_r}}{r}+\frac{1}{r}\pd{{\Omega}_\theta}{\theta}=-\avg{\pd{\omega_z\p}{z}}\label{eq:divort_avg_cyl},
\end{gather}
and introducing \eref{eq:cont_avg_cyl} and \eref{eq:divort_avg_cyl} into \eref{eq:VTE_cyl_avg} yields
\begin{gather}
\pd{{\Omega}_z}{t}+U_r\pd{{\Omega}_z}{r}+\frac{U_\theta}{r}\pd{{\Omega}_z}{\theta}={\Omega}_r\pd{U_z}{r}+\frac{{\Omega}_\theta}{r}\pd{U_z}{\theta}+\frac{1}{Re}\pars{\ddn{{\Omega}_z}{r}{2}+\frac{1}{r}\pd{{\Omega}_z}{r}+\frac{1}{r^2}\ddn{{\Omega}_z}{\theta}{2}}-\nonumber\\
-\frac{1}{r}\bracs{\pd{}{r}\pars{\avg{r u_r\p\omega_z\p}-\avg{r \omega_r\p u_z\p}}+
\pd{}{\theta}\pars{\avg{u_\theta\p\omega_z\p}-\avg{\omega_\theta\p u_z\p}}}+\nonumber\\
+{\Omega}_z\avg{\pd{u_z\p}{z}}-U_z\avg{\pd{\omega_z\p}{z}}+\frac{1}{Re}\avg{\ddn{\omega_z\p}{z}{2}}.
\end{gather}

\subsection{Relationship between SANS and spanwise-averaged VTE formulations}

Next, we show how the spanwise stresses in the SANS and the spanwise-averaged VTE frameworks are related.
First, we derive an expression for the vorticity-velocity terms of the spanwise-averaged VTE to write them as velocity-velocity terms.
Considering the following vector identity
\begin{equation}
\frac{1}{2} \nabla \left( \vect{a}\cdot\vect{a} \right) = (\vect{a} \cdot \nabla) \vect{a} + \vect{a} \times (\nabla \times \vect{a}),
\end{equation}
we can rewrite the vector identity in terms of fluctuating quantities
\begin{equation}
\frac{1}{2} \nabla \left( \vect{u}\p\cdot\vect{u}\p \right) = (\vect{u}\p \cdot \nabla) \vect{u}\p + \vect{u}\p \times \boldsymbol{\omega}\p,\label{eq:vect_identity_fluct}
\end{equation}
where the definition of the vorticity vector field, $\vect{\omega}=\nabla\times\vect{u}$, has been introduced.

Second, the conservation form of the convective forces can be expressed as
\begin{equation}
\nabla\cdot\pars{\vect u \otimes\vect u}=\pars{\vect u \cdot \nabla}\vect u + \vect u\pars{\nabla\cdot\vect u},
\end{equation}
and it can be rewritten for the fluctuating quantities as
\begin{gather}
\nabla\cdot\pars{\vect u\p \otimes\vect u\p}=\pars{\vect u\p \cdot \nabla}\vect u\p + \vect u\p\pars{\nabla\cdot\vect u\p},\\
\pars{\vect u\p \cdot \nabla}\vect u\p = \nabla\cdot\pars{\vect u\p \otimes\vect u\p}-\vect u\p\pars{\nabla\cdot\vect u\p}.\label{eq:conv_term_fluct}
\end{gather}

Third, decomposing the continuity equation for incompressible flows (\eref{eq:cont})
\begin{equation}
\partial_xU+\partial_xu\p+\partial_yV+\partial_yv\p+\partial_zw\p=0,\label{eq:continuity_decomp}
\end{equation}
and subtracting the averaged continuity equation (\eref{eq:cont_decomp_avg}) yields
\begin{equation}
\partial_xu\p+\partial_yv\p+\partial_zw\p-\avg{\partial_z w\p}=0,
\end{equation}
hence
\begin{equation}
\nabla\cdot\vect u\p =\avg{\partial_z w\p}.\label{eq:cont_fluct}
\end{equation}

We now introduce \eref{eq:cont_fluct} into \eref{eq:conv_term_fluct} obtaining
\begin{equation}
\pars{\vect u\p \cdot \nabla}\vect u\p = \nabla\cdot\pars{\vect u\p \otimes\vect u\p}-\vect u\p\avg{\partial_z w\p},\label{eq:conv_term_fluct2}
\end{equation}
and \eref{eq:conv_term_fluct2} is finally introduced into \eref{eq:vect_identity_fluct}
\begin{equation}
\frac{1}{2} \nabla \left( \vect{u}\p\cdot\vect{u}\p \right) = \nabla\cdot\pars{\vect u\p \otimes\vect u\p}-\vect u\p\avg{\partial_z w\p} + \vect{u}\p \times \boldsymbol{\omega}\p,
\end{equation}
which yields
\begin{equation}
\boldsymbol{\omega}\p\times\vect{u}\p = \nabla\cdot\pars{\vect u\p \otimes\vect u\p} - \frac{1}{2} \nabla \left( \vect{u}\p\cdot\vect{u}\p \right) - \vect u\p\avg{\partial_z w\p}.
\label{eq:vect_identity_fluct2}
\end{equation}

With this, a relation between the vorticity-velocity spanwise stresses and the velocity-velocity spanwise stresses is defined in the conservation form.
To derive a more explicit form of the spanwise-averaged stresses, let us average \eref{eq:vect_identity_fluct2} and retrieve its $\vect{e}_x$ and $\vect{e}_y$ components
\begin{gather}
\vect{e}_x: \quad \avg{\omega_y\p w\p}-\avg{\omega_z\p v\p} = \partial_x\avg{u\p u\p}+\partial_y\avg{v\p u\p}+\avg{\partial_z\pars{w\p u\p}}-\nonumber\\
-\partial_x\pars{\avg{u\p u\p}+\avg{v\p v\p}+\avg{w\p w\p}}/2,\label{eq:vec_ident_fluct_e1}\\
\vect{e}_y: \quad \avg{\omega_z\p u\p}-\avg{\omega_x\p w\p} = \partial_x\avg{u\p v\p}+\partial_y\avg{v\p v\p}+\avg{\partial_z\pars{w\p v\p}}-\nonumber\\
-\partial_y\pars{\avg{u\p u\p}+\avg{v\p v\p}+\avg{w\p w\p}}/2\label{eq:vec_ident_fluct_e2}.
\end{gather}
Note that the last term of \eref{eq:vect_identity_fluct2} vanishes because of \eref{eq:rule2}. 

The spanwise stresses of the spanwise-averaged VTE (\eref{eq:vort_e3_avg}) can be expanded using Eqs. \ref{eq:vec_ident_fluct_e1} and \ref{eq:vec_ident_fluct_e2} as follows
\begin{gather}
-\partial_x\pars{\avg{u\p\omega_z\p}-\avg{\omega_x\p w\p}}-\partial_y\pars{\avg{v\p\omega_z\p}-\avg{\omega_y\p w\p}} = \nonumber\\
=-\partial_x\pars{\partial_x\avg{u\p v\p}+\partial_y\avg{v\p v\p}+\avg{\partial_z\pars{w\p v\p}} - \partial_y\pars{\avg{u\p u\p}+\avg{v\p v\p}+\avg{w\p w\p}}/2} +\nonumber\\
+\partial_y\pars{\partial_x\avg{u\p u\p}+\partial_y\avg{v\p u\p}+\avg{\partial_z\pars{w\p u\p}} - \partial_x\pars{\avg{u\p u\p}+\avg{v\p v\p}+\avg{w\p w\p}}/2}=\nonumber\\
=-\partial_{xx}\avg{u\p v\p}-\partial_{xy}\avg{v\p v\p}-\partial_x\avg{\partial_z\pars{w\p v\p}}+\partial_{xy}\pars{\avg{u\p u\p}+\avg{v\p v\p}+\avg{w\p w\p}}/2+\nonumber\\
+\partial_{xy}\avg{u\p u\p}+\partial_{yy}\avg{v\p u\p}+\partial_y\avg{\partial_z\pars{w\p u\p}}-\partial_{xy}\pars{\avg{u\p u\p}+\avg{v\p v\p}+\avg{w\p w\p}}/2=\nonumber\\
=\pars{\partial_{yy}-\partial_{xx}}\avg{u\p v\p}+\partial_{xy}\pars{\avg{u\p u\p}-\avg{v\p v\p}}+\partial_y\avg{\partial_z\pars{w\p u\p}}-\partial_x\avg{\partial_z\pars{w\p v\p}}\label{eq:vte_spanwise_stresses}.
\end{gather}

On the other hand, the SANS equations (Eqs. \ref{eq:z-avg_X} to \ref{eq:z-avg_Z}) have yield the following spanwise stresses
\begin{equation}
-\avg{\nabla\cdot\pars{\vect u\p\otimes\vect u\p}}=
-\begin{pmatrix}
    \partial_x\avg{u\p u\p}+\partial_y\avg{u\p v\p}+\avg{\partial_z\pars{w\p u\p}} \\
    \partial_x\avg{v\p u\p}+\partial_y\avg{v\p v\p}+\avg{\partial_z\pars{w\p v\p}} \\
    \partial_x\avg{w\p u\p}+\partial_y\avg{w\p v\p}+\avg{\partial_z\pars{w\p w\p}}
  \end{pmatrix}.
\end{equation}
Taking the curl of the momentum spanwise stresses
\begin{equation}
-\nabla\times\avg{\nabla\cdot\pars{\vect u\p\otimes\vect u\p}}=-
\begin{pmatrix}
    \vect{e}_x & \partial_x & \partial_x\avg{u\p u\p}+\partial_y\avg{u\p v\p}+\avg{\partial_z\pars{w\p u\p}} \\
    \vect{e}_y & \partial_y & \partial_x\avg{v\p u\p}+\partial_y\avg{v\p v\p}+\avg{\partial_z\pars{w\p v\p}} \\
    \vect{e}_z & \partial_z & \partial_x\avg{w\p u\p}+\partial_y\avg{w\p v\p}+\avg{\partial_z\pars{w\p w\p}}
  \end{pmatrix},
\end{equation}
recovers the VTE spanwise stresses as demonstrated for the $\vect{e}_z$ component
\begin{gather}
-\nabla\times\avg{\nabla\cdot\pars{\vect u\p\otimes\vect u\p}}\vect{e}_z=\nonumber\\
=-\partial_x\pars{\partial_x\avg{v\p u\p}+\partial_y\avg{v\p v\p}+\avg{\partial_z\pars{w\p v\p}}}
+\partial_y\pars{\partial_x\avg{u\p u\p}+\partial_y\avg{u\p v\p}+\avg{\partial_z\pars{w\p u\p}}}=\nonumber\\
=-\partial_{xx}\avg{v\p u\p}-\partial_{xy}\avg{v\p v\p}-\partial_x\avg{\partial_z\pars{w\p v\p}}+\partial_{xy}\avg{u\p u\p}+\partial_{yy}\avg{u\p v\p}+\partial_y\avg{\partial_z\pars{w\p u\p}}=\nonumber\\
=\pars{\partial_{yy}-\partial_{xx}}\avg{u\p v\p}+\partial_{xy}\pars{\avg{u\p u\p}-\avg{v\p v\p}}+\partial_y\avg{\partial_z\pars{w\p u\p}}-\partial_x\avg{\partial_z\pars{w\p v\p}},\label{eq:NS_spanwise_stresses}
\end{gather}
showing that the spanwise stresses in both formulations (Eqs. \ref{eq:vte_spanwise_stresses} and \ref{eq:NS_spanwise_stresses}) are equivalent.

\section{Analytical validation and quadrature error analysis}

The aim of this exercise is to analytically demonstrate the correct derivation of the SANS equations.
The velocity vector field needs to be solenoidal and periodic, in agreement with the assumption used for the derivation of the simplified SANS equations (\eref{eq:sans_simple}). For this, the Taylor--Green vortex case is considered in a $2\pi$-periodic box.
The pressure field is taken constant.
The components of its periodic solenoidal initial velocity vector field might be written as
\begin{gather}
u = \cos (x) \sin (y) \sin (z), \label{eq:tg-u} \\
v = \sin (x) \cos (y) \sin (z), \\
w = -2 \sin (x) \sin (y) \cos (z),\label{eq:tg-w}
\end{gather}
which is indeed a solenoidal vector field
\begin{gather}
\nabla \cdot \vect{u} = \pd{u}{x}+\pd{v}{y}+\pd{w}{z}=\nonumber\\
=- \sin (x) \sin (y) \sin (z)- \sin (x) \sin (y) \sin (z)+2\sin (x) \sin (y) \sin (z)=0.
\end{gather}

\subsection{No quadrature errors and spanwise periodicity}
\label{sec:sans_analytical}

The averaging operation in the spanwise direction $\avg{\cdot}$ is defined on the $z\in\left[a,b\right]$ interval as
\begin{equation}
\avg{q}\pars{x,y,t} = \frac{1}{b-a}\int^b_a q\pars{x,y,z,t}\,dz.
\end{equation}
Numerically, this carries a quadrature error which we will neglect for the moment.
The following condition needs to hold true assuming an error-free quadrature and spanwise periodicity,
\begin{equation}
\avg{\mathcal{S}\left(\vect{u}, p\right)} = \tilde{\mathcal{S}}\left(\vect{U}, P\right) + \nabla\cdot\boldsymbol\tau_{ij}^R, \label{eq:balance}
\end{equation}
where
\begin{align}
\mathcal{S}= \vect{u}\cdot\nabla \vect{u} + \nabla p - \nu \nabla^2 \vect{u}, \qquad &\vect{u}=\pars{u,v,w},\,\,\, \vect{x}=\pars{x,y,z},\\
 \tilde{\mathcal{S}}= \vect{U}\cdot\nabla \vect{U} + \nabla P - \nu \nabla^2 \vect{U}, \qquad &\vect{U} = \pars{U,V},\,\,\, \vect{x}=\pars{x,y}.
\end{align}

The derivation to validate the SANS equations will be performed for the $\vect{e}_x$ component of \eref{eq:balance}.
In a $L=b-a=2\pi$ domain and considering the velocity vector field defined in \eref{eq:tg-u} to \eref{eq:tg-w}, we obtain $\vect{U}=\pars{0,0,0}$, so $\vect{u}=\vect{u}\p$.

We begin calculating the LHS of \eref{eq:balance},
\begin{equation}
\avg{\mathcal{S}}\vect{e}_x=\avg{u\pd{u}{x}+v\pd{u}{y}+w\pd{u}{z}+\pd{p}{x}-\nu\pars{\ddn{u}{x}{2}+\ddn{u}{y}{2}+\ddn{u}{z}{2}}}.\label{eq:LHS}
\end{equation}
The required derivatives are
\begin{gather}
\begin{align}
\pd{u}{x}&=-\sin (x) \sin (y) \sin (z), \qquad \ddn{u}{x}{2}=-\cos (x) \sin (y) \sin (z)=-u,\\
\pd{u}{y}&= \cos (x) \cos (y) \sin (z), \qquad\,\,\,\,\, \ddn{u}{y}{2}=-\cos (x) \sin (y) \sin (z)=-u,\\
\pd{u}{z}&=\cos (x) \sin (y) \cos (z), \qquad\,\,\,\,\, \ddn{u}{z}{2}=-\cos (x) \sin (y) \sin (z)=-u.
\end{align}
\end{gather}

Substituting the derivatives into \eref{eq:LHS} yields
\begin{gather}
\avg{\mathcal{S}}\vect{e}_x=\avg{u\pd{u}{x}+v\pd{u}{y}+w\pd{u}{z}+\pd{p}{x}-\nu\pars{\ddn{u}{x}{2}+\ddn{u}{y}{2}+\ddn{u}{z}{2}}}= \nonumber \\
=\bigl\langle-\sin (x) \cos (x) \sin^2 (y) \sin^2 (z) + \sin (x) \cos (x) \cos^2 (y) \sin^2 (z)-\nonumber\\
\quad\;\;-2\sin (x) \cos (x) \sin^2 (y) \cos^2 (z) + 3\nu u \bigr\rangle = \nonumber\\
=-\frac{\pi}{2\pi}\sin (x) \cos (x) \sin^2 (y) + \frac{\pi}{2\pi} \sin (x) \cos (x) \cos^2 (y) -\frac{2\pi}{2\pi} \sin (x) \cos (x) \sin^2 (y) + 0=\nonumber\\
=\frac{1}{2}\sin(x) \cos(x) \pars{\cos^2(y)-3\sin^2(y)}.
\label{eq:b21_lhs}
\end{gather}

Next, we calculate the RHS of \eref{eq:balance}. Note that $\tilde{\mathcal{S}}=0$ because $\vect{U}=0$ and the pressure is constant.
Hence, only the spanwise stresses remain, analytically
\begin{gather}
\pars{\nabla\cdot\boldsymbol\tau_{ij}^R}\vect{e}_x = \pd{}{x}\avg{u\p u\p}+\pd{}{y}\avg{u\p v\p} = \pd{}{x}\avg{u u}+\pd{}{y}\avg{u v} = \nonumber \\
=\pd{}{x}\pars{\avg{\cos^2(x) \sin^2(y) \sin^2(z)}}+\pd{}{y}\pars{\avg{\sin(x)\cos(x)\sin(y)\cos(y)\sin^2(z)}}=\nonumber \\
=\pd{}{x}\pars{ \frac{\pi }{2\pi }\cos^2(x) \sin^2(y)}+\pd{}{y}\pars{\frac{\pi }{2\pi }\sin(x)\cos(x)\sin(y)\cos(y)}=\nonumber \\
=-\sin(x)\cos(x)\sin^2(y) +\frac{1}{2}\sin(x)\cos(x)\pars{\cos^2(x)-\sin^2(x)}=\nonumber \\
=\frac{1}{2} \sin(x) \cos(x) \pars{\cos^2(y)-3\sin^2(y)}.
\label{eq:b21_rhs}
\end{gather}

Finally, we can show that \eref{eq:balance} has the correct balance between both sides by bringing all the pieces together, i.e. using \eref{eq:b21_lhs} as LHS and \eref{eq:b21_rhs} as RHS,
\begin{equation}
\frac{1}{2} \sin(x) \cos(x) \pars{\cos^2(y)-3\sin^2(y)} = 0 + \frac{1}{2} \sin(x) \cos(x) \pars{\cos^2(y)-3\sin^2(y)}.
\end{equation}

\subsection{Quadrature errors without spanwise periodicity}
\label{sec:sans_quadrature_errors1}

When quadrature errors arising from the averaging operation are considered, we have to recall the second-line terms in \eref{eq:sans_full} that were neglected in the previous section.
The quadrature error depends on the function being averaged and the quadrature itself.
From here, we will consider the composite trapezoidal quadrature which has the following associated error on the integral interval $[a,b]$
\begin{equation}
\epsilon_T(\xi) = -\frac{(b-a)h^2}{12} f''(\xi), \qquad \xi \in [a,b],
\end{equation}
where $h$ is the subinterval (spacing) length defined as $h=(b-a)/N$, and $N$ is the number of subintervals for $N+1$ quadrature points.
Note that the convergence error is $\mathcal{O}\pars{h^2}$ and the presence of $f''$, which implies that the trapezoidal rule can perfectly calculate the quadrature of zero and first degree polynomial functions.

For the Taylor--Green vortex case, only first- and second-degree trigonometric functions are required. We can quantify the error associated to its quadrature using the definition above in the $2\pi$ interval as
\begin{gather}
I_1 =\int^{2\pi}_0 \sin(z) dz, \\
\epsilon_1\pars{\xi} := I_1 - I_{1,N} = \frac{\pi h^2}{6} \sin(\xi),
\end{gather}
and
\begin{gather}
I_2 =\int^{2\pi}_0 \sin^2(z) dz, \\
\epsilon_2\pars{\xi} := I_2 - I_{2,N} = -\frac{\pi h^2}{6} \bracs{2\pars{\cos^2(\xi)-\sin^2(\xi)}}.
\end{gather} 
Next, we bound $\epsilon_1$ and $\epsilon_2$ with $|f''(\xi)|_{\max}$ within $[0,2\pi]$,
\begin{gather}
|\epsilon_1|_{\max} = \frac{\pi h^2}{6} \cdot 1, \\
|\epsilon_2|_{\max} = \frac{\pi h^2}{6} \cdot 2,
\end{gather}
hence
\begin{equation}
|\epsilon_2|_{\max}= 2|\epsilon_1|_{\max}.
\end{equation}
Defining $I_1^*$ and $I_2^*$ as $I^* = I/(2\pi)$, we obtain
\begin{gather}
I^*_1 = \frac{0+\epsilon}{2\pi}=\frac{\epsilon}{2\pi}, \\
I^*_2 = \frac{\pi+2\epsilon}{2\pi}=\frac{1}{2}+\frac{\epsilon}{\pi},
\end{gather}
where $\epsilon = \pi h^2/6$. Note that $I^*_1$ and $I^*_2$ using $\cos(z)$ instead of $\sin(z)$ are similarly calculated. 

Considering quadrature errors, \eref{eq:balance} has to be corrected with the terms previously neglected because of spanwise periodicity,
\begin{equation}
\avg{\mathcal{S}\left(\vect{u}, p\right)} = \tilde{\mathcal{S}}\left(\vect{U}, P\right) + \nabla\cdot\boldsymbol\tau_{ij}^R+W\avg{\partial_z\vect{u}\p}-\nu\avg{\partial_{zz}\vect{u}\p}+\avg{\partial_z\pars{w\p\vect{u}\p}}.\label{eq:balance_all}
\end{equation}

We proceed by calculating again the new average and fluctuating quantities using the errors described above as follows
\begin{gather}
U = \frac{\epsilon}{2\pi}\cos(x)\sin(y),\qquad u\p=\cos(x)\sin(y)\pars{\sin(z)-\frac{\epsilon}{2\pi}},\\
V = \frac{\epsilon}{2\pi}\sin(x)\cos(y),\qquad v\p=\sin(x)\cos(y)\pars{\sin(z)-\frac{\epsilon}{2\pi}},\\
W = -\frac{\epsilon}{\pi}\sin(x)\sin(y),\,\,\,\,\, w\p=2\sin(x)\sin(y)\pars{\frac{\epsilon}{2\pi}-\cos(z)}.
\end{gather}
With this, the LHS of \eref{eq:balance_all} can be written as
\begin{gather}
\avg{\mathcal{S}}\vect{e}_x=\avg{u\pd{u}{x}+v\pd{u}{y}+w\pd{u}{z}+\pd{p}{x}-\nu\pars{\ddn{u}{x}{2}+\ddn{u}{y}{2}+\ddn{u}{z}{2}}}=\nonumber \\ 
=\bigl\langle-\sin (x) \cos (x) \sin^2 (y) \sin^2 (z) + \sin (x) \cos (x) \cos^2 (y) \sin^2 (z)-\nonumber\\
\quad\;\;-2\sin (x) \cos (x) \sin^2 (y) \cos^2 (z) + 3\nu u \bigr\rangle = \nonumber\\
=-\pars{\frac{1}{2}+\frac{\epsilon}{\pi}}\sin(x)\cos(x)\sin^2(y) +\pars{\frac{1}{2}+\frac{\epsilon}{\pi}}\sin(x)\cos(x)\cos^2(y) \nonumber \\
-2\pars{\frac{1}{2}+\frac{\epsilon}{\pi}}\sin(x)\cos(x)\sin^2(y)+3\nu\frac{\epsilon}{2\pi}\cos(x)\sin(y) = \nonumber\\
=\pars{\frac{1}{2}+\frac{\epsilon}{\pi}}\sin(x)\cos(x)\pars{\cos^2(y)-3\sin^2(y)}+3\nu\frac{\epsilon}{2\pi}\cos(x)\sin(y).\label{eq:S_avg}
\end{gather}

The spatial operator of the spanwise-averaged quantities $(\tilde{\mathcal{S}})$ considering quadrature errors can be written as
\begin{gather}
\tilde{\mathcal{S}}\,\vect{e}_x=U\pd{U}{x}+V\pd{U}{y}+\pd{P}{x}-\nu\pars{\ddn{U}{x}{2}+\ddn{U}{y}{2}}=\nonumber\\
=\frac{\epsilon}{2\pi}\cos(x)\sin(y)\pd{}{x}\pars{\frac{\epsilon}{2\pi}\cos(x)\sin(y)}+\frac{\epsilon}{2\pi}\sin(x)\cos(y)\pd{}{y}\pars{\frac{\epsilon}{2\pi}\cos(x)\sin(y)}+0-\nonumber\\
-\nu\bracs{\ddn{}{x}{2}\pars{\frac{\epsilon}{2\pi}\cos(x)\sin(y)}+\ddn{}{y}{2}\pars{\frac{\epsilon}{2\pi}\cos(x)\sin(y)}}=\nonumber\\
=-\pars{\frac{\epsilon}{2\pi}}^2\sin(x)\cos(x)\sin^2(x)+\pars{\frac{\epsilon}{2\pi}}^2\sin(x)\cos(x)\cos^2(y)+2\nu\frac{\epsilon}{2\pi}\cos(x)\sin(y)=\nonumber\\
=\pars{\frac{\epsilon}{2\pi}}^2\sin(x)\cos(x)\pars{\cos^2(y)-\sin^2(x)}+\nu\frac{\epsilon}{\pi}\cos(x)\sin(y).\label{eq:S_tilde}
\end{gather}

And $\pars{\nabla\cdot\boldsymbol\tau_{ij}^R}\vect{e}_x$ is calculated as follows
\begin{gather}
\pars{\nabla\cdot\boldsymbol\tau_{ij}^R}\vect{e}_x=\pd{}{x}\avg{u\p u\p}+\pd{}{y}\avg{u\p v\p} =\nonumber\\
=\pd{}{x}\pars{\avg{\cos^2(x) \sin^2(y) \pars{\sin(z)-\frac{\epsilon}{2\pi}}^2}}+\nonumber\\
+\pd{}{y}\pars{\avg{\sin(x)\cos(x)\sin(y)\cos(y)\pars{\sin(z)-\frac{\epsilon}{2\pi}}^2}}=\nonumber \\
=\pd{}{x}\pars{\cos^2(x) \sin^2(y)\avg{ \pars{\sin^2(z)+\pars{\frac{\epsilon}{2\pi}}^2-\frac{\epsilon}{\pi}\sin(z)}}}+\nonumber\\
+\pd{}{y}\pars{\sin(x)\cos(x)\sin(y)\cos(y)\avg{\pars{\sin^2(z)+\pars{\frac{\epsilon}{2\pi}}^2-\frac{\epsilon}{\pi}\sin(z)}}}=\nonumber \\
=\bracs{\frac{1}{2}+\frac{\epsilon}{\pi}+\pars{\frac{\epsilon}{2\pi}}^2-\frac{1}{2}\pars{\frac{\epsilon}{\pi}}^2}\Bigg[\pd{}{x}\pars{\cos^2(x) \sin^2(y)}+\nonumber\\
+\pd{}{y}\pars{\sin(x)\cos(x)\sin(y)\cos(y)}\Bigg]=\nonumber\\
=\bracs{\frac{1}{2}+\frac{\epsilon}{\pi}-\frac{1}{4}\pars{\frac{\epsilon}{\pi}}^2}\sin(x)\cos(x)\pars{\cos^2(y)-3\sin^2(y)}.\label{eq:divT}
\end{gather}

Next, we calculate the terms neglected in the previous section
\begin{gather}
W\avg{\partial_z\vect{u}\p}-\nu\avg{\partial_{zz}\vect{u}\p}+\avg{\partial_z\pars{w\p\vect{u}\p}}=\nonumber \\
=-\frac{\epsilon}{\pi}\sin(x)\sin(y)\avg{\pd{}{z}\pars{\cos(x)\sin(y)\pars{\sin(z)-\frac{\epsilon}{2\pi}}}}-\nonumber \\ 
-\nu\avg{\ddn{}{z}{2}\pars{\cos(x)\sin(y)\pars{\sin(z)-\frac{\epsilon}{2\pi}}}}+\nonumber \\
+\avg{\pd{}{z}\pars{\cos(x)\sin(y)\pars{\sin(z)-\frac{\epsilon}{2\pi}}2\sin(x)\sin(y)\pars{\frac{\epsilon}{2\pi}-\cos(z)}}}=\nonumber
\end{gather}
\begin{gather}
=-\frac{\epsilon}{\pi}\sin(x)\cos(x)\sin^2(y)\avg{\cos(z)}+\nu\cos(x)\sin(y)\avg{\sin(z)}+\nonumber\\
+2\sin(x)\cos(x)\sin^2(y)\avg{\pd{}{z}\pars{\frac{\epsilon}{2\pi}\sin(z)+\frac{\epsilon}{2\pi}\cos(z)-\sin(z)\cos(z)-\pars{\frac{\epsilon}{2\pi}}^2}}=\nonumber\\
=-\frac{1}{2}\pars{\frac{\epsilon}{\pi}}^2\sin(x)\cos(x)\sin^2(y)+\nu\frac{\epsilon}{2\pi}\cos(x)\sin(y)+\nonumber\\
+2\sin(x)\cos(x)\sin^2(y)\avg{\frac{\epsilon}{2\pi}\cos(z)-\frac{\epsilon}{2\pi}\sin(z)+\sin^2(z)-\cos^2(z)}=\nonumber\\
=-\frac{1}{2}\pars{\frac{\epsilon}{\pi}}^2\sin(x)\cos(x)\sin^2(y)+\nu\frac{\epsilon}{2\pi}\cos(x)\sin(y)+\nonumber\\
+2\sin(x)\cos(x)\sin^2(y)\cancelto{0}{\bracs{\pars{\frac{\epsilon}{2\pi}}^2-\pars{\frac{\epsilon}{2\pi}}^2+\frac{1}{2}+\frac{\epsilon}{\pi}-\frac{1}{2}-\frac{\epsilon}{\pi}}}=\nonumber\\
=-\frac{1}{2}\pars{\frac{\epsilon}{\pi}}^2\sin(x)\cos(x)\sin^2(y)+\nu\frac{\epsilon}{2\pi}\cos(x)\sin(y).\label{eq:extra}
\end{gather}
We note that $\avg{\partial_z\pars{w\p\vect{u}\p}}$ automatically cancels out.

Putting all the pieces together (Eqs. \ref{eq:S_avg}, \ref{eq:S_tilde}, \ref{eq:divT} \& \ref{eq:extra} into \eref{eq:balance_all}) we obtain the error between the balance of its LHS and RHS as follows
\begin{gather}
\mathcal{E} = \left|\avg{\mathcal{S}\left(\vect{u}, p\right)} - \tilde{\mathcal{S}}\left(\vect{U}, P\right) - \nabla\cdot\boldsymbol\tau_{ij}^R - W\avg{\partial_z\vect{u}\p}+\nu\avg{\partial_{zz}\vect{u}\p} - \avg{\partial_z\pars{w\p\vect{u}\p}}\right|=\nonumber\\
=\Bigg|\cancel{\pars{\frac{1}{2}+\frac{\epsilon}{\pi}}\sin(x)\cos(x)\pars{\cos^2(y)-3\sin^2(y)}}+\bcancel{3\nu\frac{\epsilon}{2\pi}\cos(x)\sin(y)} - \nonumber\\
- \pars{\frac{\epsilon}{2\pi}}^2\sin(x)\cos(x)\pars{\cos^2(y)-\sin^2(x)}-\bcancel{\nu\frac{\epsilon}{\pi}\cos(x)\sin(y)} - \nonumber \\
- \bracs{\cancel{\frac{1}{2}+\frac{\epsilon}{\pi}}-\frac{1}{4}\pars{\frac{\epsilon}{\pi}}^2}\sin(x)\cos(x)\pars{\cos^2(y)-3\sin^2(y)}+ \nonumber \\
+\frac{1}{2}\pars{\frac{\epsilon}{\pi}}^2\sin(x)\cos(x)\sin^2(y)-\bcancel{\nu\frac{\epsilon}{2\pi}\cos(x)\sin(y)}\Bigg| \nonumber=\\
=\Bigg|-\frac{1}{4}\pars{\frac{\epsilon}{\pi}}^2\sin(x)\cos(x)\pars{\cos^2(y)-\sin^2(x)}+\nonumber\\
+\frac{1}{4}\pars{\frac{\epsilon}{\pi}}^2\sin(x)\cos(x)\pars{\cos^2(y)-3\sin^2(y)}+\frac{1}{2}\pars{\frac{\epsilon}{\pi}}^2\sin(x)\cos(x)\sin^2(y)\Bigg| \nonumber= \\
=\Bigg|-\frac{1}{2}\pars{\frac{\epsilon}{\pi}}^2\sin(x)\cos(x)\sin^2(x)+\frac{1}{2}\pars{\frac{\epsilon}{\pi}}^2\sin(x)\cos(x)\sin^2(y)\Bigg|= 0.
\end{gather}

With this, we can conclude that the quadratic errors arising from the RHS of \eref{eq:balance_all} cancel out with each other.
Also, the linear errors arising on both LHS and RHS also cancel each other.

\subsection{Quadrature error with spanwise periodicity}
\label{sec:sans_quadrature_errors2}

The SANS derivation employed in \cref{chapter:sans} and \cref{chapter:zanspy} (\eref{eq:balance}) neglects errors arising from the quadratures of spanwise derivatives (second line of \eref{eq:sans_full}).
Because of this, \eref{eq:balance_all} is not balanced.
Here, we derive the error associated to this practice.
Inserting the previously derived terms into \eref{eq:balance} carries the following error
\begin{gather}
\mathcal{E} = \avg{\mathcal{S}\left(\vect{u}, p\right)} - \tilde{\mathcal{S}}\left(\vect{U}, P\right) - \nabla\cdot\boldsymbol\tau_{ij}^R=\nonumber\\
=\cancel{\pars{\frac{1}{2}+\frac{\epsilon}{\pi}}\sin(x)\cos(x)\pars{\cos^2(y)-3\sin^2(y)}}+3\nu\frac{\epsilon}{2\pi}\cos(x)\sin(y) - \nonumber\\
- \pars{\frac{\epsilon}{2\pi}}^2\sin(x)\cos(x)\pars{\cos^2(y)-\sin^2(x)}-{\nu\frac{\epsilon}{\pi}\cos(x)\sin(y)} - \nonumber \\
- \bracs{\cancel{\frac{1}{2}+\frac{\epsilon}{\pi}}-\frac{1}{4}\pars{\frac{\epsilon}{\pi}}^2}\sin(x)\cos(x)\pars{\cos^2(y)-3\sin^2(y)} \nonumber=\\
=\frac{\epsilon}{\pi}\nu\cos(x)\sin(y) - \frac{1}{4}\pars{\frac{\epsilon}{\pi}}^2\sin(x)\cos(x)\pars{\cos^2(y)-\sin^2(x)} + \nonumber \\
+\frac{1}{4}\pars{\frac{\epsilon}{\pi}}^2\sin(x)\cos(x)\pars{\cos^2(y)-3\sin^2(y)} \nonumber=\\
=\frac{\epsilon}{\pi}\nu\cos(x)\sin(y)-\frac{1}{2}\pars{\frac{\epsilon}{\pi}}^2\sin(x)\cos(x)\sin^2(y),
\end{gather}
which corresponds indeed to the derivation of \eref{eq:extra}
\begin{equation}
\mathcal{E}=|\avg{w}\avg{\partial_z\vect{u}\p}-\nu\avg{\partial_{zz}\vect{u}\p}|.
\end{equation}
It can be observed that, given $\epsilon$ is a small positive number, the viscous term has a larger contribution to the model error.

It is important to note that the midpoint and trapezoidal quadratures obtain a higher convergence rate when the integrand is periodic and smooth \citep{Weideman2002,Trefethen2014}, precisely from $\mathcal{O}(h^2)$ to $\mathcal{O}(e^{-1/h})$.
Applying the spanwise-periodicity constraint in our problem and considering incompressible flow (smooth functions), the quadrature error should converge faster than the spatial discretisation error of the governing equations $\pars{\mathcal{O}(h^2)}$.
% ---------------------------------------------------------------- 

\chapter{Machine-learning model hyper-parameters tuning}

\begin{figure}[!h]
	\centering
	\begin{subfigure}[t]{\linewidth}		
		\centering
		\includegraphics[width=0.7\linewidth]{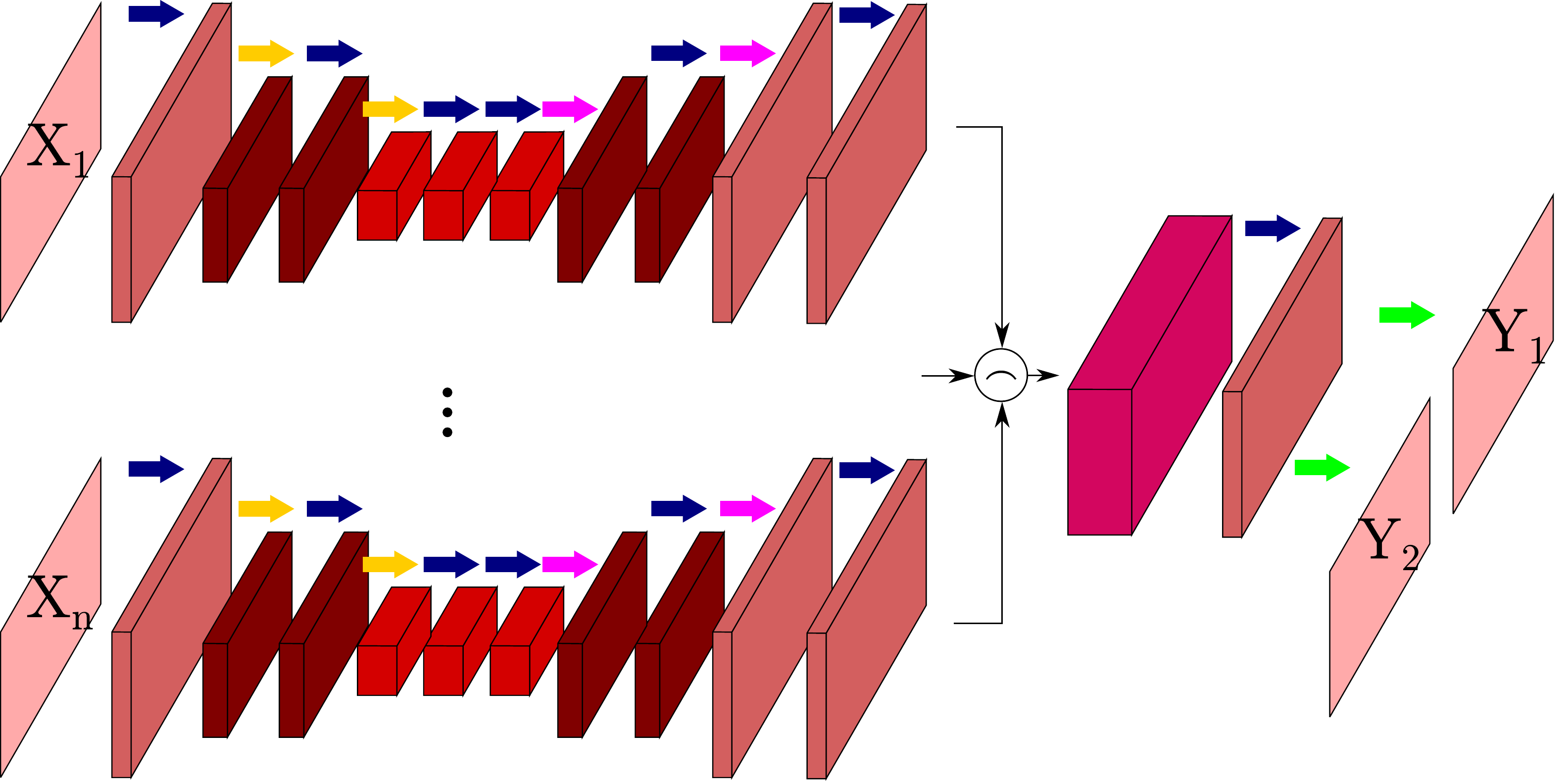}%
		\caption{Model 1 (M1)}\vspace{0.7cm}
	\end{subfigure}
	\begin{subfigure}[t]{\linewidth}
		\centering
		\vspace{0.5cm}\includegraphics[width=0.95\linewidth]{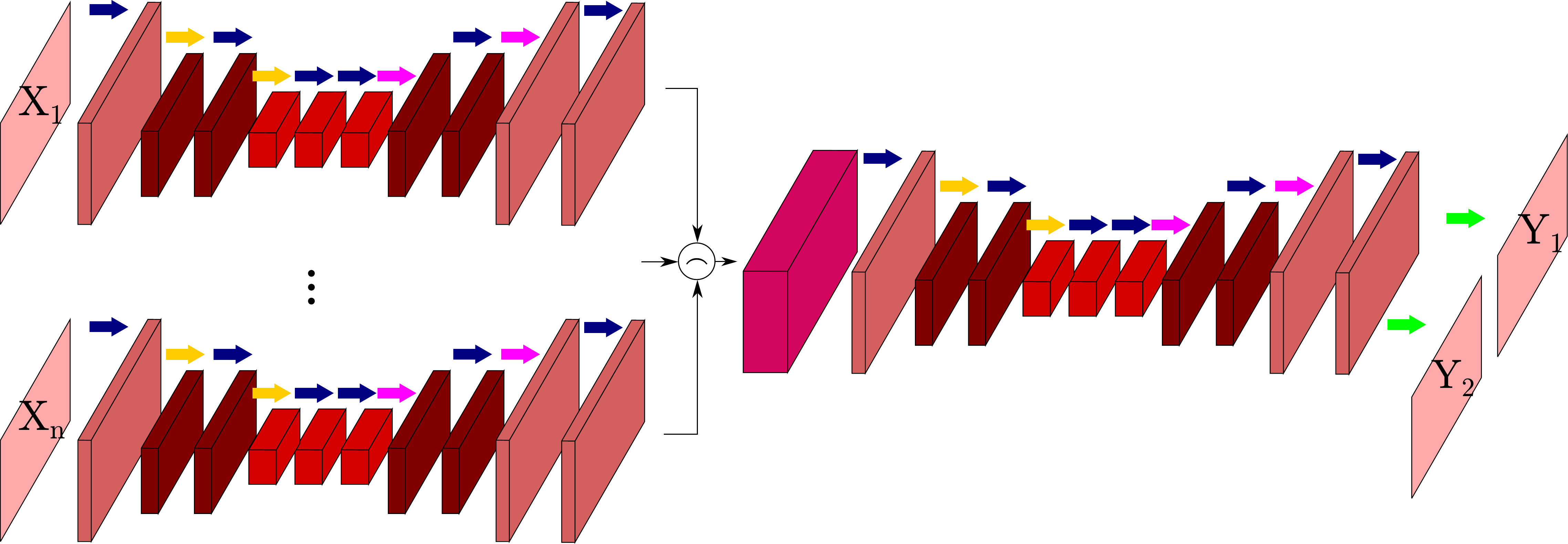}%
		\caption{Model 2 (M2)}\vspace{0.9cm}
	\end{subfigure}
	\begin{subfigure}[t]{\linewidth}
		\centering
		\includegraphics[width=0.6\linewidth]{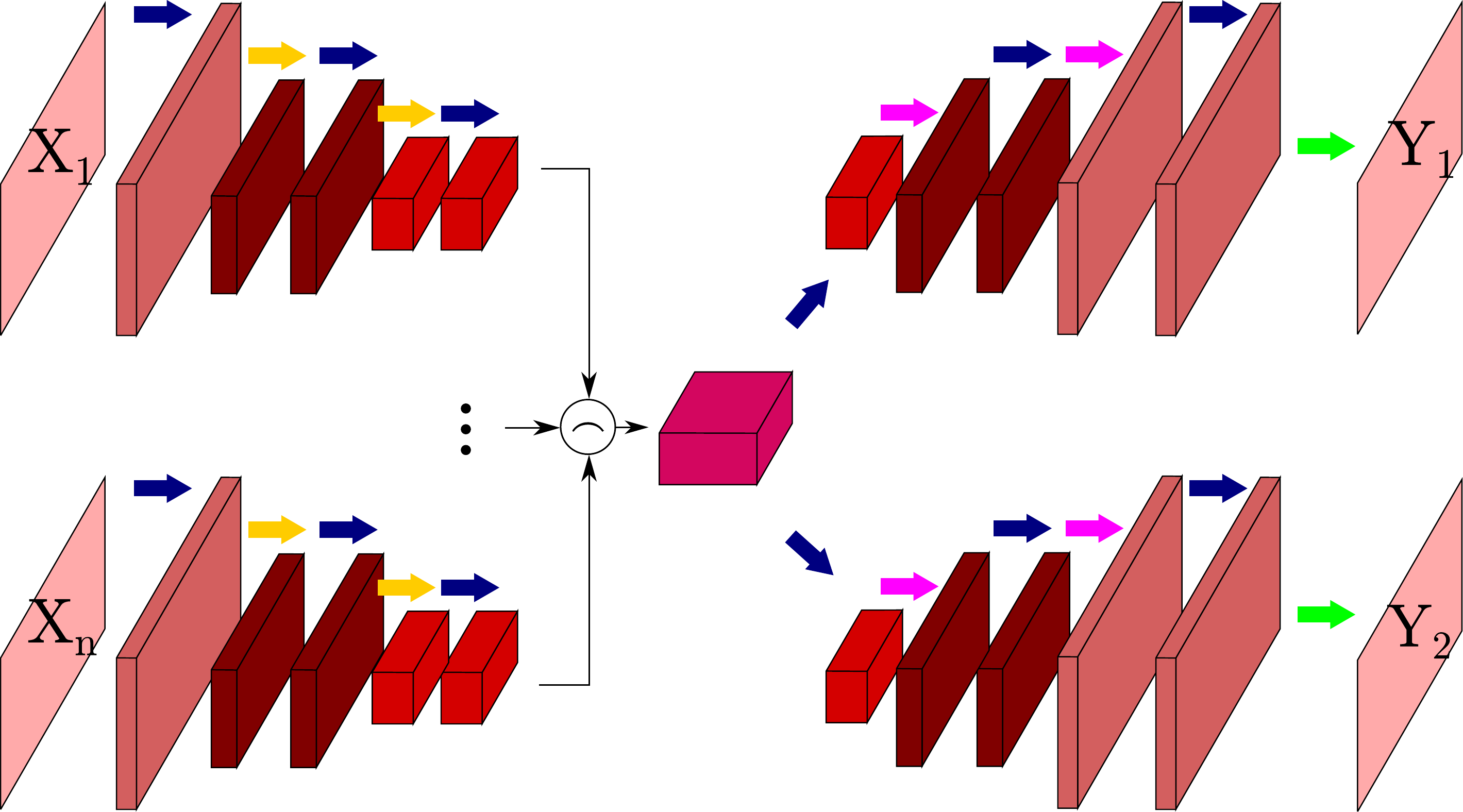}%
		\caption{Model 3 (M3)}
	\end{subfigure}
	\captionsetup[subfigure]{width=\linewidth}
	\caption{CNN architectures investigated.
The input $(\mathrm{X}_n)$ and output $(\mathrm{Y}_n)$ fields size is $1216\times540$.
Arrows color indicate layers of:
Dark blue: Conv2D $+$ batch normalisation $+$ activation function.
Yellow: MaxPooling $2\times2$.
Magenta: UpSampling $2\times2$.
Green: Conv2D $1\times1$ $+$ batch normalisation $+$ linear function.
The thickness of each block (noted on the last dimension) represents the number of filters.
On the encoding blocks, the number of filters is doubled at every convolutional layer, reaching a maximum number of filters just before the decoding block, where the operation is reversed halving the number of filters at every convolutional layer.
The merging operator $\frown$ indicates concatenation in the last dimension, hence stacking filters from multiple blocks.}\label{fig:cnn}
\end{figure}

A hyper-parametric study of the proposed ML model for the SANS equations is performed next.
Different MIMO-CNN architectures are considered, as shown in \fref{fig:cnn}.
All of the different CNNs are based on encoding the input information into some lower-dimensional latent space followed by a decoding procedure that maps such latent space into the target outputs, similarly to SegNet \citep{Badrinarayanan2015} or U-net \citep{Ronneberger2015} architectures.

For the hyper-parametric study, a dataset of 10152 snapshots containing the resolved quantities $\left\lbrace U, V, P \right\rbrace$ and the residual closure terms $\lbrace \mathcal{S}^R_x,\mathcal{S}^R_y\rbrace$ is generated throughout approximately 600 wake cycles.
The size of the recorded 2-D flow fields is $1216\times540$.
The data is split 80-20 for training and testing, respectively.
The training data is further split 75-25 for optimisation and validation, respectively.
The test dataset is generated after the 600 wake cycles, hence this data is completely hidden from the CNN during the training stage.
Only the inner uniform domain (see \fref{fig:computational_domain}) is stored during the data generation process.
For the training stage, an early-stopping approach is considered as the convergence criteria when the validation loss does not improve for 8 consecutive epochs (dataset iterations), or a maximum of 200 epochs is reached.
Success is measured by the Pearson correlation coefficient between the target $({\mathcal{S}}^{R})$ and predicted $(\mathcal{S}^{R,\,\mathrm{ML}})$ output fields, respectively
\begin{gather}
\mathcal{CC}_x=\mathcal{CC}\pars{\mathcal{S}^R_x,\mathcal{S}_x^{R,\,\mathrm{ML}}}=\frac{\mathrm{cov}\pars{\mathcal{S}^R_x,\mathcal{S}_x^{R,\,\mathrm{ML}}}}{\sqrt{\mathrm{cov}\pars{\mathcal{S}^R_x,\mathcal{S}^R_x}\mathrm{cov}\pars{\mathcal{S}_x^{R,\,\mathrm{ML}},\mathcal{S}_x^{R,\,\mathrm{ML}}}}},\\
\mathcal{CC}_y=\mathcal{CC}\pars{\mathcal{S}^R_y,\mathcal{S}_y^{R,\,\mathrm{ML}}}=\frac{\mathrm{cov}\pars{\mathcal{S}^R_y,\mathcal{S}_y^{R,\,\mathrm{ML}}}}{\sqrt{\mathrm{cov}\pars{\mathcal{S}^R_y,\mathcal{S}^R_y}\mathrm{cov}\pars{\mathcal{S}_y^{R,\,\mathrm{ML}},\mathcal{S}_y^{R,\,\mathrm{ML}}}}}.
\end{gather}

A mini-batch gradient descend method (4 to 6 samples depending on the GPU memory) using the Adam optimiser \citep{Kingma2014} is employed.
The learning rate is set to $10^{-4}$ with a decay rate of $10^{-5}$.
The full model is developed under Keras \citep{Chollet2015}, a deep-learning Python library using the TensorFlow \citep{Tensorflow} backend.

Because of the network weights are initialised randomly and stochastic gradient descent backpropagation is used to optimise the network, there is an irreducible level of randomness in the final results.
Hence, the training process is repeated for cases converged into a very shallow local minima and only the best results have been reported.

\begin{figure}
	\centering
	\captionsetup[subfigure]{width=1\linewidth}
	\begin{subfigure}[t]{0.9\linewidth}		
		\centering
		\includegraphics[width=\linewidth]{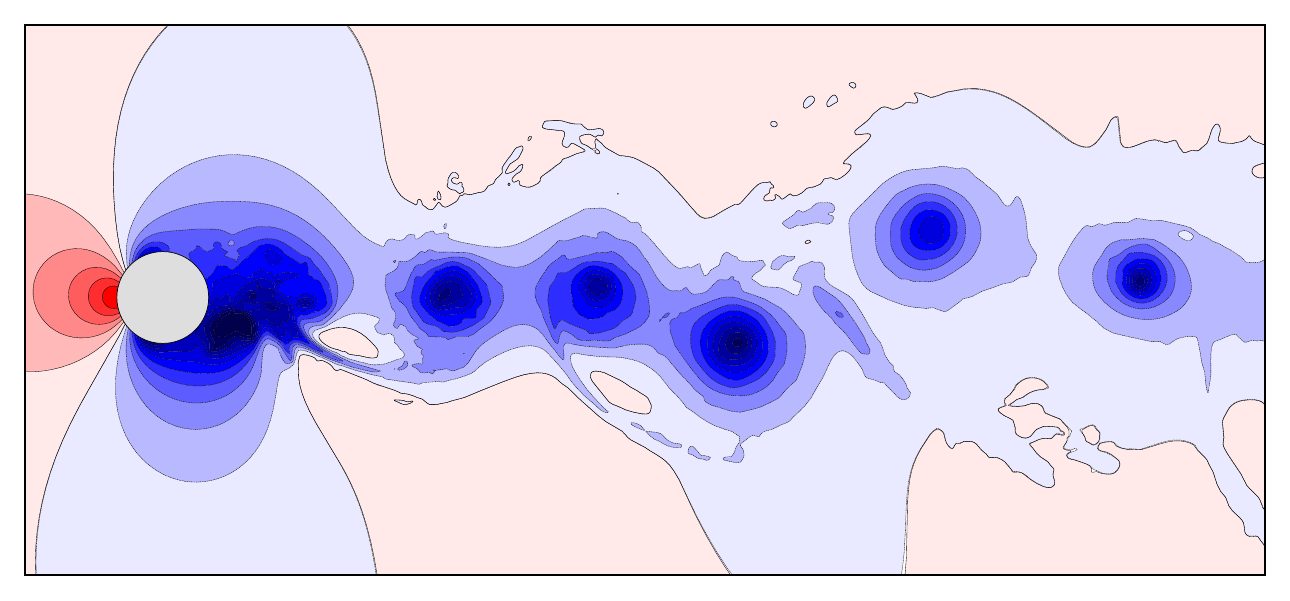}%
	\caption{$P$}
	\end{subfigure}
	\begin{subfigure}[t]{\linewidth}		
		\centering
		\includegraphics[width=0.9\linewidth]{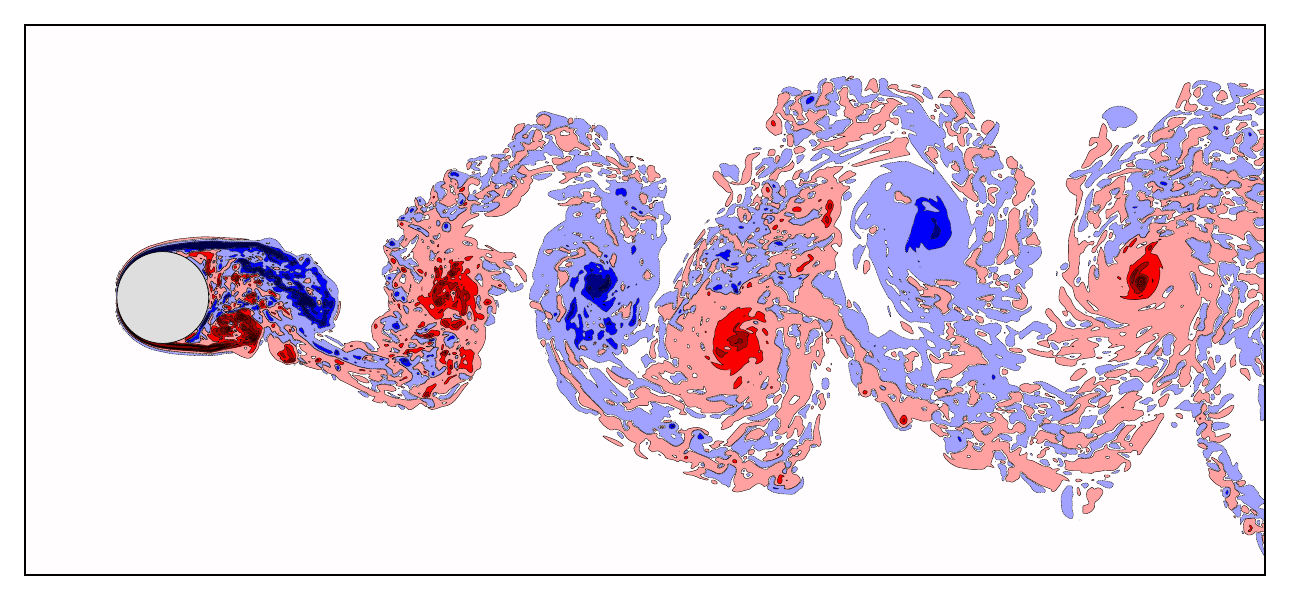}%
	\caption{$\Omega_z$}
	\end{subfigure}
	\begin{subfigure}[t]{\linewidth}
		\centering
		\includegraphics[width=0.9\linewidth]{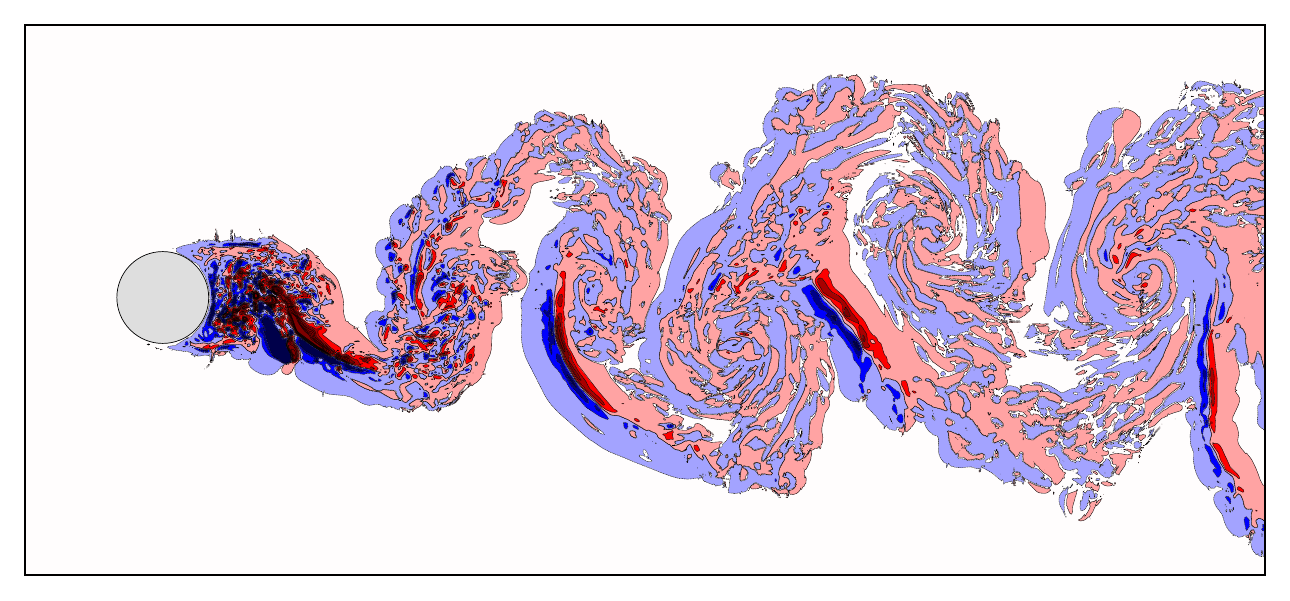}%
		\caption{$\mathcal{S}^R_x$}
	\end{subfigure}
	\caption{Flow snapshot of resolved (top, middle) and target closure (bottom) fields.} \label{fig:flow_field}
\end{figure}

A flow field snapshot of resolved quantities and a target output is displayed in \fref{fig:flow_field}.
The complex flow structures at the shear layer roll-up region behind the cylinder can be particularly challenging to model.
This can indicate that a large CNN in terms of number of trainable parameters (weights $\matr{W}$ and bias $\vect{b}$) is required.
Also, beyond exploring different CNN architectures, we test different loss functions, activation functions, number of filters and kernel size, effect of input data normalisation, and different input sets.
A summary of the parametric study is detailed in \tref{tab:parameters} and the base-case parameters are shown in \tref{tab:models_fixed_parameters}.

\begin{table}
\begin{center}
\begin{tabular}{ll}
\toprule
Architecture & M1, M2, M3\\
Loss function & sse, sae \\
Activation function & ReLU, tanh, sigmoid \\
Max. filters & 16, 32, 64 \\
Kernel size & $3\times3, \,\,\, 5\times5, \,\,\,7\times7$ \\
Normalised input data & False, True \\
Input set & $\left\lbrace U, V, P \right\rbrace$, $\left\lbrace \Omega_z, d \right\rbrace$,  $\left\lbrace \nabla\vect{U}, \nabla P \right\rbrace$ \\
\bottomrule
\end{tabular}
\end{center}
\caption{Parametric study.
sse: sum square error.
sae: sum absolute error.
ReLU: rectified linear unit.
$\Omega_z$: Spanwise-averaged vorticity.
$d$: Distance function to the cylinder.}
\label{tab:parameters}
\end{table} 

\begin{table}
\begin{center}
\begin{tabular}{ll}
\toprule
Architecture & M3 \\
Loss function & sse \\
Activation function & ReLU \\
Max. filters & 64 \\
Kernel size & $3\times3$ \\
Normalised input data & True \\
Input set & $\left\lbrace U, V, P \right\rbrace$ \\
\bottomrule
\end{tabular}
\end{center}
\caption{Base-case parameters of the parametric investigation.}
\label{tab:models_fixed_parameters}
\end{table}

\clearpage

\section{Results}

\subsection{Model architecture and data normalisation}

We start investigating the CNN architecture based on three different models, namely M1, M2 and M3 (see \fref{fig:cnn}).
The base-case parameters are detailed in \tref{tab:models_fixed_parameters}.
We also study the effect of normalising each input data sample with its standard score, i.e. $\hat{x} = \pars{x-\bar{x}}/\sigma_x$.

\begin{figure*}[!ht]
\centering
\vspace{1cm}
\begin{multicols}{2}
	\begin{subfigure}[t]{\linewidth}		
		\includegraphics[width=\linewidth]{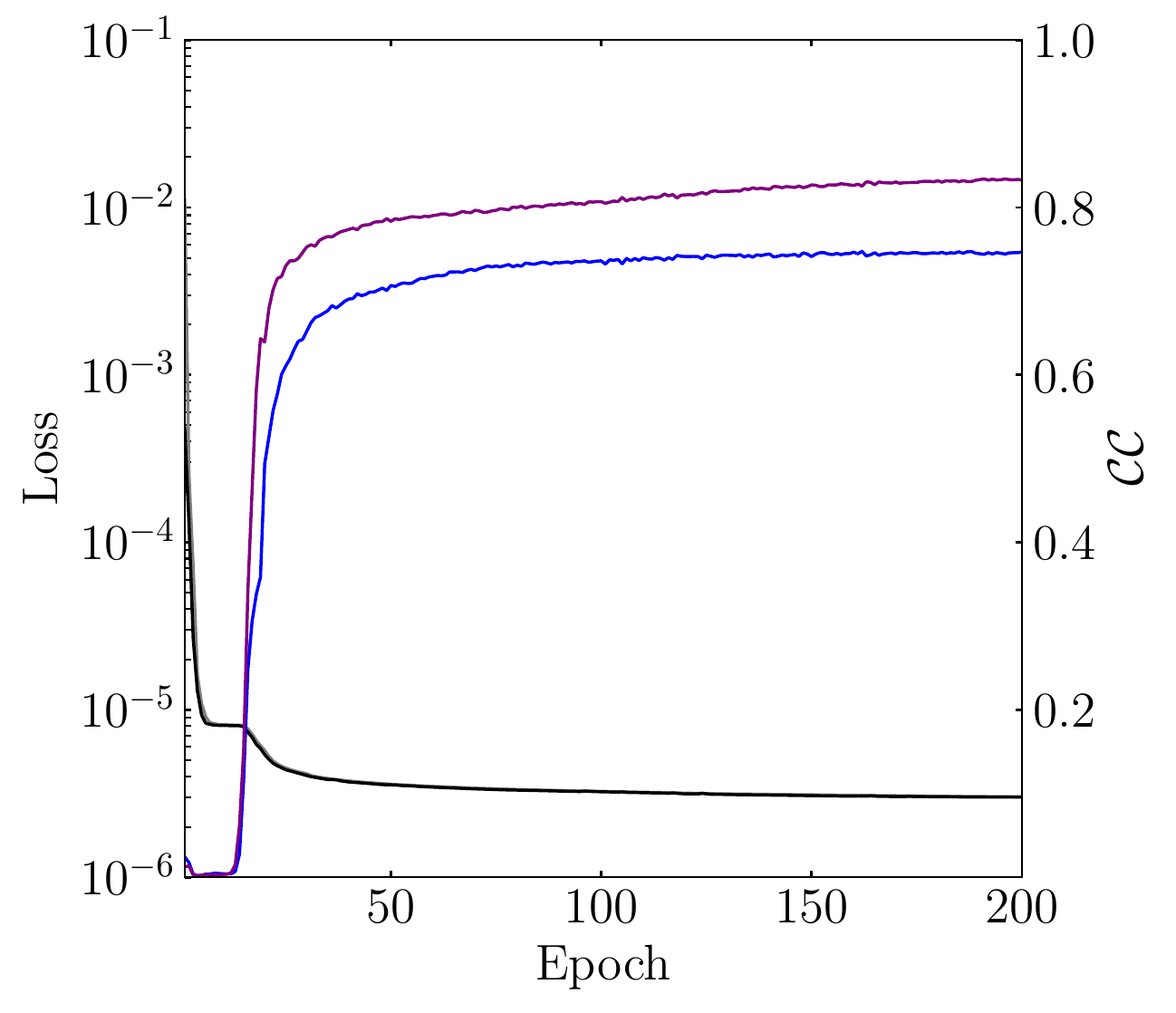}%
		\caption{M1}
	\end{subfigure}\vspace{0.3cm}
	\begin{subfigure}[t]{\linewidth}		
		\includegraphics[width=\linewidth]{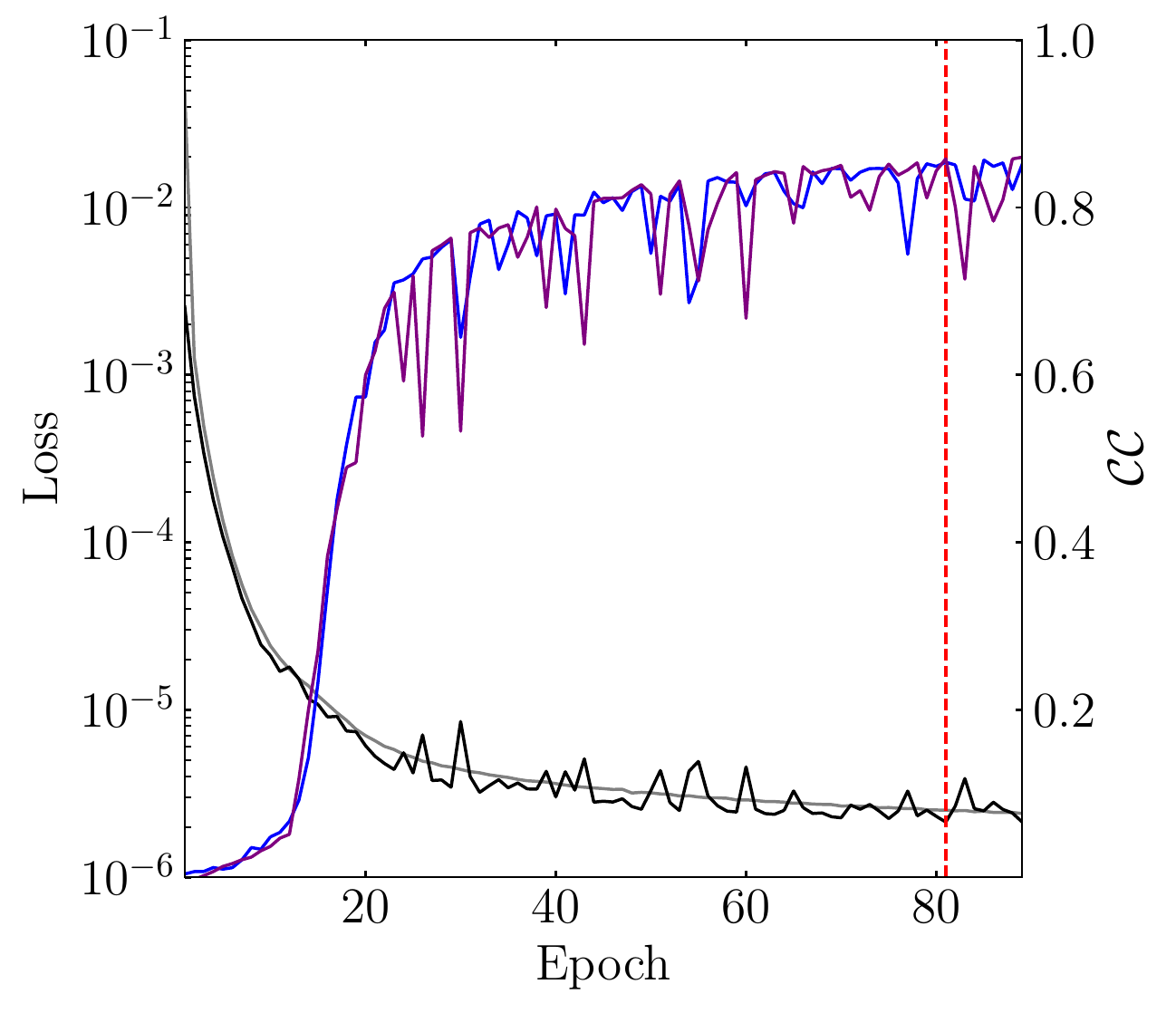}%
		\caption{M2}
	\end{subfigure}\vspace{0.3cm}
	\begin{subfigure}[t]{\linewidth}
		\includegraphics[width=\linewidth]{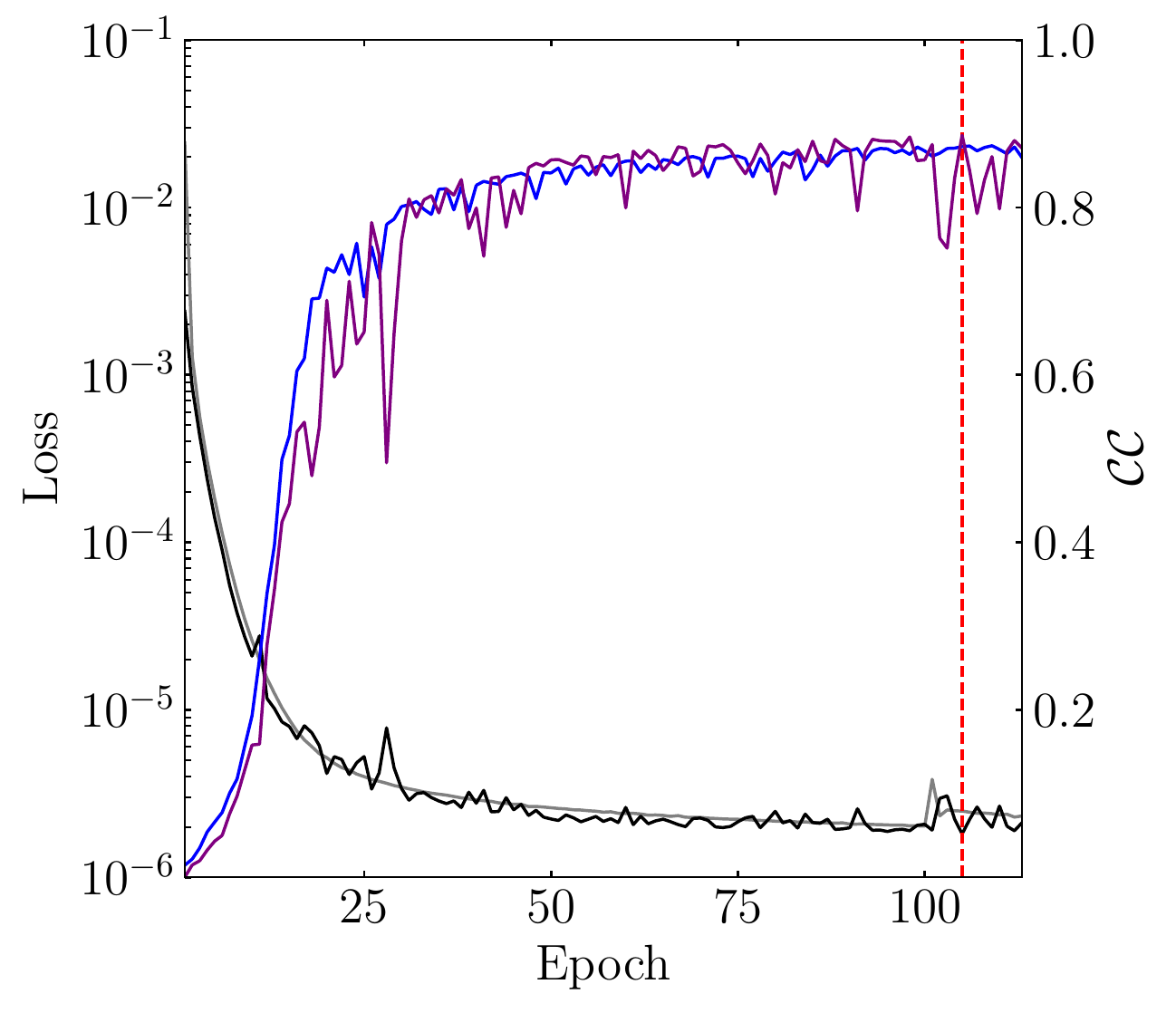}
		\caption{M3}
	\end{subfigure}
		\begin{subfigure}[t]{\linewidth}		
		\includegraphics[width=\linewidth]{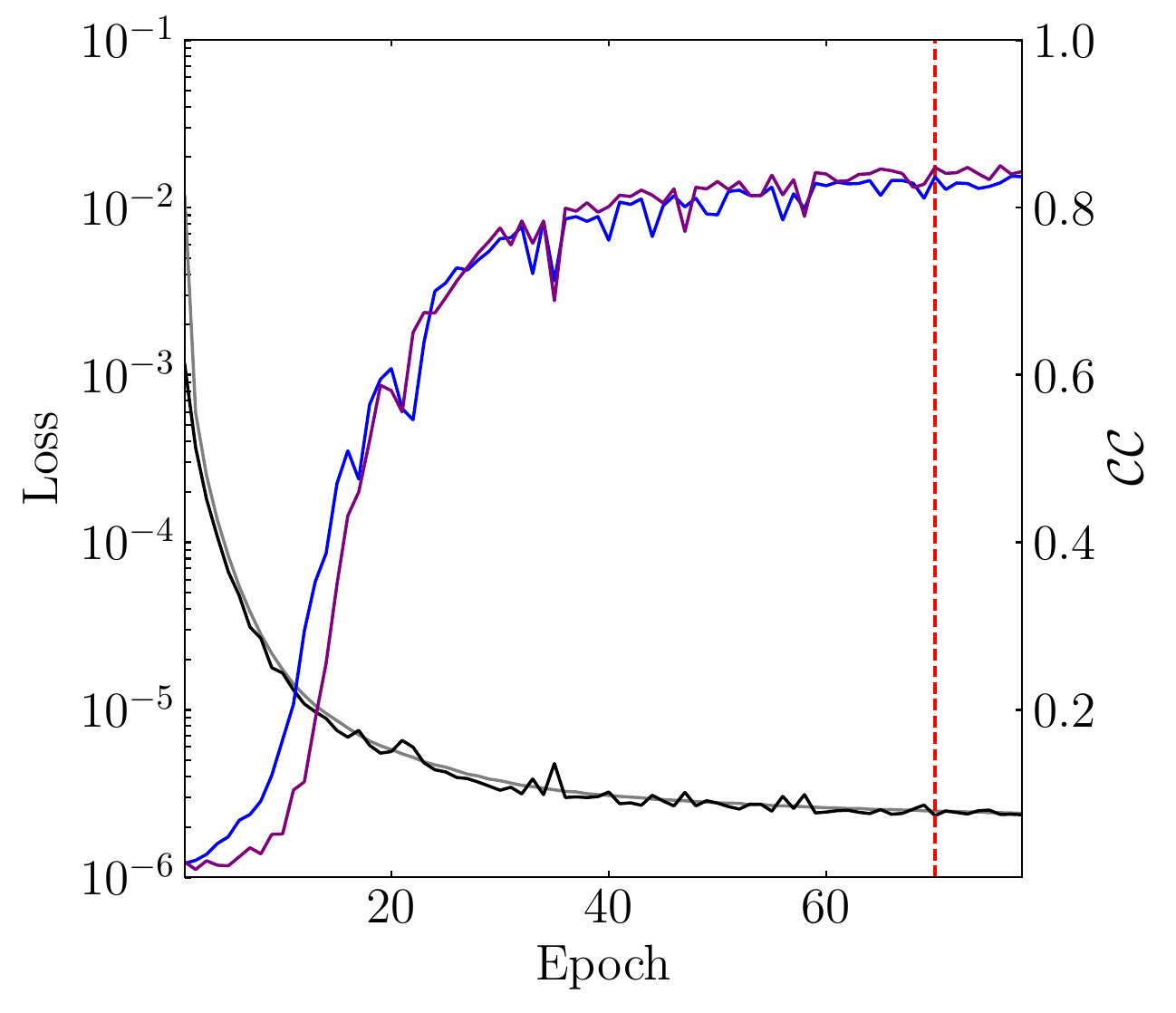}%
		\caption{M1, norm.}
	\end{subfigure}\vspace{0.3cm}
		\begin{subfigure}[t]{\linewidth}		
		\includegraphics[width=\linewidth]{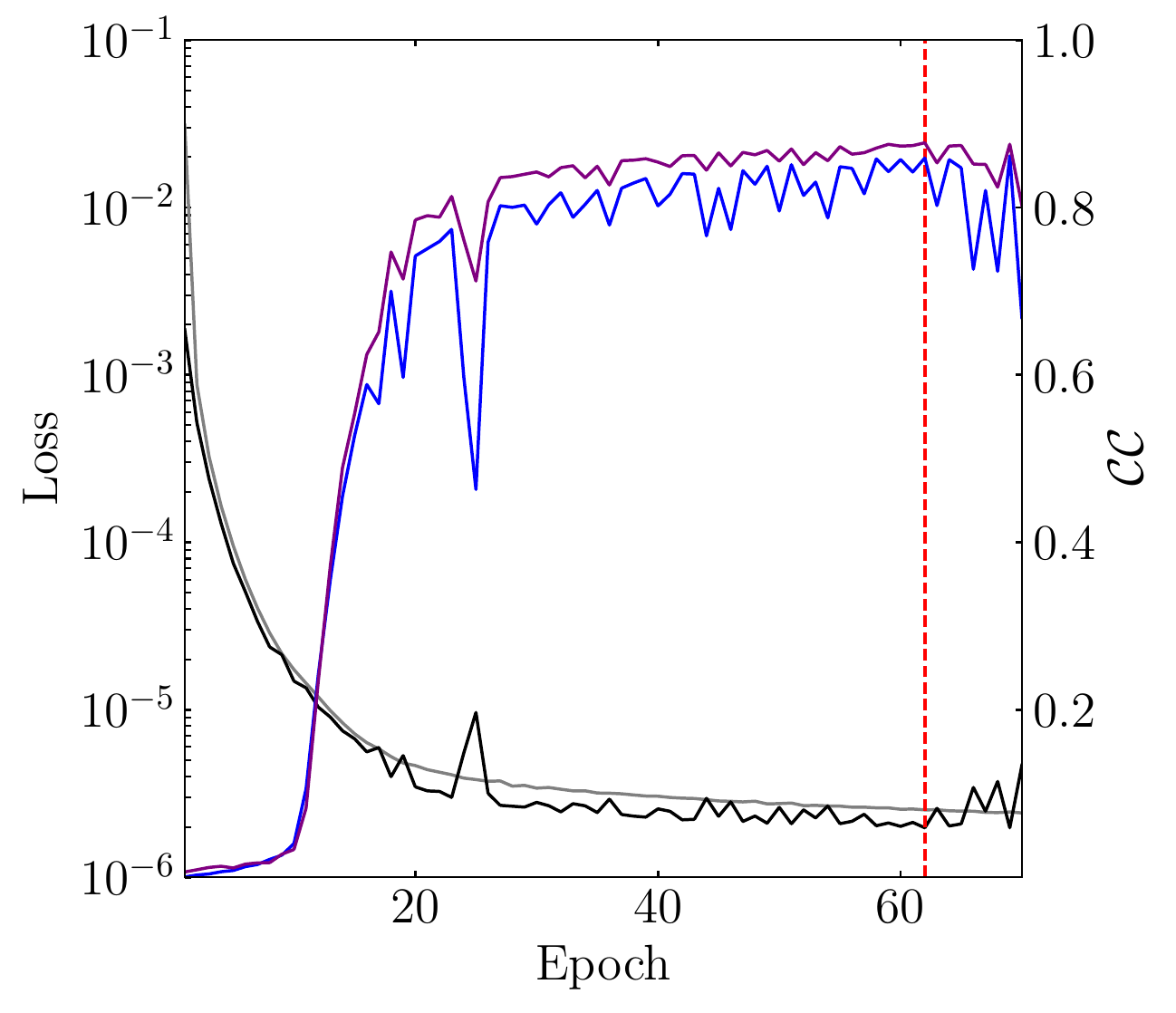}%
		\caption{M2, norm}
	\end{subfigure}\vspace{0.3cm}
	\begin{subfigure}[t]{\linewidth}		
		\includegraphics[width=\linewidth]{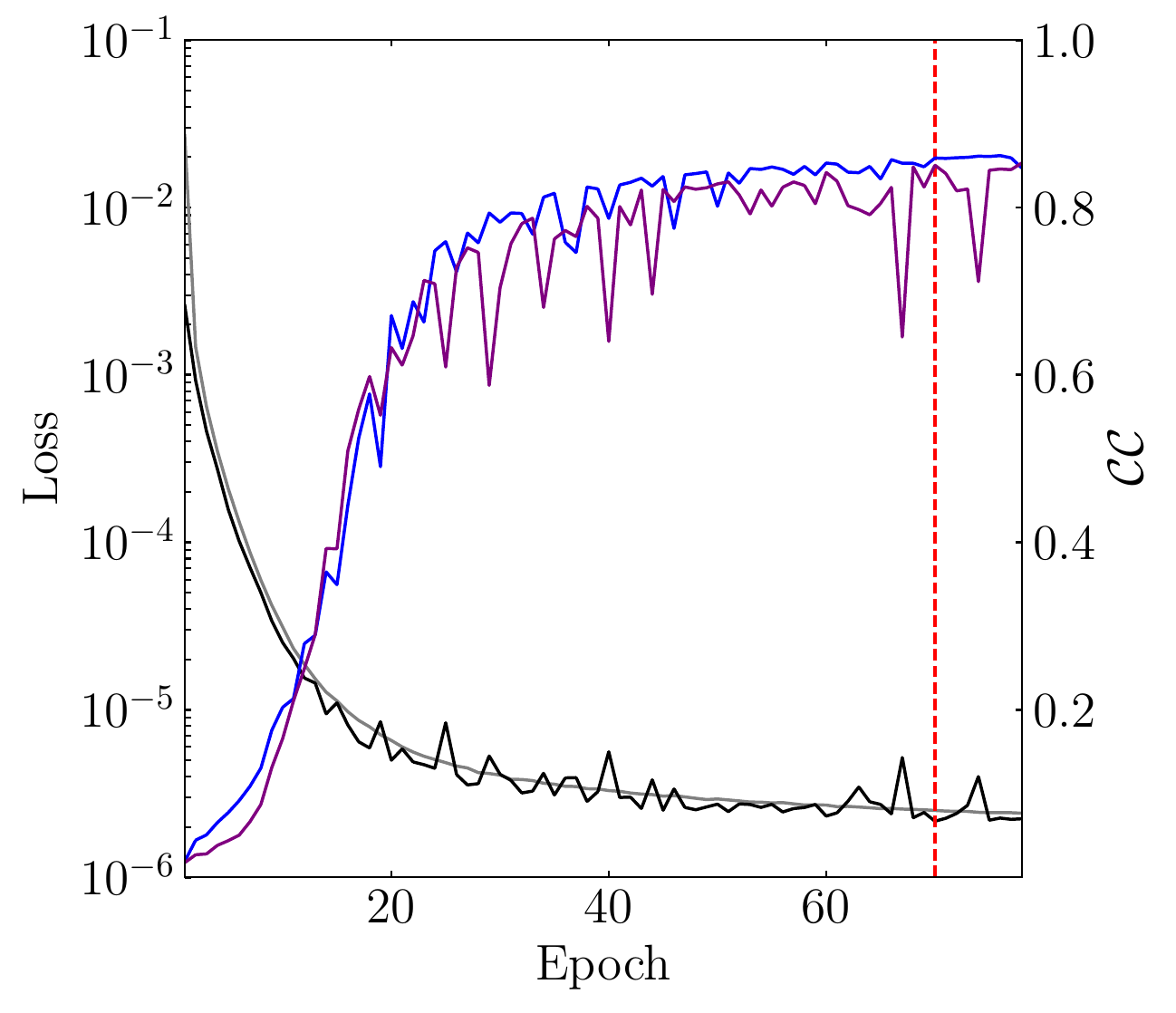}%
		\caption{M3, norm.}
	\end{subfigure}
\end{multicols}
	\caption{Training history of the different model architectures.
Left: Input data not normalised.
Right: Input data normalised.
Legend: Grey: fit loss.
Black: val. loss.
Blue: $\mathcal{CC}_x$ (val. dataset).
Purple: $\mathcal{CC}_y$ (val. dataset).
Red: Early stop mark.} 
	\label{fig:models_convergence}
\end{figure*}

Results for the training stage of each model are displayed in \fref{fig:models_convergence}.
It can be observed that normalising the input data provides a faster convergence, noting that the non-normalised M1 case hits the maximum number of epochs limit.
The correlation coefficient results are given in \tref{tab:models_correlation}.
All models provide a high correlation coefficient for both validation and test datasets (hence no over-fitting is detected).
The general trend is that accuracy increases with the network size.
It is expected that output decoding branches having access to a larger latent space (number of filters encoding information on the encoder output), such as M3, are more flexible and perform better than networks with a smaller latent space.
The M3 model with normalised input data is selected for all the subsequent analysis, unless specified otherwise.
The choice is motivated by the M3 model training consistency and its slightly faster convergence when compared to other models.

{\renewcommand{\arraystretch}{1.2}
\begin{table}[t]
\begin{center}
\begin{tabular}{ccccc}
\toprule
Case & \specialcell{Train.\\params.} & val.: $\mathcal{CC}_x,\,\mathcal{CC}_y$ & test: $\mathcal{CC}_x,\,\mathcal{CC}_y$ & Best epoch\\
\midrule
M1 & 258242 & 0.75, 0.83 &  0.74, 0.83 & 200\\
\specialcell{M1\\norm.} & 258242 & 0.84, 0.85 & 0.83, 0.84 & 70\\
M2 & 300194 & 0.85, 0.86 & 0.84, 0.84 & 81\\
\specialcell{M2\\norm.} & 300194 & {0.86, 0.88} & {0.85, 0.88}& 62\\
\specialcell{M2\\norm.\\stacked} & 58538 & 0.76, 0.77 &  0.74, 0.76 & 63\\
M3 & 338498 & 0.87, 0.88 & 0.86, 0.87 & 105\\
\specialcell{M3\\norm.} & 338498 & 0.86, 0.85 & 0.85, 0.84 & 70\\
\bottomrule
\end{tabular}
\end{center}
\caption{Validation and test results for different model architectures.}
\label{tab:models_correlation}
\end{table}}

\clearpage

\clearpage
Additionally, a simple auto-encoder concatenating the resolved quantities in a single input branch (stacking them in the filter dimension, noted in \tref{tab:models_correlation} as ``stacked'') has been tested.
Even though this model is much smaller than the other ones in terms of number of trainable parameters, it still manages to provide acceptable predictions.

\subsection{Loss function}

The loss function between predicted and target fields is investigated next.
The same previous base-case parameters are chosen and results are summarised in \tref{tab:loss_correlation}.

{\renewcommand{\arraystretch}{1.2}
\begin{table}
\begin{center}
\begin{tabular}{ccccc}
\toprule
Case & \specialcell{Train.\\params.} & val.: $\mathcal{CC}_x,\,\mathcal{CC}_y$ & test: $\mathcal{CC}_x,\,\mathcal{CC}_y$ & Best epoch\\
\midrule
sse & 338498 & 0.86, 0.85 & 0.85, 0.84 & 70 \\
sae & 338498 & 0.86, 0.57 & 0.83, 0.48 & 45\\
\bottomrule
\end{tabular}
\end{center}
\caption{Validation and test results for different loss functions.}
\label{tab:loss_correlation}
\end{table}}

It can be observed that using the absolute error yields to a local minima where one of the outputs has not converged successfully.
The squared error function being an easier metric to differentiate (convex parabola) than the absolute error might be causing such difference.

\subsection{Activation function}

Activation functions ($h$) are used to dictate the output of a neuron based on some input value $x$ (the model input or a previous layer output), a weight and a bias: $a = h(wx+b)$.
Activation functions widely used in ANNs are tested here.
We consider the ReLU, tanh, and sigmoid activation functions.
Results are displayed in \tref{tab:activation_correlation}.

It can be observed that the ReLU activation function provides a clear advantage in comparison with the other tested functions.
The ReLU function allows for faster trainings of deep ANNs because it preserves the neuron relative intensity across layers (maximum value is not capped) \citep{Nair2010}.
This is also due to the fact that the ReLU function helps mitigating vanishing-gradient caveats which arise more naturally when using the sigmoid function.

{\renewcommand{\arraystretch}{1.2}
\begin{table}
\begin{center}
\begin{tabular}{ccccc}
\toprule
Case & \specialcell{Train.\\params.} & val.: $\mathcal{CC}_x,\,\mathcal{CC}_y$ & test: $\mathcal{CC}_x,\,\mathcal{CC}_y$ & Best epoch\\
\midrule
ReLU & 338498 & 0.86, 0.85 & 0.85, 0.84 & 70 \\
tanh & 338498 & 0.78, 0.76 & 0.77, 0.76 & 62\\
sigmoid & 338498 & 0.81, 0.82 & 0.80, 0.82 & 36\\
\bottomrule
\end{tabular}
\end{center}
\caption{Validation and test results for different activation functions.}
\label{tab:activation_correlation}
\end{table}}

\subsection{Number of filters and kernel size}

The number of filters per convolutional layer and the convolution kernel size is explored.
The results for the maximum number of filters per convolutional layer are displayed in \tref{tab:filters_correlation}, where the number of trainable parameters jumps by a factor of 4.
The results observed in this particular study are strongly biased by the number of training epochs.
It should be noted that the case using a maximum of 16 filters only converged after almost 400 epochs (the maximum epochs limit has been lifted here because of the network size), whereas the other cases are both converged under 100 epochs.
The slow convergence of the 16 filters might be due to its lack of flexibility (model degrees of freedom) compared to the models with more filters available.
Still, the 64 filters case displays correlation coefficients relatively close to the 16 filters case with only 70 epochs of training.
Hence it can be argued that increasing the number of filters allows the model to be optimised based on a wider range of spatial features thus achieving a faster convergence.
Simultaneously, increasing the number of filters does not automatically yield better accuracy.

From a computational cost standpoint, the training time for the 64 maximum filters case is approximately 26 min./epoch, whereas the 32 and 16 filters cases are both around 23 min./epoch (the latter only measured during its first 100 epochs).
Hence, the computational cost does not increase linearly with the network size.
On the other hand, the computational training time of the 16 filters case gradually increases with every new epoch.
For the 378 epochs until convergence, a total of 180 hours of computational time is required, averaging to roughly 28 min./epoch.
Therefore, the computational time per epoch increases with the number of epochs, making the 16 maximum filters case less efficient compared to the larger networks.

{\renewcommand{\arraystretch}{1.2}
\begin{table}
\begin{center}
\begin{tabular}{ccccc}
\toprule
Case & \specialcell{Train.\\params.} & val.: $\mathcal{CC}_x,\,\mathcal{CC}_y$ & test: $\mathcal{CC}_x,\,\mathcal{CC}_y$ & Best epoch\\
\midrule
64 & 338498 & 0.86, 0.85 & 0.85, 0.84  & 70\\
32 & 85154 & 0.81, 0.72 & 0.81, 0.71 & 34\\
16 & 21554 & 0.87, 0.87 & 0.87, 0.86 & 386\\
\bottomrule
\end{tabular}
\end{center}
\caption{Validation and test results for different maximum number of filters.}
\label{tab:filters_correlation}
\end{table}}

{\renewcommand{\arraystretch}{1.2}
\begin{table}
\begin{center}
\begin{tabular}{ccccc}
\toprule
Case & \specialcell{Train.\\params.} & val.: $\mathcal{CC}_x,\,\mathcal{CC}_y$ & test: $\mathcal{CC}_x,\,\mathcal{CC}_y$ & Best epoch\\
\midrule
$3\times3$ & 338498 & 0.85, 0.86 & 0.86, 0.84 & 70 \\
$5\times5$ & 937282 & 0.89, 0.89 & 0.89, 0.89 & 101\\
$7\times7$ & 1835458 & 0.87, 0.86 & 0.86, 0.86 & 34\\
\bottomrule
\end{tabular}
\end{center}
\caption{Validation and test results for different kernel sizes.}
\label{tab:kernels_correlation}
\end{table}}

The results for the kernel size effect are displayed in \tref{tab:kernels_correlation}.
The kernel size is important for the detection of small- and large-scale structures.
The kernel size to capture the smallest scale in the wake should be, at least, a $3\times3$ filter since our iLES solver is resolving spatial scales down to grid level.
Structures larger than the grid-level scale might be better identified by slightly larger kernels.
However, the nonlinear combination of feature maps across the CNN layers allows small kernels to reconstruct such large-scale structures, hence the implied advantage of slightly larger kernels is not as relevant.
On the other hand, large kernels will filter out information of the subkernel-scale structures.
This is why the current accepted practice in CNNs is to use either $3\times3$ or $5\times5$ kernels and our results agree with that practice.
The $5\times5$ kernel configuration provides the best prediction among the kernels set, where a compromise between the detection of small- and large-scale structures is achieved.

\subsection{Input data}

Last, different input sets are evaluated.
Here, we consider the M2 model without normalising the inputs.
This is motivated by the fact that the M3 model becomes very large when 6 different inputs are used, as it is the case with the $\left\lbrace \nabla\vect{U}, \nabla P \right\rbrace$ input set.
The remaining base-case parameters are employed.
By testing different inputs we seek to include features more closely related to the expected outputs so that the regression problem solved by the CNN becomes easier \citep{Gamahara2017, Beck2019}.

As summarised in \tref{tab:inputs_correlation}, it is found that the primitive quantities set and the gradient set provide the highest correlation between target and predicted closure terms.
On the other hand, the vorticity plus the distance function input set does not perform as well.
The closure terms are, in their core, convective terms of the spanwise-fluctuating velocity field.
Hence, it is reasonable to argue that spatial derivatives of the resolved spanwise-averaged velocity field carry more relevant information to the closure terms than the other input sets.
Still, the primitive quantities set performs very similarly to the gradient set.
This can be an indicator of the network being able to find the gradients on its own.
Also, compared to the vorticity input set, it seems that information on the pressure field is important although this has not been tested in direct comparison.

{\renewcommand{\arraystretch}{1.2}
\begin{table}
\begin{center}
\begin{tabular}{ccccc}
\toprule
Case & \specialcell{Train.\\params.} & val.: $\mathcal{CC}_x,\,\mathcal{CC}_y$ & test: $\mathcal{CC}_x,\,\mathcal{CC}_y$ & Best epoch\\
\midrule
$\left\lbrace U, V, P \right\rbrace$& 338498 & 0.86, 0.85 & 0.85, 0.84 & 70\\
$\left\lbrace \Omega_z, d \right\rbrace$& 255746 & 0.57, 0.52  & 0.56, 0.52 & 110\\
$\left\lbrace \nabla\vect{U}, \nabla P \right\rbrace$ & 433538 & 0.86, 0.87 & 0.85, 0.86 & 112\\
\bottomrule
\end{tabular}
\end{center}
\caption{Validation and test results for different input sets.}
\label{tab:inputs_correlation}
\end{table}}

\section{Conclusion}

The M2 model has provided the highest correlations across the tested models for a given set of base-case parameters, and, on the other hand, the M3 model offered the best consistency while training.
Also, normalising the input data allows for faster training convergence without significantly compromising accuracy.
The sse loss function, a maximum of 64 filters, and a kernel size of $5\times5$ proved to be the best possible parameters.
Finally, the gradient input set and the primitive quantities set performed similarly providing the highest correlation coefficients (see \tref{tab:best_model}).

\begin{figure*}[!ht]
\begin{multicols}{2}
\begin{subfigure}[t]{\linewidth}
    \includegraphics[width=\linewidth]{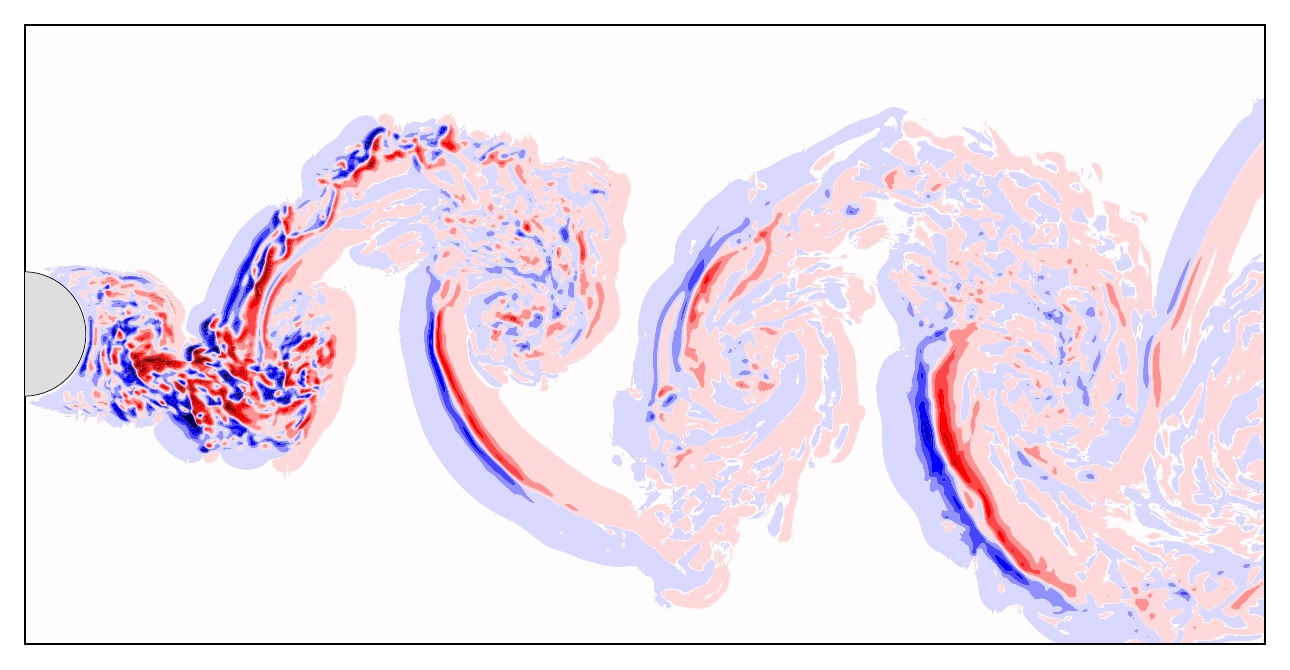}\caption{$\mathcal{S}^R_x$} 
\end{subfigure}\vspace{0.5cm}
\begin{subfigure}[t]{\linewidth}
    \includegraphics[width=\linewidth]{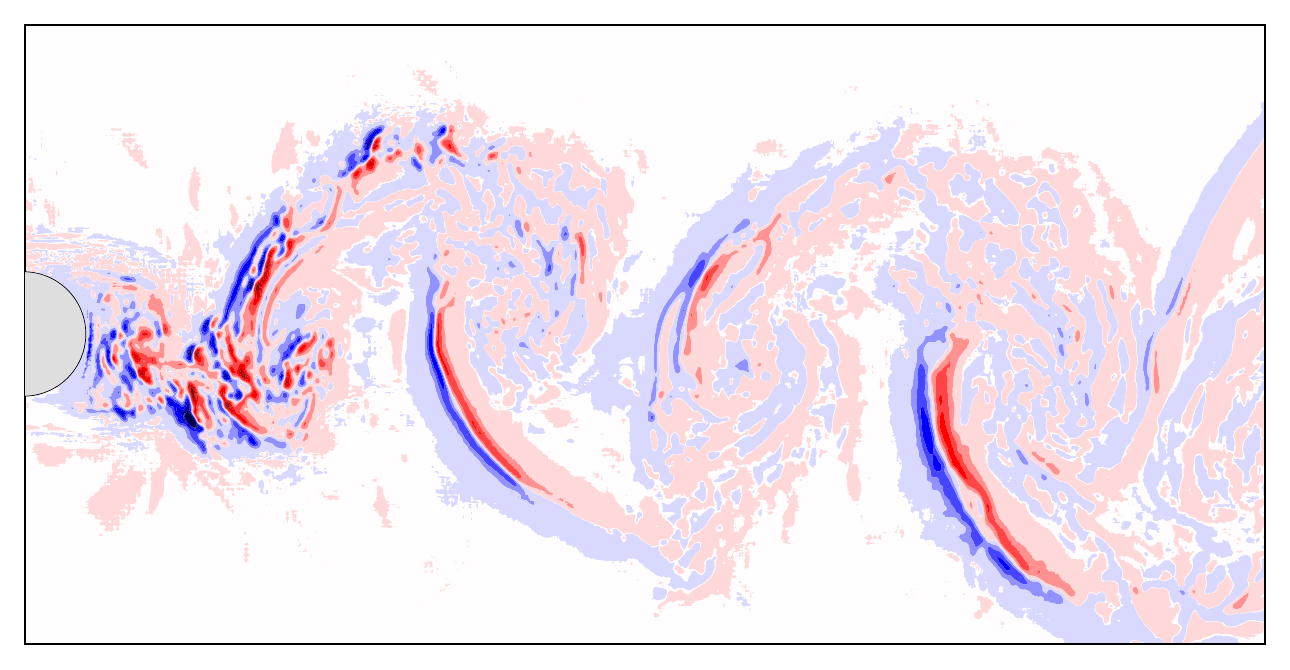} \caption{$\mathcal{S}^{R,\,\mathrm{ML}}_x$ (raw)}
\end{subfigure}\vspace{0.5cm}
\begin{subfigure}[t]{\linewidth}
    \includegraphics[width=\linewidth]{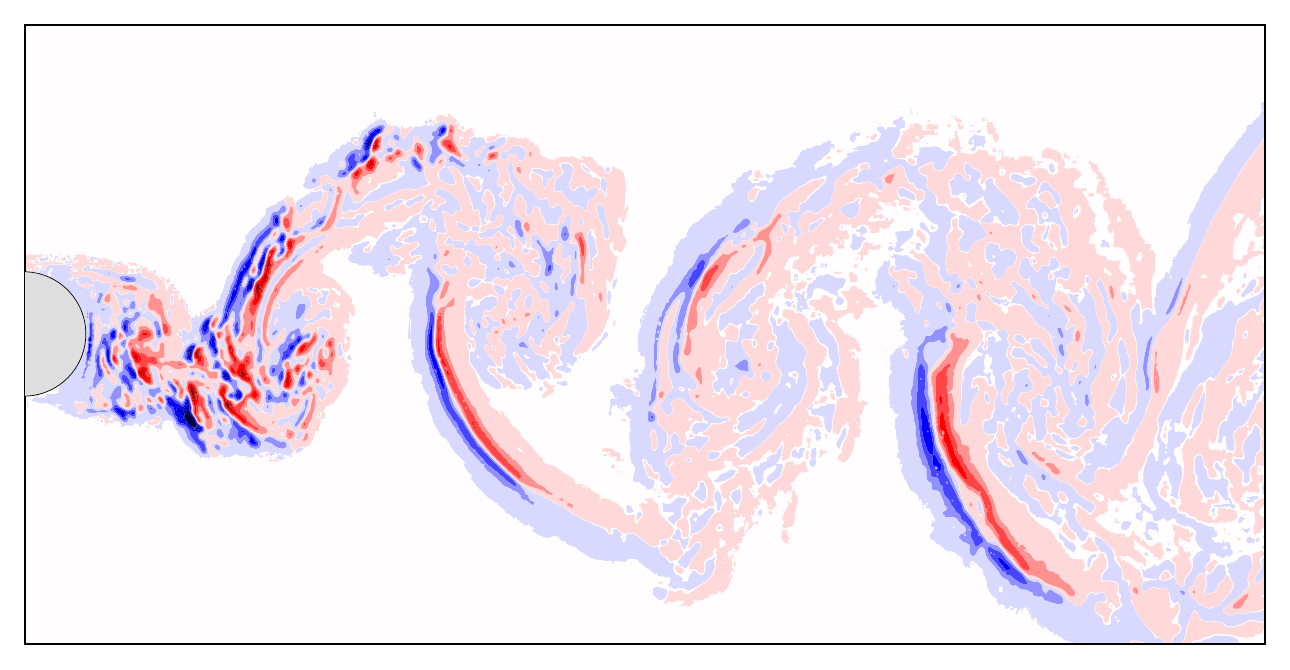} \caption{$\mathcal{S}^{R,\,\mathrm{ML}}_x$ (post-processed)}
\end{subfigure}
\begin{subfigure}[t]{\linewidth}
    \includegraphics[width=\linewidth]{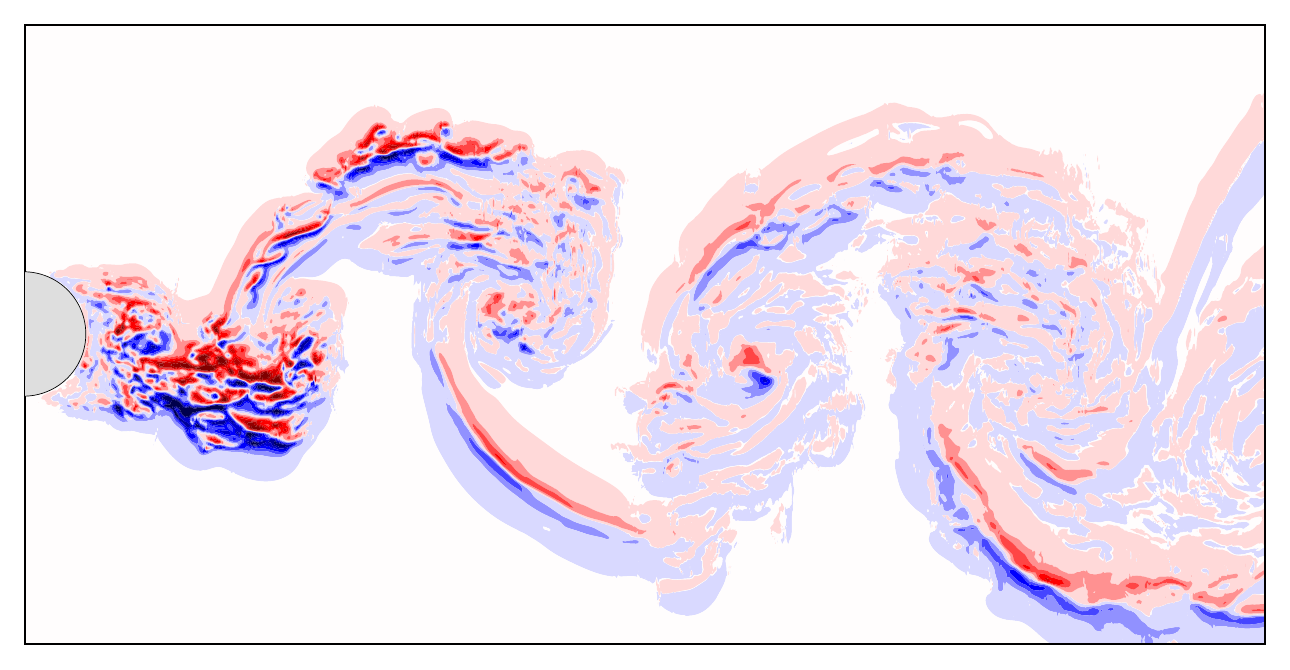}\caption{$\mathcal{S}^R_y$}
\end{subfigure}\vspace{0.5cm}
\begin{subfigure}[t]{\linewidth}
    \includegraphics[width=\linewidth]{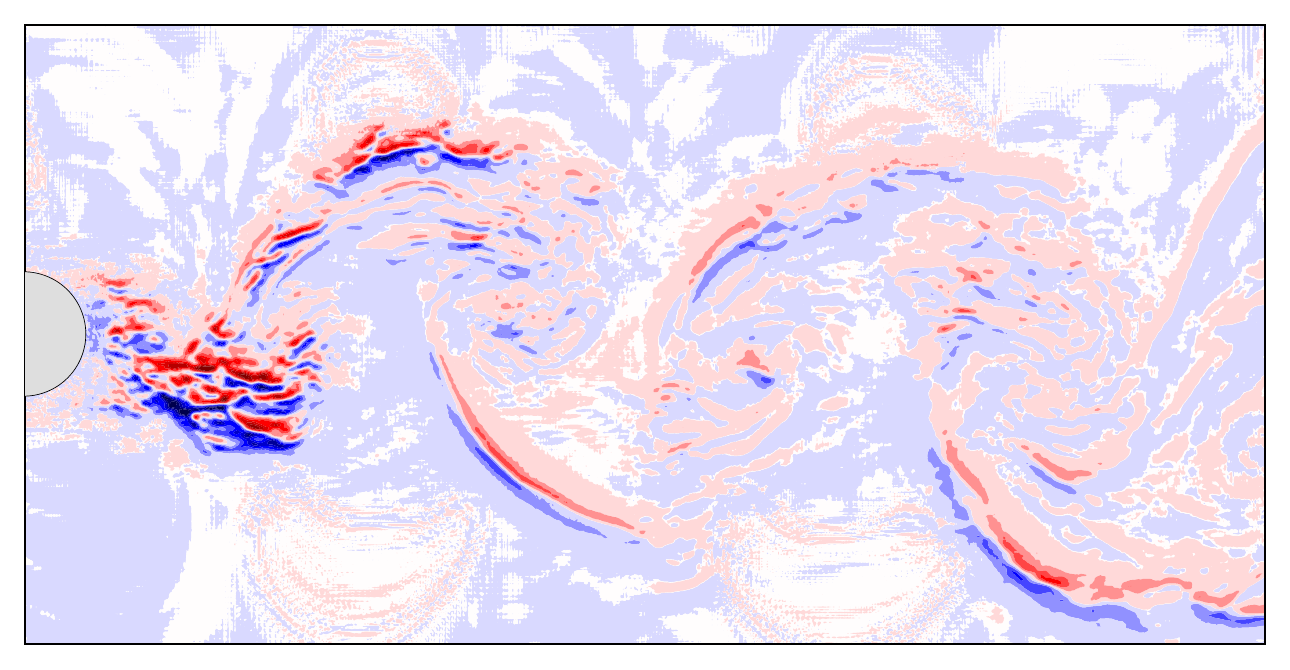}\caption{$\mathcal{S}^{R,\,\mathrm{ML}}_y$ (raw)}
\end{subfigure}\vspace{0.5cm}
\begin{subfigure}[t]{\linewidth}
    \includegraphics[width=\linewidth]{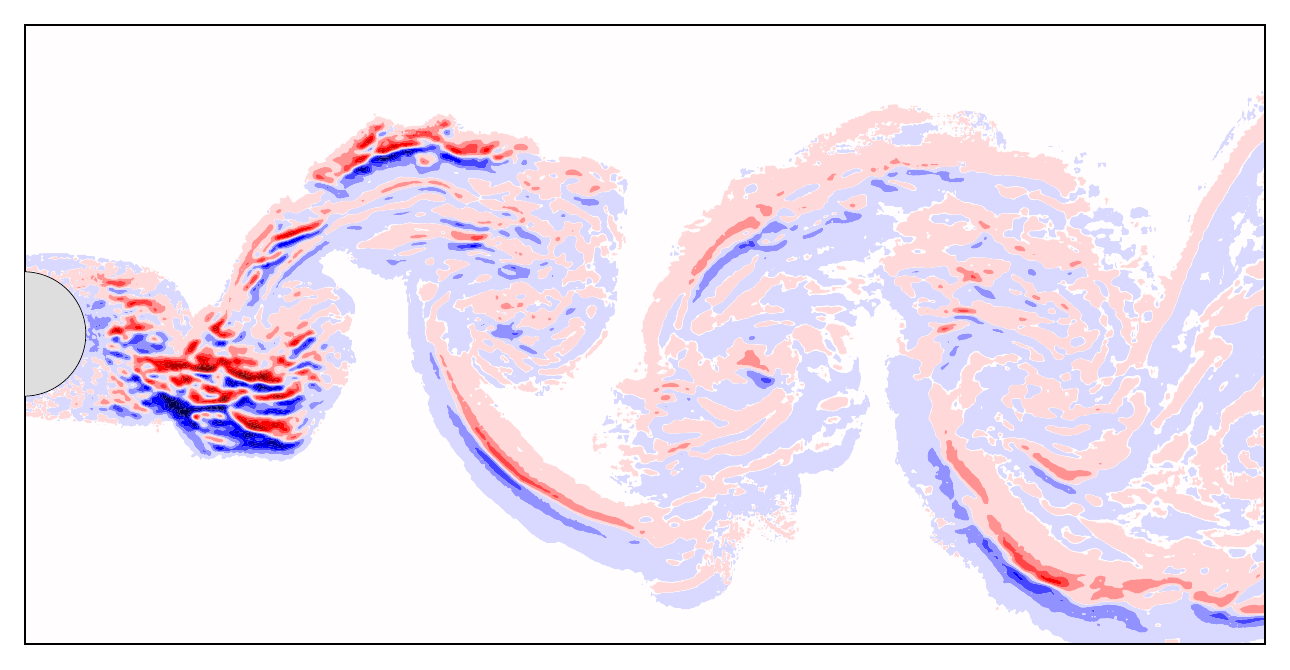}\caption{$\mathcal{S}^{R,\,\mathrm{ML}}_y$ (post-processed)}
\end{subfigure}
\end{multicols}
\caption{Qualitative comparison between a snapshot of target fields (top), CNN raw predictions (middle), and CNN post-processed predictions (bottom).
The post-processing step removes high-frequency oscillations with a Gaussian filter and applies a wake detection filter constructed from the resolved vorticity field.}
\label{fig:predictions}
\end{figure*}

From a qualitative standpoint, predictions and target closure terms are displayed in \fref{fig:predictions}.
The M3 model using the normalised primitive quantities input set, a kernel size of $5\times5$, the sse loss function, and 64 maximum filters has been employed to generate the predicted closure terms (highest $\mathcal{CC}$ observed: $\mathcal{CC}(\mathcal{S}^R_x)=0.89,\, \mathcal{CC}(\mathcal{S}^R_y)=0.89$).
The predicted fields have been post-processed including a wake mask step constructed from the resolved vorticity field and a Gaussian filter to mitigate the noise in the raw CNN output (note that all the presented correlation values do not include this post-processing step).
This post-processing step is similar to process described in \sref{sec:ML_model} although in that case it is embedded in the CNN.

\fref{fig:predictions} shows that both large- and mid-scale structures are well captured by the model.
On the other hand, small-scale structures are not so well identified.
Similar issues are encountered in \cite{Lee2019}, where deep-learning models employed for recursively generating snapshots of flow past a circular cylinder at high $Re$ failed to fully reproduce small-scale structures.
Also, the noise observed in the $\mathcal{S}^{R,\,\mathrm{ML}}_y$ output motivates the inclusion of a wake mask, as described in \sref{sec:ML_model}.

{\renewcommand{\arraystretch}{1.2}
\begin{table}
\begin{center}
\begin{tabular}{ll}
\toprule
Architecture &\noindent \specialcell{M2 (highest correlation)\\M3 (best training consistency)} \\
Loss function & sse \\
Activation function & ReLU \\
Max. filters & 64 \\
Kernel size & $5\times5$ \\
Normalise input data & True \\
Input dataset & $\left\lbrace U, V, P \right\rbrace$, $\left\lbrace \nabla\vect{U}, \nabla P \right\rbrace$\\
\bottomrule
\end{tabular}
\end{center}
\caption{Summary of best results.}
\label{tab:best_model}
\end{table}}

% ---------------------------------------------------------------- 

\chapter{Hardware and software details}

The 3-D simulations employed throughout the thesis have been performed in the Iridis4 supercomputer at the University of Southampton.
The fluid solver is run in 128 parallel processes across 8 different nodes.
Each node is equipped with 16x 2.6 GHz Intel\textsuperscript\textregistered $ $ Sandybridge\texttrademark $ $ processors. With this set-up, 205 hours of computational time was spent for the dataset generation used in \cref{chapter:zanspy}.
A NVIDIA\textsuperscript\textregistered $ $ Tesla V100 GPU is used for training the ML model in the Iridis5 supercomputer.
It takes approximately 30 minutes to iterate through the training dataset (i.e. 30 min./epoch).

In the a-posteriori analysis, the fluid solver (Fortran) and the ML model (Python) need to run in sync.
At every time step, the resolved quantities data is transferred to the ML model in order to compute the closure terms.
After the prediction is completed, the closure terms are sent back to the fluid solver to compute the solution of the next time step.
This has been accomplished using network sockets, a data transfer protocol between processes living in the same network.
In particular, internet sockets have been employed allowing to transfer data across a network port.
An example of Fortran-Python sockets data transfer can be found in \cite{Font2019-f2py}.
Since the SANS simulation is purely 2-D, this has been carried in a 4.0 GHz Intel\textsuperscript\textregistered $ $ Core\texttrademark $\,\,$i7-6700K processor using serial computations.
It takes approximately 40 minutes to complete $\Delta t^*=10$ time units in this a-posteriori configuration.
% ---------------------------------------------------------------- 

\backmatter

\bibliographystyle{etc/apalike-refs}
\bibliography{main}
%% ---------------------------------------------------------------
\end{document}